\documentclass{aa}  

\usepackage{color}
\usepackage{graphicx}
\usepackage[fleqn]{amsmath}	
\usepackage{amssymb}	
\usepackage{mathtools}
\usepackage{txfonts}
\usepackage{booktabs}
\usepackage{natbib,twoopt}
\defcitealias{nardini2019}{Paper I}
\defcitealias{lusso2020}{L20}
\defcitealias{rl19}{RL19}
\defcitealias{timlin2020}{T20}

\usepackage{float}
\usepackage[caption = false]{subfig}
\usepackage[breaklinks=true]{hyperref}
\hypersetup{
  colorlinks=true,   
  citecolor=blue,    
  urlcolor=blue,     
  linkcolor=blue,     
}
\bibpunct{(}{)}{;}{a}{}{,} 
\usepackage{lscape}

\makeatletter
\newcommandtwoopt{\citeads}[3][][]{\href{http://adsabs.harvard.edu/abs/#3}%
{\def\hyper@linkstart##1##2{}%
\let\hyper@linkend\@empty\citealp[#1][#2]{#3}}}
\newcommandtwoopt{\citepads}[3][][]{\href{http://adsabs.harvard.edu/abs/#3}%
{\def\hyper@linkstart##1##2{}%
\let\hyper@linkend\@empty\citep[#1][#2]{#3}}}
\newcommandtwoopt{\citetads}[3][][]{\href{http://adsabs.harvard.edu/abs/#3}%
{\def\hyper@linkstart##1##2{}%
\let\hyper@linkend\@empty\citet[#1][#2]{#3}}}
\newcommandtwoopt{\citeyearads}[3][][]%
{\href{http://adsabs.harvard.edu/abs/#3}
{\def\hyper@linkstart##1##2{}%
\let\hyper@linkend\@empty\citeyear[#1][#2]{#3}}}
\makeatother

\newcommand{\lo}{\log L_{\rm UV}}
\newcommand{\lx}{\log L_{\rm X}}
\newcommand{\fo}{\log F_{\rm UV}}

\newcommand{\Lo}{L_{\rm UV}}
\newcommand{\Fo}{F_{\rm UV}}
\newcommand{\Lx}{L_{\rm X}}
\newcommand{\Fx}{F_{\rm X}}
\newcommand{\Fh}{F_{2-10\,\rm keV}}
\newcommand{\Lh}{L_{2-10\,\rm keV}}

\newcommand{\aox}{\alpha_{\rm ox}}
\newcommand{\daox}{\Delta\alpha_{\rm ox}}

\newcommand{\lbol}{L_{\rm bol}}

\newcommand{\gammax}{\Gamma_{\rm X}}

\newcommand{\ebv}{E(B-V)}
\newcommand{\vp}{\upsilon_{\rm peak}}

\newcommand{\kms}{km s$^{-1}$}

    \newcommand{\xmm}{\emph{XMM--Newton}\xspace}

    \newcommand{\chandra}{\emph{Chandra}\xspace}

    \newcommand{\qsfit}{\texttt{QSFit}\xspace}

\DeclareRobustCommand{\ion}[2]{%
\relax\ifmmode
\ifx\testbx\f@series
{\mathbf{#1\,\mathsc{#2}}}\else
{\mathrm{#1\,\mathsc{#2}}}\fi
\else\textup{#1\,{\mdseries\textsc{#2}}}%
\fi}

\newcommand{\rev}[1]{{ #1}}
\DeclareCaptionFormat{cont}{#1 (cont.)#2#3\par}

\begin{document} 

\title{The most luminous blue quasars at 3.0\,$<$\,\textit{z}\,$<$\,3.3}
\subtitle{II. \ion{C}{iv}/X-ray emission and accretion disc physics}

   \author{
   E. Lusso\inst{1,2}\thanks{\email{elisabeta.lusso@unifi.it}},
    E.~Nardini\inst{2,1}, 
    S.~Bisogni\inst{3}, 
    G.~Risaliti\inst{1,2},
	R.~Gilli\inst{4},
    G.~T.~Richards\inst{5},
	F.~Salvestrini\inst{2},
	C.~Vignali\inst{6,4}
    G.~Bargiacchi\inst{7},
	F.~Civano\inst{8},
	M.~Elvis\inst{8},
	G.~Fabbiano\inst{8},
	A.~Marconi\inst{1,2}, 
    A.~Sacchi\inst{1},
    M.~Signorini\inst{1,2}
          }
          
\institute{
$^{1}$Dipartimento di Fisica e Astronomia, Universit\`a di Firenze, via G. Sansone 1, 50019 Sesto Fiorentino, Firenze, Italy\\
$^{2}$INAF -- Osservatorio Astrofisico di Arcetri, Largo Enrico Fermi 5, I-50125 Firenze, Italy\\
$^{3}$INAF -- Istituto di Astrofisica Spaziale e Fisica Cosmica Milano, via Corti 12, 20133 Milano, Italy \\
$^{4}$INAF -- Osservatorio di Astrofisica e Scienza dello Spazio di Bologna, via Gobetti 93/3, I-40129 Bologna, Italy\\
$^{5}$Department of Physics, 32 S. 32nd Street, Drexel University, Philadelphia, PA 19104\\
$^{6}$Dipartimento di Fisica e Astronomia, Universit\`a degli Studi di Bologna, via Gobetti 93/2, I-40129 Bologna, Italy\\
$^{7}$Scuola Superiore Meridionale, Largo S. Marcellino 10, I-80138, Napoli\\
$^{8}$Center for Astrophysics | Harvard \& Smithsonian, 60 Garden Street, Cambridge, MA 02138, USA
}

  \titlerunning{The most luminous blue quasars at $z \simeq 3$ II. \ion{C}{iv}/X-ray emission and accretion disc physics}
  \authorrunning{E. Lusso et al.}
   \date{\today}

 
  \abstract{We analyse the properties of the high-ionisation \ion{C}{iv}\,$\lambda$1549 broad emission line in connection with the X-ray emission of 30 bright, optically selected quasars at $z$\,$\simeq$\,3.0--3.3 with pointed \xmm observations, which were selected to test the suitability of active galactic nuclei as cosmological tools. \rev{In our previous work, we found that a large fraction ($\approx$\,25\%) of the quasars in this sample are X-ray underluminous by factors of $>$\,3--10. As absorbing columns of $\gtrsim$\,10$^{23}$ cm$^{-2}$ can be safely ruled out, their weakness is most likely intrinsic.}
  Here we explore possible correlations between the UV and X-ray features of these sources to investigate the origin of X-ray weakness with respect to X-ray normal quasars at similar redshifts. We fit the UV spectra from the Sloan Digital Sky Survey of the quasars in our sample and analyse their \ion{C}{iv} properties (e.g., equivalent width, EW; line peak velocity, $\vp$) as a function of the X-ray photon index and 2--10 keV flux. We confirm the statistically significant trends of \ion{C}{iv} $\vp$ and EW with UV luminosity at 2500 \AA\ for both X-ray weak and X-ray normal quasars, as well as the correlation between X-ray weakness (parametrised through $\daox$) and \ion{C}{iv} EW. In contrast to some recent work, we do not observe any clear relation between the 2--10 keV luminosity and $\vp$. We find a statistically significant correlation between the hard X-ray flux and the integrated \ion{C}{iv} flux for X-ray normal quasars, which extends across more than 3 (2) decades in \ion{C}{iv} (X-ray) luminosity, whilst X-ray weak quasars deviate from the main trend by more than 0.5 dex. We argue that X-ray weakness might be interpreted in a starved X-ray corona picture associated with an ongoing disc-wind phase. If the wind is ejected in the vicinity of the black hole, the extreme-UV radiation that reaches the corona will be depleted, depriving the corona of seeds photons and generating an X-ray weak quasar. Nonetheless, at the largest UV luminosities ($>$\,10$^{47}$ erg s$^{-1}$), there will still be an ample reservoir of ionising photons that can explain the `excess' \ion{C}{iv} emission observed in the X-ray weak quasars with respect to normal sources of similar X-ray luminosities.}

   \maketitle
%

\section{Introduction}
Active galactic nuclei (AGN) represent a phase that almost all galaxies will undergo in their lifetime, where the supermassive black hole (SMBH) located in the galaxy centre starts to efficiently accrete matter in the form of a disc \citep[e.g.][]{soltan1982}. The flow of matter towards the SMBH dissipates angular momentum through viscous friction, heating the disc that will shine at ultraviolet (UV) wavelengths \citep[e.g.][]{Salpeter1964,LB1969}. The spectral energy distribution (SED) of AGN also presents significant emission ($\sim$\,10--30\% of their total luminosity) at high energies ($>$\,0.1 keV), which cannot be produced directly from the accretion disc but is rather interpreted as Comptonised radiation originating in a plasma of hot relativistic electrons in the vicinity of the SMBH, the so-called X-ray corona. The intrinsic nature of this corona is still uncertain, but the existence of a non-linear correlation between the continuum UV (at 2500 \AA, $\Lo$) and X-ray (at 2 keV, $\Lx$) emission implies a physical connection between the disc and the corona \citep[e.g.][]{avnitananbaum79,zamorani81,vignali03,steffen06,just07,lusso2010,martocchia2017}. The observation that this non-linear correlation is very tight \citep[$\leq$\,0.2 dex of scatter;][]{lr16}, with a slope independent of redshift, indicates that a strong disc/corona synergy must subsist in AGN across cosmic time \citep[e.g.][]{nicastro2000,merloni2003,lr17,arcodia2019}.

The tight $\Lx-\Lo$ relation also allows us to accurately estimate the X-ray luminosity of a quasar for any given UV luminosity, in order to define a standard range of soft X-ray emission for typical (i.e., non-broad absorption line, non-jetted, with minimal deviation due to absorption) quasars, or vice-versa to easily identify peculiar objects (e.g. intrinsically X-ray weak, with strong radio jets, or extremely red). The former case is important in designing clean quasar samples with cosmological value, where systematics and biases are minimised \citep[][L20 hereafter]{lusso2020}. In our recent analysis of a sample of 30 quasars at $z$\,$\simeq$\,3.0--3.3, we found that a significant fraction of these high-redshift quasars depart from the $\Lx-\Lo$ relation towards lower X-ray fluxes, when compared to quasars with similar UV emission \citep[][Paper I]{nardini2019}.
This quasar sample was selected from the Sloan Digital Sky Survey (SDSS) to be representative of the most luminous, intrinsically blue quasar population at $z$\,$\sim$\,3, for which we have obtained good-quality X-ray spectra from a dedicated {\it XMM-Newton} campaign. The main aim of this observational campaign was to investigate the evolution of the quasar Hubble diagram at high redshift making use of a highly homogeneous sample in terms of UV properties (for the details on its cosmological application, see \citealt[][RL19 hereafter]{rl19}). We thus presumed that all the 30 quasars would follow the $\Lx-\Lo$ relation. Nonetheless, we discovered that $\approx$\,25\% of the targets present an X-ray emission much weaker than expected, by factors of $>$\,3. This fraction is significantly larger than those previously reported for radio-quiet, non-BAL quasars at lower redshift and luminosity ($\approx$\,10\%, e.g. \citealt{Gibson2008}; see also \citealt{brandt2000}). 

In this paper, the second of the series dedicated to the study of the physical properties of these 30 high-redshift quasars, we focus on the connection between their UV (i.e., \ion{C}{iv} and continuum) and X-ray emission to better understand the origin X-ray weakness for the sub-sample of objects that depart from the $\Lx-\Lo$ relation. 

The paper is organised as follows: in Section~\ref{dataset} we describe the data set and its selection, whilst Section~\ref{uvspectralfit} is dedicated to the UV observations and spectral analysis. Results and relative discussion are presented in Sections~\ref{results} and \ref{discussion}, respectively. Conclusions are drawn in Section~\ref{conclusions}.

\begin{figure*}
\centering
 \resizebox{\hsize}{!}{\includegraphics{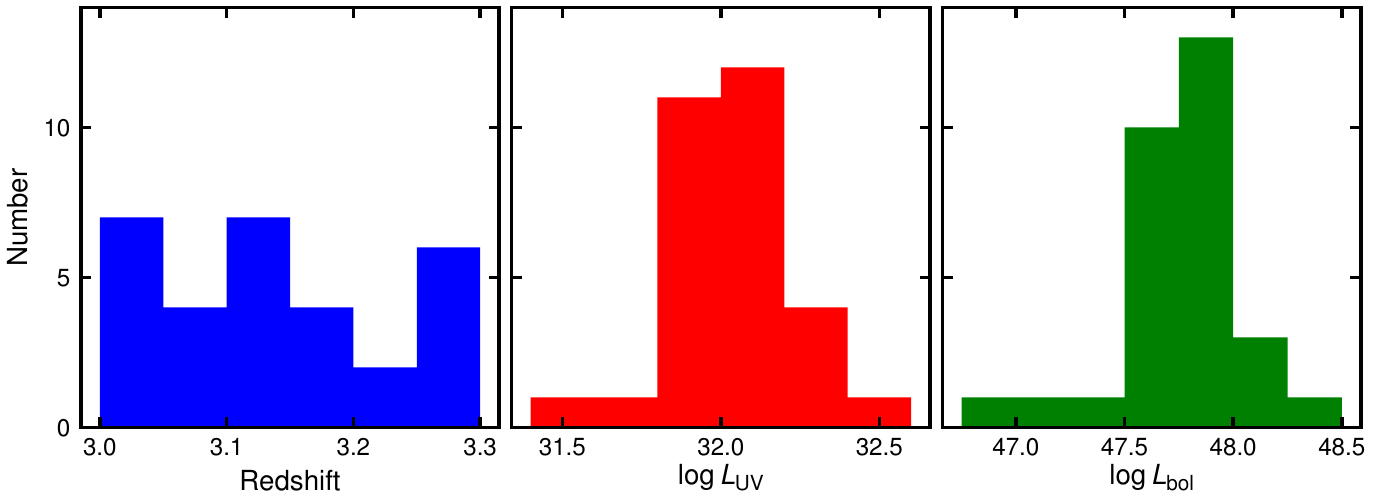}}
\caption{From left to right: distribution of redshift, rest-frame monochromatic luminosity at 2500 \AA, and bolometric luminosity (as listed in \citealt{shen2011}) for the 30 sources in our $z$\,$\simeq$\,3 quasar sample.}
\label{distrall}
\end{figure*}

\section{The data set}
\label{dataset}
The sample analysed here consists of 30 luminous ($\lbol$\,$>$\,10$^{47}$ erg s$^{-1}$) quasars in a narrow redshift interval, $z$\,$=$\,3.0--3.3, for which X-ray observations were obtained through an extensive campaign performed with \xmm (cycle 16, proposal ID: 080395, PI: Risaliti), for a total exposure of 1.13 Ms. This sample, selected in the optical from the SDSS Data Release 7 to be representative of the most luminous, intrinsically blue radio-quiet quasars, boasts by construction a remarkable degree of homogeneity in terms of optical/UV properties. Figure~\ref{distrall} shows the distribution of redshift\footnote{\rev{We considered the improved redshifts for SDSS quasars computed by \citet{hw2010}.}}, rest-frame monochromatic luminosity at 2500 \AA\ \rev{(derived from the fit of the SDSS spectra, as described in \S~\ref{uvspectralfit} and appendix~\ref{Ultraviolet SDSS spectra})} and bolometric luminosity (computed from the UV luminosity at 1350 \AA, as listed in \citealt{shen2011}) for the 30 quasars. 
\rev{All the sources were selected to be radio-quiet (with radio loudness parameter $R=L_{2500\AA}/L_{1.4\rm GHz}<10$). The only quasar in the sample that was conservatively flagged as radio-bright is SDSS~J090033.50$+$421547.0 ($R$\,$\simeq$\,2--2.5; more details on the sample selection in the Supplementary Material of \citetalias{rl19}). 
We thus exclude J0900$+$42 from any general consideration regarding the sample.}

\subsection{A sample of bright blue quasars}
\label{A blue quasar sample}
The $z$\,$\simeq$\,3 quasar sample was filtered to retain minimum levels of both dust reddening and host-galaxy contamination. To select the targets, we followed a similar approach to the one presented in our previous works (\citealt{rl15,lr16}; \citetalias{rl19}). We briefly summarise below the main points. 
We built the rest-frame photometric SEDs to compute, for each object, the slopes $\Gamma_1$ and $\Gamma_2$ of a $\log(\nu)-\log(\nu L_\nu)$ power law in the rest-frame ranges 0.3--1 $\mu$m and 1450--3000~\AA, respectively 
(see also \citeads{hao2013}). 
We discuss how we constructed the SEDs in Appendix~\ref{Photometric spectral energy distributions}, whilst Figure~\ref{seds} presents the full SEDs of the $z$\,$\simeq$\,3 quasars from the near-infrared to the X-rays. 

The wavelength intervals for $\Gamma_1$ and $\Gamma_2$ were chosen based on the fact that the SED of an {\it intrinsically} blue quasar displays a different shape from the one of an inactive galaxy or a dust-reddened AGN. On the one hand, the intrinsic quasar SED is characterised by a dip around 1\,$\mu$m, where the galaxy has the peak of the emission from the passive/old stellar population (e.g. \citeads{1994ApJS...95....1E,2006AJ....131.2766R,shang2011,2012ApJ...759....6E,2013ApJS..206....4K}). 
On the other hand, as dust reddening is wavelength dependent, the UV part of the quasar SED will be attenuated differentially. These two concurrent factors impact on the quasar SED shape, so we can define a set of slopes that single out quasars with minimum levels of both host-galaxy emission and dust reddening (see also figure~1 in \citeads{hao2013}).

\begin{figure}
\centering
 \resizebox{\hsize}{!}{\includegraphics{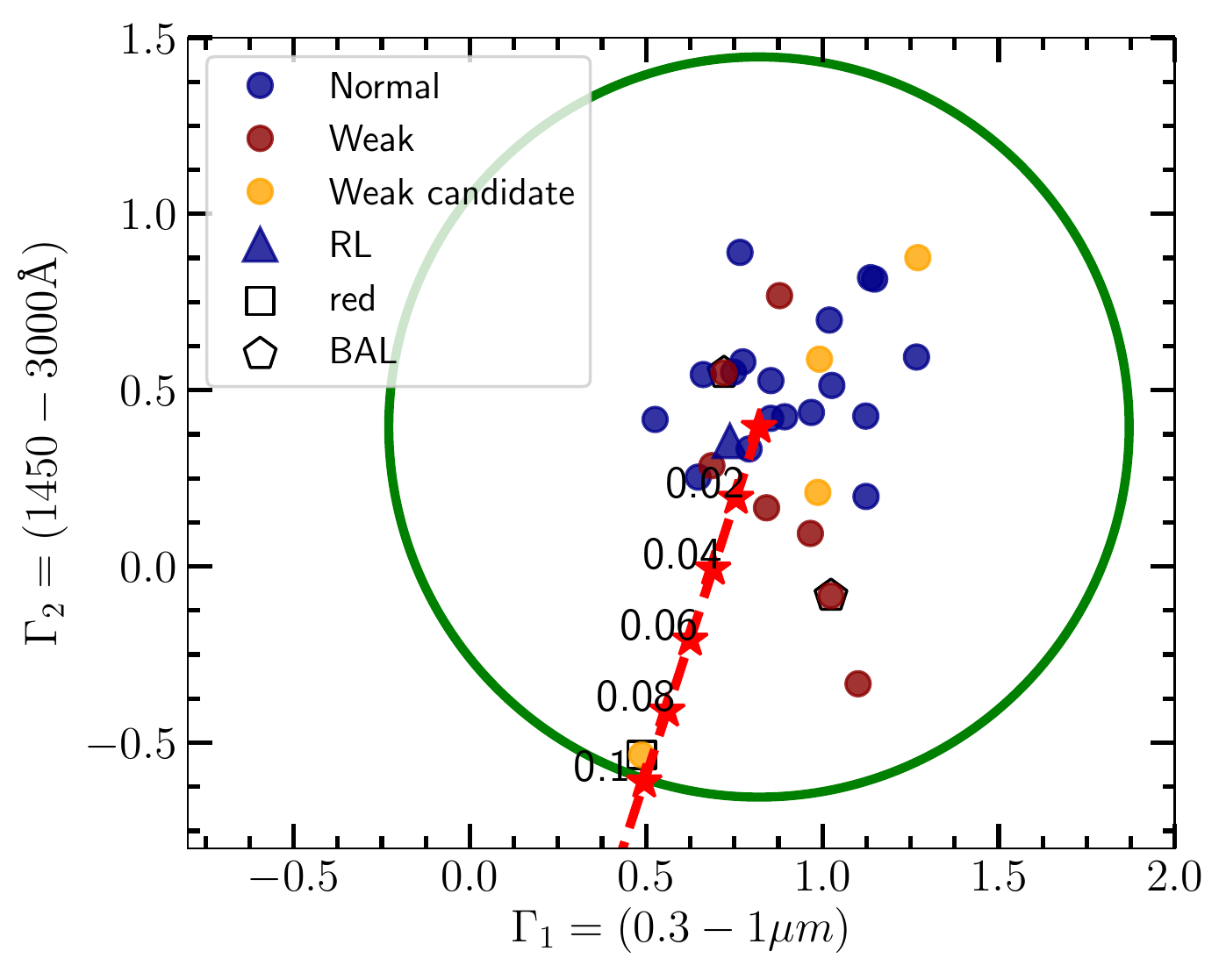}}
\caption{Distribution of the $z$\,$\simeq$\,3 quasar sample in the $\Gamma_{1}-\Gamma_{2}$ plot, where $\Gamma_{1}$ and $\Gamma_{2}$ are the slopes of a power law in the $\log(\nu)-\log(\nu F_\nu)$ plane in the 0.3--1~$\mu$m and 1450--3000~\AA\ intervals (see \S\,\ref{A blue quasar sample}), respectively. The red stars represent the $\Gamma_{1}-\Gamma_{2}$ values of the quasar SED by \citet{2006AJ....131.2766R} with increasing dust reddening (following the extinction law of \citealt{prevot84}), with $\ebv$ in the range 0--0.1. All quasars are within $\ebv$\,$=$\,0.1, as marked by the green circle. Blue, brown and orange symbols represent X-ray normal, weak and weak candidates, respectively, following the definition in \S\,\ref{The X-ray weak quasar fraction}. The triangle, pentagon and square symbols mark the radio-bright (J0900+42), BALs (J1148+23, J0945+23), and reddest (J1459+00) quasars in the sample.}
\label{g1g2plot}
\end{figure}

The $\Gamma_1-\Gamma_2$ distribution for the $z$\,$\simeq$\,3 quasars is shown in Figure~\ref{g1g2plot}. We assumed a standard SMC extinction law $k(\lambda)$ after \citet{prevot84}, with $R_V$\,$=$\,$A(V)/E(B-V)$\,$=$\,3.1 (as appropriate for unobscured AGN; \citealt{2004AJ....128.1112H,salvato09}), to estimate the $\Gamma_1-\Gamma_2$ correlation as a function of extinction, parametrised by the colour excess $\ebv$. The red dashed line in Figure~\ref{g1g2plot} shows how $\ebv$ varies in the $\Gamma_1-\Gamma_2$ plane, where the green circle (with a radius corresponding to a reddening $\ebv$\,$\simeq$\,0.1) is centred at the reference values for a standard quasar SED of \citet[i.e., $\Gamma_1=0.82$, $\Gamma_2=0.40$]{2006AJ....131.2766R} with zero extinction. 
The distribution of $\Gamma_1-\Gamma_2$ 
values along the red dashed line is indicative of possible dust reddening, growing in the outward direction, whilst sources with markedly smaller (i.e., negative) $\Gamma_1$ values are objects with possible host-galaxy contamination (see \citealt{rl15,lr16}; \citetalias{rl19} for further details). 
It is clear from Figure~\ref{g1g2plot} that the majority of the $z$\,$\simeq$\,3 quasars are blue with negligible absorption by dust, with $\ebv$\,$<$\,0.08. Only one quasar, J1459+00, has $\ebv$\,$\simeq$\,0.1, and it was selected to test our threshold value in the $\Gamma_1-\Gamma_2$ plane. 

\subsection{The X-ray data}
\label{The X-ray data}
The X-ray spectroscopic data have been extensively analysed in the first paper on this sample; the interested reader should refer to \citetalias{nardini2019} 
for details. Here we briefly summarise the main results we obtained from our X-ray analysis.

Twenty-five quasars in the sample are found to be extremely X-ray luminous, with rest-frame 2--10 keV luminosities of 0.5--7\,$\times$\,10$^{45}$ erg s$^{-1}$. Their continuum photon index distribution shows an average $\gammax$\,$\sim$\,1.85, in excellent agreement with objects at lower redshift, luminosity and black-hole mass (e.g. \citeads{just07,bianchi2009}).
Three sources turned out to be very faint (J1159$+$31, J1425$+$54, and J1507$+$24), but we can at least rely on a marginal detection with the EPIC/pn (see table~1 in \citetalias{nardini2019}). Only one target (J0945$+$23) \rev{can be formally considered as undetected, given a} spurious detection level of 4\%. \rev{In the following, we consider for the latter four quasars the more conservative flux value computed by assuming a fixed $\gammax$\,$=$\,1.8 and free $N_{\rm H}$, thus taking into account intrinsic absorption (see Section 4.3 in \citetalias{nardini2019} for more details; the resulting upper limits are listed in their Table~2).} 

\begin{figure}
\centering
 \resizebox{\hsize}{!}{\includegraphics{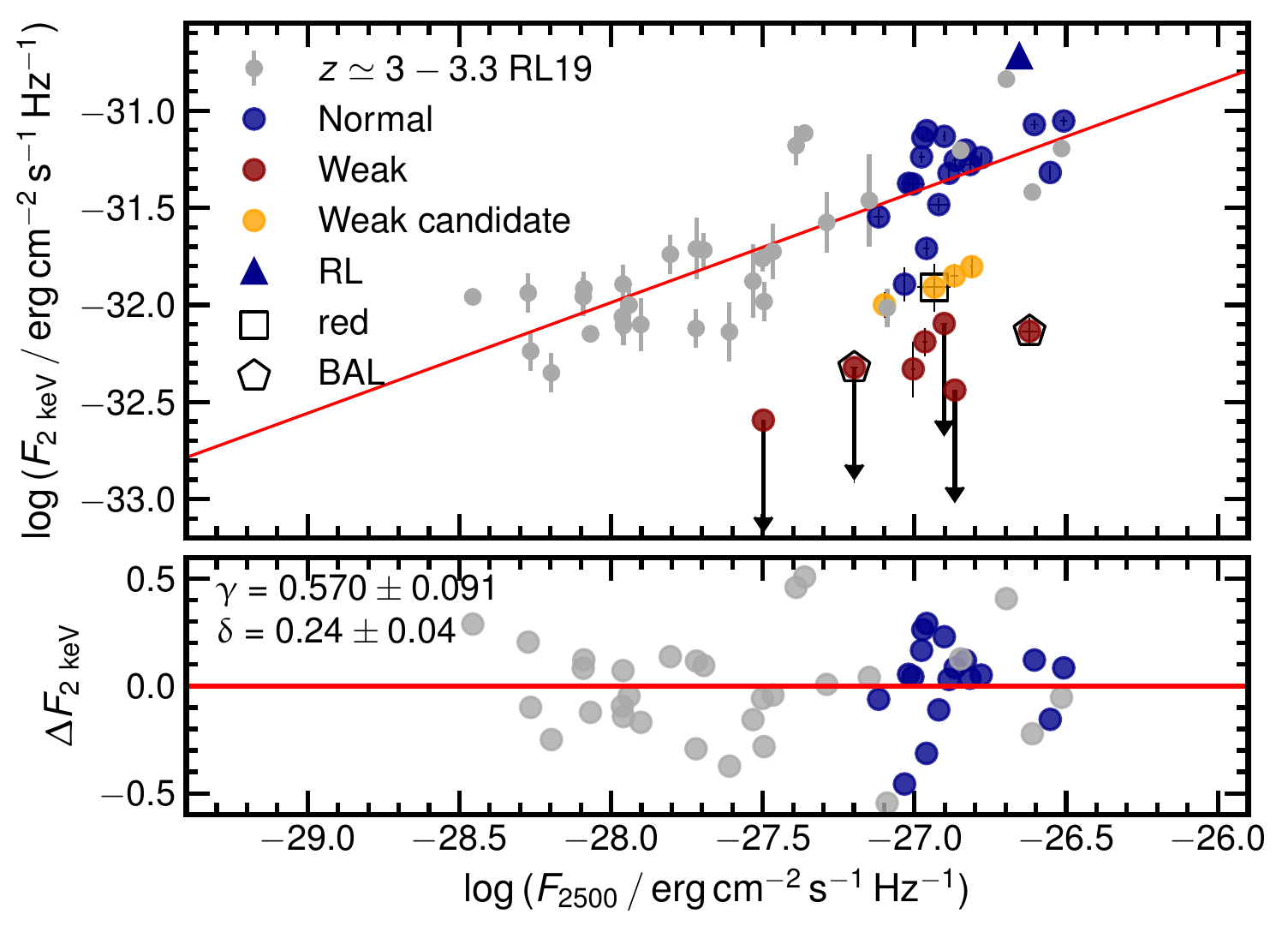}}
\caption{Relation between the rest-frame monochromatic X-ray (2 keV) and UV (2500 \AA) fluxes for quasars in the redshift range $z$\,$=$\,3.0--3.3. 
The colour code is the same adopted in figure~\ref{g1g2plot} for the different sub-samples defined in \S\,\ref{The X-ray weak quasar fraction}. Grey points represent the other sources in the same redshift range from the cleaned quasar sample of \citetalias{rl19} (30 objects). The red line is the regression best fit for the $z$\,$=$\,3.0--3.3 \citetalias{rl19} sources only. Slope and dispersion values are reported with the residuals from the best fit in the bottom panel (from which X-ray weak, in brown, and weak candidates, in orange, are omitted for clarity).}
\label{fofx}
\end{figure}

Yet, our quasars show an unexpectedly diverse behaviour when we plot them in the $\lx-\lo$ plane.
In Figure~\ref{fofx} we illustrate the relation between the rest-frame monochromatic fluxes in the X-rays at 2 keV and in the UV at 2500 \AA\ for quasars at $z$\,$=$\,3.0--3.3,\footnote{Given the narrow redshift range, we can equivalently deal with fluxes or luminosities.} where the grey points represent the remaining quasars in the redshift range of interest from the cleaned sample of \citetalias{rl19} (30 objects, see their supplementary figure 4). 
We refer the reader to \citetalias{rl19} (see also \citetalias{lusso2020}) 
for a detailed discussion of their selection criteria. 
The red line in Figure~\ref{fofx} is the regression best fit of the $z$\,$=$\,3.0--3.3 RL19 quasars (grey points only). The observed slope is $\gamma$\,$=$\,0.570\,$\pm$\,0.090, and the intercept is $\beta$\,$=$\,$-$0.318\,$\pm$\,0.048\footnote{The $\beta$ value incorporates a normalisation (in logarithm) of $-$27.7 and $-$31.5 on the monochromatic UV and X-ray fluxes, respectively.}, with a dispersion $\delta$\,$=$\,0.24\,$\pm$\,0.03.
While most of the $z$\,$\simeq$\,3 sources analysed here are clustered around the best-fit relation, a significant fraction ($\approx$\,25\%) of them appear to be X-ray underluminous by factors of $>$\,3--10. Interestingly, the reddest quasar in our sample is not amongst the most deviating objects, suggesting that our selection in the $\Gamma_1-\Gamma_2$ plane does not play a major role in driving this deviation.

\subsection{The X-ray weak quasar fraction}
\label{The X-ray weak quasar fraction}
In \citetalias{nardini2019} we reported on the surprisingly high fraction ($\approx$\,25\%) of X-ray weak sources among our $z$\,$\simeq$\,3 blue quasars, much larger than previously found for radio-quiet, non-BAL quasars at lower redshift and luminosity.
The definition of {\it X-ray weakness} is discussed at length in Section 5.6 of \citetalias{nardini2019}, and it is based on the non-linearity of the X-ray to UV correlation. This produces the well-known anti-correlation between $\aox$, \rev{defined as $\aox$\,$=$\,$0.384 \log(\Fx/\Fo)$,} and $\Lo$ \citep[e.g.][and references therein]{lr16,lr17}, as the UV-to-X-ray part of the quasar SED becomes steeper with higher UV luminosities. The $\Lx-\Lo$ relation is thus utilised as a natural benchmark to determine the extent of any {\it intrinsic} X-ray weakness. 

We distinguish X-ray {\it normal} from X-ray {\it weak} quasars in our sample by computing the differences $(\daox)$ between the observed and predicted values of $\aox$. To obtain the expected $\aox$ to be compared to the observed values, we assumed as the reference $\Fx-\Fo$ relation the one shown in Figure~\ref{fofx}, determined from the 30 objects in the clean sample of \citetalias{rl19} in the same redshift range as the $z$\,$\simeq$\,3 quasars . 
\begin{figure}
\centering
 \resizebox{\hsize}{!}{\includegraphics{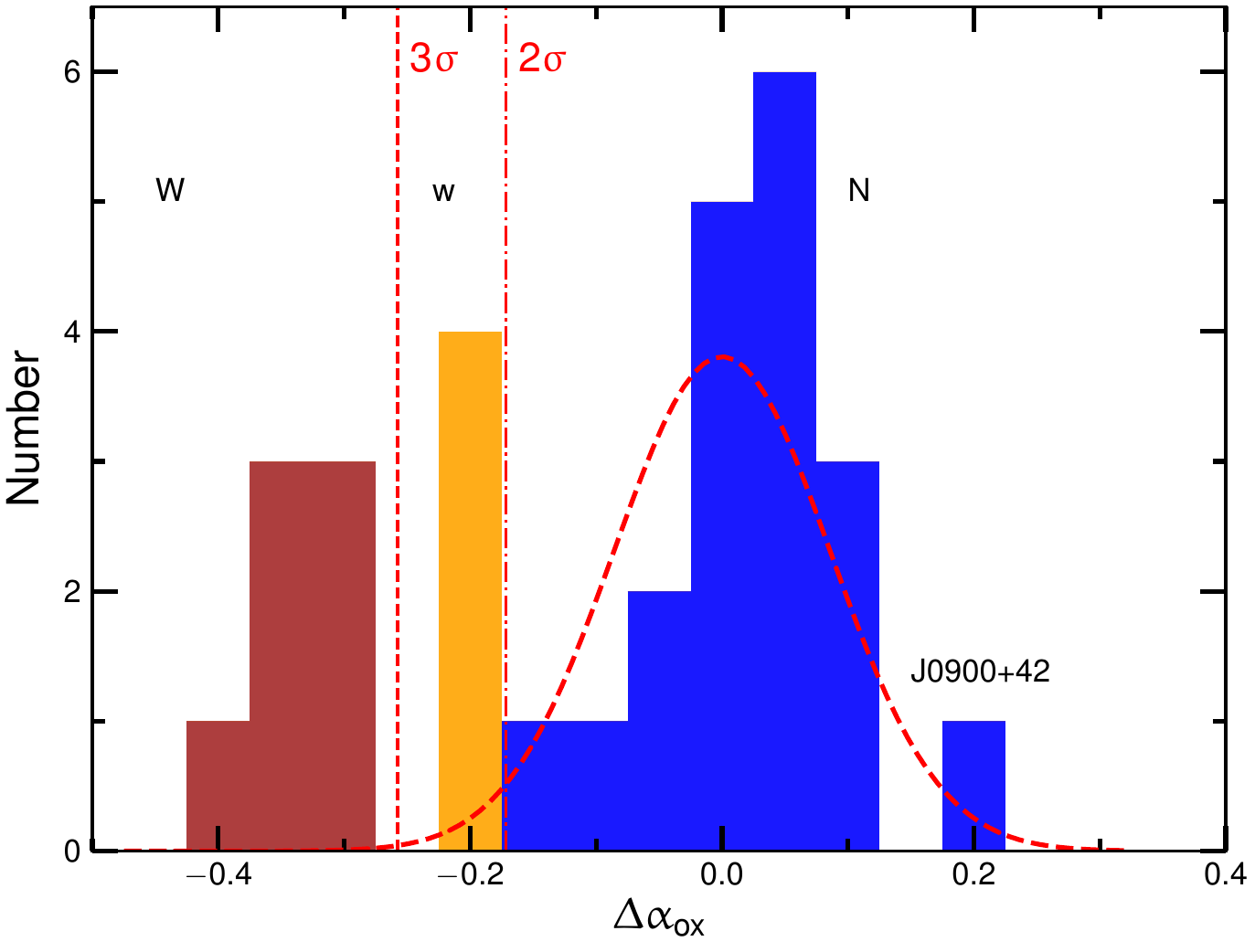}}
\caption{Distribution of $\daox$ in the $z$\,$\simeq$\,3 sample (see \S\,\ref{The X-ray weak quasar fraction} for details). The dashed curve is the best-fitting normal distribution for the 25 radio-quiet sources with unquestionable X-ray detection (we also neglected the radio-bright quasar J0900$+$42), where the peak position is forced to be at $\daox$\,$=$\,0. 
We define as X-ray {\it weak} ($W$) and X-ray {\it weak candidates} ($w$) all the sources at more than 3\,$\sigma$ (7 objects, brown histogram) and 2\,$\sigma$ (4 objects, orange histogram) from the peak, respectively. All the other quasars are considered as X-ray {\it normal} ($N$, blue histogram).}
\label{histdaox}
\end{figure}
Figure~\ref{histdaox} presents the distribution of the differences $(\daox)$ between the observed and predicted values of $\aox$ for all the 30 
$z$\,$\simeq$\,3 quasars. The dashed curve is the best-fitting normal distribution, with the peak position forced at $\daox$\,$=$\,0, for the 25 radio-quiet sources with solid X-ray detection (the radio-bright quasar J0900$+$42 is conservatively excluded from the fit).
We define as X-ray {\it weak} ($W$) and X-ray {\it weak candidates} ($w$) all the sources that fall at more than $3\,\sigma$ (7 objects) and $2\,\sigma$ (4 objects) from the peak of the distribution, respectively. The remaining 18 quasars are then labelled as X-ray {\it normal} ($N$).
Figure~\ref{weakf} reproduces the continuum photon index against the intrinsic flux density at rest-frame 2 keV and integrated 2--10 keV flux, as presented in \citetalias{nardini2019}. The dashed line in the left panel indicates the minimum $\gammax$ value adopted by \citetalias{rl19} to define the clean quasar sample for cosmology. Eighteen sources (15 $N$ and 3 $w$) are located above this limit.
The majority of the $z$\,$\simeq$\,3 quasars have a photon index consistent with (within the uncertainties) or higher than 1.7, with only 4 objects (all classified as $W$) displaying a much flatter ($\gammax$\,$\la$\,1.4) continuum. 


\begin{figure}
\centering
 \resizebox{\hsize}{!}{\includegraphics{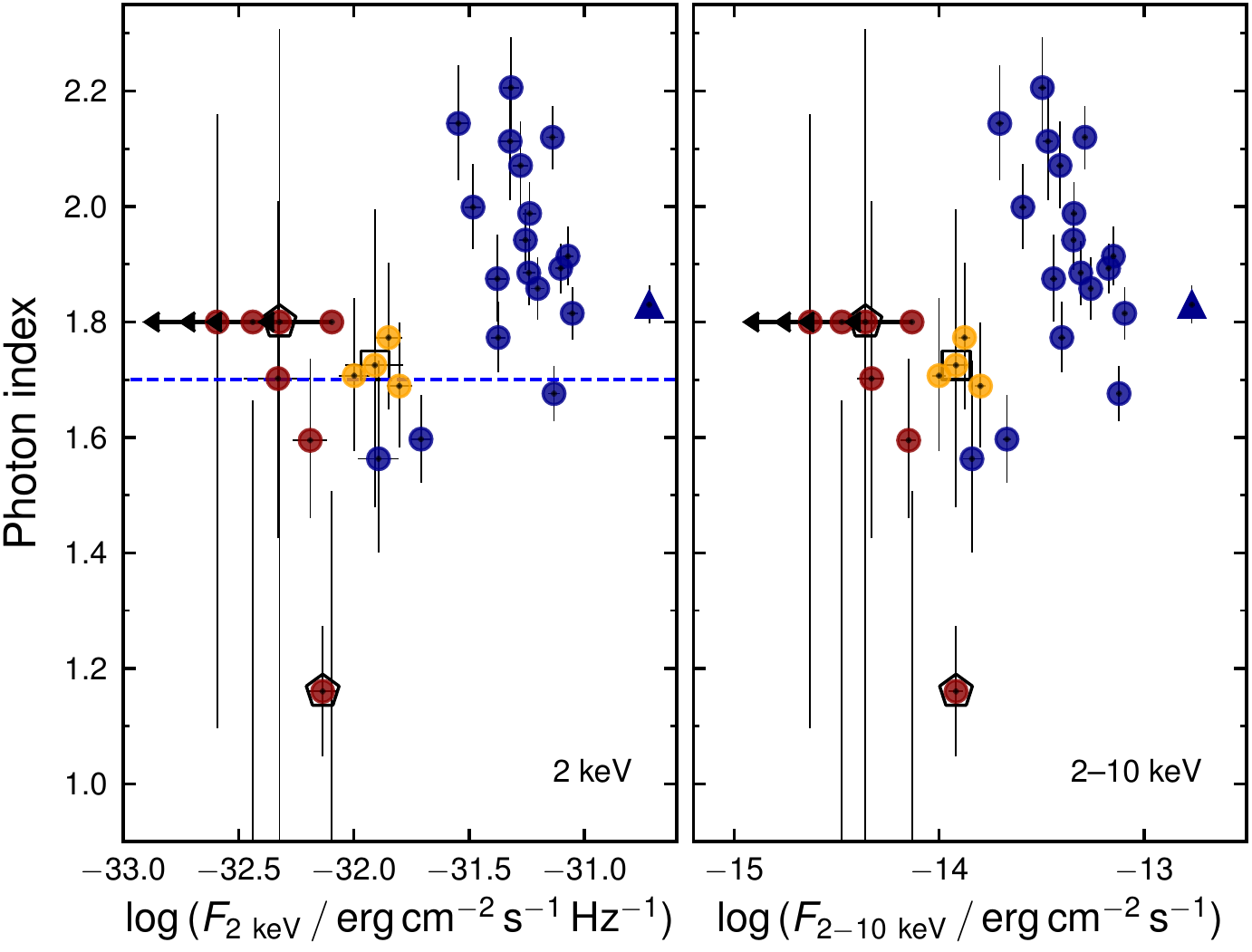}}
\caption{{\it Left panel}: Continuum photon index against intrinsic flux density at rest-frame 2 keV, as measured in \citetalias{nardini2019}. The colour code is the same adopted in 
the previous figures. The dashed line represents the minimum photon index value ($\gammax$\,$=$\,1.7) adopted by \citetalias{rl19} to select the clean quasar sample for cosmology. {\it Right panel}: Same for the intrinsic rest-frame flux integrated in the 2--10 keV band.}
\label{weakf}
\end{figure}

\begin{figure}
\centering
 \resizebox{\hsize}{!}{\includegraphics{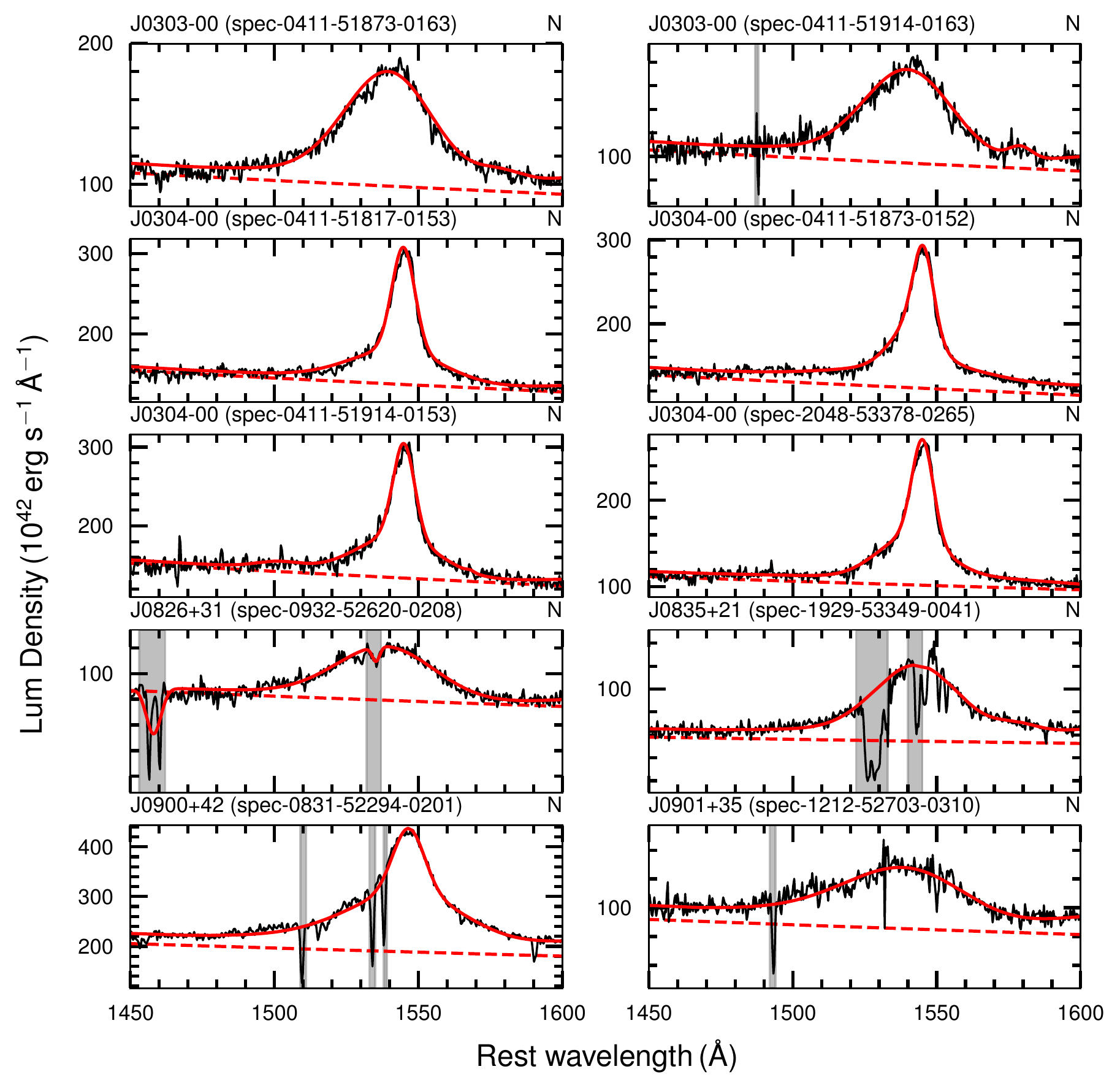}}
\caption{Fits to the \ion{C}{iv} emission line of the 38 SDSS quasar spectra. The data are shown in black, the best-fitting model in red, and the continuum with the red dashed line. The source ID, the name of the SDSS spectrum FITS file and a flag for the X-ray normal ($N$), weak ($W$), and weak candidates ($w$) are shown at the top of each panel. Grey shaded areas highlight bad pixels and/or strong absorption features that are either masked in the fit or considered in the fit to improve the least-$\chi^2$.}
\label{sfit}
\end{figure}
\section{Ultraviolet spectral fit}
\label{uvspectralfit}
The main aim of this paper is to connect the X-ray properties of the $z$\,$\simeq$\,3 sample with the UV ones (e.g. emission-line strength, continuum luminosity) to provide an interpretation of the observed X-ray weakness. We thus carried out a detailed spectral fitting of all the SDSS spectra, also considering multiple observations, using both a custom-made code and the publicly available package for spectral fitting \qsfit \citepads{calderone2017}. The spectral fitting was performed independently with each tool, 
and the relative results were then compared.

Our custom-made code is based on a procedure that uses the IDL MPFIT package \citepads{Markwardt2009}, written with the purpose of a simultaneous fitting of continuum, \ion{Fe}{ii} complex (both optical and UV), and several emission lines. The spectra were corrected for Galactic extinction using the $\ebv$ values from \citetads{sf2011} and the \citetads{F99} reddening law with $R_V$\,$=$\,3.1.
Broad lines were fitted with a broken power law, convolved with a Gaussian function to avoid the presence of a cusp at the peak \citepads{nagao2006}. This profile is particularly useful when dealing with emission-line complexes, and it helps in
limiting the degeneracy in the fits. Narrow lines were fitted with a Gaussian profile, as well as blue/red asymmetries if required by the data. For more details, the interested reader should refer to \citet[][see their Section 3]{bisogni2017}.  

We also independently fitted the SDSS data making use of \qsfit. We modelled each spectrum as follows: the \ion{C}{iv} emission line was reproduced by a broad (FWHM\,$>$\,2,000 \kms) profile and, whenever required, an additional narrow component (FWHM\,$<$\,2,000 \kms), whilst the continuum includes the contributions from the iron UV complex and the AGN continuum. The spectra were corrected for Galactic extinction using the $\ebv$ values from \citetads{schlegel98} and the parametrisation by \citetads{CCM1989} and \citetads{odonnel1994}, with a total to selective extinction $R_V$\,$=$\,3.1 \citepads{calderone2017}. 

Making use of different parametrisations for the \ion{Fe}{ii} complex, reddening values, extinction laws, and line profiles allows us to check their effect on the output parameters.

\begin{figure}
\addtocounter{figure}{-1}
\centering
 \resizebox{\hsize}{!}{\includegraphics{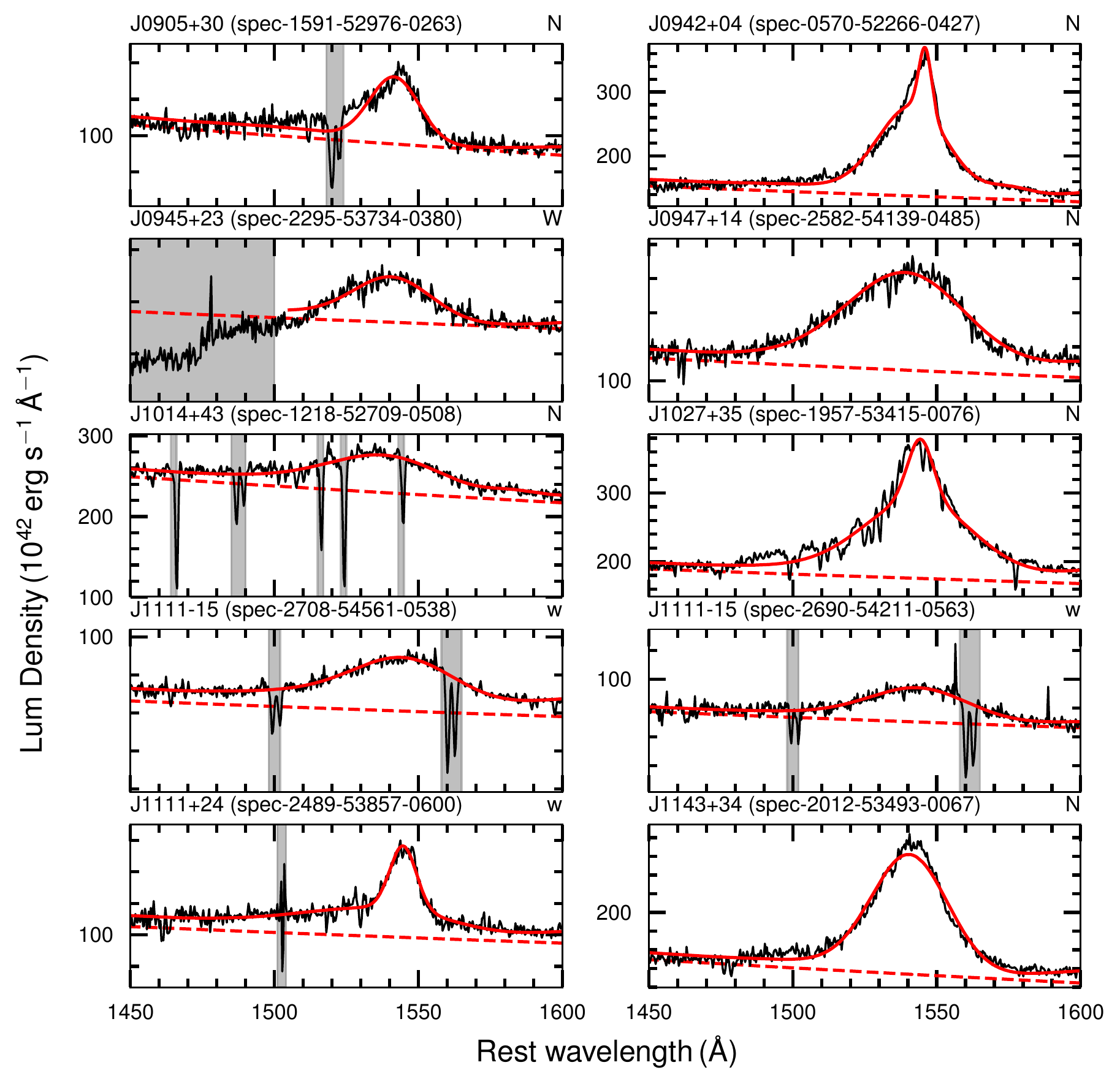}}
\caption{{\it Continued}}
\label{sfit}
\end{figure}
\begin{figure}
\addtocounter{figure}{-1}
\centering
 \resizebox{\hsize}{!}{\includegraphics{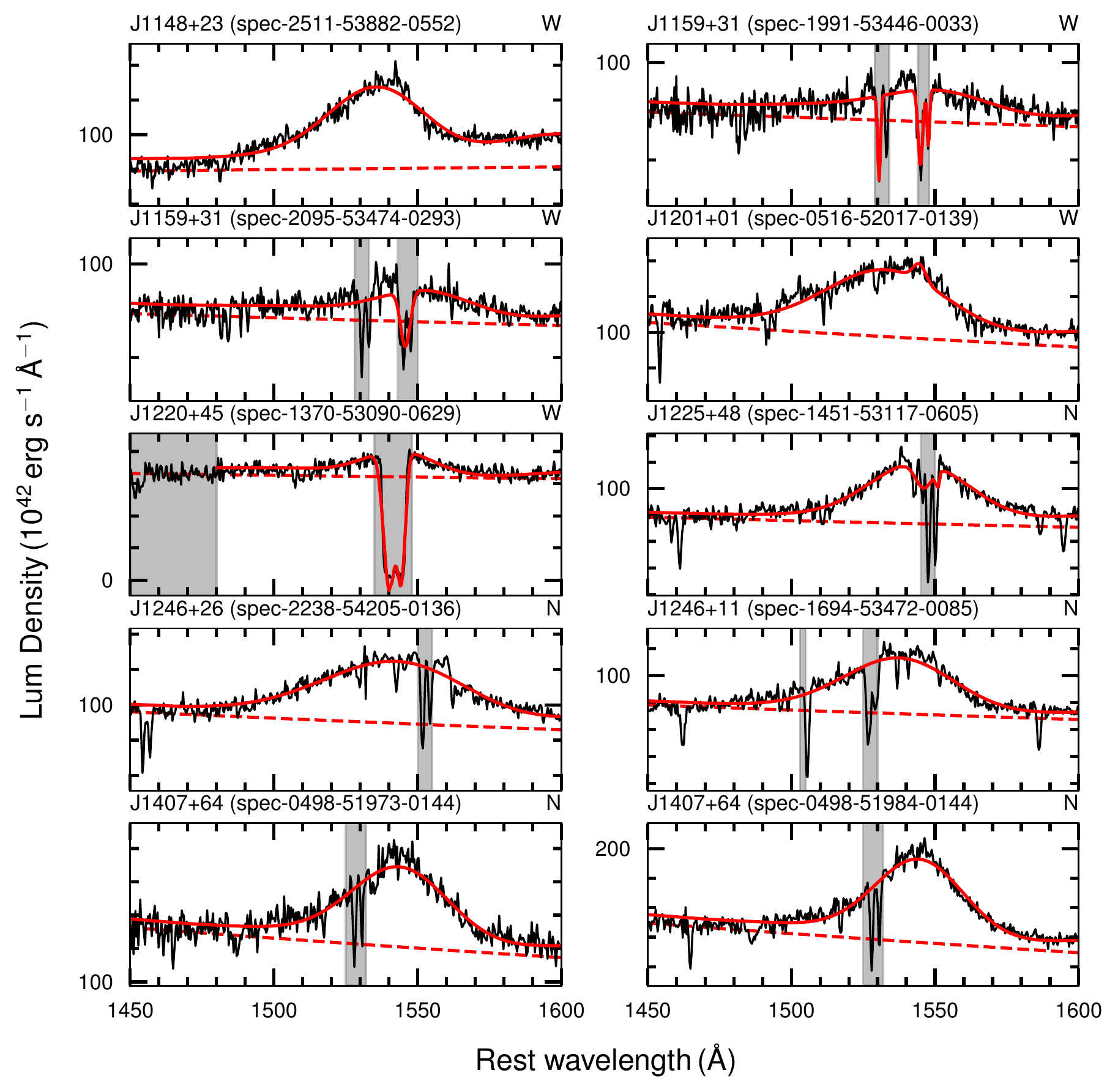}}
\caption{{\it Continued}}
\label{sfit}
\end{figure}
\begin{figure}
\addtocounter{figure}{-1}
\centering
 \resizebox{\hsize}{!}{\includegraphics{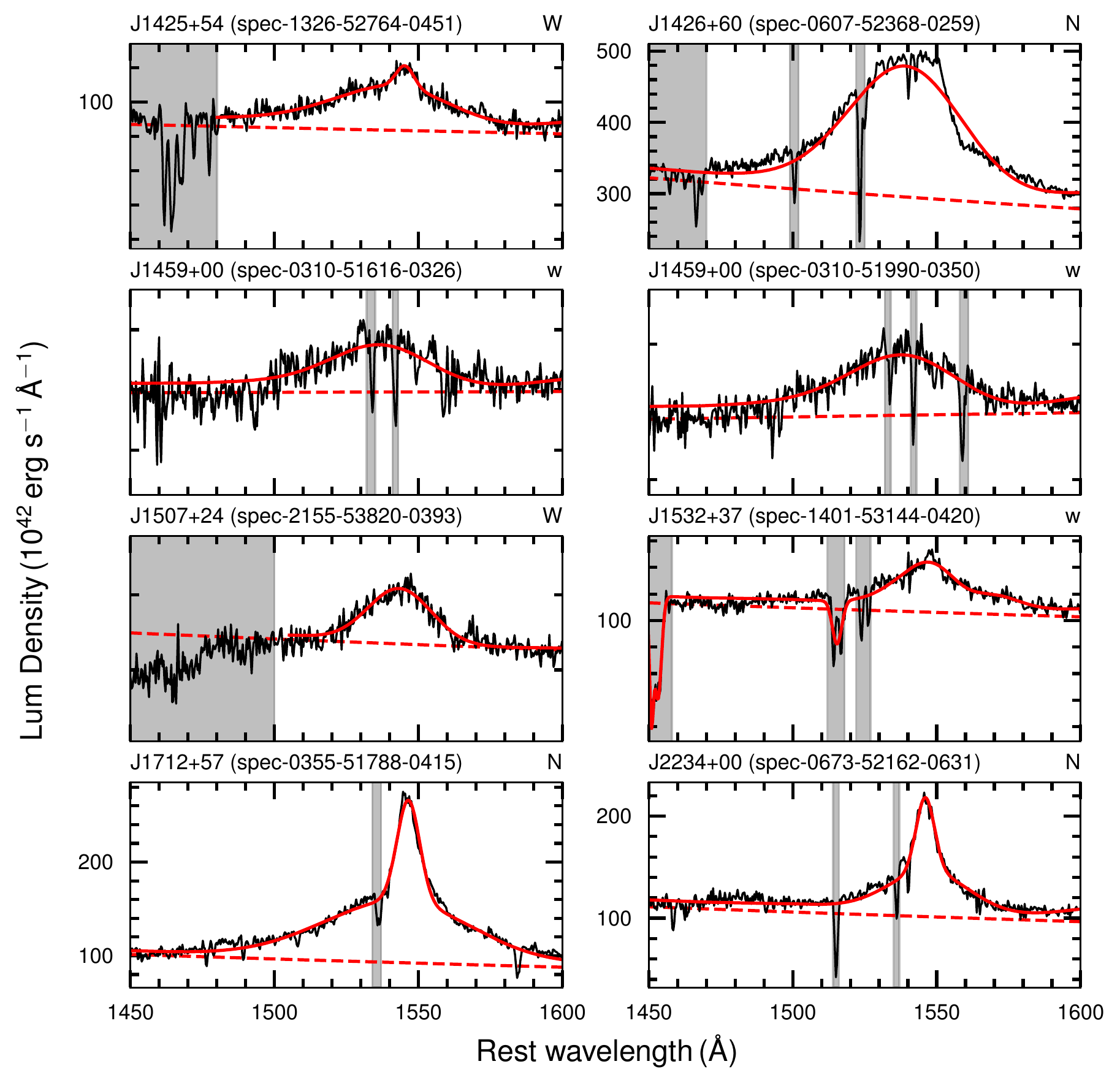}}
\caption{{\it Continued}}
\label{sfit}
\end{figure}

\subsection{The SDSS data}
\label{The SDSS data}
We searched for all the spectroscopic data in the SDSS archive finding 38 observations, with 6 quasars having 2 or more spectra. In our analysis we excluded additional BOSS spectroscopy, as its flux calibration may be uncertain\footnote{See http://www.sdss3.org/dr9/spectro/caveats.php\#qsoflux for a list of issues in the BOSS flux calibration.}.
We then started a systematic analysis of the SDSS spectroscopic data of these 38 spectra, focusing mainly on the \ion{C}{iv} emission line and the nuclear continuum.
Figure~\ref{sfit} shows the best-fit models of the \ion{C}{iv} emission line (in red), whilst the continuum is plotted with the red dashed line. The full UV spectra are shown in Figure~\ref{spec}. All the output parameters (i.e., \ion{C}{iv} FWHM and EW, continuum slope and luminosity) are in good agreement (within a factor of two) between the two different codes. 
Table~\ref{tbl1} summarises the UV properties, output of the spectral analysis.

\section{Results}
\label{results}
\subsection{UV composite of X-ray normal versus weak quasars}
\label{UV composite}
\begin{figure}
\centering
 \resizebox{\hsize}{!}{\includegraphics{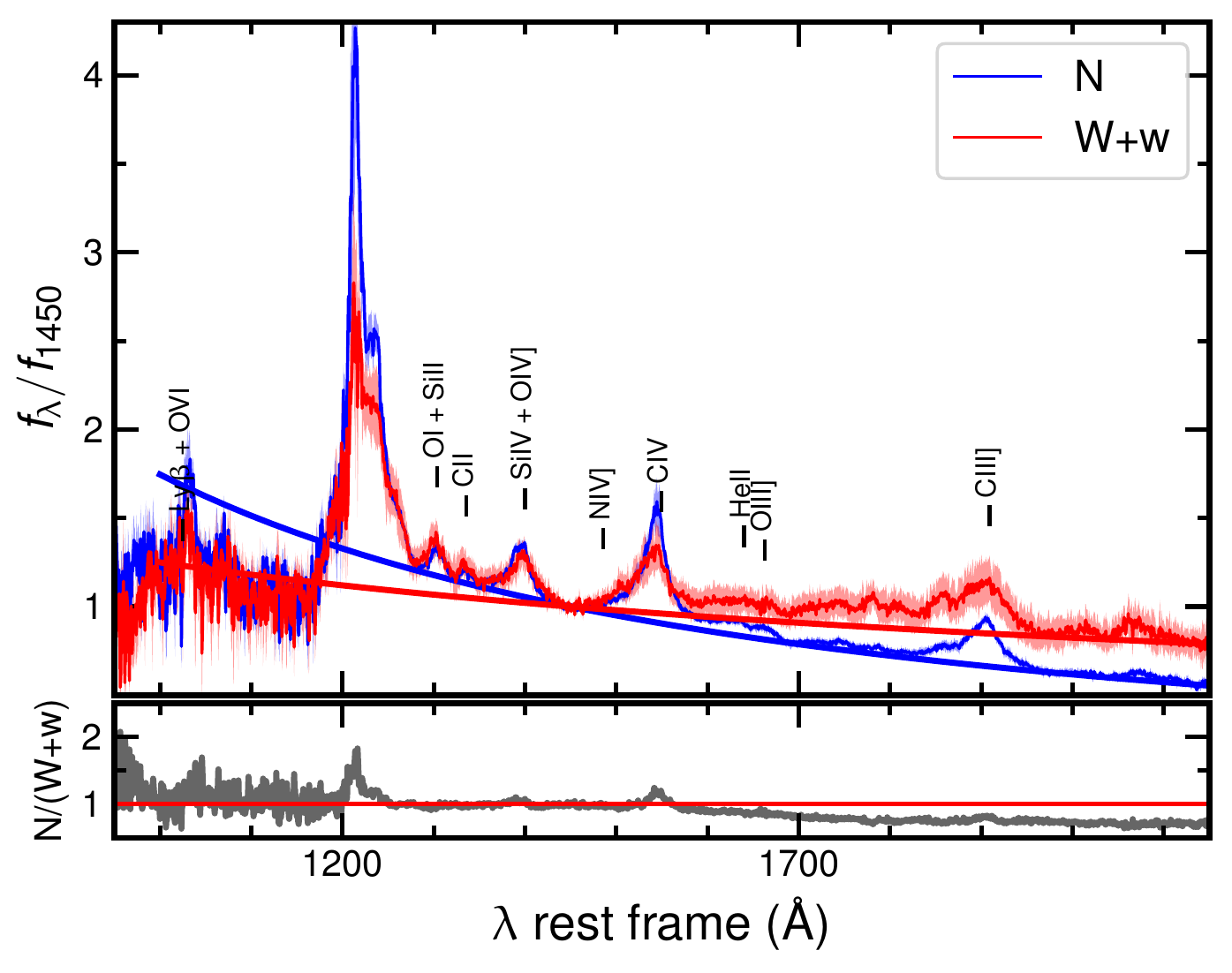}}
 \resizebox{\hsize}{!}{\includegraphics{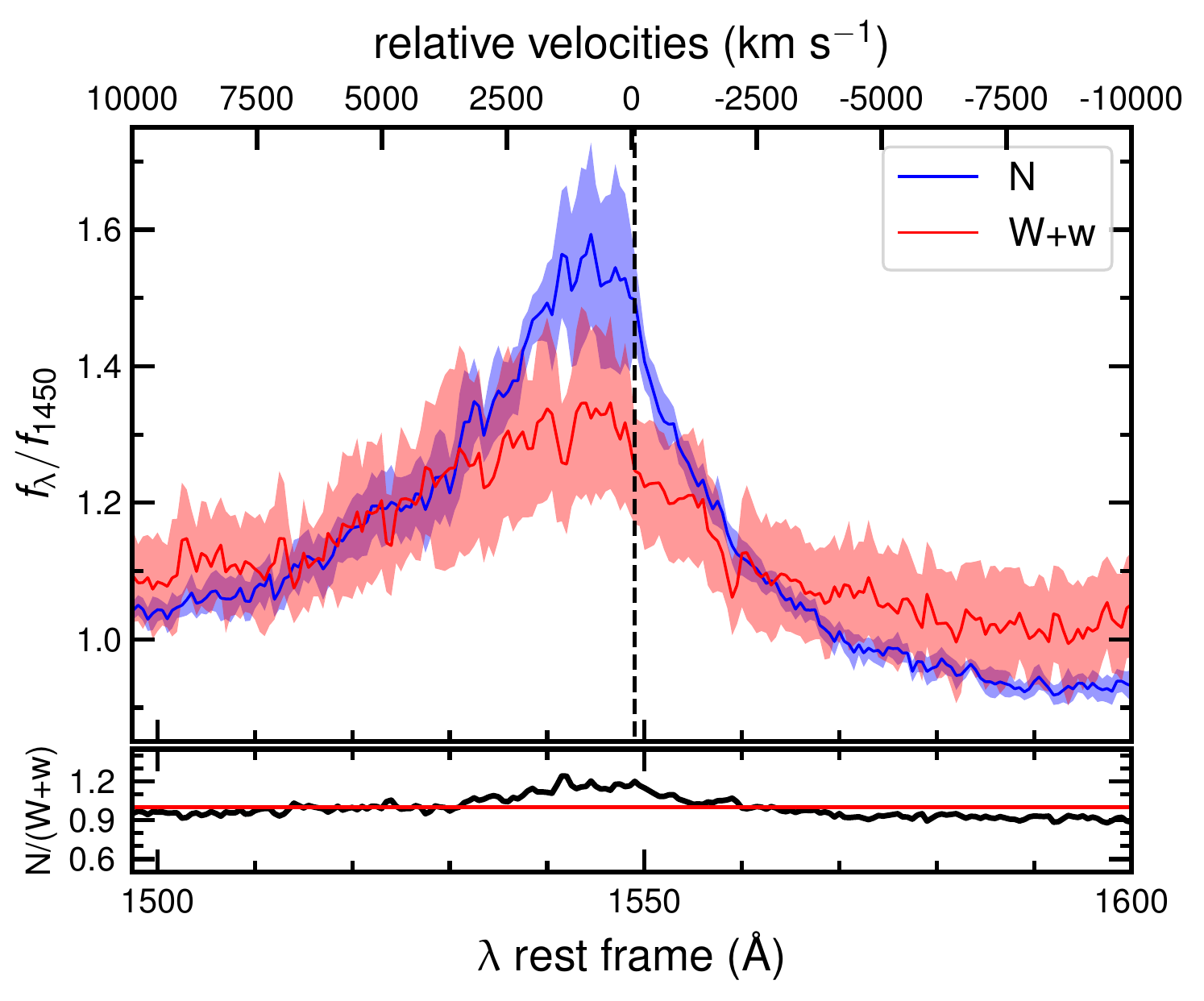}}
\caption{{\it Top panel}: Mean observed quasar spectra for the X-ray normal ($N$, blue) and X-ray weak plus weak candidates (\textit{W+w}, red). These composites are estimated from a stack of the SDSS spectra (23 $N$ and 10 \textit{W+w} spectra), each normalised to unit flux at 1450 \AA\ with uncertainties from bootstrap (shaded area). The UV continuum slopes for the $N$ and \textit{W+w} composites are shown with the blue and red lines, respectively. {\it Bottom panel}: zoom-in of the \ion{C}{iv} emission line for the $N$ and \textit{W+w} composites. The velocity shift from the rest-frame transition wavelength (1549 \AA) is shown on the top $x-$axis. In both panels, the ratio of the $N$ over the \textit{W+w} mean spectra is also shown at the bottom with a black line.}
\label{stacks}
\end{figure}
The quasars in the $z$\,$\simeq$\,3 sample are characterised by highly homogeneous UV spectra by construction, with an intrinsically blue continuum. This homogeneity is clearly shown in \citetalias{nardini2019} (see their figure~2), where the $z$\,$\simeq$\,3 spectral composite is compared with the composite of SDSS AGN of \citet{vandenberk2001}, and with the one of \citet{lusso2015}, based on 53 quasars at $z$\,$\sim$\,2.4 and corrected for intervening absorption by neutral hydrogen in the intergalactic medium. 
A mild decrease in the emission-line strength due to the Baldwin effect \citep{baldwin1977} is observed, but it is clear that both the UV continuum and the overall spectral properties of our sources are in good agreement with the expected intrinsic quasar spectrum. 

Following the same methodology for the composite construction as in \citetalias{nardini2019} (see also \citealt{lusso2015}), we also built two separate stacks for the $N$ and the \textit{W+w} sub-samples. 
Figure~\ref{stacks} (top panel) shows the resulting mean quasar spectra. These stacks are obtained from a mean of the SDSS spectra, each normalised to unit flux at 1450 \AA, with uncertainties estimated from bootstrap (shaded areas). The spectral ratio between the $N$ and \textit{W+w} sub-samples is also shown with a black line. 
The radio-bright (J0900+42), the two BALs (J1148+23, J0945+23), and the reddest quasar (J1459+00) are excluded when building the stacks, since we want to compare possible different spectral features between $N$ and \textit{W+w} sources that may not be connected to their BAL/red nature (e.g. redder/flatter UV slope, broad absorption lines).
The stack of the $N$ sub-sample is thus composed by 23 spectra, whilst the \textit{W+w} one contains 10 spectra. 

There are two main differences between the quasar composites, although the statistics are such that uncertainties are significant. First, the \textit{W+w} stack displays a slightly redder/flatter UV continuum slope ($\alpha_\lambda$\,$\simeq$\,$-0.6$)\footnote{The relation between the flux densities in wavelength ($f_\lambda$\,$\propto$\,$\lambda^{\alpha_\lambda}$ and frequency ($f_\nu$\,$\propto$\,$\nu^{\alpha_\nu}$) space is given by $\alpha_\nu$\,$=$\,$-(2+\alpha_\lambda)$.} with respect to the $N$ one ($\alpha_\lambda$\,$\simeq$\,$-1.5$). Second, emission lines in the \textit{W+w} stack are broader and fainter than in the $N$ one. The latter property is also evident in the bottom panel of Figure~\ref{stacks}, where we show the velocity shift from the rest-frame wavelength of the \ion{C}{iv} emission-line transition (1549 \AA) for the $N$ and \textit{W+w} composites. The emission line presents a mild blueshift\footnote{In the general sign convention, blueshifts correspond to `negative' velocity shifts. Here, for simplicity, we assign to blueshift a positive velocity value, which is equivalent to adopting the natural quasar frame (see \citealt{richards2002}).} of $\simeq$\,1,000 \kms\ with respect to the reference wavelength\footnote{The average \ion{C}{iv} emission-line shift for radio-quiet quasars in the SDSS is $\sim$\,810 \kms\ \citep{richards2011}.}, and an asymmetric profile towards the blue side in both stacks. Yet, the \ion{C}{iv} emission line is broader (FWHM\,$\simeq$\,10,000 \kms) in the \textit{W+w} than in the $N$ composite ($\simeq$\,7,000 \kms). Within the uncertainties, however, these features are consistent with quasars at similar redshifts (e.g. \citealt{richards2002,shen2011}) and luminosities (e.g. \citealt{vietri2018}) for both the $N$ and \textit{W+w} stacks\footnote{We also note that the redshifts for the $z$\,$\simeq$\,3 sample could still be uncertain by about 600--800 km s$^{-1}$, since they are based on UV lines \citep{hw2010}.}.
\rev{We stress that the $z$\,$\simeq$\,3 sample was selected to be `blue', with minimal levels of UV absorption (as shown in Figure~\ref{g1g2plot}), and to span a narrow range of UV fluxes (see Figs.~\ref{distrall} and \ref{fofx}), therefore any difference between the $N$ and \textit{W+w} sub-samples in the UV are minimised by construction.}

\rev{Even if the difference in the $N$ and \textit{W+w} spectral properties is not statistically significant, the mild variation between these two sub-sample may hint at a higher incidence of outflowing gas in the \textit{W+w} sub-sample with respect to the $N$ one (the former displaying a higher blueshift and a broader profile of the \ion{C}{iv} emission line).}
To confirm this interpretation, we have embarked upon a near-IR observational campaign (covering the rest-frame optical) with the Large Binocular Telescope (LBT) located in Mount Graham, Arizona, which will provide a better estimate of the systemic redshift from narrow emission lines (e.g. [\ion{O}{iii}]). Moreover, broad emission components in the rest-frame optical with blueward asymmetry, such as that observed in the [\ion{O}{iii}] forbidden line (e.g. \citealt{bischetti2017,vietri2018}, and references therein), also probe ionised gas that may be in an outflowing phase at much larger (kpc) scales. We will discuss this point further in Section~\ref{discussion}.

\subsection{X-ray parameters as a function of the UV spectral properties}
\label{xrayuv}
\begin{figure*}
\centering
 \resizebox{\hsize}{!}{\includegraphics{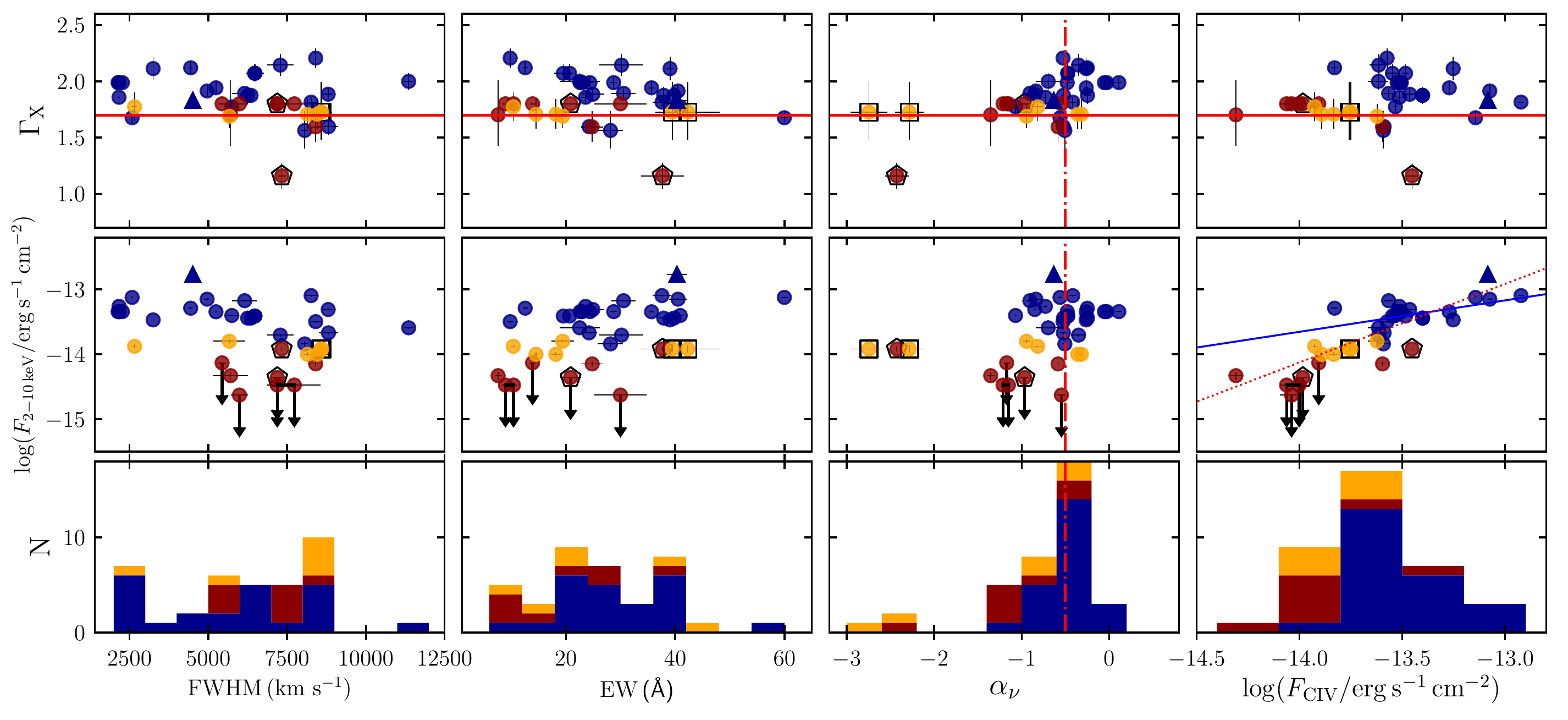}}
 \resizebox{\hsize}{!}{\includegraphics{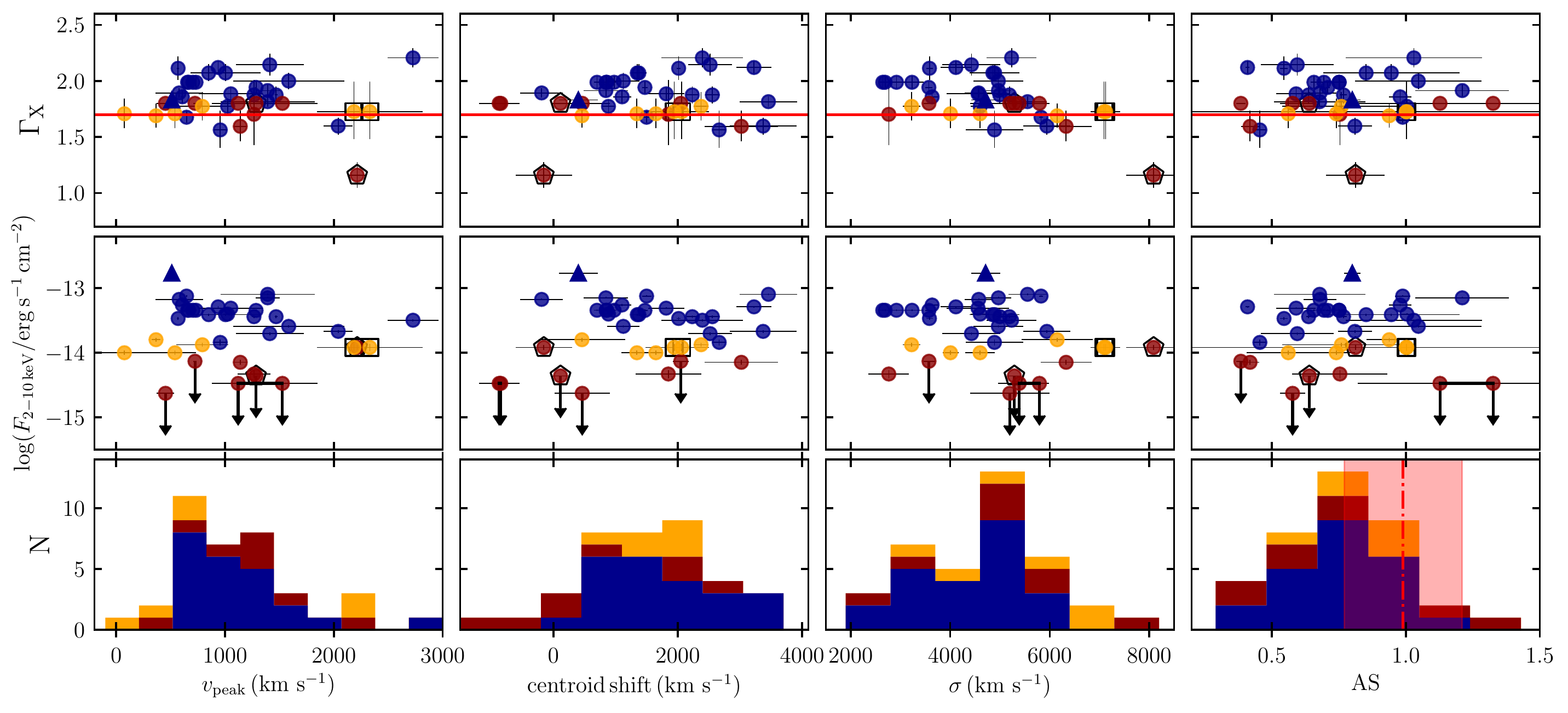}}
\caption{X-ray versus \ion{C}{iv} emission-line properties \rev{for the 38 SDSS spectra}. In the first two rows from top, the continuum photon index and the 2--10 keV flux (in erg s$^{-1}$ cm$^{-2}$) are plotted against the \ion{C}{iv} FWHM (in \kms), EW (in \AA), the UV continuum slope ($\alpha_\nu$) and the total \ion{C}{iv} emission-line flux (in erg s$^{-1}$ cm$^{-2}$), the distributions of which are shown in the third row. The red solid line at $\gammax$\,$=$\,1.7 represents the minimum value adopted by \citetalias{rl19} to select the clean quasar sample for cosmology. The red dot-dashed line in the $\alpha_\nu$ panel marks an UV continuum slope of $-0.5$, typical for $z$\,$>$\,2 quasars \citep{lusso2015}. Only $\Fh$ shows a significant correlation with $F_{\rm C\,IV}$, where the blue solid and red dotted lines represent the best-fit regression solutions for $N$ and {\it N+W+w} sources (radio-bright, BALs and red quasars are excluded, see \S\,\ref{A blue quasar sample}). In the fourth and fifth rows $\gammax$ and $\Fh$ are plotted as a function of the \ion{C}{iv} peak velocity $\vp$ (with respect to the nominal wavelength), the velocity shift of the line centroid, the velocity dispersion $\sigma$, and the asymmetry index (AS). Redward (blueward) asymmetries correspond to AS values $>$\,1 ($<$\,1), where the dot-dashed line marks AS\,$=$\,1 (as generally found at lower UV luminosity and redshift) and the red shaded area the typical dispersion on this parameter of 0.22 dex.}
\label{civxprop}
\end{figure*}
Evidence of correlations involving the \ion{C}{iv} and X-ray spectral properties have been reported and widely discussed in the context of the Eigenvector 1 formalism \citep[e.g.][and references therein]{sulentic2000,sulentic2007,marzianisulentic2014}. Recent studies have found that the connection between \ion{C}{iv} and X-ray parameters also apply to extremely luminous ($\lbol$\,$>$\,10$^{47}$ erg s$^{-1}$), high-redshift ($z$\,$\sim$\,2--4) quasars \citep{martocchia2017,vietri2018,zappacosta2020}. Nonetheless, previously analysed samples are characterised by a rather high level of inhomogeneity regarding their classification, often comprising a mixed bag of various different types such as BALs, radio-loud and UV-red quasars. By contrast, here we can leverage on the uniform selection of the $z$\,$\simeq$\,3 sample.
 
Figure~\ref{civxprop} presents the key X-ray properties (i.e., continuum slope and intensity) as a function of \ion{C}{iv} emission-line characteristics \rev{computed for the 38 SDSS spectra}. In the first two rows from top, the X-ray photon index and the 2--10 keV flux values are plotted against the FWHM, the rest-frame EW, the UV continuum slope ($\alpha_\nu$) and the total \ion{C}{iv} flux. 
The distribution of each UV parameter is shown in the third row. The red dot-dashed line in the $\alpha_\nu$ panel marks an UV continuum slope of $-0.5$, typical for $z$\,$>$\,2 quasars \citep{lusso2015}, while the red solid line is the $\gammax$ threshold adopted by \citetalias{rl19} to select the clean quasar sample for cosmology. The cut at $\gammax$\,$=$\,1.7 clearly singles out quasars with a typical blue SED in the UV.
The fourth and fifth rows show how $\gammax$ and $\Fh$ are distributed as a function of the velocity of the \ion{C}{iv} line peak ($\vp$, computed with respect to the nominal wavelength), the velocity shift of the centroid of the \ion{C}{iv} line (first momentum of the flux distribution), the second momentum of the flux distribution ($\sigma$), and the asymmetry index (AS\footnote{The AS parameter is defined as $\ln(\lambda_{\rm red}/\lambda_{\rm peak})/\ln(\lambda_{\rm peak}/\lambda_{\rm blue})$, where $\lambda_{\rm red}$ e $\lambda_{\rm blue}$ are the wavelengths at half-width of the line profile in the red and blue part, respectively.}). Following the definition in \citet{shenliu2012}, redward and blueward asymmetries correspond to AS values $>$\,1 and $<$\,1, respectively. The dot-dashed line marks ${\rm AS}=1$, which is usual for sources of lower UV luminosities and redshifts, while the red shaded area is the typical dispersion on this parameter of 0.22 dex.

\begin{figure}
\centering
 \resizebox{\hsize}{!}{\includegraphics{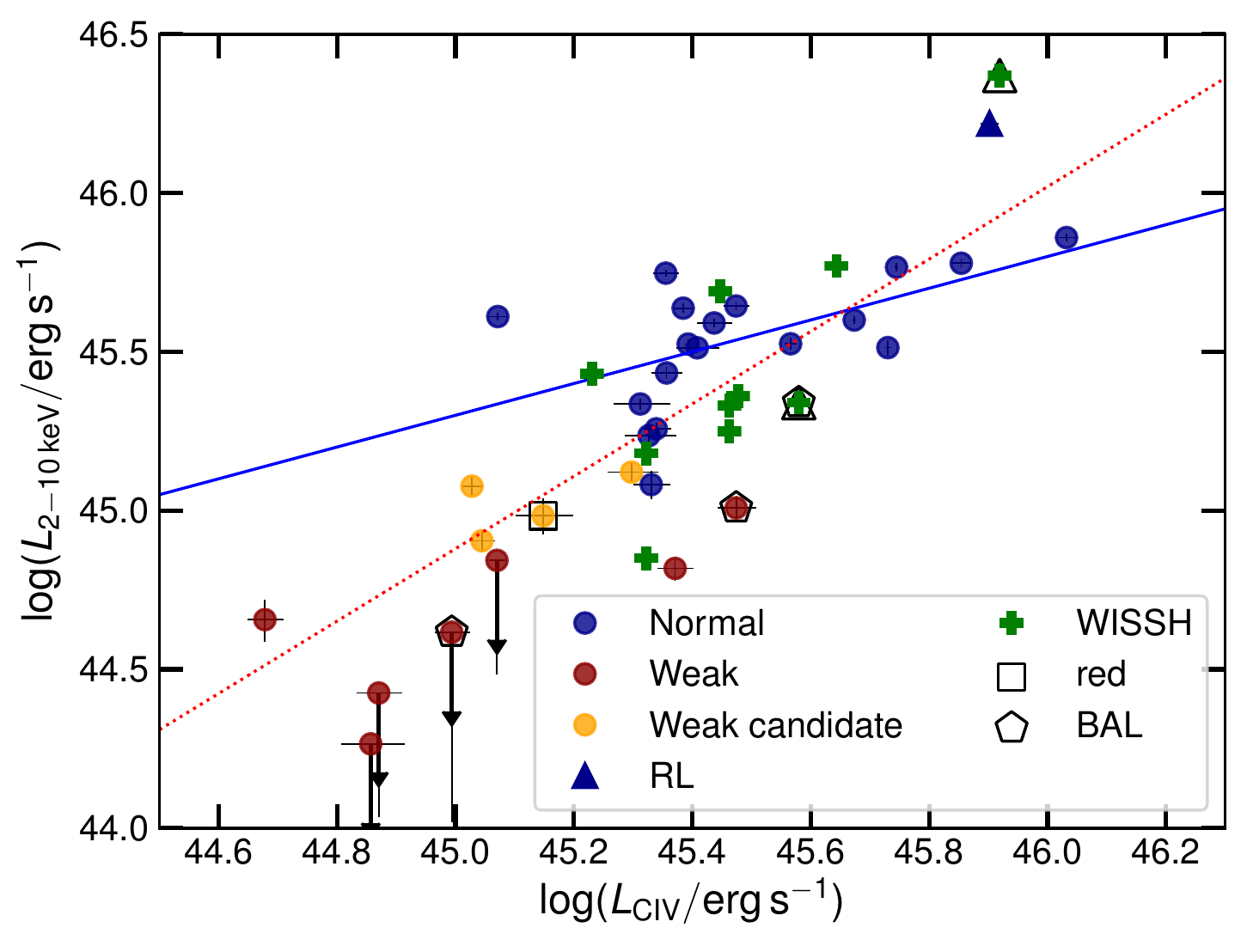}}
\caption{X-ray luminosity at 2--10 keV (erg s$^{-1}$) as a function of the total integrated \ion{C}{iv} line luminosity (erg s$^{-1}$). Green plus symbols represent 10 quasars from the WISSH sample \citep{martocchia2017}.}
\label{fxfciv}
\end{figure}
The only statistically significant correlation is revealed between $\Fh$ and $F_{\rm C\,IV}$, with the blue solid and red dotted lines representing the best-fit regression solutions considering the $N$ and {\it N+W+w} objects, where radio-bright, BALs and the reddest quasar in our sample are excluded. 
Figure~\ref{fxfciv} illustrates the same correlation in luminosities. We performed a regression analysis finding a statistically significant correlation at the $\sim$\,3\,$\sigma$ level in the case of the $z$\,$\simeq$\,3 $N$ quasars, for which 
\begin{equation}
\label{lxlciv}
\log (\Lh - 45) = (0.50\pm0.18)(\log L_{\rm C\,IV}-45) + 0.30\pm0.10,
\end{equation}
with an estimated dispersion of 0.17\,$\pm$\,0.04 dex.
If we perform the regression analysis on the entire $z$\,$\simeq$\,3 sample, but excluding the radio-bright quasar, the correlation becomes
\begin{equation}
\label{lxlcivall}
\log (\Lh - 45) = (1.14\pm0.18)(\log L_{\rm C\,IV}-45)- 0.12\pm0.08.
\end{equation}
The observed $\Lh-L_{\rm C\,IV}$ correlation is more significant in the latter case ($\simeq$\,9\,$\sigma$), when the $W$ and $w$ quasars are also included. However, the dispersion is much larger (0.28\,$\pm$\,0.05 dex) with respect to the same relation for the $N$ objects only. All the sources at $\Lh$\,$<$\,10$^{45}$ erg s$^{-1}$ are either $W$ or $w$, and also show lower emission-line luminosities, consistent with what we observe in their UV spectral composite in Figure~\ref{stacks}.

Figure~\ref{fxfciv} also includes 11 quasars from the WISSH sample analysed by \citet{martocchia2017}, for which both $L_{\rm C\,IV}$ and $\Lh$ are available\footnote{We excluded two quasars with a $L_{\rm C\,IV}$ measurement but an upper limit in the hard X-ray band. Amongst the 11 WISSH sources, the radio-bright quasar is in common with our sample.}. The WISSH quasars include 3 radio emitting objects and one BAL, which are marked accordingly in Figure~\ref{fxfciv}. The WISSH subset follows the $L_{\rm C\,IV}-\Lh$ relation, and its inclusion does not significantly change our results when all the quasars are considered.


Summarising, we have explored the hard X-ray band flux and the X-ray photon index as a function of several UV spectral features, finding no significant correlation apart from the one we observe between the X-ray flux and the integrated flux of the \ion{C}{iv} emission line.

\subsection{Comparison with other samples: on the relations with the \ion{C}{iv} EW}
\label{civew}
As already pointed out by \citet{Gibson2008}, there is a possible correlation between $\daox$ and log\,(\ion{C}{iv} EW), significant at the $>$\,99.99\% level. 
We observe the same correlation in the $z$\,$\simeq$\,3 sample, significant at the $\sim$\,98.5\% despite the lower statistics, according to both Pearson's $\rho$ and Kendall's $\tau$ tests.
Such a correlation is interpreted in the context of the quasar disc-wind scenario \citep[e.g.][]{elvis2000}, where high-ionisation lines are associated with the presence of a wind component exposed to the quasar UV continuum. 
Since the $z$\,$\simeq$\,3 quasars are probing the bright end of the quasar luminosity function, we want to compare their properties with those of a lower redshift/luminosity sample selected in a similar way.
The work of \citet{Gibson2008} was recently expanded by \citet[T20 hereafter]{timlin2020}, who examined 2,106 quasars selected from the SDSS-DR14 in the redshift range 1.7\,$\leq$\,$z$\,$\leq$\,2.7, with archival \chandra serendipitous observations. We singled out quasars with blue colours (i.e., $\Delta(g-i)$\,$\leq$\,0.45, see their Section~3), flagged as non-BAL, and with a radio-loudness value (estimated from the ratio of the 2500-\AA\ and the 6-cm flux densities) lower than 10. These criteria to a sample of 1,106 `typical' quasars, of which 637 
have a measurement of the rest-frame EW for the \ion{C}{iv} emission line.

\begin{figure}
\centering
 \resizebox{\hsize}{!}{\includegraphics{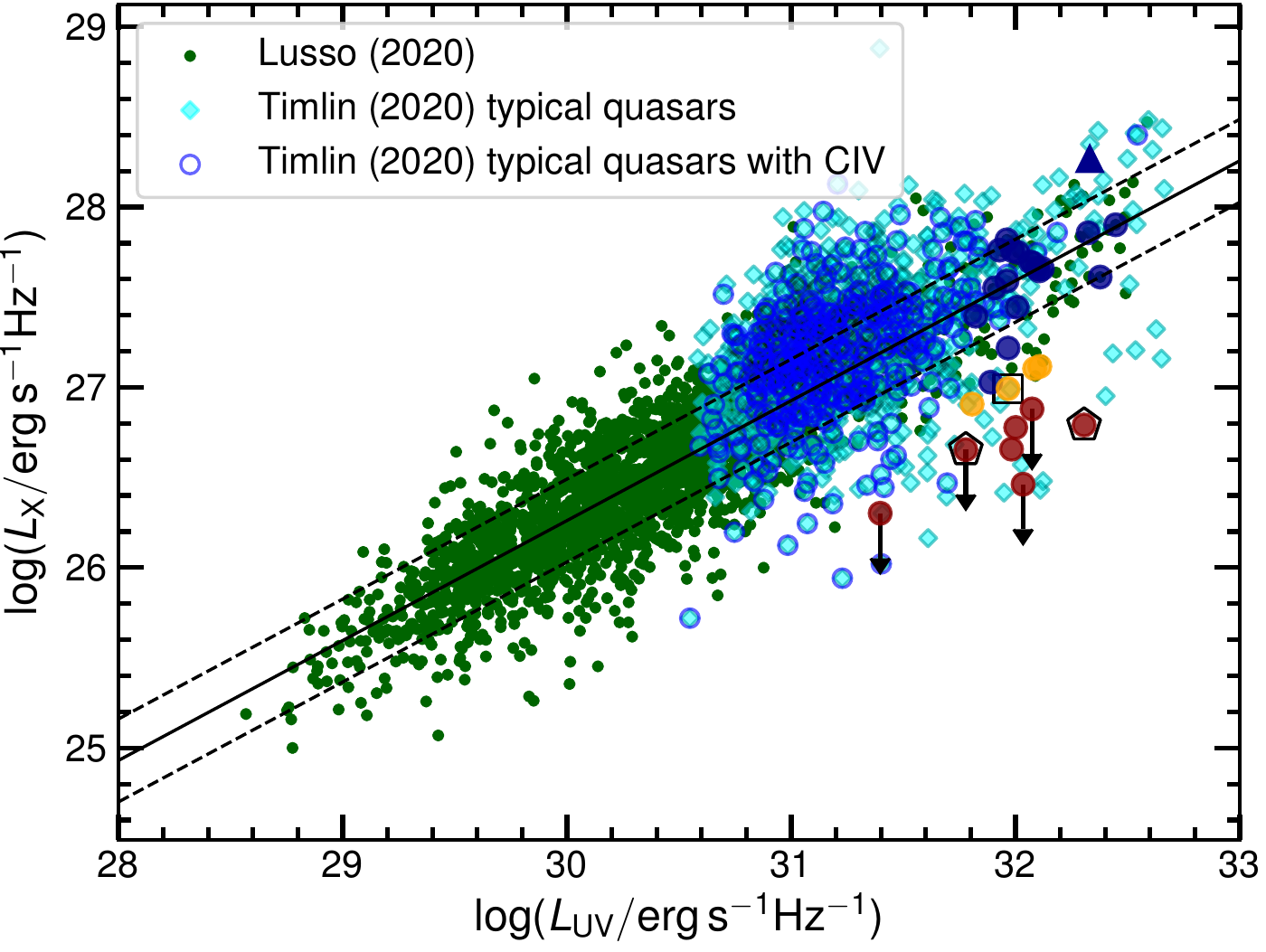}}
\caption{Rest-frame monochromatic luminosities $\lx$ against $\lo$ for the 30 quasars in our $z$\,$\simeq$\,3 sample. The colour coding is the same as in the previous figures. The green dots represent the sample of $\sim$\,2,400 quasars from \citetalias{lusso2020}, with the relative regression line. The dashed lines trace the 1\,$\sigma$ dispersion, 0.23 dex. We overplotted the sample of typical quasars from \citet{timlin2020}, where sources with an available \ion{C}{iv} measurement are highlighted with a blue circle.}
\label{lolx}
\end{figure}

To visualise possible differences between the lower and higher redshift/luminosity samples, we compare their rest-frame monochromatic luminosities $\lx$ against $\lo$ in Figure~\ref{lolx}. Sources with an available \ion{C}{iv} measurement (T20/\ion{C}{iv}) are highlighted with an open circle and span about 2 decades in both UV and X-ray luminosities, and they are preferentially located at lower values than the $z$\,$\simeq$\,3 quasars, as expected. There is no significant difference in either the UV or X-ray luminosity distributions of the sub-sample with \ion{C}{iv} measurements with respect to typical quasars. In the following we will thus focus on the sample with the \ion{C}{iv} emission-line properties available.
Figure~\ref{lolx} also shows the sample of $\sim$\,2,400 quasars from \citetalias{lusso2020}, with the relative regression line (with a slope $\gamma$\,$=$\,0.665\,$\pm$\,0.007) for comparison. The dashed lines trace the 1\,$\sigma$ dispersion, 0.23 dex. 
The \citetalias{lusso2020} sample is composed by 2,421 optically selected quasars (mostly from SDSS) with X-ray data from \xmm and \chandra, and it spans a redshift interval 0.009\,$\leq$\,$z$\,$\leq$\,7.541, with a mean (median) redshift of 1.442 (1.295). These sources were selected to have minimal contamination from the host galaxy emission (this is especially important for the $z$\,$<$\,0.7 AGN) and minimal dust/gas absorption. The Eddington bias is also taken into account. Details about the sample selection are provided in their Section~5. Most of the $z$\,$\simeq$\,3 $N$ quasars are included in the \citetalias{lusso2020} sample. 
As a whole, the T20/\ion{C}{iv} quasar sample is located at relatively bright X-ray and UV luminosities, and is characterised by a much higher dispersion in the $\lx-\lo$ relation (0.32 dex). The slope $\gamma$\,$=$\,0.540\,$\pm$\,0.060 is flatter with respect to \citetalias{lusso2020}, but consistent within a $2\sigma$ statistical level. 
The \citetalias{lusso2020} sample was built by applying more stringent constraints than the one considered in \citetalias{timlin2020} (e.g. possibly X-ray absorbed AGN with $\gammax$\,$<$\,1.7 were excluded). However, since our focus here is to verify the presence of possible correlations between the X-ray and UV properties of quasars at high redshift, we do not need to reduce the dispersion in the \citetalias{timlin2020} sample further.

\begin{figure*}
\centering
\includegraphics[width=0.495\linewidth]{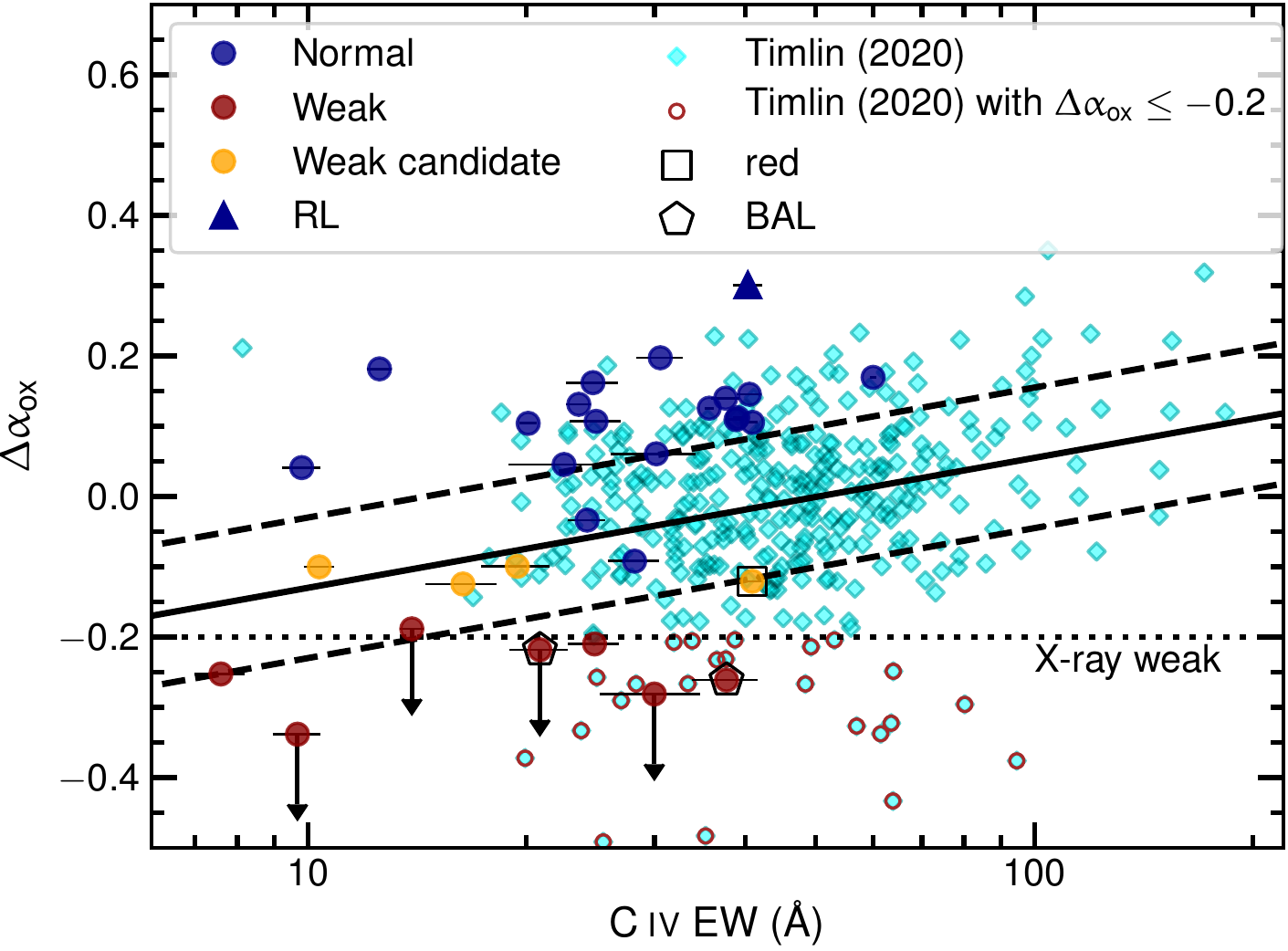}
\includegraphics[width=0.495\linewidth]{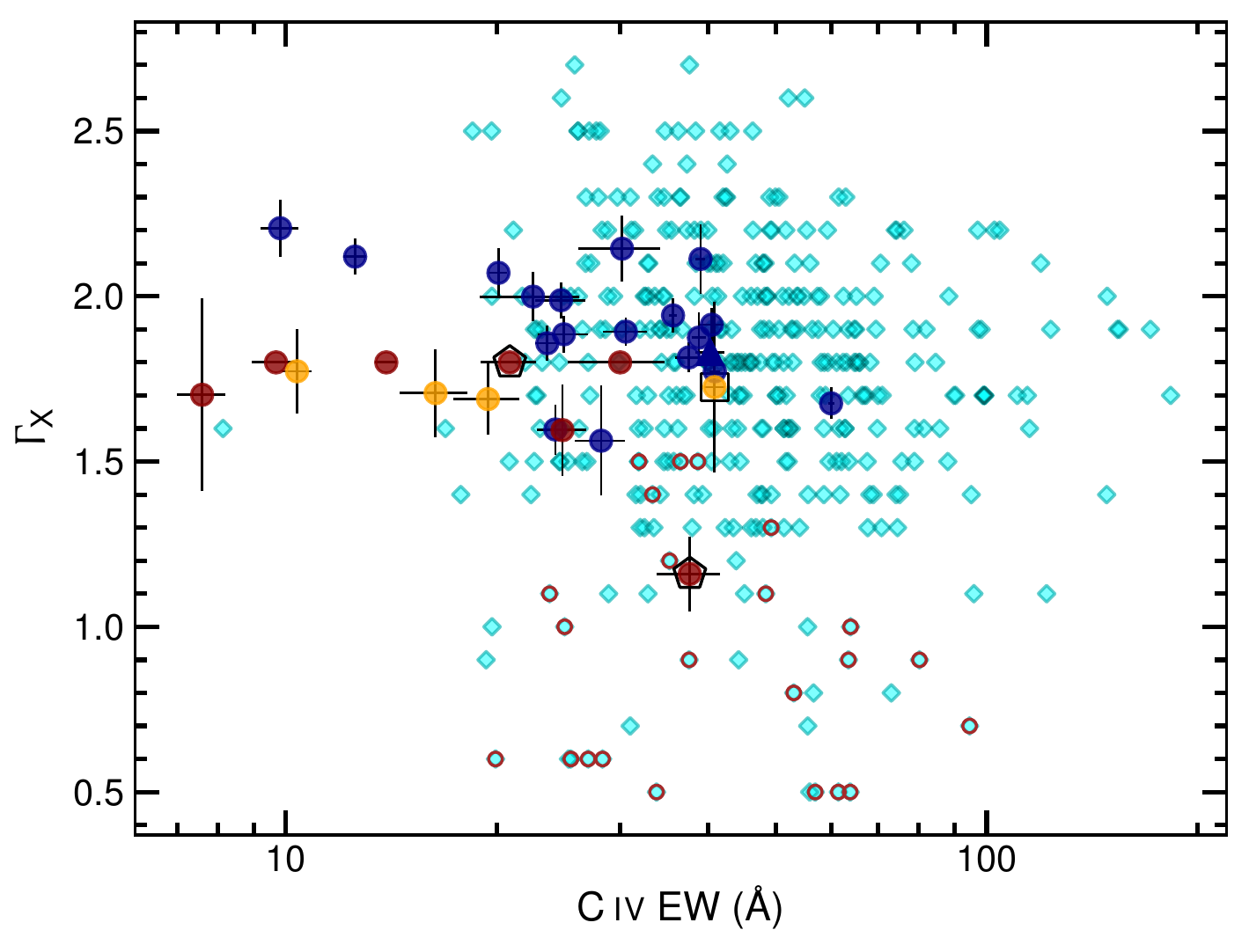}
\caption{$\daox$ (left) and photon index $\gammax$ (right) as a function of the rest frame \ion{C}{iv} EW for the $z$\,$\simeq$\,3 and T20/\ion{C}{iv} samples. The $\daox$ values for the $z$\,$\simeq$\,3 sample have been adjusted to be consistent with \citetalias{timlin2020} (see text for details). The solid line in the left panel is the $\daox$\,$-$\,\ion{C}{iv} EW relation reported by \citetalias{timlin2020} (see their equation~4). X-ray weak quasars in \citetalias{timlin2020} ($\daox$\,$\leq$\,$-0.2$) are marked with brown open circles.}
\label{ewciv1}
\end{figure*}
\begin{figure*}
\centering
\includegraphics[width=0.495\linewidth]{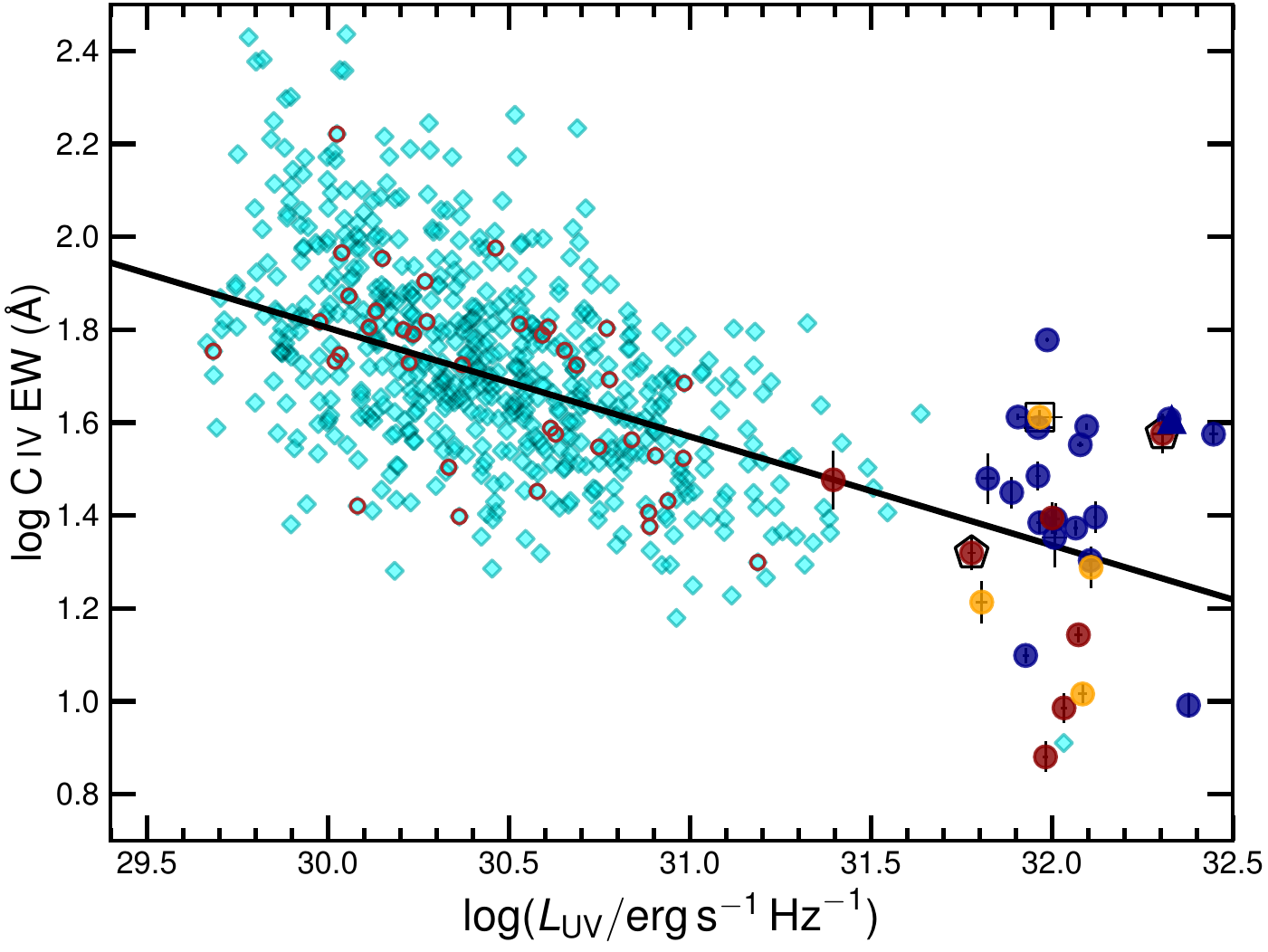}
\includegraphics[width=0.495\linewidth]{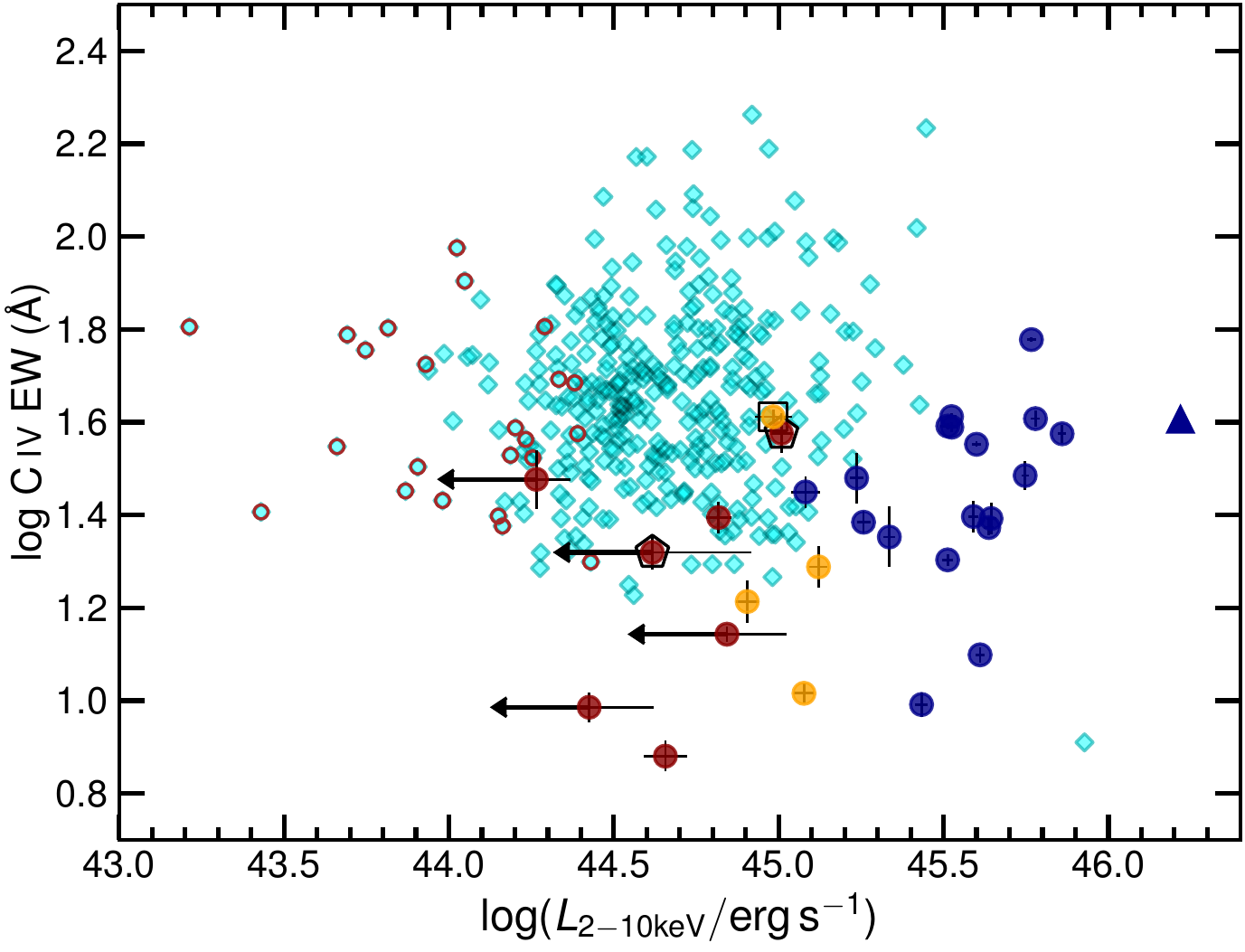}
\caption{$\Lo$ (left) and $\Lh$ (right) as a function of the rest frame \ion{C}{iv} EW for the $z$\,$\simeq$\,3 and T20/\ion{C}{iv} samples. The solid line in the left panel is the \ion{C}{iv} EW\,$-$\,$\Lo$ relation reported by \citetalias{timlin2020} (see their equation~8). X-ray weak quasars in \citetalias{timlin2020} ($\daox$\,$\leq$\,$-0.2$) are marked with brown open circles.}
\label{ewciv2}
\end{figure*}

Figures~\ref{ewciv1} and \ref{ewciv2} show the comparison of the X-ray weakness parameter $\daox$, photon index $\gammax$, $\Lh$, and $\lo$, as a function of the rest-frame \ion{C}{iv} EW between the $z$\,$\simeq$\,3 and T20/\ion{C}{iv} samples. Since the $\daox$ values for the $z$\,$\simeq$\,3 sample were computed assuming a different relation (see Section~\ref{The X-ray weak quasar fraction}) with respect to the one employed by \citetalias{timlin2020} (i.e., $\aox=-0.199\lo+4.573$, see their eq. (3) and related discussion in their Section~4.1), we decided to modify our $\daox$ by assuming for consistency their $\aox-\lo$ relation as a benchmark. 
The difference between our original $\daox$ values for the $z$\,$\simeq$\,3 sample and the new ones is about $-$0.08, thus the applied correction shifts all the data points to slightly higher values.

The $z$\,$\simeq$\,3 sample follows a similar trend to the one published by \citetalias{timlin2020}, i.e., $\daox=0.185\log ($\ion{C}{iv}\,{\rm EW}$) -0.315$ (with a reported dispersion of 0.10 dex), which is plotted in Figure~\ref{ewciv1} with the solid line. 
Even with the modified $\daox$ values, the $W$ quasars will be still considered as such ($\daox$\,$\leq$\,$-0.2$).
The \citetalias{timlin2020} sample extends to larger \ion{C}{iv} EW than the $z$\,$\simeq$\,3 quasars, which is expected given the relative ranges of UV luminosities and the Baldwin effect. The $z$\,$\simeq$\,3 quasars are selected in a very narrow $\Lo$ range, as shown in Figure~\ref{ewciv2}, whilst the \citetalias{timlin2020} sample is located at much lower UV luminosities. For this reason, the $z$\,$\simeq$\,3 $N$ quasars also have higher hard X-ray luminosities than the \citetalias{timlin2020} sample, but do not show any statistical trend with \ion{C}{iv} EW. 
The $\gammax$ values of our high-redshift $N$ quasars are consistent with the lower redshift/luminosity \citetalias{timlin2020} sample, 
and again no statistical trend with \ion{C}{iv} EW is present. 

We also report the best-fit regression line published by \citetalias{timlin2020} between \ion{C}{iv} EW and $\Lo$ (their equation~8, see also \citealt{green2001}). The $z$\,$\simeq$\,3 quasars follow the same relation as observed for the \citetalias{timlin2020} AGN sample. This negative trend is due to the well-known Baldwin effect of the \ion{C}{iv} line, i.e., the brighter the ionising source the fainter (in terms of EW) the emission line \citep[e.g.][]{green1996,dietrich2002}. 
Together, these two samples cover almost three decades in UV luminosity, and the brightest quasars tend to be located at low \ion{C}{iv} EW values. However, X-ray weak quasars show no preference of \ion{C}{iv} EW values, covering a range in both UV luminosity and \ion{C}{iv} EW similar to that of X-ray normal quasars. 


\subsection{Comparison with other samples: on the relations with the \ion{C}{iv} velocity shift}
\label{civvshift}
\begin{figure*}
\centering
\includegraphics[width=0.46\linewidth]{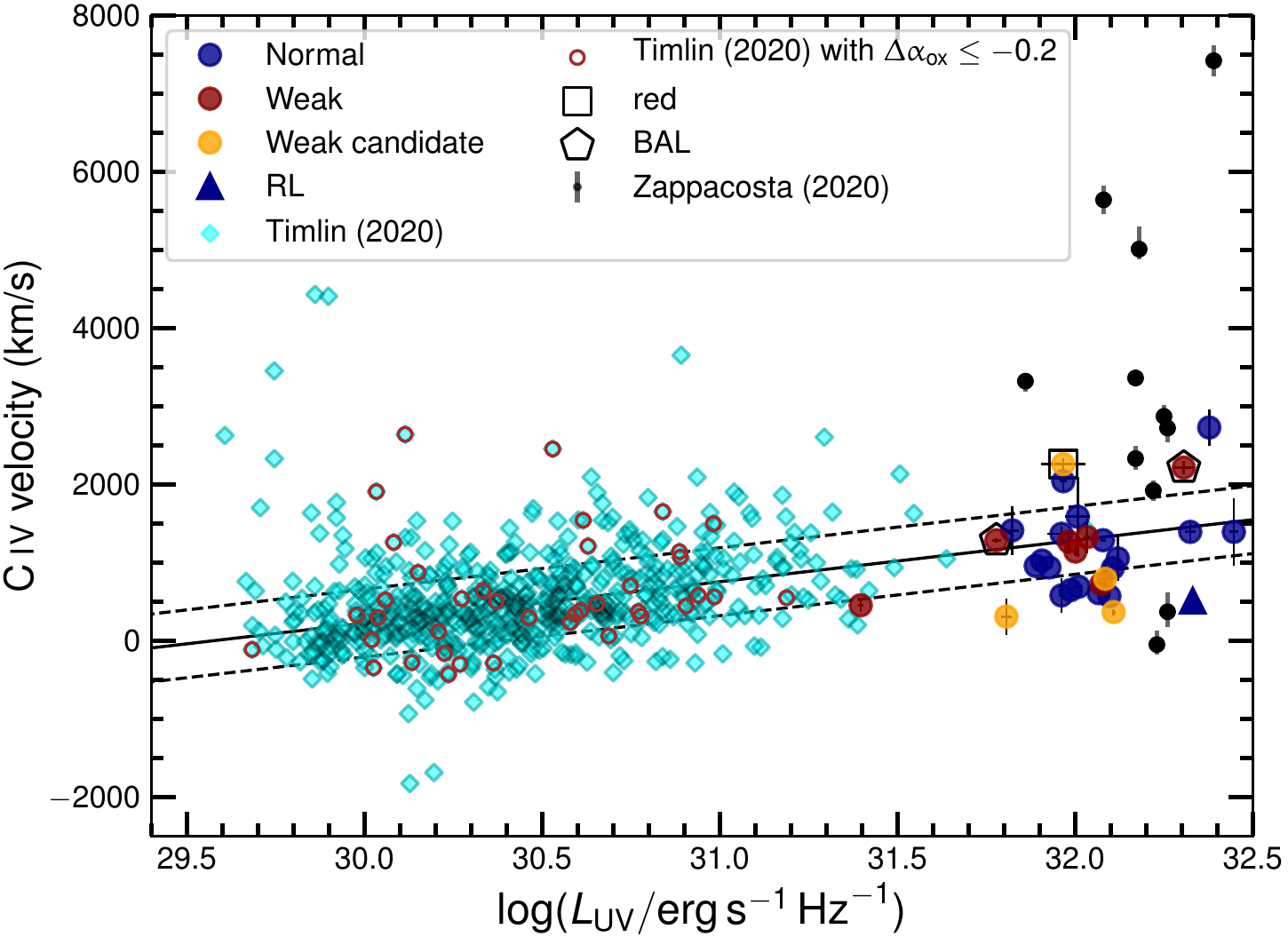}
\includegraphics[width=0.45\linewidth]{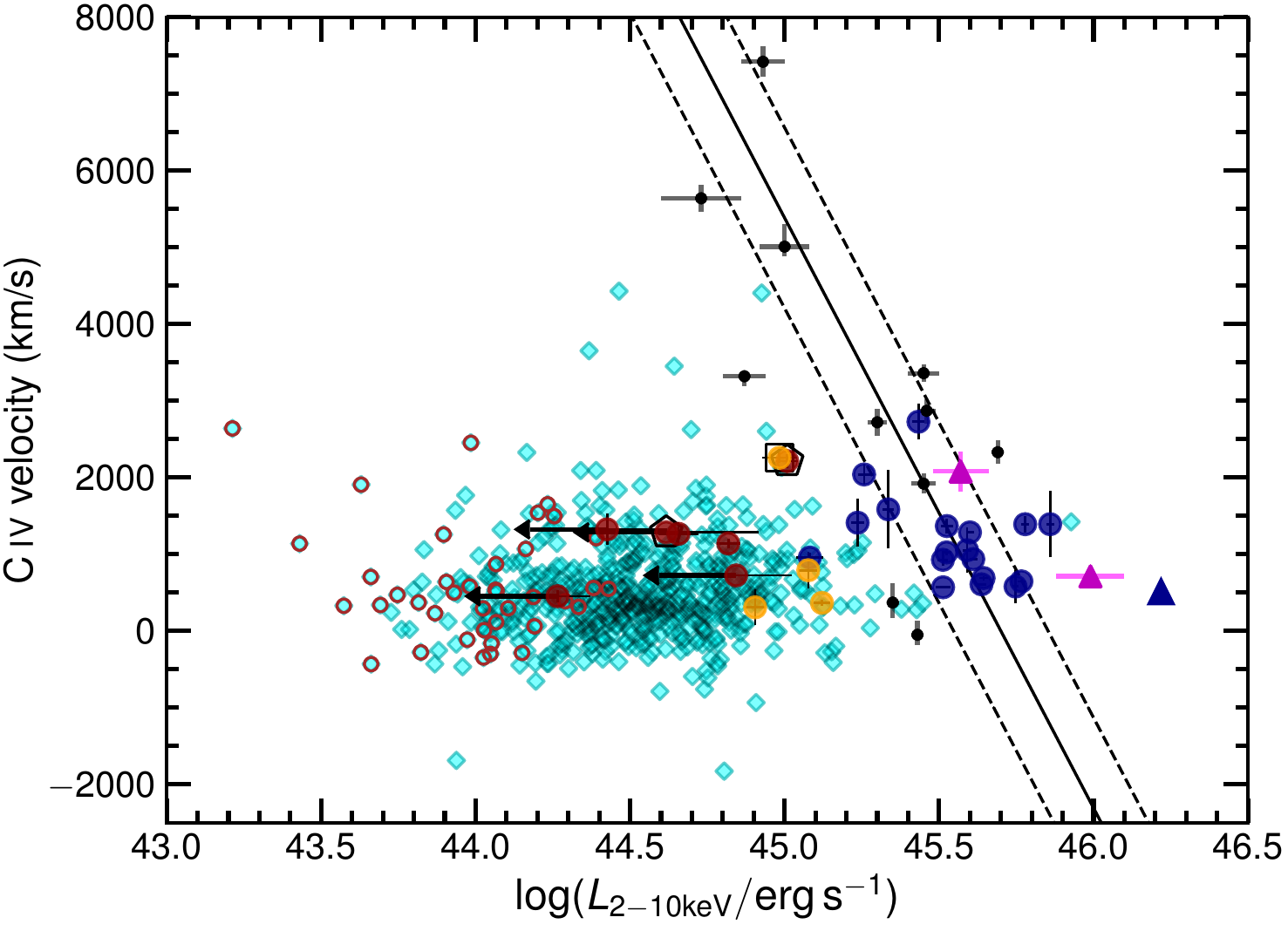} 
\caption{$\Lo$ (left), and $\Lh$ (right), as a function of the rest frame \ion{C}{iv} peak velocity shift for the $z$\,$\simeq$\,3, T20/\ion{C}{iv}, WISSH \citep{zappacosta2020} samples (we excluded 2 quasars that overlap with our $z\simeq3$ sample), and five additional $z\simeq3.2$--3.7 quasars with available $\vp$ 
(published by \citealt[magenta triangles]{coatman2017}, we neglected 3 overlaps with our $z\simeq3$ sample). The solid line in the left panel is the \ion{C}{iv} $\vp-\Lo$ best-fit relation by \citetalias{timlin2020}, whilst the one in the right panel is the $\Lh$\,$-$\,\ion{C}{iv} $\vp$ best fit relation reported by \citet[see their table~1]{zappacosta2020}. 
X-ray weak quasars in \citetalias{timlin2020} ($\daox$\,$\leq$\,$-0.2$) are marked with brown open circles.}
\label{vcivlolx}
\end{figure*}

The \ion{C}{iv} emission line centroid usually shows a blueshift (up to about 10,000 \kms) with respect to the laboratory wavelength or the systemic redshift measured from the narrow emission lines \citep[e.g.][]{gaskell1982,wilkes1984}. Such a property is often observed in bright quasars and it is usually ascribed to the presence of strong winds \citep[e.g.][]{richards2002,2005MNRAS.356.1029B,richards2011}. 

Recently, \citet{zappacosta2020} reported on a strong correlation between $\vp$ and the unabsorbed 2--10 keV luminosity $\Lh$ for a sample of 13 WISSH quasars at 2.075\,$\leq$\,$z$\,$\leq$\,3.490, with X-ray observations from \xmm and/or \chandra and \ion{C}{iv} emission-line properties available \citep{vietri2018}. Figure~\ref{vcivlolx} (right) shows such correlation and the relative dispersion, where we have included our $z$\,$\simeq$\,3 sample. The $N$ quasars occupy the top right corner of the $\Lh-\vp$ plane and are in very good agreement with the observed trend, whilst the $w$ and $W$ objects have $\vp$ values consistent with the $N$ quasars in spite of their fainter $\Lh$. 

Figure~\ref{vcivlolx} also shows the location in the $\Lh-\vp$ plane of the lower redshift/luminosity \citetalias{timlin2020} sample. Given their $\vp$ in the range $\sim$\,500--2000 \kms\ and fainter $\Lh$ values, the \citetalias{timlin2020} quasars extends the coverage to low $\Lh$ values, with no observed trend of an decreasing $\vp$ (or blueshift) with increasing $\Lh$. 
The absence of a relation between $\Lh$ and $\vp$ in the \citetalias{timlin2020} sample could be due to the lack of strong outflows in these sources, since also their UV luminosity is much lower than that observed in the higher redshift samples, as shown in the left panel of Figure~\ref{vcivlolx}. There is, however, a strong correlation between $\vp$ and $\Lo$ for the \citetalias{timlin2020} sample, and our $z$\,$\simeq$\,3 quasars are located at the high end of this $\vp-\Lo$ relation. 
\citet{zappacosta2020}, instead, observed no statistically significant correlation between $\vp$ and $\Lo$, which can be due to the combination of low sample statistics and the fact that the WISSH quasars (likewise ours) probe too narrow a range of $\Lo$ values. 
Since the \citetalias{timlin2020} data are covering a much wider range of $\Lo$, 
we believe that such a correlation is real, and both our $z$\,$\simeq$\,3 and the WISSH sources are sampling its brighter end. Given the scatter of the relation, and the presence of a few outliers (with $\vp$ higher than a few thousands \kms) also in the \citetalias{timlin2020} sample, we argue that the WISSH data alone are not sufficient to determine if a relation between $\Lh$ and $\vp$ exists, since it can be simply driven by a few peculiar objects.\footnote{The three most deviating objects with high blueshifts ($\vp$\,$>$\,5,000 \kms) in the $\vp-\Lo$ plane, which are the main drivers of the $\Lh-\vp$ correlation seen in the WISSH sample, are J0958+2827 ($z$\,$=$\,3.434, $\gammax$\,$=$\,1.87$^{+0.79}_{-0.64}$), J1421+4633 ($z$\,$=$\,3.454, $\gammax$\,$=$\,1.30$^{+0.39}_{-0.36}$) and J1521+5202 ($z$\,$=$\,2.218, $\gammax$\,$=$\,1.71$^{+0.41}_{-0.39}$). The $\daox$ parameter reported by \citet{zappacosta2020} for these objects would place them at the boundary between $W$ and $w$ (they all have $\daox$\,$\simeq$\,$-0.25$).}

It is worth noting that different investigations use slightly different definitions of $\vp$. 
\citetalias{timlin2020} defines the \ion{C}{iv} blueshift (in units of \kms) as $c(1549.06-\lambda_{\rm peak})/1549.06$, where $\lambda_{\rm peak}$ is the measured peak of the modelled \ion{C}{iv} emission line (in \AA), 1549.06 \AA\ is the \ion{C}{iv} laboratory wavelength and $c$ is the speed of light.
Also the \ion{C}{iv} velocity shifts in \citet{zappacosta2020} are measured from the modelling of the emission line, but they refer to the line centroid offset with respect to the systemic redshift (see figure~2 in \citealt{vietri2018} and relative discussion).
Our velocity shifts are computed with respect to the \ion{C}{iv} vacuum wavelength as in \citetalias{timlin2020}. Nonetheless, we note that differences of even several hundreds of \kms\ amongst the different analyses would not change our conclusions.

Summarising, we confirm a relation between $\vp$ and $\Lo$ across three decades in $\Lo$ from $z$\,$\simeq$\,1.7 up to $z$\,$\simeq$\,3.5, whilst we do not observe a statistically significant correlation between $\vp$ and $\Lh$, in contrast to \citet{zappacosta2020}. We argue that the $\vp-\Lh$ correlation reported in the latter study 
is driven by 
the small sample statistics, as it is not confirmed when including the lower luminosity \citetalias{timlin2020} sample.

\subsection{On the relation between X-ray and \ion{C}{iv} fluxes}
\label{fxfciv_extended}
\begin{figure}
\centering
 \resizebox{\hsize}{!}{\includegraphics{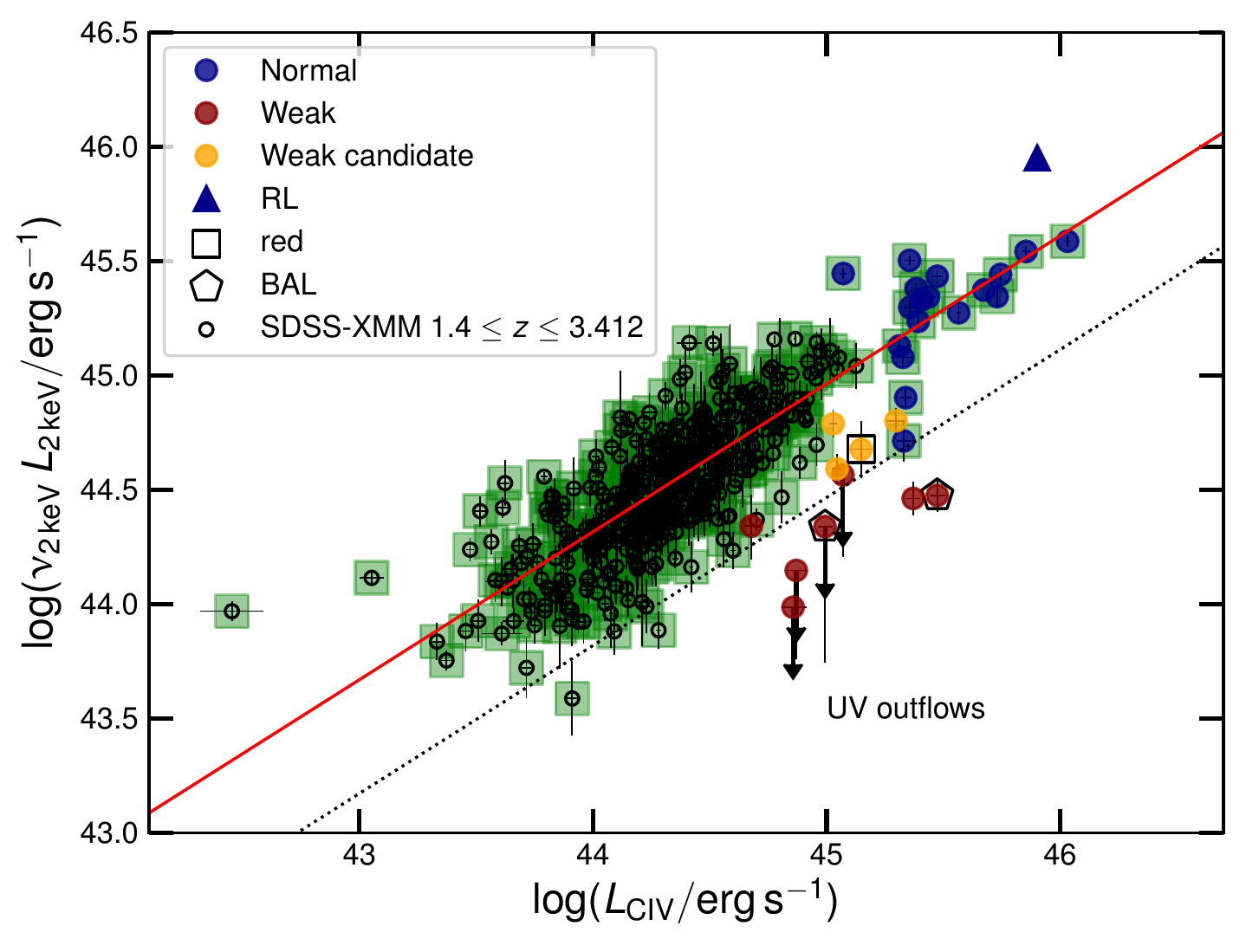}}
\caption{X-ray luminosity at $2$ keV as a function of the total integrated \ion{C}{iv} line luminosity. Green squares highlight the sources used in the MCMC regression line analysis. Black circles describe a lower redshift/luminosity SDSS-XMM sample with a robust measurement of the integrated \ion{C}{iv} line luminosity (Signorini et al., in preparation). The dotted line represents 3$\times$ the intrinsic dispersion on the relation, $\sim$\,0.5 dex.}
\label{fxfcivextended}
\end{figure}

We have also checked whether the relation we observe between $\Lh$ and $L_{\rm C\,IV}$ (Figure~\ref{fxfciv}) is still present when including lower redshift/luminosity data. We cannot consider at this stage the \citetalias{timlin2020} and WISSH samples as their integrated \ion{C}{iv} fluxes are not publicly available. We thus examined a sample of SDSS quasars with the \ion{C}{iv} emission line covered by their spectra and an X-ray observation in the latest 4XMM catalogue. 
The sample is composed by 1,761 quasars in the redshift range 1.703\,$\leq$\,$z$\,$\leq$\,2.697, selected to fulfil all the quality criteria discussed by \citetalias{lusso2020}. All the spectra were fitted with \qsfit, and here we consider the quasars with a robust measurement of the \ion{C}{iv} spectral properties, leading to 444 sources. 
All the details will be provided in a forthcoming publication (Signorini et al., in preparation).

We performed a MCMC regression analysis of the low redshift/luminosity sample, where we included the $z$\,$\simeq$\,3 $N$ quasars (462 sources in total). We considered the relation between the luminosity at 2 keV and $L_{\rm C\,IV}$, but the results do not change when considering $\Lh$ instead. The choice of using $\Lx$ instead of $\Lh$ allows us to directly interpret this relation in the context of the $\Lx-\Lo$ correlation. 
We find 
\begin{equation}
\label{lxlcivextended}
\log \Lx = (0.647\pm0.001)\log L_{\rm C\,IV} + (15.850\pm0.001),
\end{equation}
with an estimated intrinsic dispersion of 0.17\,$\pm$\,0.01 dex (observed dispersion 0.19 dex).
Figure~\ref{fxfcivextended} presents the X-ray luminosity at $2$ keV as a function of the total integrated \ion{C}{iv} line luminosity. We overplotted the $w$, $W$ and the radio-bright quasars, not considered in the fit. $W$ quasars deviate from the observed relation, being located at faint X-fluxes at a given $L_{\rm C\,IV}$. The slope is similar (slightly steeper\footnote{The slope of the $\Lx-\Lo$ relation on the \ion{C}{iv} sample is 0.510\,$\pm$\,0.018.}) to the one characterising the $\Lx-\Lo$ relation, whilst the dispersion of $\simeq$\,0.17 dex is akin, if not tighter, with respect to that observed in the $\Lx-\Lo$ relation. 

\section{Discussion}
\label{discussion}
\begin{figure}
\centering
 \resizebox{\hsize}{!}{\includegraphics{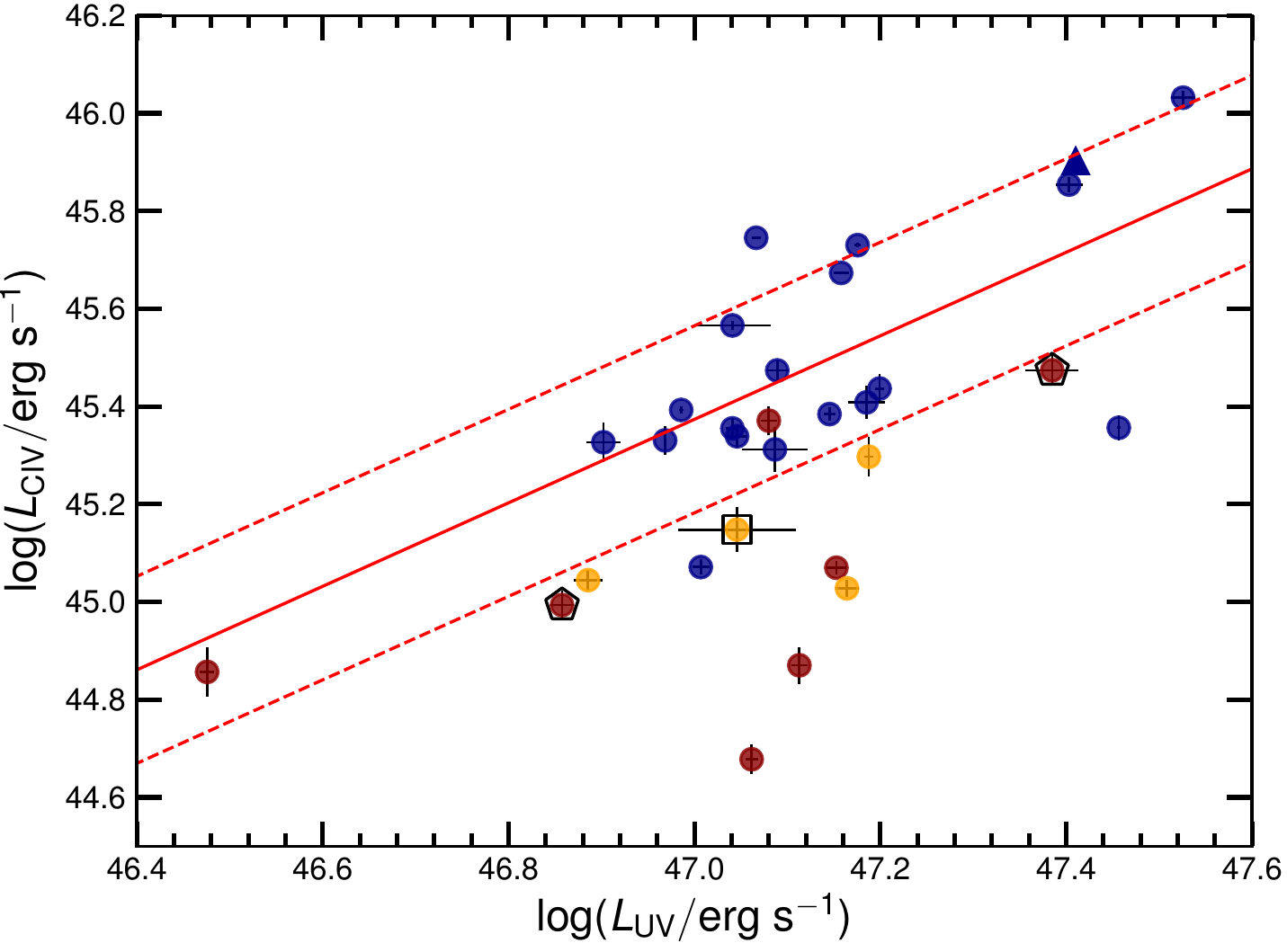}}
\caption{\rev{Integrated \ion{C}{iv} emission line luminosities as a function of the UV luminosities. The regression line (red solid line) is estimated by considering the $N$ quasars only. We observed a slope of 0.855\,$\pm$\,0.095 with a standard deviation of 0.19 dex (red dashed lines).}}
\label{lolciv}
\end{figure}

We report on a tight relation between X-ray and \ion{C}{iv} emission-line fluxes in a homogeneously selected sample of blue quasars at redshift $z$\,$\simeq$\,3. 
The predicted dependence between $L_{\rm C\,IV}$ and $\Lo$ can be derived directly from the observed Baldwin effect of the \ion{C}{iv} emission line, i.e., $\log\,($\ion{C}{iv}\,{\rm EW}$)$\,$\propto$\,$-$0.2\,$L_{1450}$, where $L_{1450}$ is the continuum luminosity at the rest-frame 1450 \AA\ (see table 1 in \citealt{dietrich2002}), which implies a slope of roughly 0.8 between $L_{\rm C\,IV}$ and $\Lo$\footnote{The rest frame EW can be written as the ratio between the integrated luminosity of the \ion{C}{iv} line and the continuum luminosity at 1549 \AA, i.e., \ion{C}{iv} EW\,$\simeq$\,$L_{\rm C\,IV}/L_{1549}$. We then assume that $\Lo \propto L_{1549} \propto L_{1450}$.}.
We observe a relation between $L_{\rm C\,IV}$ and $\Lo$ for the $N$ quasars with a slope of 0.855\,$\pm$\,0.095 \rev{(see Figure~\ref{lolciv})}, in agreement with the expectations.
The $\Lx-L_{\rm C\,IV}$ correlation suggests a strong connection between the relative strength of the X-rays with respect to the UV continuum and the \ion{C}{iv} line emission. Specifically, bright high-redshift quasars that appear 
weaker in the X-rays with respect to the expected emission (if they were to follow the $\Lx-\Lo$ relation), also show a weaker \ion{C}{iv} line flux compared to objects at similar redshifts. 
Whilst it is well-known that the \ion{C}{iv} emission line is subject to an anti-correlation between its equivalent width and the UV luminosity (i.e., the Baldwin effect), which is more generally associated with the Eigenvector 1 analysis (e.g. \citealt{bg1992}), it is less clear how to connect the \ion{C}{iv} properties to the X-ray emission in the X-ray weak quasars. 
The $\aox-\Lo$ relationship, thus in turn the $\Lx-\Lo$ one, as observed in X-ray normal, bright blue quasars, is most likely the main driver for the \ion{C}{iv} (and overall) Baldwin effect, since the steepening of $\aox$ with respect to the values observed in quasars of lower UV luminosity leads to a deficit of ionising photons in the far-UV/soft X-rays (e.g. \citealt{wu2009}).
The $\Lx - L_{\rm C\,IV}$ correlation could just be a by-product of the $\aox-\Lo$ relationship, yet this does not entirely explain the excess of \ion{C}{iv} line flux in the X-ray weak quasar subset compared to the X-ray normal quasars of similar X-ray luminosity (see Figure~\ref{fxfcivextended}). 
\citet[see also e.g. \citealt{richards2011}]{2011AJ....142..130K} found that radio-quiet quasars with both strong \ion{C}{iv} line emission and small \ion{C}{iv} blueshifts 
have a flatter $\aox$ and present significant X-ray emission as compared to the UV, whilst quasars with both weak \ion{C}{iv} emission and large \ion{C}{iv} blueshifts have a 
have steeper $\aox$ and weaker X-rays as compared to the UV. 
Moreover, the former (latter) preferentially exhibit harder (softer) X-ray spectra, with $\gammax<2$ ($\gammax>2$). This is interpreted in the Eigenvector 1 context, which was
developed after the study by \citet[][see also \citealt{wem1987}]{bg1992}, 
who discussed the correlations between the optical, radio, and X-ray continuum (i.e., radio-loudness and $\aox$) and the rest-frame optical emission-line properties 
(e.g. \ion{He}{ii}, H$\beta$, [\ion{O}{iii}], \ion{Fe}{ii} complex)
through a principal components analysis. A similar approach, with the inclusion of X-ray spectral information and UV lines such as \ion{C}{iv}, was then widely utilised to identify the main factors that explain the variance in the AGN spectra (e.g. \citealt{wang1996,sulentic2000,sulentic2007,richards2011,marzianisulentic2014,rivera2020}). 
In particular, steep soft X-ray ($E$\,$<$\,2 keV) spectra and large \ion{C}{iv} blueshifts have been connected to the original Eigenvector 1 ingredients (i.e., H$\beta$ FWHM and \ion{Fe}{ii} strength), and are considered as diagnostics of high accretion rates \citep{wang1996,sulentic2007,shenho2014}.
These trends, however, do not explain the flatter $\gammax$ observed in the \textit{W+w} quasars, which do not seem to show much different blueshifts with respect to the `normal' sources (at a similar UV emission). 
The mild contrast between the emission-line properties of the \textit{W+w} and $N$ subsets could be due to intrinsic differences in their SED in the region where the photons that affect the \ion{C}{iv} line are emitted, i.e., above the ionisation potential of $\simeq$\,48 eV (corresponding to a wavelength of 258 \AA\footnote{The continuum photons principally responsible for \ion{C}{iv} line emission likely cover a wider energy range, also including the Lyman continuum (13.6--24.5 eV) and the soft (300--400 eV) X-rays \citep{KK1988}.}). 

A deficit of continuum photons that ionise elements like \ion{C}{iv} could be interpreted in the context of the \textit{shielding gas} scenario (e.g. \citealt{leighly2004b,wu2011,luo2015,ni2018}). 
As the Eddington ratio increases ($L_{\rm bol}/L_{\rm Edd}$\,$\geq$\,0.3), 
the inner disc is expected to become geometrically thick (i.e., a slim disc; \citealt{Abramowicz1988,Mineshige2000,Ohsuga_2011,nt2013,wang2014,cn2016}).  
This 
puffed up disc could partly prevent ionising photons to reach the broad line region. Moreover, also X-ray weakness would take place when the line of sight of the observer intercepts the thick disc, leading to an apparently X-ray weak quasar where broad, high-ionisation 
emission lines are also faint. Conversely, when the line of sight to the innermost regions is not obstructed by the disc, the quasar would still present weak high-ionisation 
lines, but the X-ray emission is consistent with the one from a standard X-ray corona.

As mentioned, substantial \ion{C}{iv} blueshifts (with $\vp$\,$\ga$\,1,000 \kms) and large broadening (FWHM\,$>$\,5,000 \kms) are often observed in weak-line AGN, implying that \ion{C}{iv} mostly (if not only) originates in a wind. 
%
Also radiation-driven winds \citep[e.g.][]{murray1995} could affect the strength of the emission lines through self-shielding, and depend upon X-ray illumination, as the latter can over-ionise the surrounding gas preventing the establishment of UV-line driving \citep[][]{proga1998,proga2000,nomura2020}. Simulations of such radiation-driven winds from bright accretion discs are consistent with the observed anti-correlation between the relative strength of the soft X-ray emission and the \ion{C}{iv} absorption equivalent width \citep[see e.g.][]{brandt2000}, or the X-ray weakness of BALs in general \citep{gibson2009}. The wind material should have high column densities ($N_{\rm H}$\,$\sim$\,10$^{23-24}$ cm$^{-2}$) for an efficient shield, especially the slow part of the wind closer to the disc, which can prevent any other radiation (both UV and X-ray) to reach the broad line region. 
The gas flow geometry strongly depends upon the geometry of the radiation field, where a bright disc/corona produces dense, warm and fast equatorial outflows in which the \ion{C}{iv} emission line originates \citep{proga2000}. 
Also in this case, orientation would play a role in explaining the UV/X-ray properties of X-ray weak quasars when the observer's line of sight passes through the dense outflow, but a standard quasar will be perceived at more polar viewing angles. 

In both disc and wind shielding, we should then expect to observe hard and weak X-ray spectra in a fraction of sources, suggesting high levels of X-ray absorption, with $N_{\rm H}$\,$\geq$\,10$^{23}$ cm$^{-2}$ or higher.
Since the X-ray spectra of our quasars exclude such a strong gas absorption  and their bolometric luminosities are in excess of $10^{47}$ erg s$^{-1}$ (see Figure~\ref{distrall}), in most of the $z$\,$\simeq$\,3 sample the disc should be seen almost face-on (see below), as larger disc inclinations would imply that these luminosities, already very high, are underestimated.
\rev{Gas shielding could still be a viable interpretation of the X-ray weakness for the two BALs in the \textit{W+w} subset (J1148+23, J0945+23), whilst for another couple of objects (J1201+01, J1459+0024) the data quality is not sufficient to rule out some degree of X-ray absorption.
In \citetalias{nardini2019}, we investigated through both a detailed spectral analysis and a hardness ratio analysis whether the \textit{W+w} quasars could suffer from high levels (i.e., $N_{\rm H}$\,$>$\,10$^{23}$ cm$^{-2}$) of X-ray absorption. Since soft emission is detected in the majority of the \textit{W+w} quasars, significant X-ray absorption appears unlikely.}
Indeed, the \textit{W+w} quasars show lower emission by a median (mean) factor of $\sim$\,6 (8) in the soft X-ray band (0.5--2 keV, observed frame) compared to $N$ quasars, which would require a (nearly) neutral shielding column of the order of $\sim$\,5\,$\times$\,$10^{23}$ cm$^{-2}$. \rev{We point out that column densities of the order of $10^{23}$ cm$^{-2}$ would produce a clear cut-off around 1 keV, which is not seen even though it would be easily detectable also in the spectra of lower quality.} 
\rev{Absorption due to} shielding would completely obliterate the observed flux density at rest-frame 2 keV, whilst no evidence of such a spectral turnover is found. 
\rev{Even with a slightly steeper intrinsic photon index, moving from an unabsorbed $\gammax$\,$=$\,1.8 to a moderately absorbed $\gammax$\,=\,2.0, the difference in terms of the 2-keV flux density would be of 40\% only. As the average intrinsic photon index is likely intermediate ($\langle\gammax\rangle$\,$\simeq$\,1.91 for the $N$ quasars), such a correction should be taken as an upper limit, and none of our results are substantially affected.}
Therefore, the current data do not favour the shielding scenario to explain the X-ray weakness observed in the \textit{W+w} sources. Nonetheless, X-ray absorption could be variable, so multiple X-ray observations of the \textit{W+w} $z$\,$\simeq$\,3 quasars can provide deeper insights into the origin of X-ray weakness. \rev{Summarising, in the absence of additional data, the low observed gas absorption (i.e., $N_{\rm H}$\,$<$\,10$^{23}$ cm$^{-2}$) suggests that the observed X-ray weakness is \textit{intrinsic} to the source.}

In this framework, most of the possible explanations provided by \citet[][see the discussion in their Section 5.1]{Leighly2007a} for the prototypical intrinsically X-ray weak quasar PHL\,1811 remain valid. In particular, in high-Eddington sources a significant fraction of photons can be trapped in the accretion flow towards the SMBH and accreted before they can actually escape, so starving both the corona and the broad line region from photons \citep{begelman1978}. 
Here, we want to elaborate in some more depth on the possible connection between X-ray weakness and disc winds.

In a \textit{slab} or \textit{patchy} geometry for the accretion-disc corona (see the top and bottom panels of figure 6 in \citealt{rn2003}), the radiation pressure due to the UV photons from the disc can push a dense flow into the corona, thus intertwining the processes that give rise to the UV and X-ray emission. This coupling may lead to a quenching of the corona, as the region above the disc will become too dense, opaque, hence too cold to produce sufficient X-rays, as noticed by \citet{proga2005}. 
Intrinsic quenching (rather than obscuration) of the corona by a  \textit{failed} wind that stalls because of over-ionisation could control the wind duty cycle, and act as a self-regulating mechanism of the accretion/ejection process. Moreover, it can also explain a variety of observations, such as 
the spectral properties of several emission and absorption features 
\citep{Leighly2007a,Leighly2007b,wu2011}. 
This scenario can also account for the flat (hard) $\gammax$ observed in the \textit{W+w} quasars, as an increase of the optical depth naturally leads to a flatter photon index in the Comptonization process \citep{Zdziarski1996,Beloborodov1999}, and for the fact that we do not observe strong \ion{C}{iv} blueshifts.

We argue that X-ray weakness might also be interpreted in a starved X-ray corona picture, associated with an ongoing disc-wind phase. Although we do not observe significant \ion{C}{iv} blueshifts, a wind could still be present given the low disc inclination (i.e., mostly face-on) of our sources and the prevalent equatorial nature of line-driven disc winds \citep[][but see also \citealt{rivera2020} for the relation between the \ion{C}{iv} properties and orientation]{proga2000}. These winds can significantly change the local accretion rate, and should alter the SED shape and the overall disc emission \citep[as well as the SMBH growth rate; e.g.][]{sn2012,ld2014,nomura2020}. 
The consequences on the corona have not been theoretically explored, but are likely to be important. 
High accretion rates, such as the ones characterising the $z$\,$\simeq$\,3 sample, generate large bolometric luminosities and shift the peak of the disc emission towards higher frequencies ($>$\,10$^{15.5}$ Hz, see e.g. \citealt{sn2012}). 
The closer the high accretion rate reaches in with respect to the inner disc radius, 
the stronger is the emitted ionising UV radiation\footnote{Here we refer to the ionising photons that arise in the extreme-UV part of the quasar SED, which we cannot observe.}. Yet, if an outflow is ejected in the vicinity of the SMBH, the extreme-UV radiation that `feeds' the corona will be depleted, starving the corona from the needed seed photons and generating an X-ray weak quasar. 
Nonetheless, at large UV luminosities ($>$\,10$^{47}$ erg s$^{-1}$, Figure~\ref{distrall}), there could still be a reservoir of ionising photons 
that can explain the `excess' \ion{C}{iv} line emission observed in the \textit{W+w} quasars with respect to the sources at similar X-ray luminosities that follow the $\Lx-\Lo$ relation. 
The strength of the \ion{C}{iv} line, in fact, 
does not simply depend on the amount of ionising photons in the high energy tail of the disc spectrum. The excited level (lying only 8 eV above the ground state) is primarily populated via electron--ion collisions, and X-ray photons are key contributors to the gas heating. In X-ray weak sources, the integrated \ion{C}{iv} flux presumably suffers from the inefficient line excitation due to the scarcity of hot electrons, rather than the limited production of \ion{C}{iv} due to the lack of ionising photons \citep[see also][]{timlin2021}. In normal quasars of comparable X-ray luminosity, but lower $\Lo$, the flux of ionising photons is smaller, hence the \ion{C}{iv} emission line is fainter.

For the above reasons, the \ion{C}{iv} line is not a good indicator of the total budget of ionising photons and of the extreme-UV SED shape. A more suitable line in this sense is the \ion{He}{ii}\,$\lambda$1640 line. The \ion{He}{ii} EW values of the $z$\,$\simeq$\,3 quasars are in the range $\sim$\,0.04--0.90 \AA, consistent with the expectations from the Baldwin effect of the \ion{He}{ii} line at the UV luminosities of our sample (see the recent analysis by \citealt{timlin2021} and \citealt{rankine2020}). We were not able to achieve a good fit to the \ion{He}{ii} line for 7 quasars (4 $N$, 2 $W$ and 1 $w$) as the line is significantly blended with the \ion{C}{iv} red wing and the \ion{Fe}{ii} UV complex. Given the intrinsic faintness of the \ion{He}{ii} line and the moderate spectral resolution and S/N of the SDSS spectra, the uncertainties in the measure of the \ion{He}{ii} EW do not allow us to reveal any statistically significant difference between the $N$ and \textit{W+w} sub-samples. 
A more detailed investigation of \ion{He}{ii} emission is beyond the scope of the present paper, and requires spectra with higher S/N than the SDSS ones.


Other emission lines could be affected by the presence of a wind and by a shortage of extreme-UV photons, in the first place the [\ion{O}{iii}]\,$\lambda$5007 line, which has an ionisation potential of $\simeq$\,35 eV (corresponding to $\simeq$\,354 \AA) and is produced at larger distances from the SMBH (where the gas density is lower). To explore the [\ion{O}{iii}] properties, 
we are currently analysing the near-IR spectra (probing both emission lines and continuum at 1 $\mu$m) 
of our $z$\,$\simeq$\,3 quasars obtained with LUCI, the near-IR spectrograph and imager at the LBT. Whilst we have continuum observations for all the sources, medium resolution spectroscopy ($R_\lambda$\,$\simeq$\,1100 in the $K_{\rm s}$ band for 1 arcsec slit) was obtained only for the 9 (4 $N$, 4 $W$ and 1 $w$) quasars with redshift between $z$\,$\simeq$\,3.2--3.3\footnote{For the other quasars in the sample, the redshift is such that the H$\beta$--[\ion{O}{iii}] spectral range is not entirely observable, as it falls in an atmospheric window with poor or no transmission.}. 
Besides probing the [\ion{O}{iii}] line emission, the LBT/LUCI spectra will also allow us to measure with better accuracy the rest-frame 2500 \AA\ fluxes without relying on the continuum extrapolation of the SDSS spectra, and to estimate more reliable black-hole masses from the H$\beta$ and/or \ion{Mg}{ii} lines. Better estimates of black-hole mass will also allow us to check for possible differences in the accretion disc temperature \citep[e.g.][]{ld2011}.
The details on the data reduction and spectral analysis will be the subject of a forthcoming publication. It is worth anticipating, however, that by using the [\ion{O}{iii}] EW as an orientation indicator (e.g. \citealt{2017MNRAS.464..385B}), the data suggest that the accretion disc is \rev{preferentially} seen at low inclination (i.e., likely to be face-on) in a \rev{large fraction of} the $z$\,$\simeq$\,3 quasars \rev{(all but one have [\ion{O}{iii}] EW $<$\,25~\AA), which is not surprising given their huge UV luminosity}. \rev{Moreover, all the \textit{W+w} quasars observed with LUCI (5 sources) show weak [\ion{O}{iii}] ($<$\,10 \AA), which is also expected in objects with high Eddington ratios (e.g. \citealt{boroson2002}).}
All this, together with the lack of a clear evidence of X-ray absorption in the \xmm spectra, argues against the coronal shielding interpretation, irrespective of its origin (either failed wind or inflated inner disc).
 


Summarising, several scenarios may explain the origin of intrinsic X-ray weakness in connection with the \ion{C}{iv} emission-line properties in the $z$\,$\simeq$\,3 sample. Following \citet[][see also \citealt{Casebeer2006}]{KK1988}, intrinsic X-ray weakness can affect \ion{C}{iv} emission in two ways, that is through a deficit of ionising photons produced in the high energy tail of the disc and/or of X-ray photons that contribute to gas heating, since collisional excitation of the line is critical. The latter could affect \ion{C}{iv} emission in X-ray weak quasars, but since we still observe significant \ion{C}{iv} emission in the \textit{W+w} subset (see Figure~\ref{sfit}), there must be a sufficient amount of ionising photons with energies $>$\,48 eV that can produce \ion{C}{iv} emission in excess with respect to what is expected from X-rays only.


\section{Conclusions}
\label{conclusions}
We presented the UV analysis of 30 quasars at 3.0\,$<$\,$z$\,$<$\,3.3 observed as part of an \xmm Large Programme in 2017--2018. This sample was selected in the optical/UV from the SDSS-DR7 to be representative of the most luminous, intrinsically blue quasars at high redshift to further test the potential of quasars as cosmological standard candles. Despite the UV homogeneity of the whole sample, the study of the $\Lx-\Lo$ relation in \citetalias{nardini2019} revealed two distinct X-ray populations. About two thirds of our quasars (X-ray normal, $N$) cluster around the relation, with a minimal dispersion of $\sim$\,0.1 dex. The remaining one third (X-ray weak and weak candidates, \textit{W+w}) appear to be X-ray underluminous by factors of $>$\,3--10. Moreover, the X-ray weakness fraction among our $z$\,$\simeq$\,3 quasars ($\approx$\,25\%) is larger than previously reported for radio-quiet, non-BAL quasars at lower redshift and luminosity.
While the \textit{W+w} quasars are a miscellaneous subset, we speculated in \citetalias{nardini2019} that, in some cases, the X-ray corona might be in a radiatively inefficient state owing to the presence of an accretion-disc wind.

In the second paper of this series we focus on the analysis of the rest-frame UV spectra, in particular on the \ion{C}{iv} emission line properties (proxy of winds/outflows) and on the UV continuum (proxy of the accretion disc), in connection with the X-ray emission (proxy of the corona).
Our main results are summarised as follows:
\begin{enumerate}
    \item The analysis of the spectral composites \rev{(see Figure~\ref{stacks})} for the $N$ and the \textit{W+w} subsets shows that the UV continuum slope is slightly redder ($\alpha_\lambda$\,$\simeq$\,$-0.6$) for the \textit{W+w} stack compared to the $N$ one ($\alpha_\lambda$\,$\simeq$\,$-1.5$). Emission lines in the \textit{W+w} stack are broader and fainter than in the $N$ one. The \ion{C}{iv} emission line presents only a mild blueshift of $\simeq$\,600--800 \kms\ with respect to the reference emission-line wavelength (1549 \AA). The line profile, asymmetric towards the blue side, is broader ($\simeq$\,10,000 \kms) than that observed in the $N$ composite ($\simeq$\,7,000 \kms). Yet, uncertainties on the stacks are such that these features are consistent with quasars at similar redshifts and luminosities for both the $N$ and \textit{W+w} composites, further confirming the homogeneity of the sample by construction.
    
    \item We recover a relation between $\vp$ and $\Lo$ across three decades in $\Lo$ in the redshift range $z$\,$\simeq$\,1.7--3.5 \rev{(Figure~\ref{vcivlolx}, left panel)}, whilst we do not observe a statistically significant correlation between $\vp$ and $\Lh$ \rev{(Figure~\ref{vcivlolx}, right panel)}, differently from previous claims. We argue that the $\vp-\Lh$ correlation recently reported in high-redshift quasars is mostly driven by the relatively small sample statistics.
    
    \item We report on a tight ($\simeq$\,0.17 dex) log-linear relation between $\Lx$ and the integrated \ion{C}{iv} line emission $L_{\rm C\,IV}$ across about four orders of magnitude in luminosity \rev{(Figure~\ref{fxfcivextended})}. The slope value is similar to the one already observed in the $\Lx-\Lo$ relation ($\gamma$\,$\sim$\,0.6), yet slightly steeper, thus preserving the \ion{C}{iv} Baldwin effect. Consequently, the observed $\Lx-L_{\rm C\,IV}$ correlation could be in part a by-product of the $\Lx-\Lo$ relationship. X-ray weak quasars deviate from the main $\Lx-L_{\rm C\,IV}$ relation, occupying the low X-ray/\ion{C}{iv} luminosity part of the plane, thus displaying an `excess' of \ion{C}{iv} line emission with respect to normal sources of similar X-ray luminosities.
    
    \item We interpret the X-ray weak quasars deviating from the $\Lx-L_{\rm C\,IV}$ relation in the context of the disc-wind quasar scenarios. Whilst the shielding model may explain X-ray weakness for only a couple of X-ray weak quasars in our sample, for the other X-ray weak objects we consider absorption unlikely. We thus argue that their X-ray weakness might be interpreted in terms of a starved (or quenched) corona associated with an ongoing disc-wind phase. Although collisional excitation of the line might be less efficient, the observed UV luminosities ($>$\,10$^{47}$ erg s$^{-1}$) characterising the $z$\,$\simeq$\,3 quasars ensure a significantly large reservoir of ionising photons 
    that can explain the `excess' \ion{C}{iv} emission observed in the X-ray weak sources with respect to normal quasars at similar X-ray luminosities. 
\end{enumerate}

Our work on the $z$\,$\simeq$\,3 blue quasar sample demonstrates that a systematic study of X-ray spectra, optical to X-ray SEDs, and optical/UV emission-line properties in the context of accretion-disc models is mandatory to better understand the observable effects of disc winds. In the future, it would be interesting to extend this kind of analysis to quasars that deviate from the $\Lx - L_{\rm C\,IV}$ relation at different redshifts and luminosities.

\begin{acknowledgements}
We acknowledge financial contribution from the agreement ASI-INAF n.2017-14-H.O. EL acknowledges the support of grant ID: 45780 Fondazione Cassa di Risparmio Firenze.
\end{acknowledgements}

%
   \bibliographystyle{aa} 
   \bibliography{bibl} 
%
\begin{appendix} 
\section{Photometric spectral energy distributions of the $z$\,$\simeq$\,3 quasar sample}
\label{Photometric spectral energy distributions}
\begin{figure}[t!]
\centering
\includegraphics[width=0.7\linewidth]{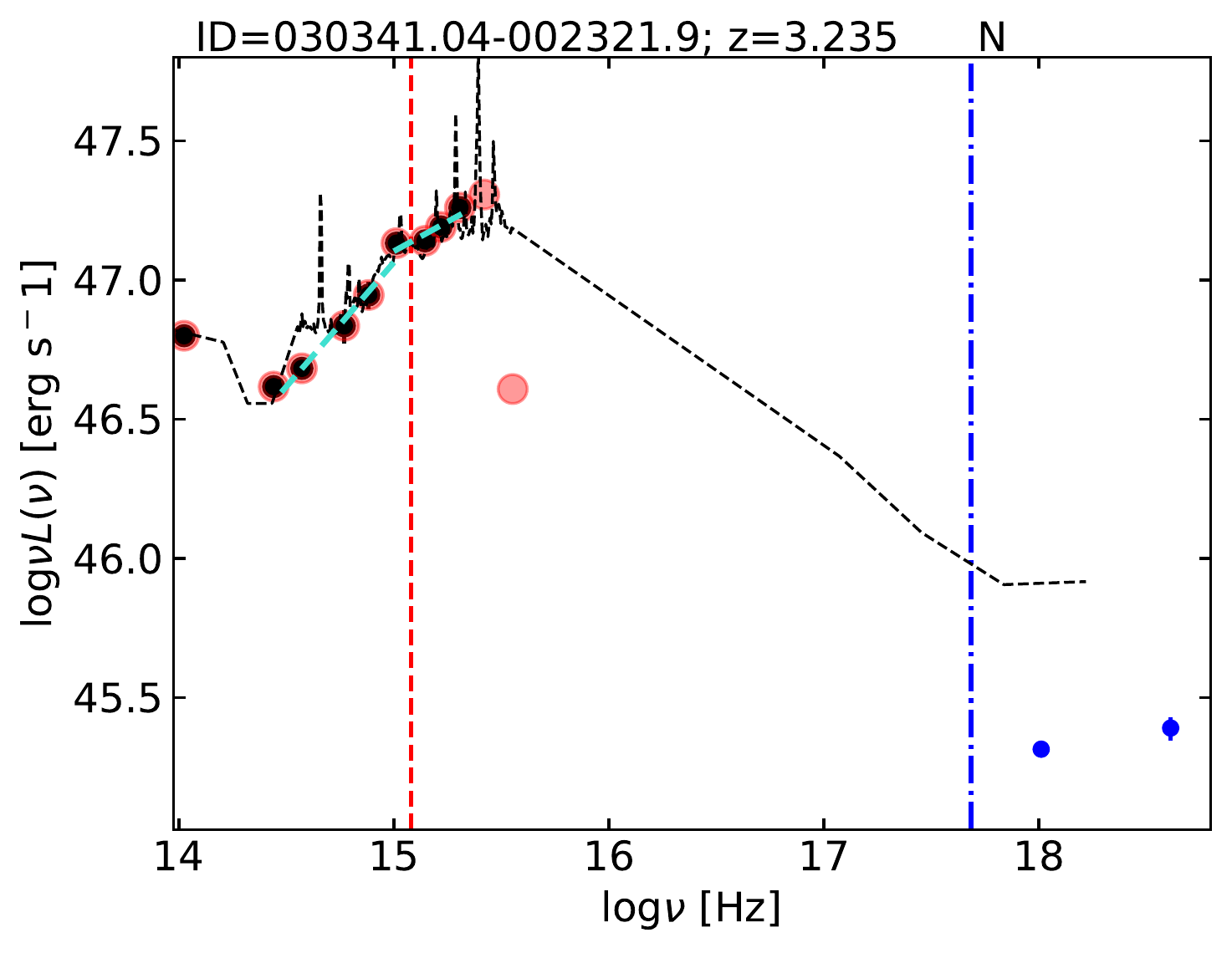}
\includegraphics[width=0.7\linewidth]{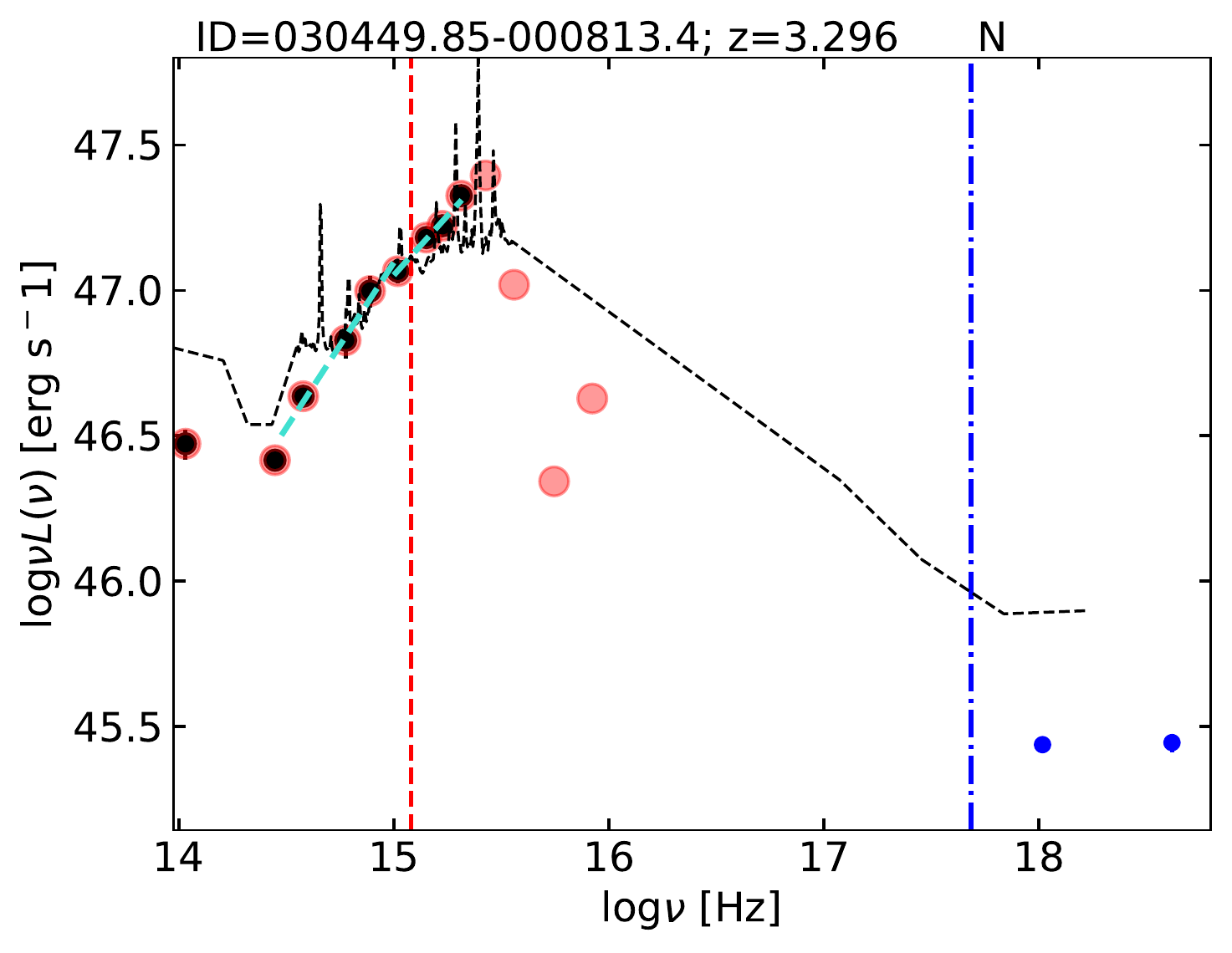}
\includegraphics[width=0.7\linewidth]{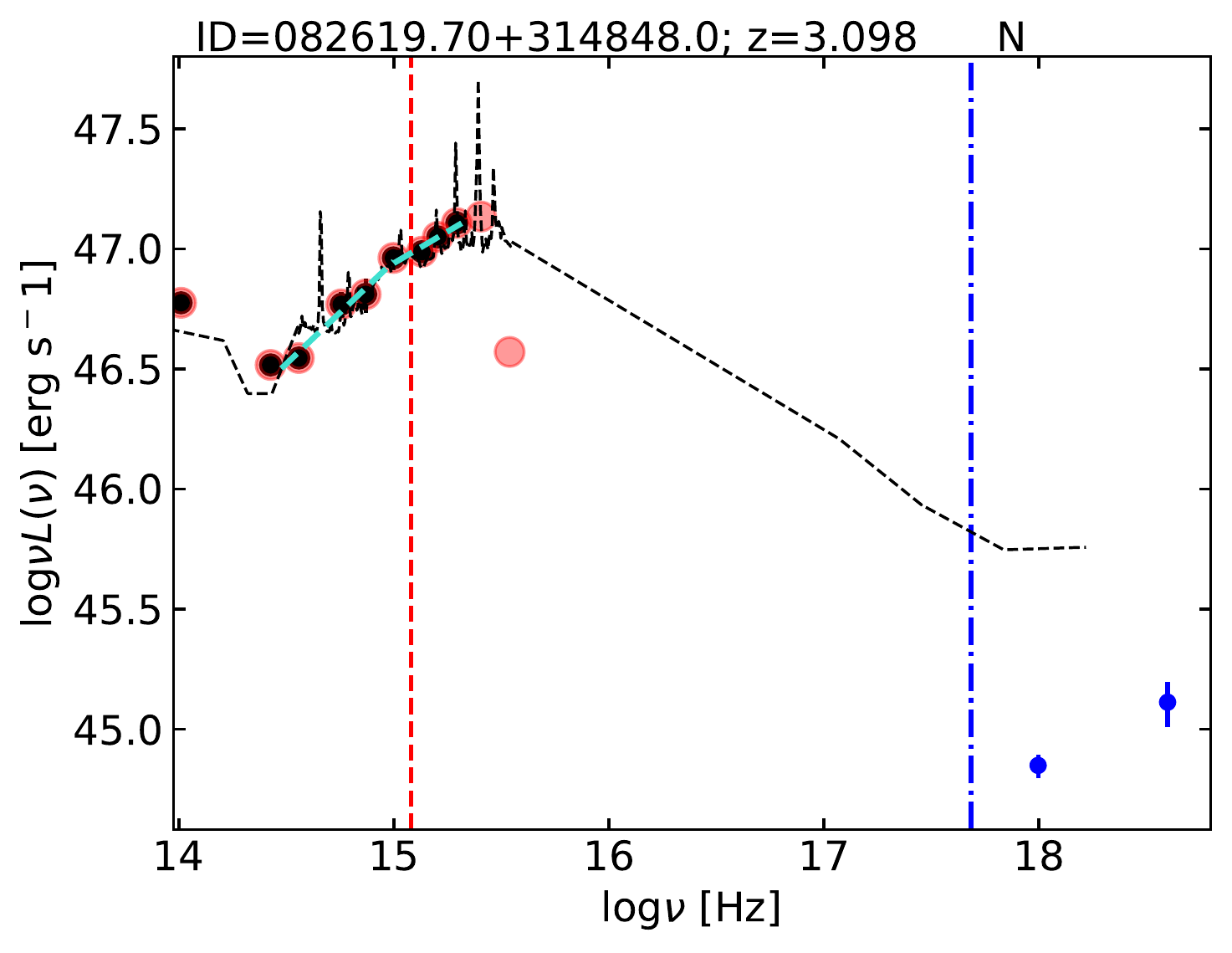}
\caption{SEDs of the $z$\,$\simeq$\,3 quasars in the rest-frame $\nu-\nu L(\nu)$ logarithmic plane. The red circles mark all the available photometry from the SDSS-DR14 catalogue, but only the ones highlighted with black circles were used to construct the SEDs. The cyan dashed lines represent the slopes $\Gamma_1$ and $\Gamma_2$ in the 0.3--1 $\mu$m and 1450--3000~\AA\ range (rest frame), respectively. The composite SDSS radio-quiet quasar spectrum from \citetads{shang2011} is also shown for reference with the black-dashed line, normalised at 2500 \AA\ (indicated by the red dashed line). We included the X-ray data with blue points for the soft and the hard band. The blue dot-dashed line locates the 2-keV point.}
\label{seds}
\end{figure}
Figure~\ref{seds} presents the full SEDs of the $z$\,$\simeq$\,3 quasars from near-IR to the X-rays. Photometry from the near-IR to the UV is drawn from SDSS-DR14 quasar catalogue \citep[][see their Section~2.1]{paris2018}: the FIRST survey in the radio \citepads{becker1995}, the Wide-Field Infrared Survey (WISE, \citeads{write2010}) in the mid-IR, the Two Micron All Sky Survey (2MASS, \citeads{cutri2003,2006AJ....131.1163S}) and the UKIRT Infrared Deep Sky Survey (UKIDSS; \citeads{lawrence2007}) in the near-IR, and the Galaxy Evolution Explorer (GALEX, \citeads{martin2005}) survey in the UV. To construct the SEDs, we followed the same approach as in \citetalias{lusso2020} (see their Section~3). Briefly, after collecting all the photometry available in the SDSS catalogue for each quasar in the sample, we corrected it for Galactic reddening by utilising the selective attenuation of the stellar continuum $k(\lambda)$ from \citetads{F99}. Galactic extinction is estimated from \citetads{schlegel98} for each object in the sample. We expect significant absorption by the intergalactic medium (IGM) in the continuum ($\sim$\,10\% between the Ly$\alpha$ and \ion{C}{iv} emission lines, see \citeads{lusso2015} for details) at wavelengths bluer than about 1400 \AA. Hence, when computing the relevant parameters, we excluded from the SED all the rest-frame data at $\lambda$\,$<$\,1500 \AA.
The red circles in Figure~\ref{seds} mark all the available photometry from the SDSS-DR14 catalogue, whilst the ones used to construct the SEDs are highlighted with black circles. The cyan dashed lines represent the two near-IR/optical slopes $\Gamma_1$ and $\Gamma_2$ in the 0.3--1 $\mu$m and 1450--3000~\AA\ range (rest frame), respectively. 
The blue points at high energies represent the flux at the rest-frame 1 keV and 4 keV derived from the \xmm data obtained in our campaign. We computed these flux values following the same approach as in \citet[][see also \citeads{lusso2013}]{lusso2010}. Briefly, we first computed the integrated unabsorbed (considering only the Galactic column density at the source location) fluxes in the observed 0.5--2 keV and 2--8 keV bands from our \xmm data. These fluxes are then converted into monochromatic X-ray fluxes in the observed frame at 1 keV and 4 keV for the soft and hard band respectively, assuming a power-law spectrum with the observed photon index slope. These X-ray fluxes are converted into luminosities and finally redshifted.
We also show for reference the composite SDSS quasar spectrum from \citetads{shang2011} with the black-dashed line, which is built from a sample of 27 low-redshift ($z$\,$<$\,0.5) optical/UV bright radio-quiet quasars. 
On average, the X-ray part of the \citetads{shang2011} SED shows relatively higher X-ray emission than our data and previous composites as well, indicating that their sample is not representative of the SDSS quasars at high energies and probably does not follow the $\Lx-\Lo$ relationship (see discussion in their Section~5.3.3).
Nonetheless, the SED shape of the $z$\,$\simeq$\,3 quasars at UV frequencies is in excellent agreement with the one obtained in the literature for quasars of lower redshift/luminosity and, thus, lower black-hole mass. This corroborates the notion that the physical mechanism responsible for the intrinsic X-ray emission (tightly connected with the disc emission) of quasars does not evolve with cosmic time and is scale-invariant.

\rev{We have compared the 2500 \AA\ fluxes obtained as described above from the broadband photometry to the ones computed from the spectral fits, as presented in Section~\ref{uvspectralfit} and appendix~\ref{Ultraviolet SDSS spectra}. The top panel of Figure~\ref{pscomp} shows such a comparison, whilst the distribution of the differences between photometric and spectroscopic UV fluxes, $\Delta \fo$\,(photo\,--\,spectro), is shown in the botton panel. The photometric $\Fo$ is in very good agreeement with the spectroscopic one, with a mean (median) $\Delta \fo$\,(photo\,--\,spectro)\,$=$\,$-0.002\,(0.010)$, with a standard deviation of 0.05. Even the photometric UV flux of the most deviating object in the $\Delta \fo$\,(photo\,--\,spectro) distribution, J1148+23, is consistent within a factor of 1.6 with the continuum UV flux from spectral fitting.
}

\begin{figure*}[!]
\addtocounter{figure}{-1}
\includegraphics[width=0.33\linewidth]{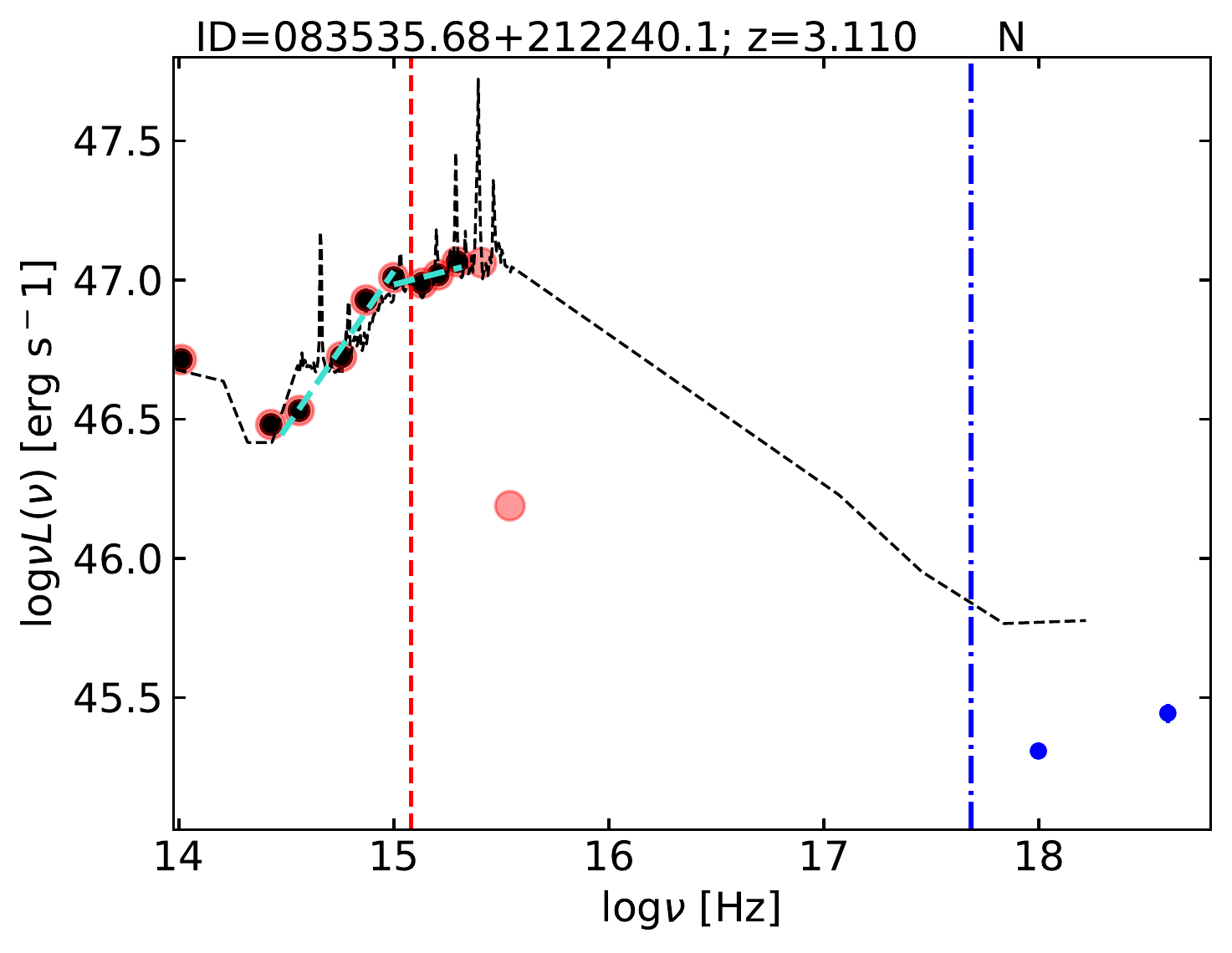}
\includegraphics[width=0.33\linewidth]{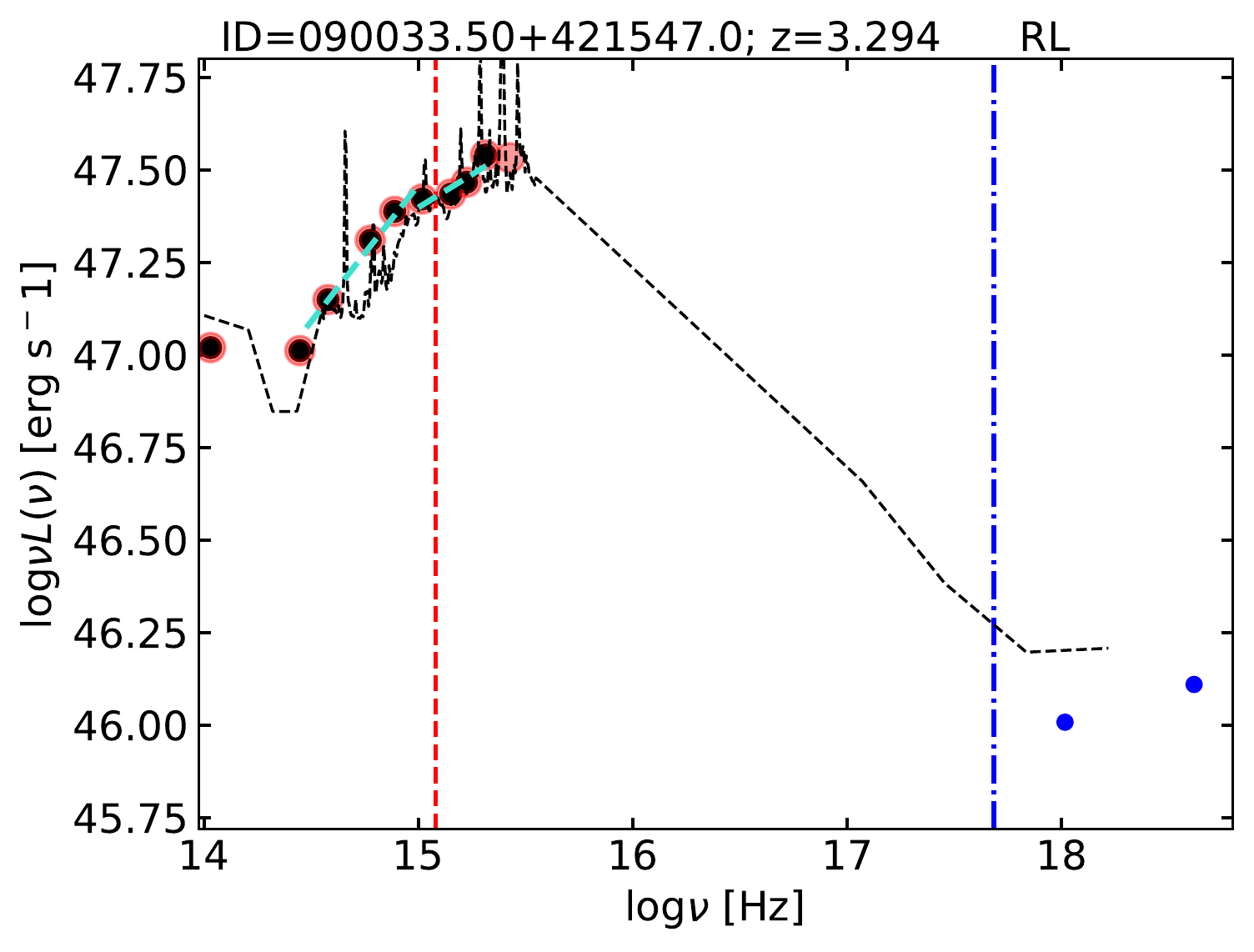}
\includegraphics[width=0.33\linewidth]{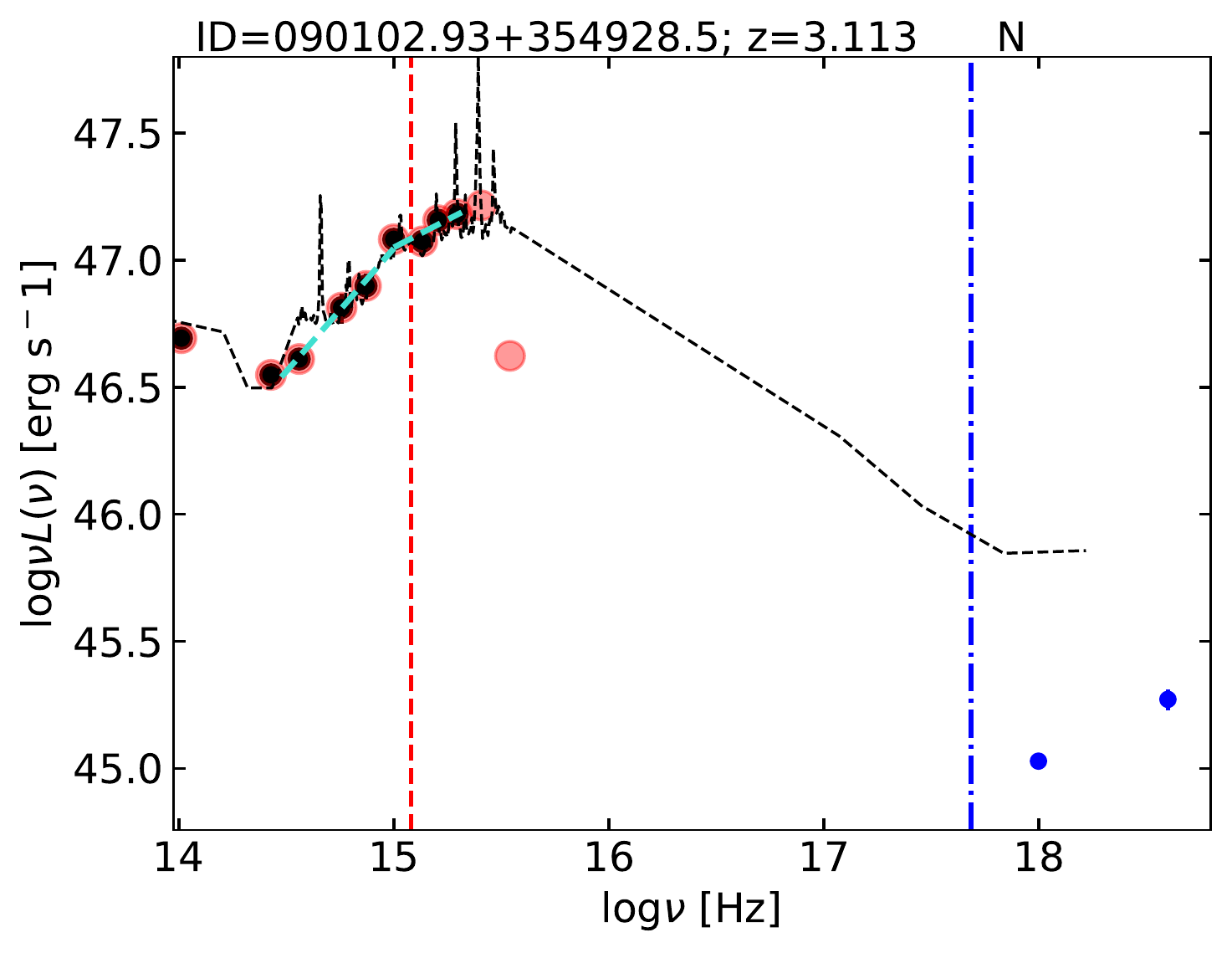}
\includegraphics[width=0.33\linewidth]{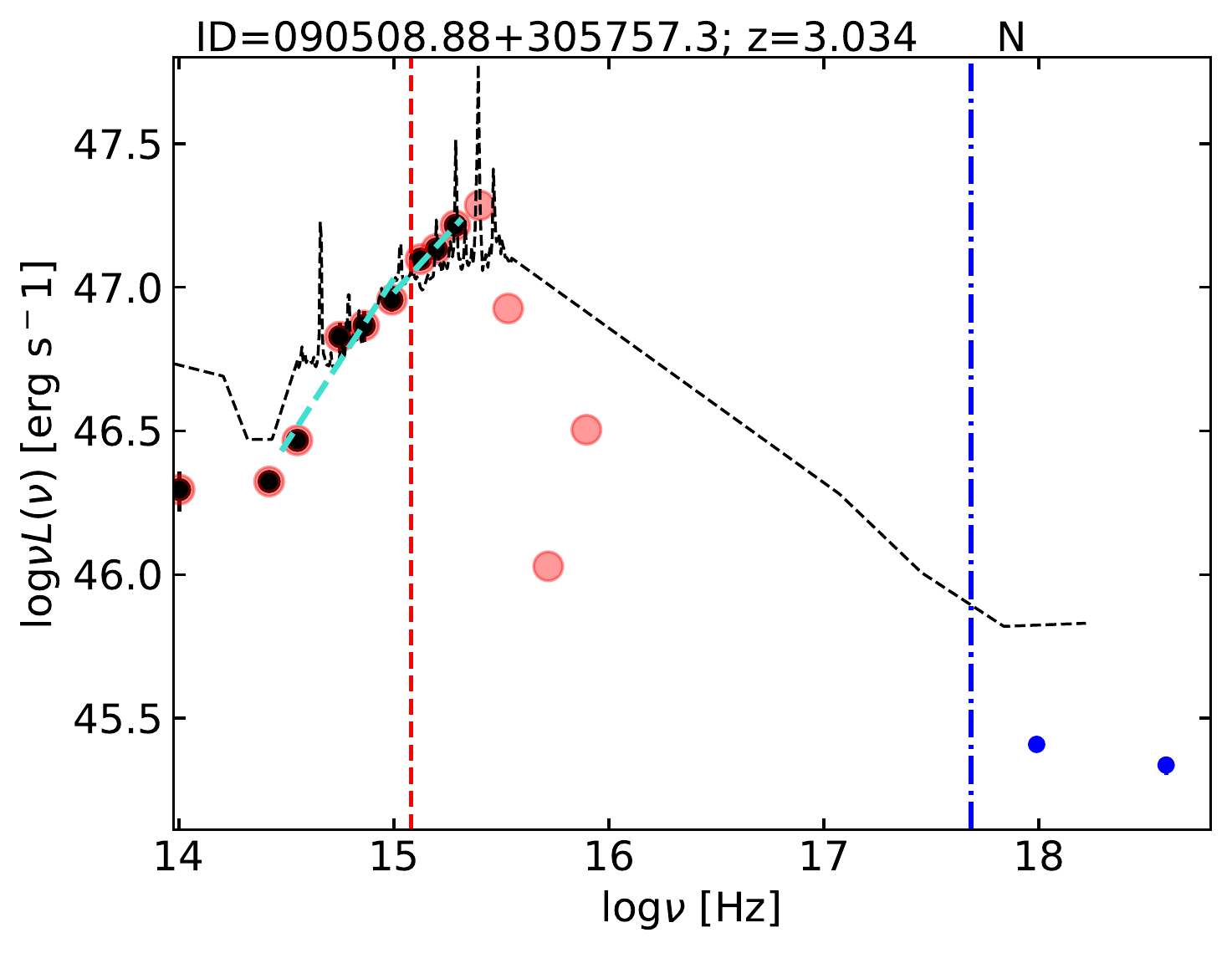}
\includegraphics[width=0.33\linewidth]{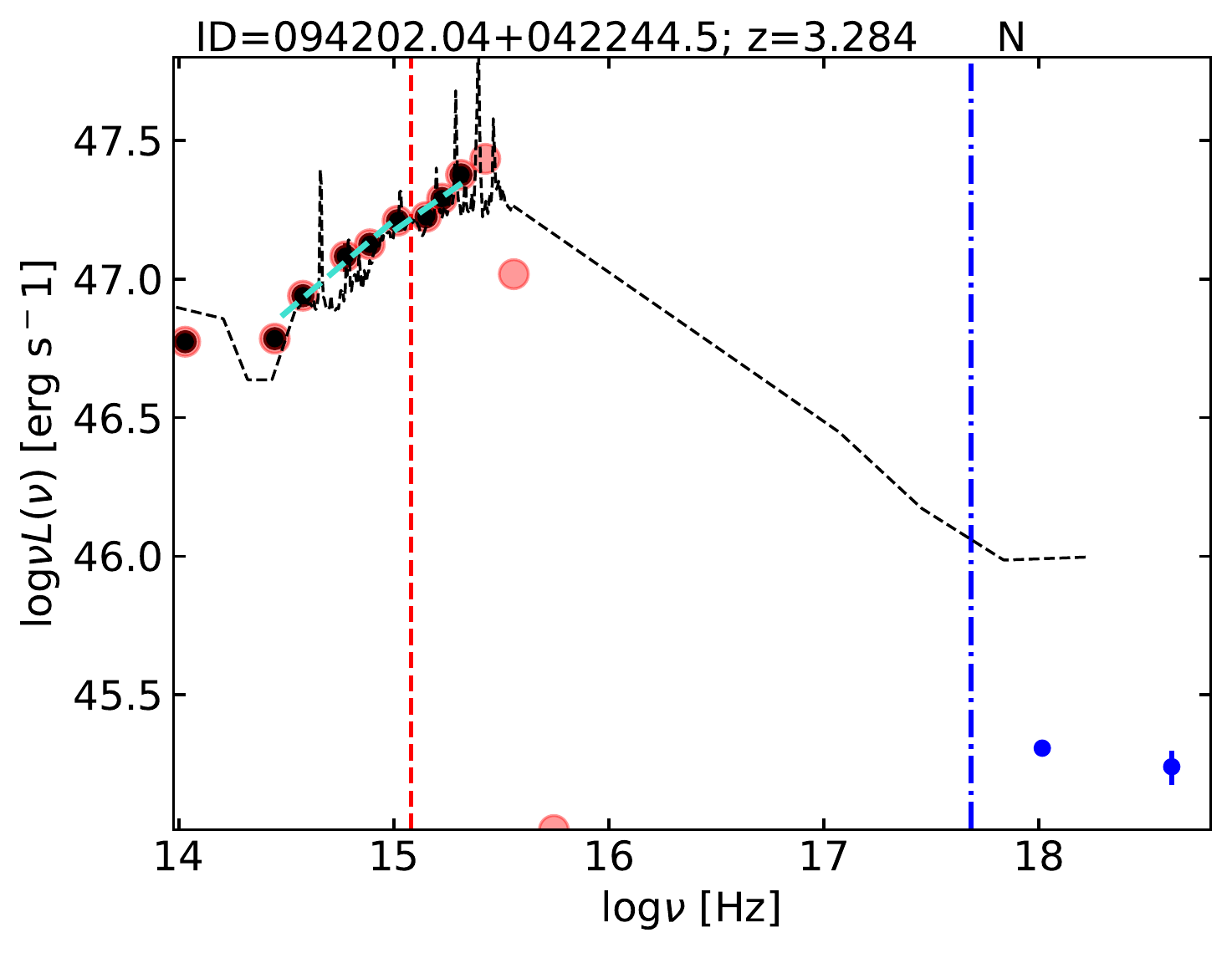}
\includegraphics[width=0.33\linewidth]{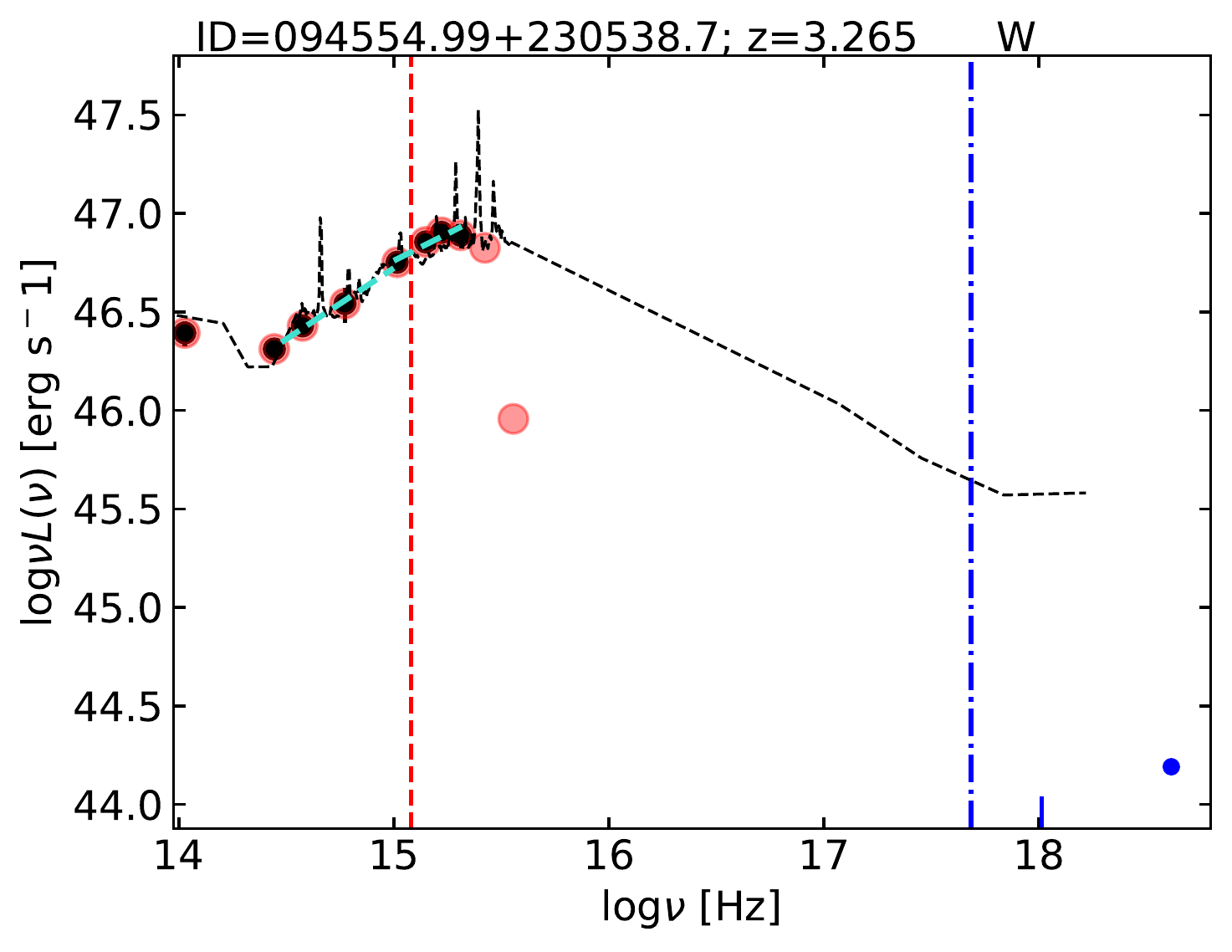}
\includegraphics[width=0.33\linewidth]{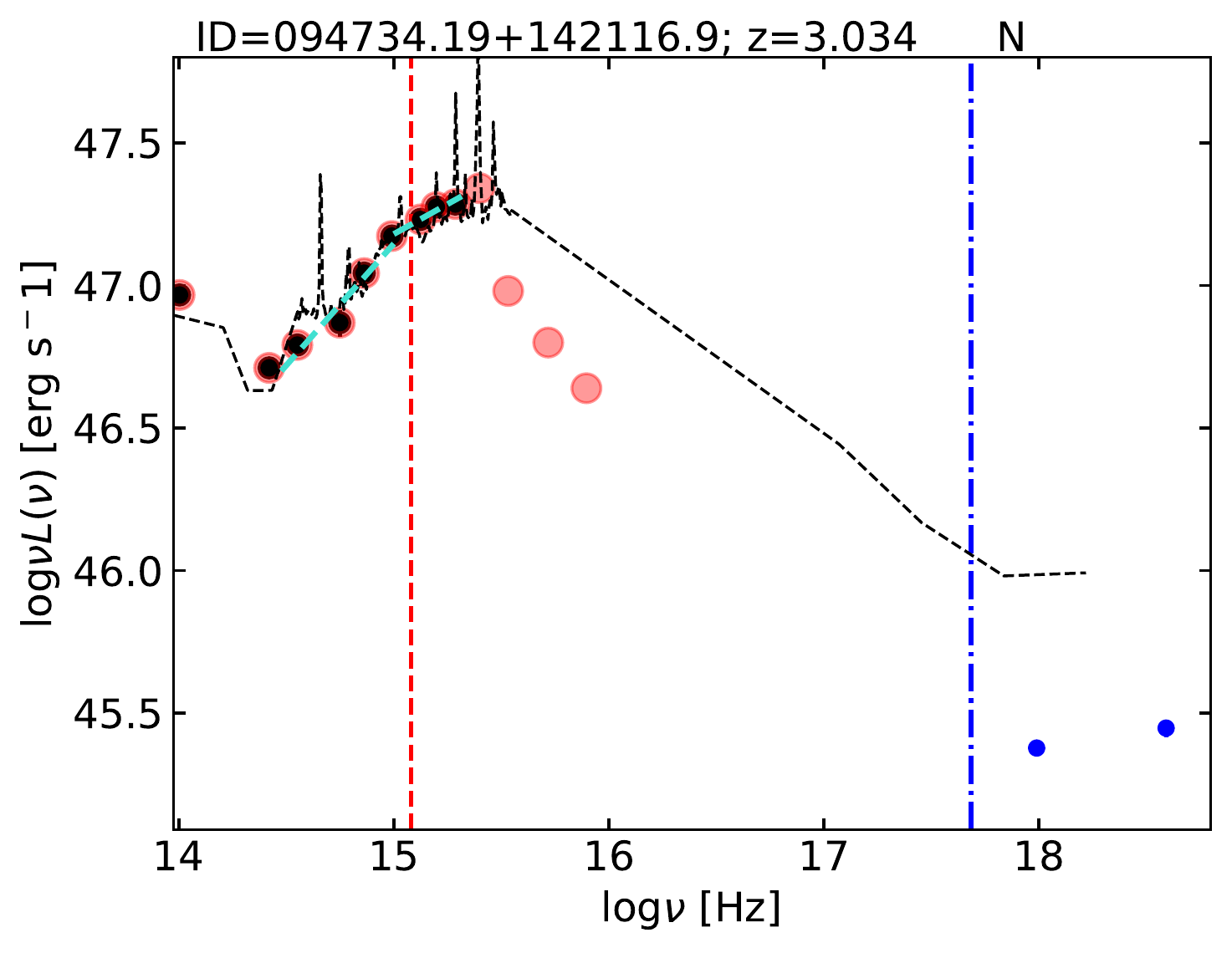}
\includegraphics[width=0.33\linewidth]{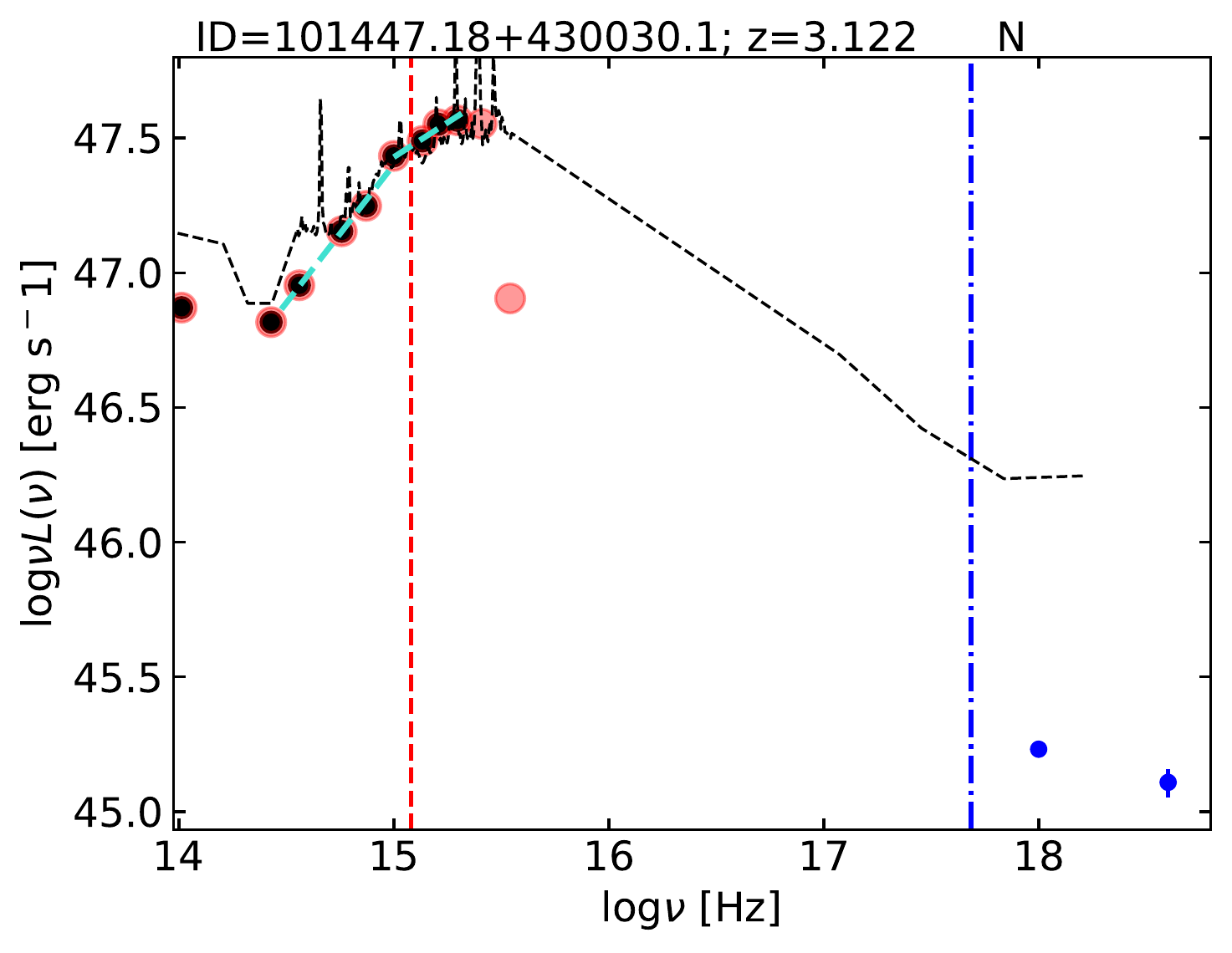}
\includegraphics[width=0.33\linewidth]{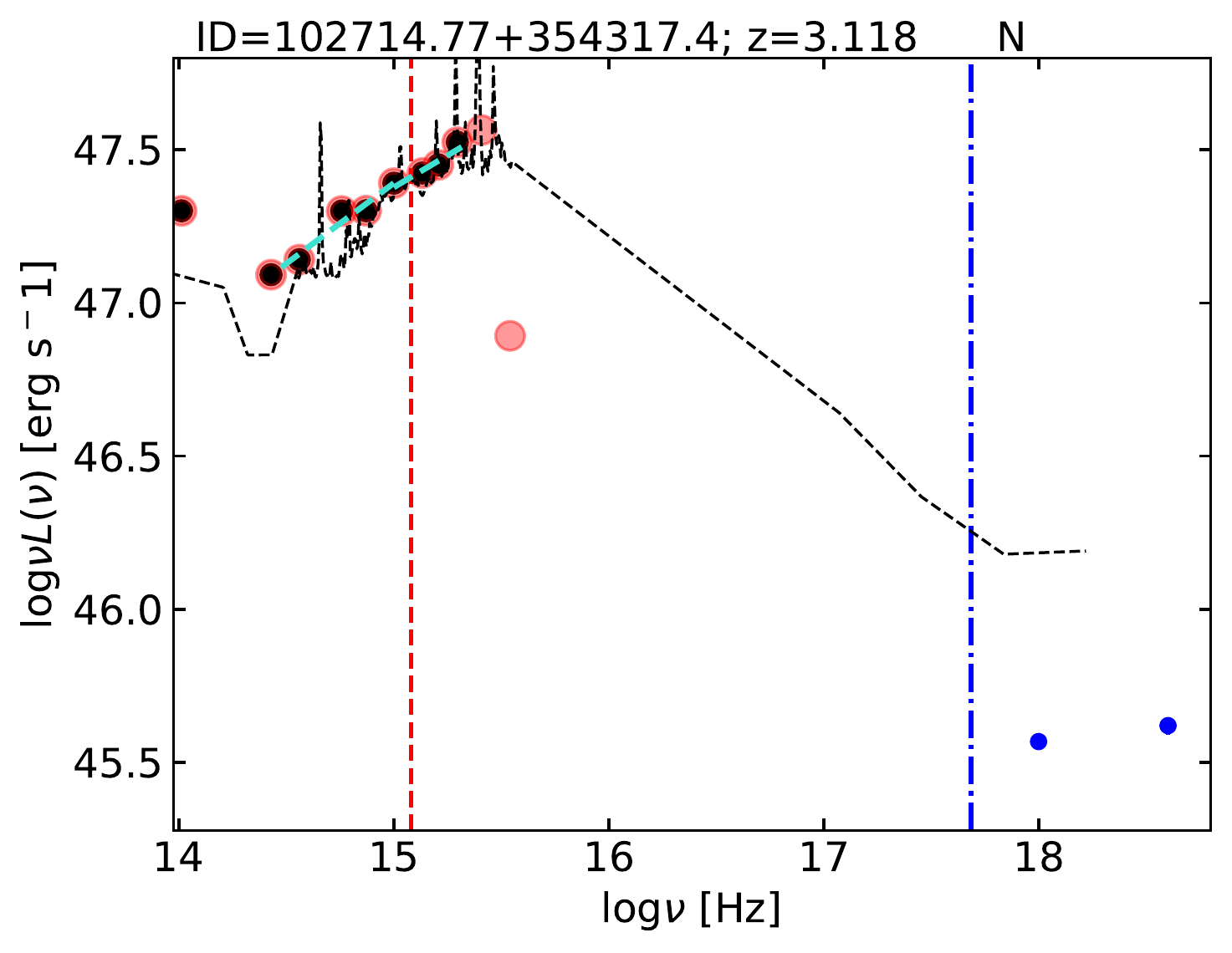}
\includegraphics[width=0.33\linewidth]{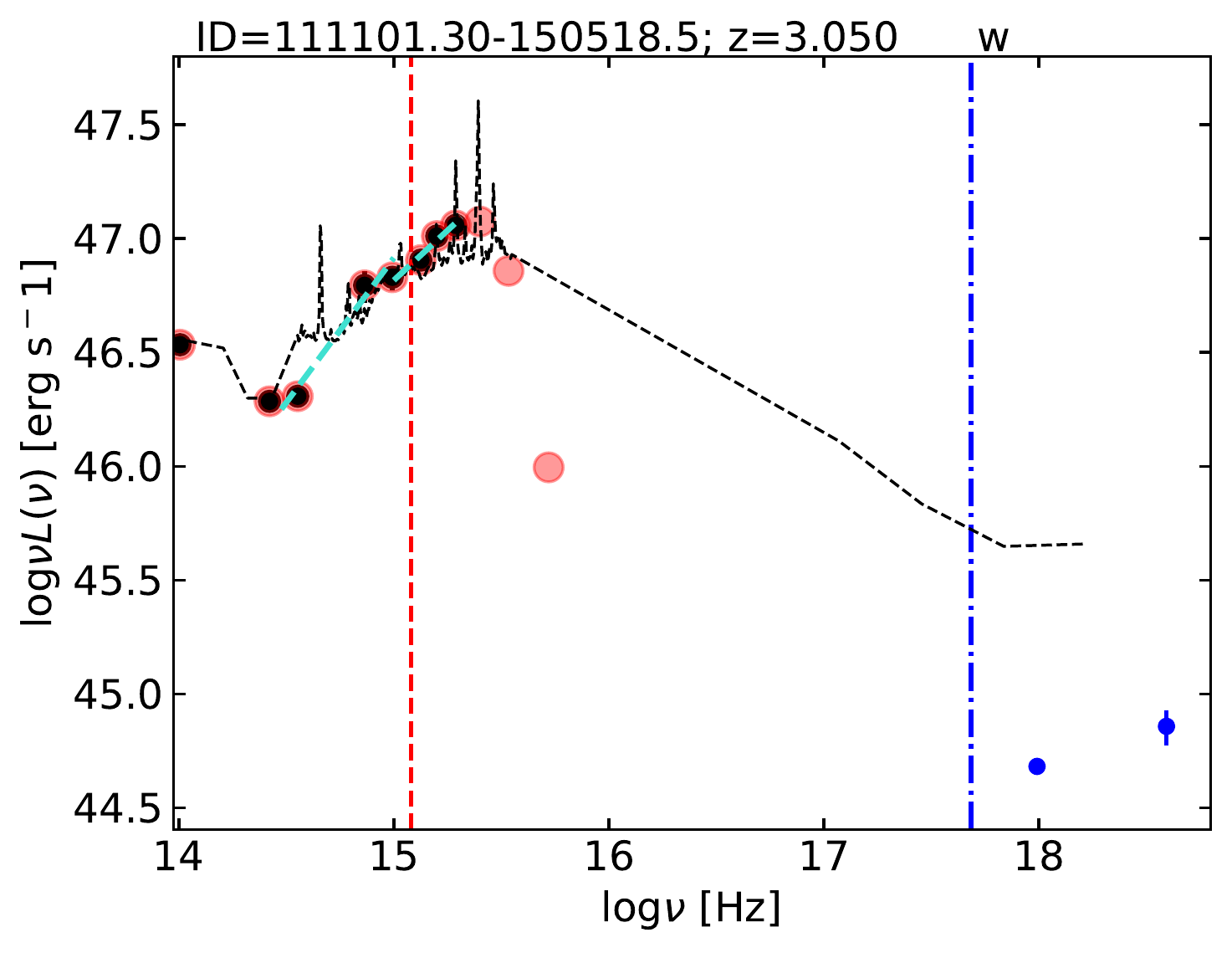}
\includegraphics[width=0.33\linewidth]{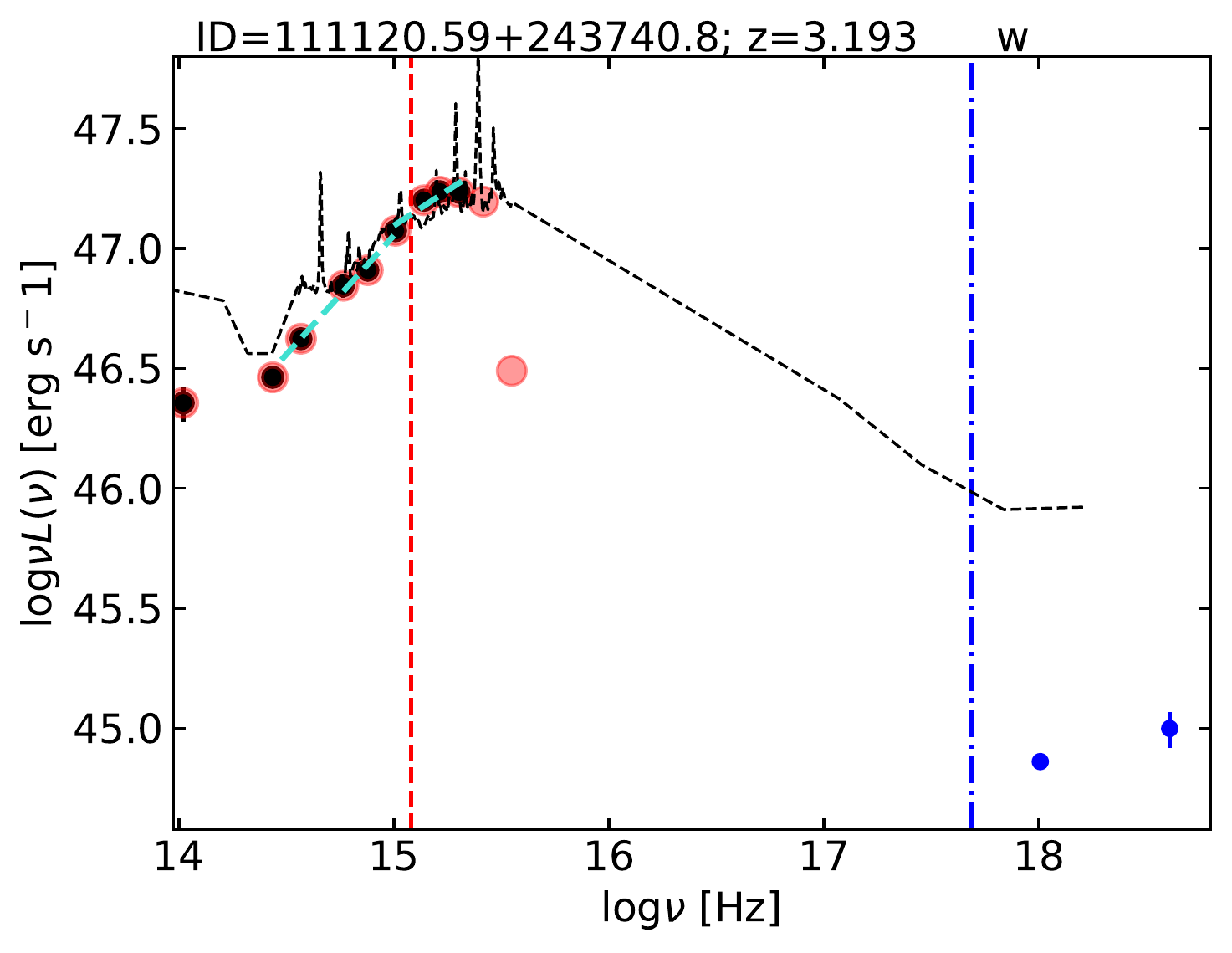}
\includegraphics[width=0.33\linewidth]{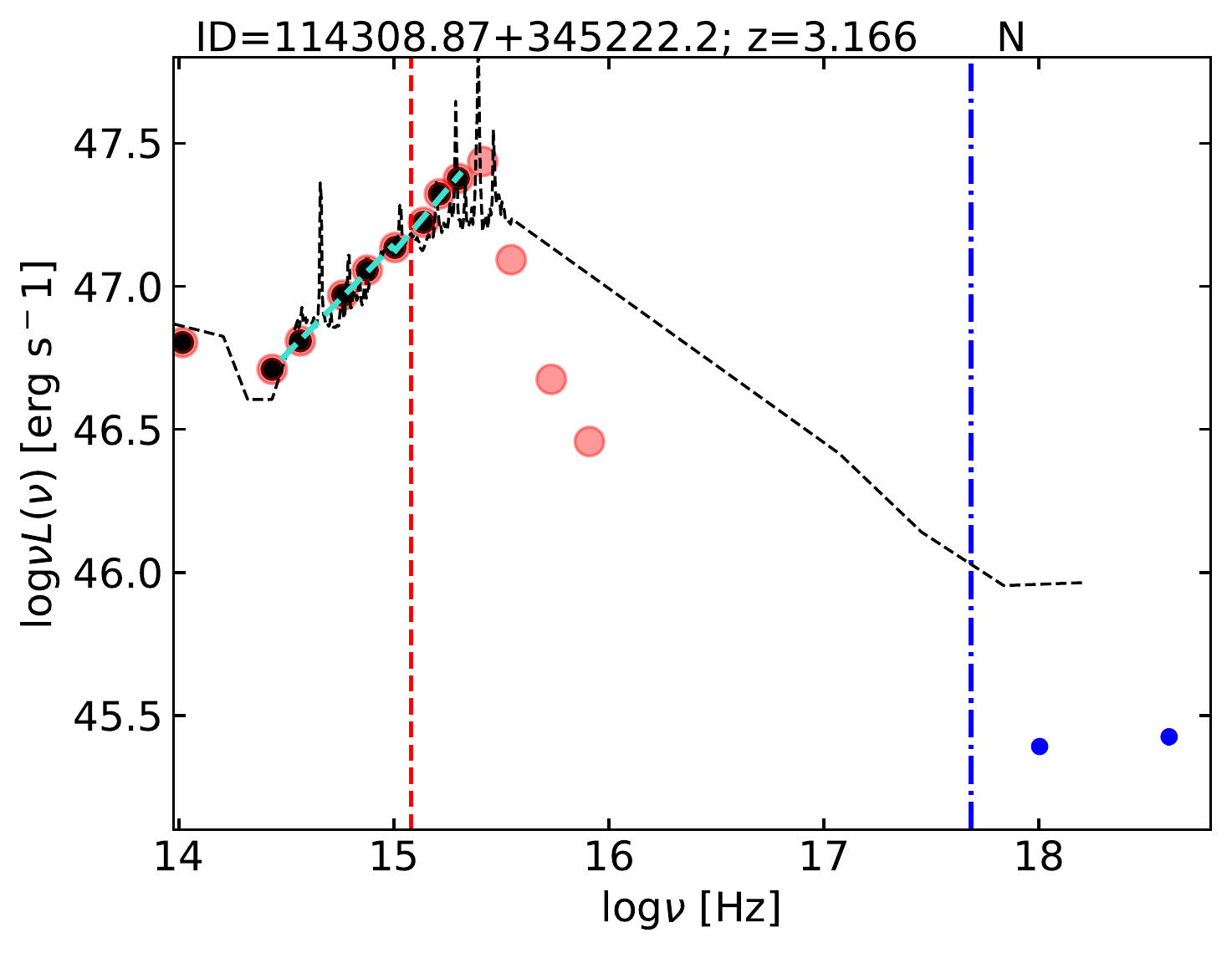}
\includegraphics[width=0.33\linewidth]{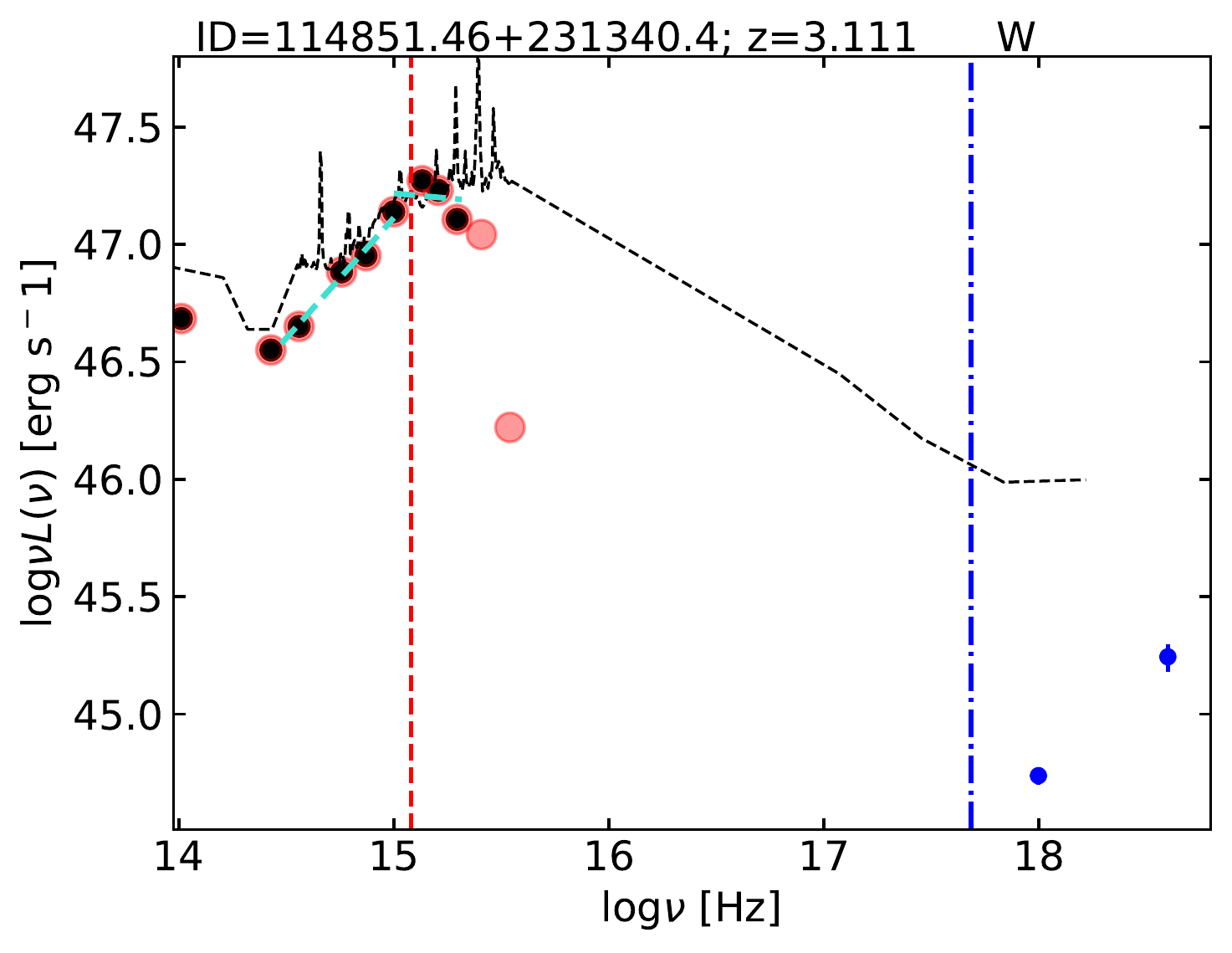}
\includegraphics[width=0.33\linewidth]{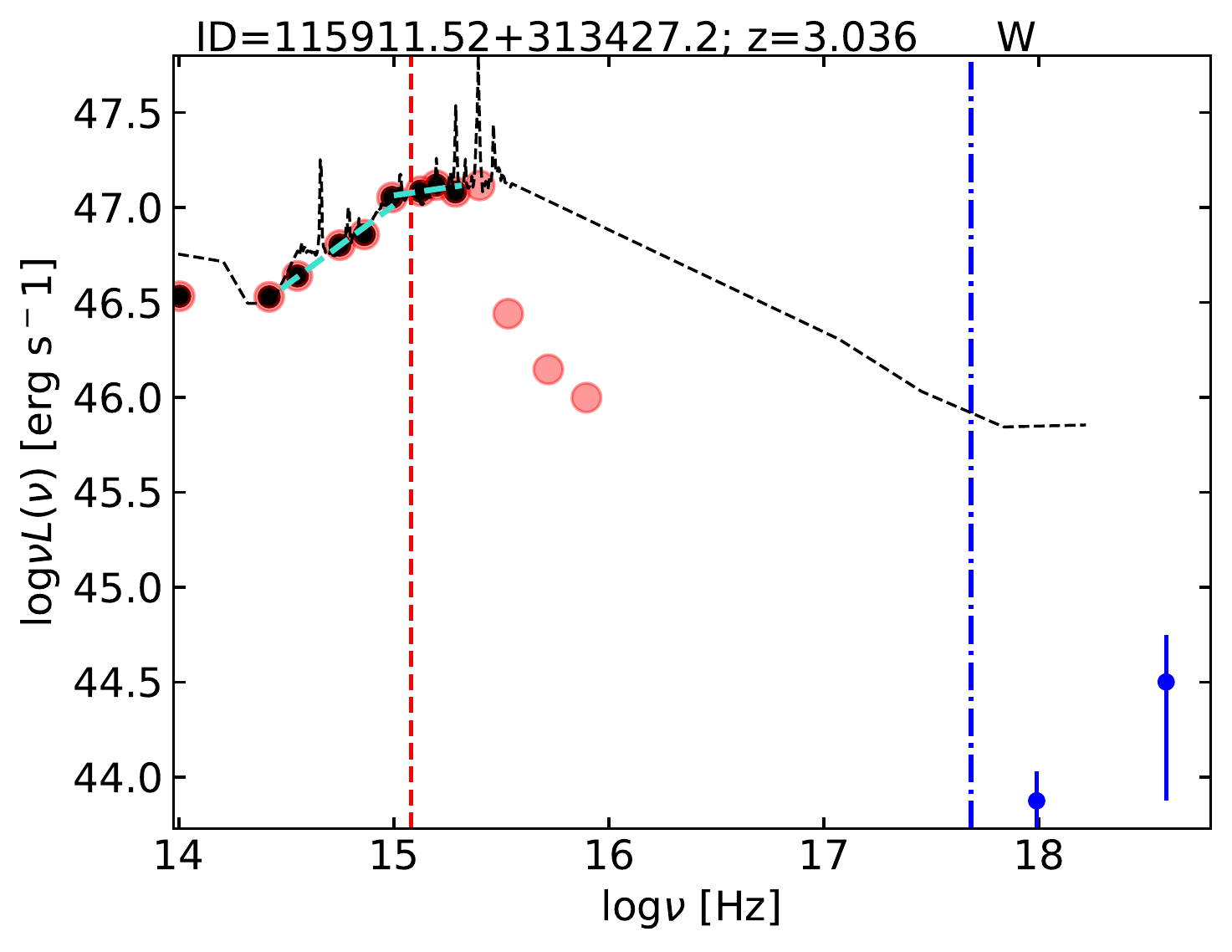}
\includegraphics[width=0.33\linewidth]{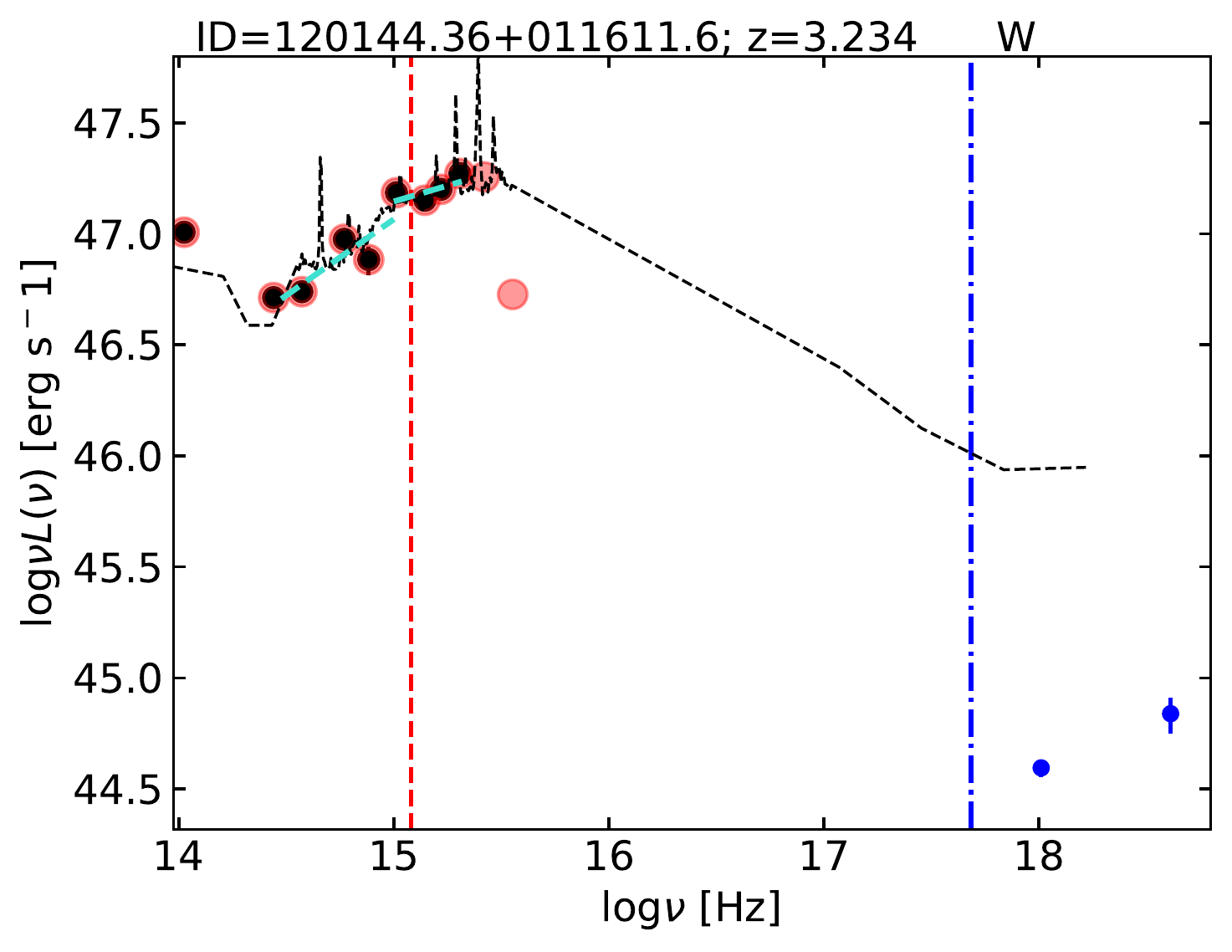}
\caption{Continued.}
\label{seds}
\end{figure*}

\begin{figure*}
\addtocounter{figure}{-1}
\includegraphics[width=0.33\linewidth]{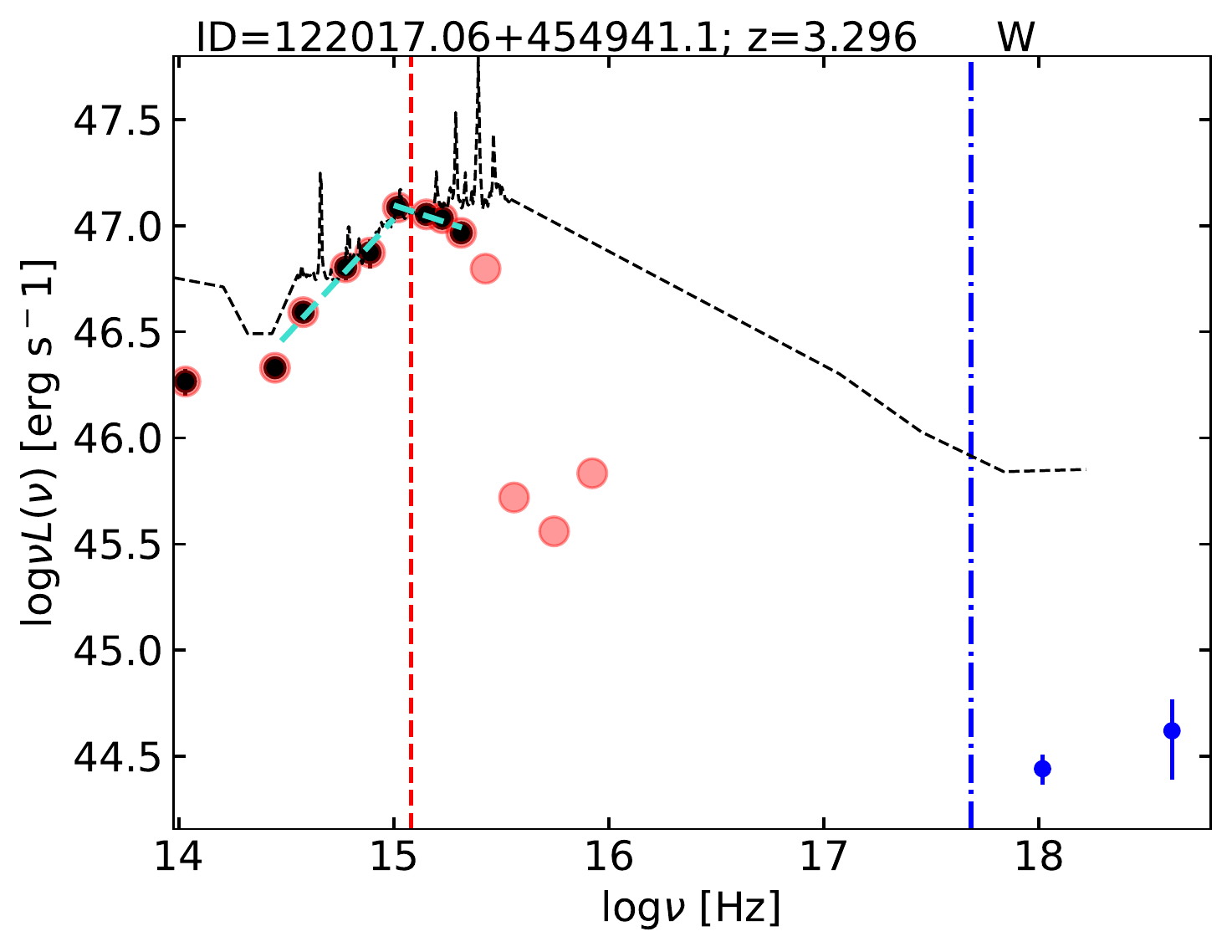}
\includegraphics[width=0.33\linewidth]{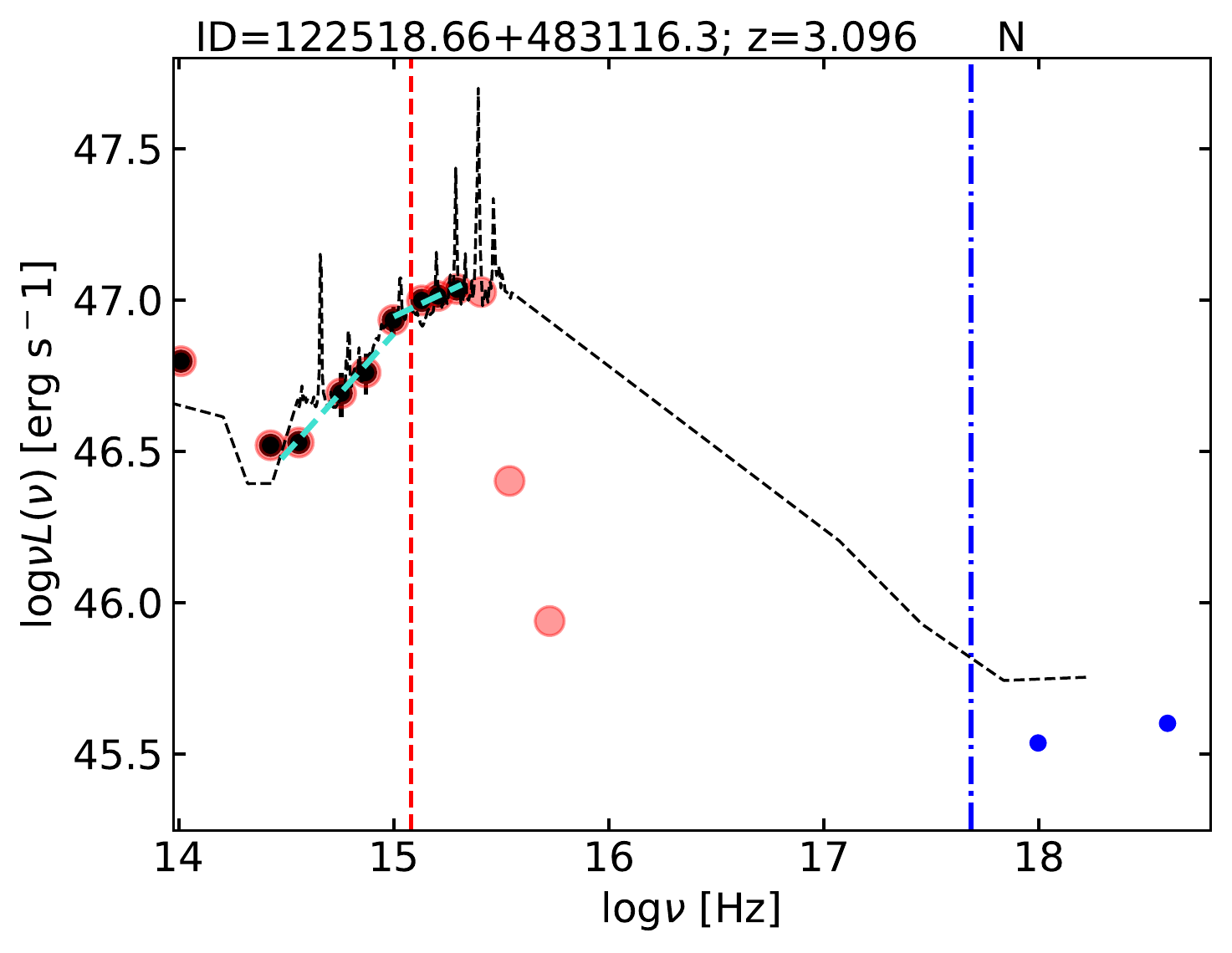}
\includegraphics[width=0.33\linewidth]{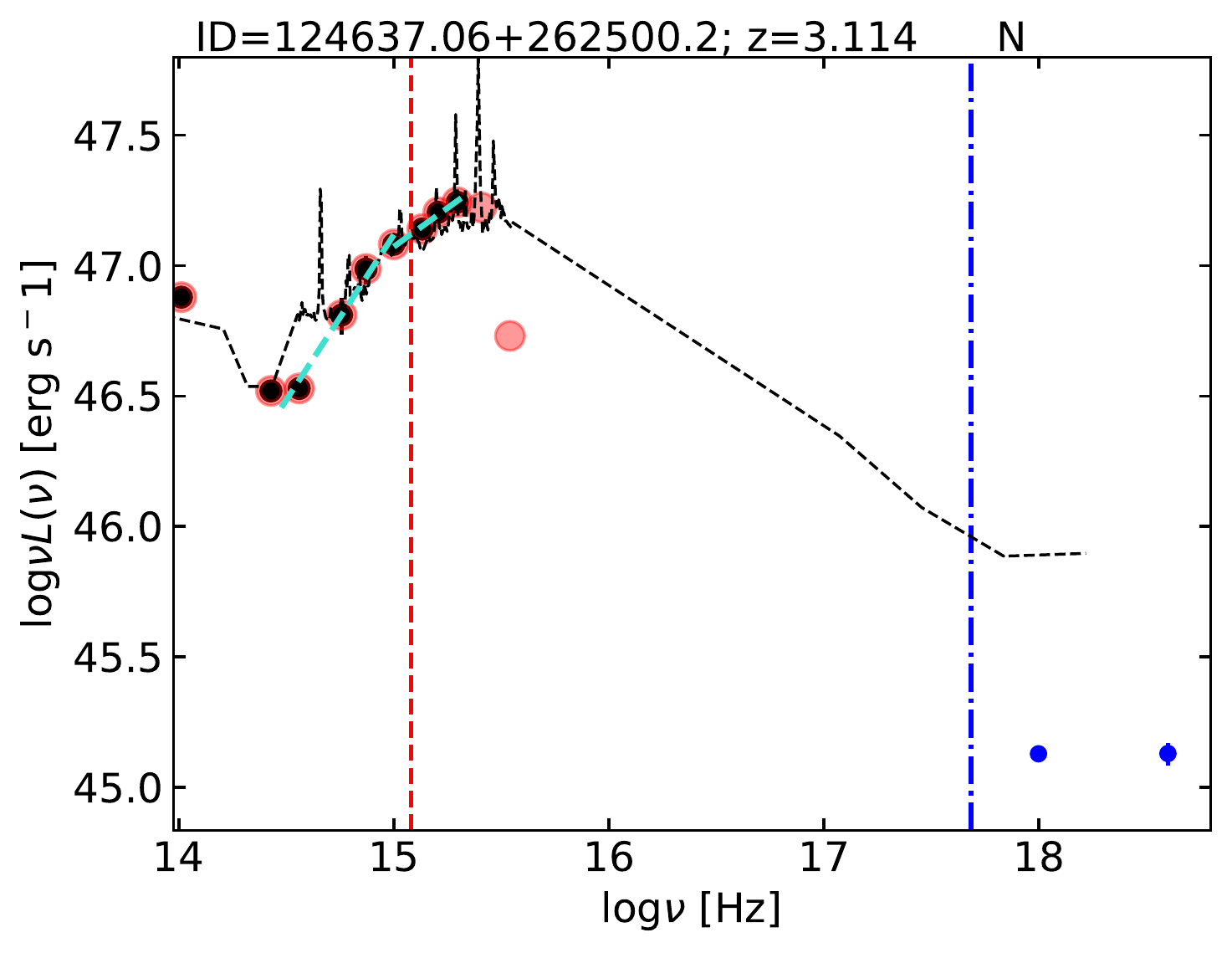}
\includegraphics[width=0.33\linewidth]{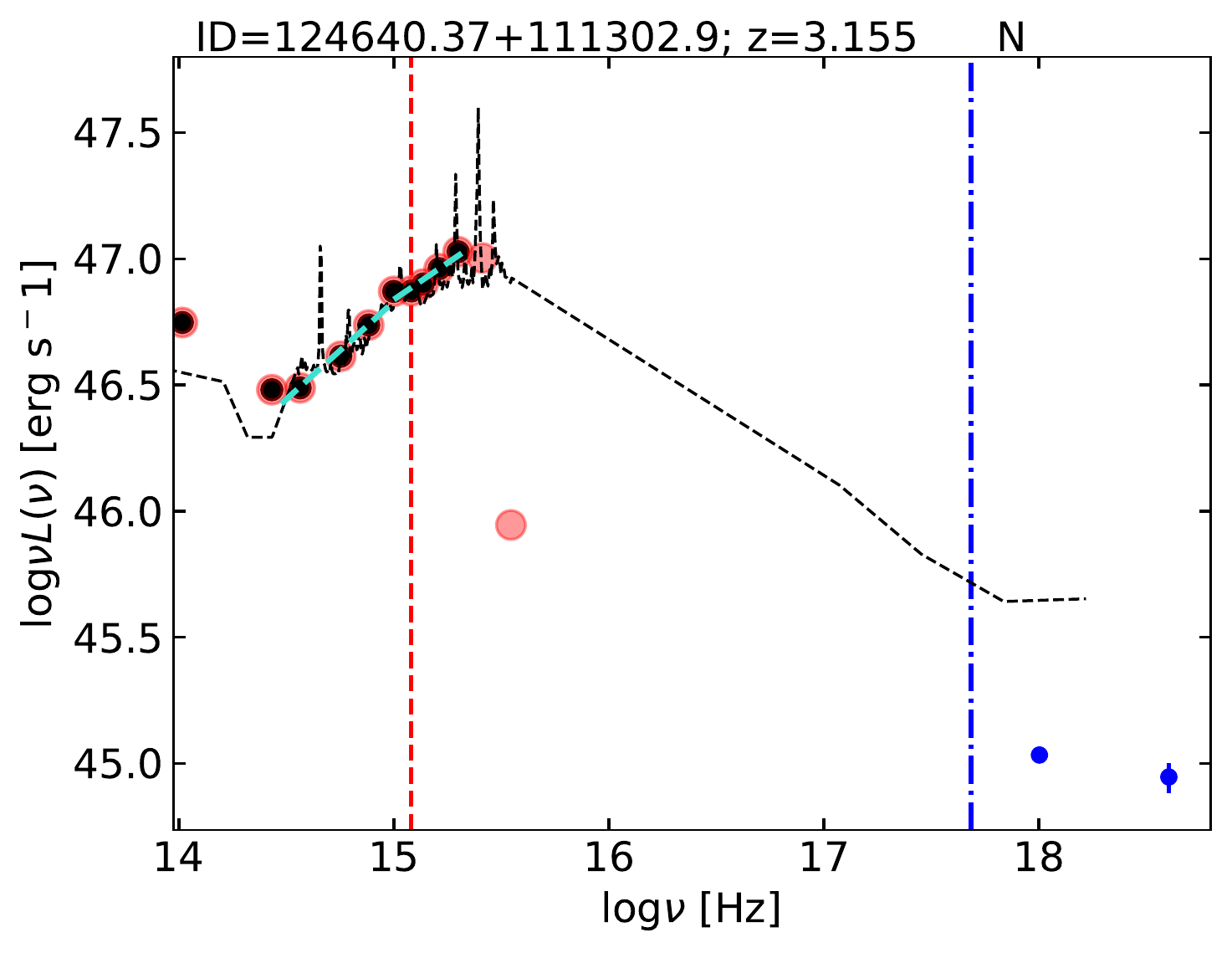}
\includegraphics[width=0.33\linewidth]{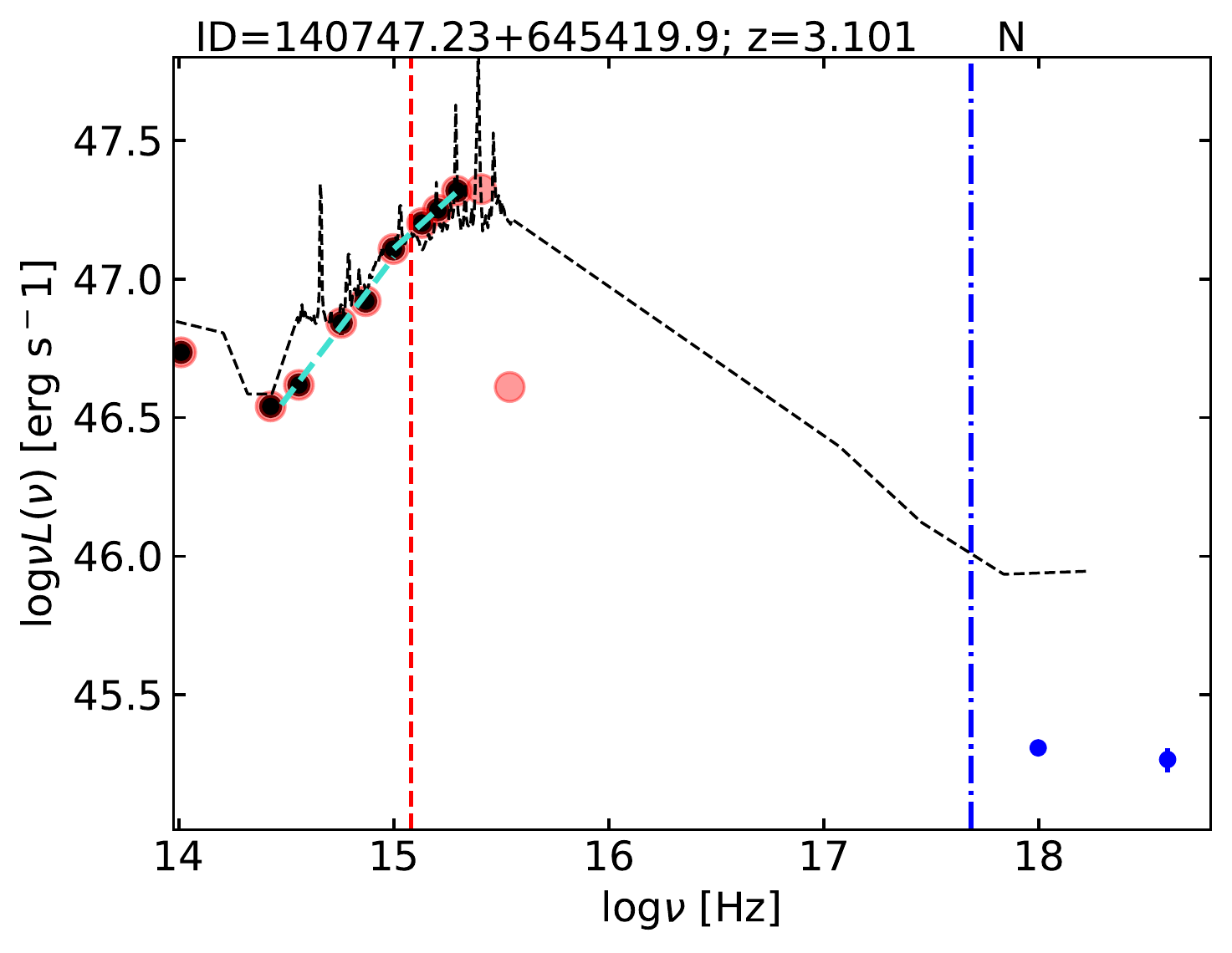}
\includegraphics[width=0.33\linewidth]{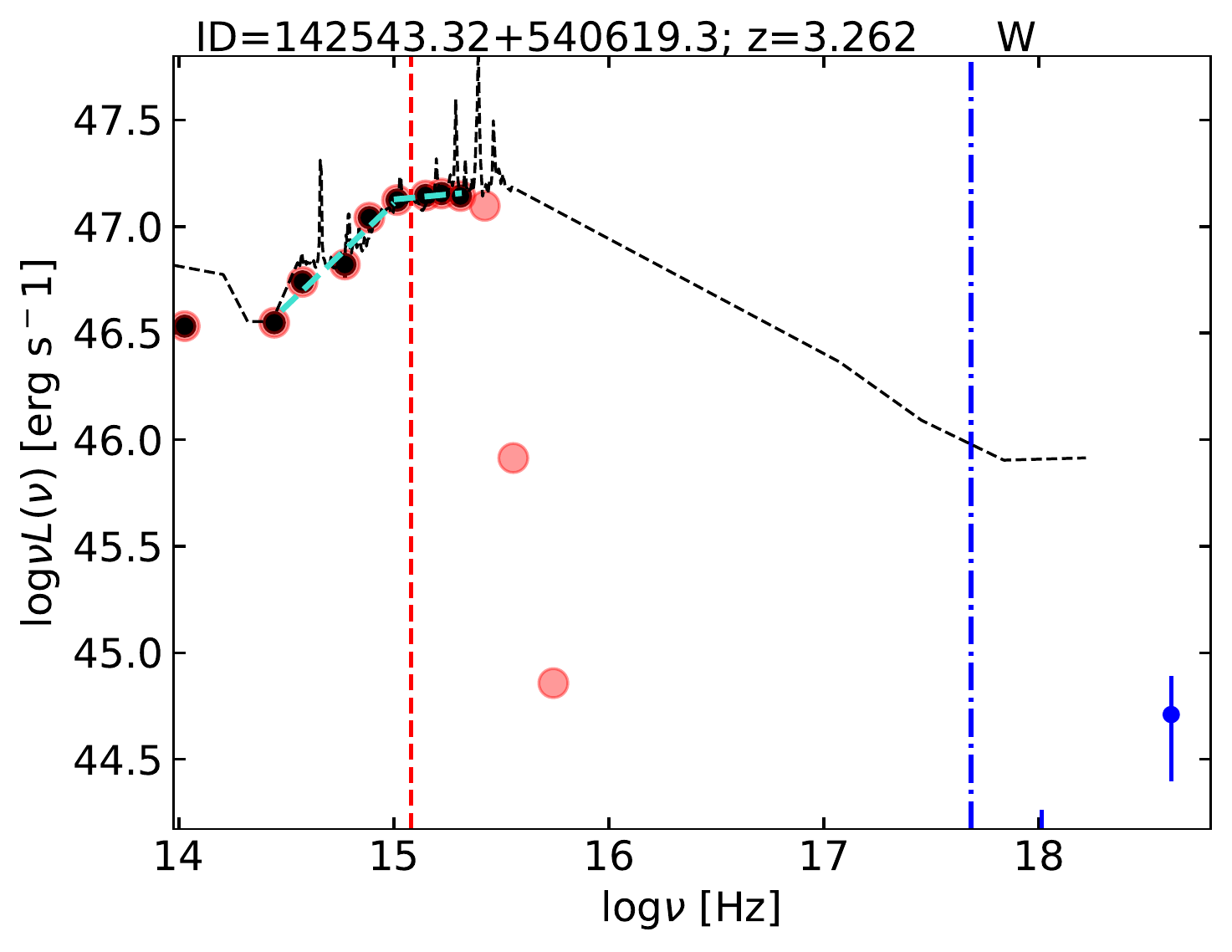}
\includegraphics[width=0.33\linewidth]{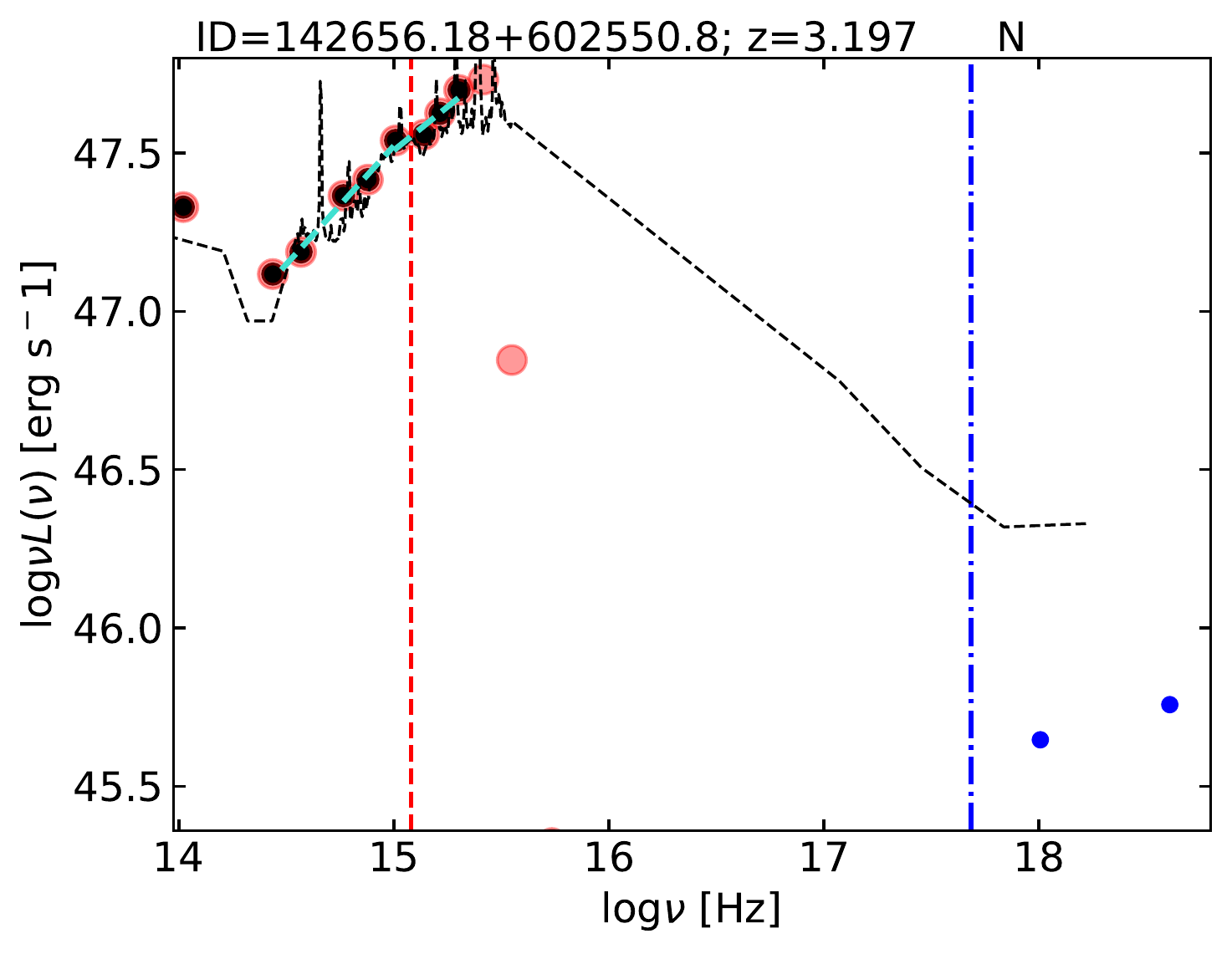}
\includegraphics[width=0.33\linewidth]{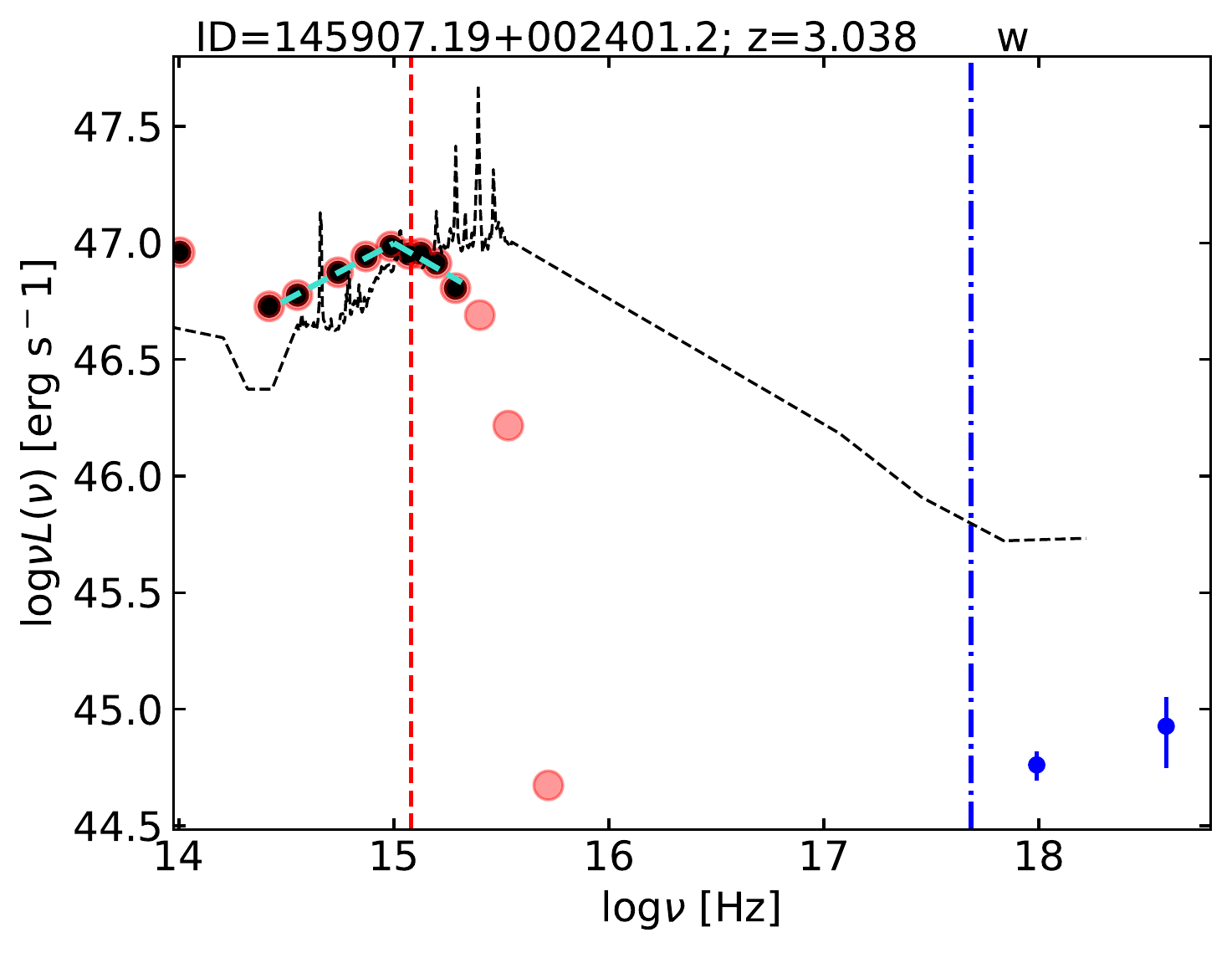}
\includegraphics[width=0.33\linewidth]{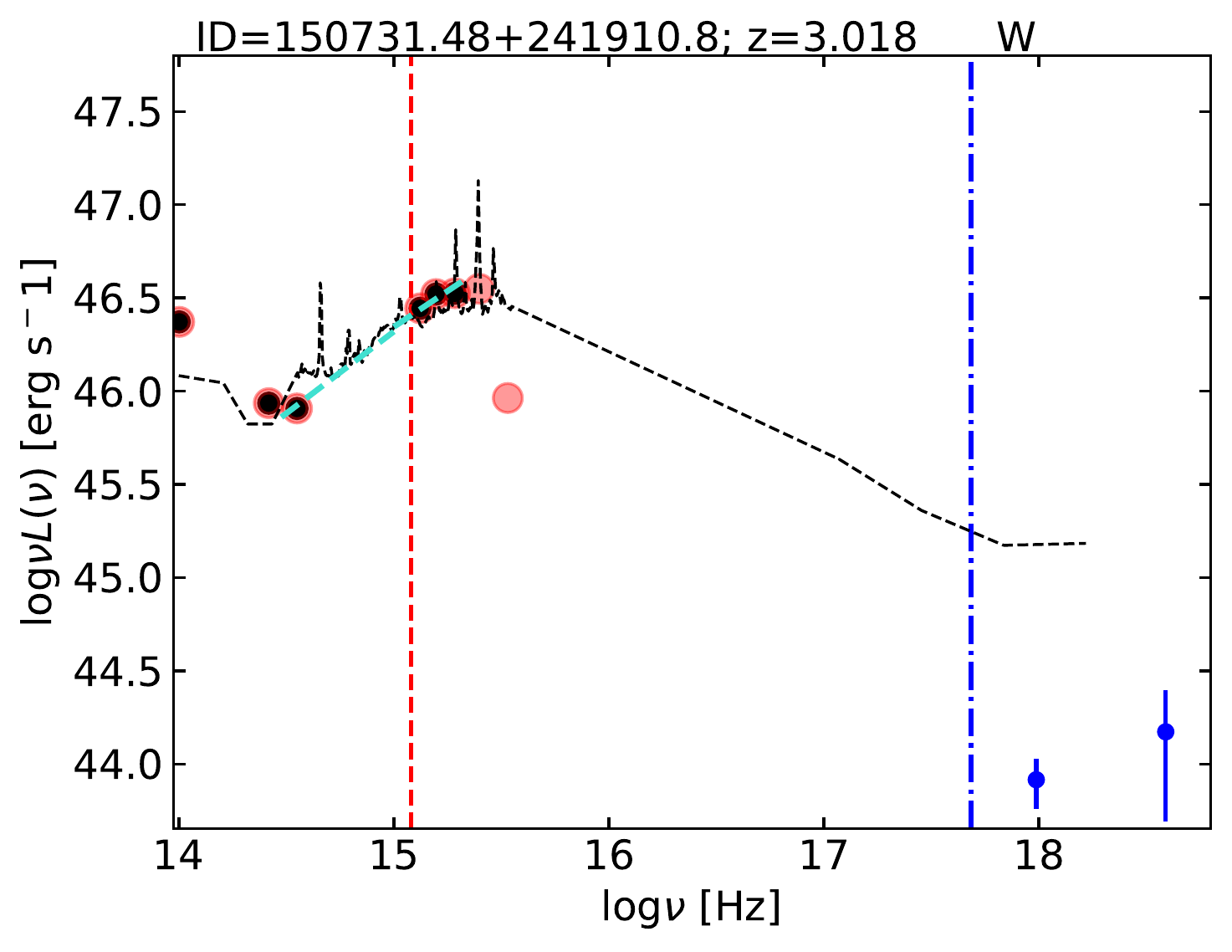}
\includegraphics[width=0.33\linewidth]{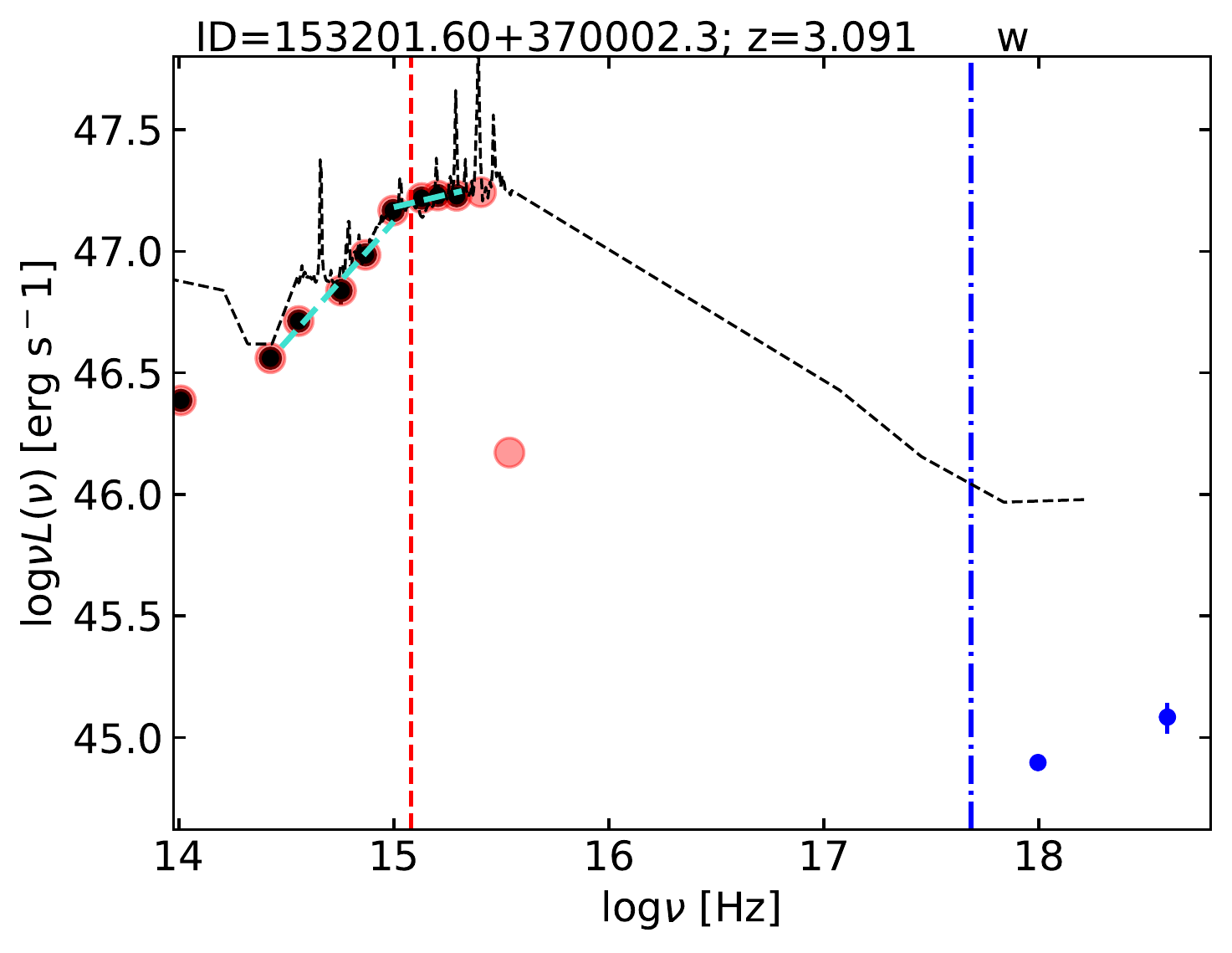}
\includegraphics[width=0.33\linewidth]{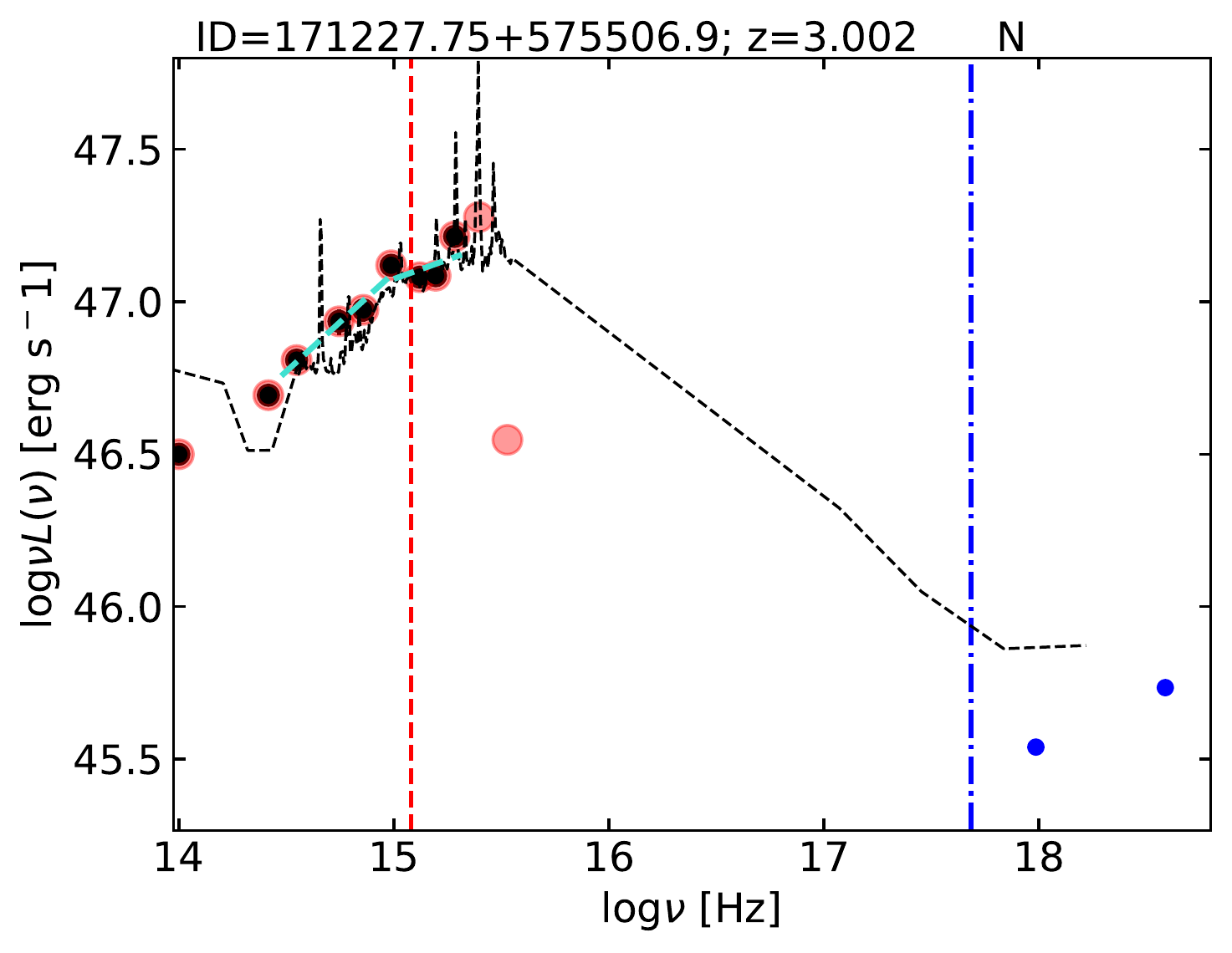}
\includegraphics[width=0.33\linewidth]{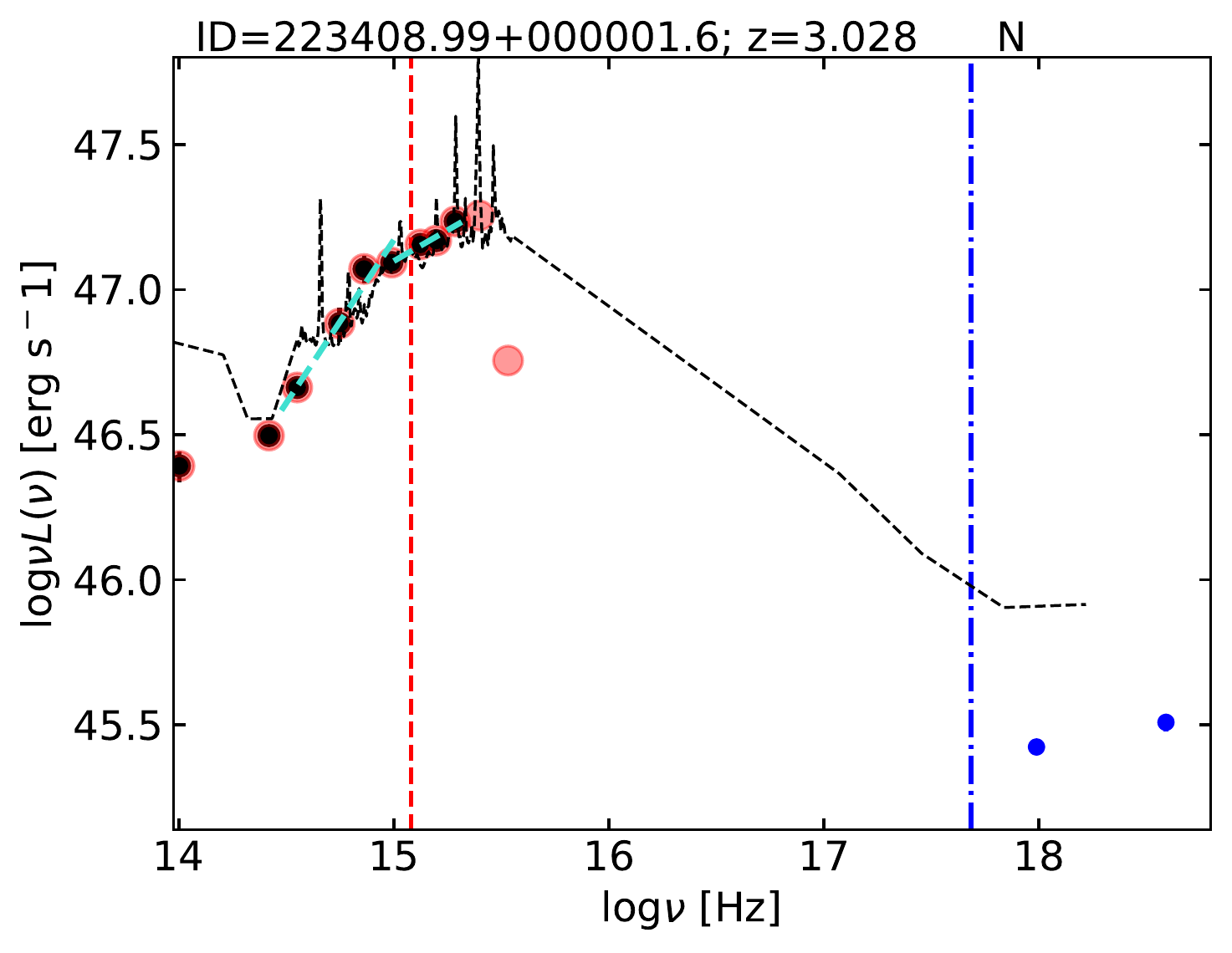}
\caption{{\it Continued}}
\label{seds}
\end{figure*}

\begin{figure}
\centering
 \resizebox{\hsize}{!}{\includegraphics{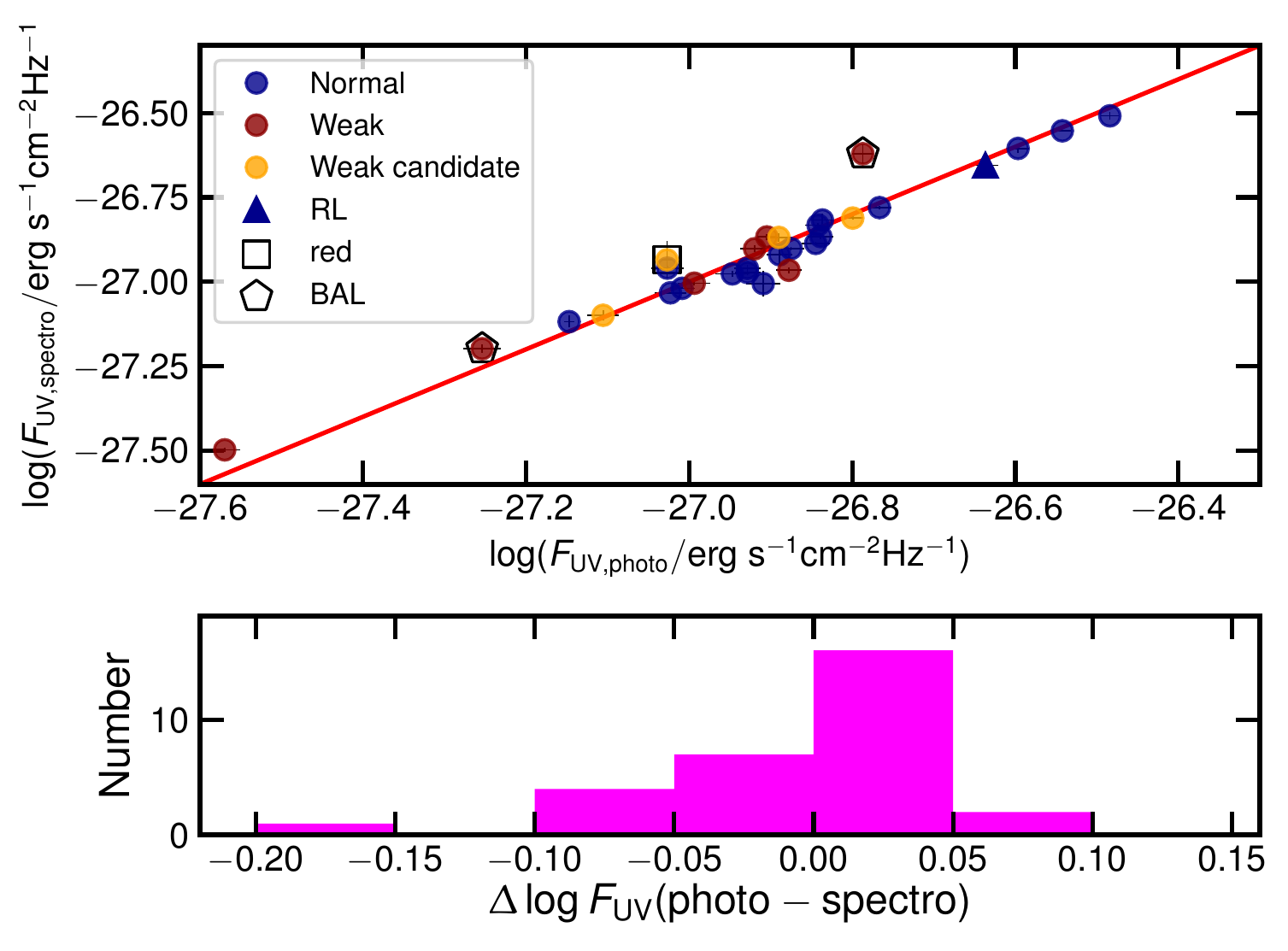}}
\caption{Comparison between the 2500 \AA\ monochromatic fluxes obtained from the broad band SEDs (\S~\ref{Photometric spectral energy distributions}) and the ones from spectral fitting (\S~\ref{uvspectralfit} and appendix~\ref{Ultraviolet SDSS spectra}). The red represent the 1:1 relation. The distribution of the differences between photometric and spectroscopic UV fluxes, $\Delta \fo$(photo-spectro), is shown in the botton panel.}
\label{pscomp}
\end{figure}

\section{Ultraviolet SDSS spectra of the $z$\,$\simeq$\,3 quasar sample}
\label{Ultraviolet SDSS spectra}
Figure~\ref{spec} shows all the ultraviolet SDSS spectra used in this analysis (see Table~\ref{tbl1}). 
The red line represents the best-fit model of the data using \qsfit. The red square symbols are the continuum luminosities estimated by \qsfit, whilst the green and blue squares at 1350 \AA\ and 2500 \AA, respectively, represent the location where the nuclear continuum luminosities are extrapolated from the best-fit of the continuum (red dotted line). For both the fitting procedure (\qsfit or our custom fitting code) we follow the strategy below.

We performed the least$-\chi^2$ fits to the SDSS spectra for each object, where the spectrum was de-reddened for Galactic extinction using the \citet{CCM1989} Galactic reddening law, assuming a total to selective extinction $R_V$\,$=$\,3.1, where $\ebv$ are derived from the \citet{schlegel98} dust map. The spectrum was then shifted to rest frame using the improved redshifts provided by \citet{hw2010} for SDSS quasars. We masked out narrow absorption line features
imprinted on the spectrum, which could bias both the continuum and emission line fits (intrinsic reddening is neglected). For a few spectra, absorption features are considered in the fit to improve the least-$\chi^2$.
The \ion{C}{iv} and \ion{C}{iii}] are fitted with one and two Gaussian profiles for the broad component, respectively (in the case of our custom-made code, all the broad lines are fitted with a broken power-law convolved with a Gaussian function in order to avoid a singularity at the peak, see \citealt{bisogni2017} for details). A Lorenzian profile is also used in the cases where the peak of the line is not well modelled by a single Gaussian profile (e.g. J0304$-$0008). We also fit the \ion{He}{ii} feature at 1640 \AA\ and the excess emission at 1600--1650 \AA\ with a pseudo-continuum that is attributed with \ion{Fe}{ii} emission. We do not impose any constrain between the broad and narrow-line component.
We measure line widths for \ion{C}{iv} both with and without an additional narrow-line subtraction to test whether it is required to improve the fit. For the majority of the cases, a single Gaussian or Lorenzian profile (or a single broken power-law profile for the custom-made code) is enough to model the entire line.
Uncertainties on the relevant parameters (e.g. FWHM, EW, $\sigma$, $\vp$) are computed from the standard deviation of 10 fits where different spectral channels are masked to increase the noise with respect to the original data. 

\begin{figure*}[t!]
\includegraphics[width=6cm]{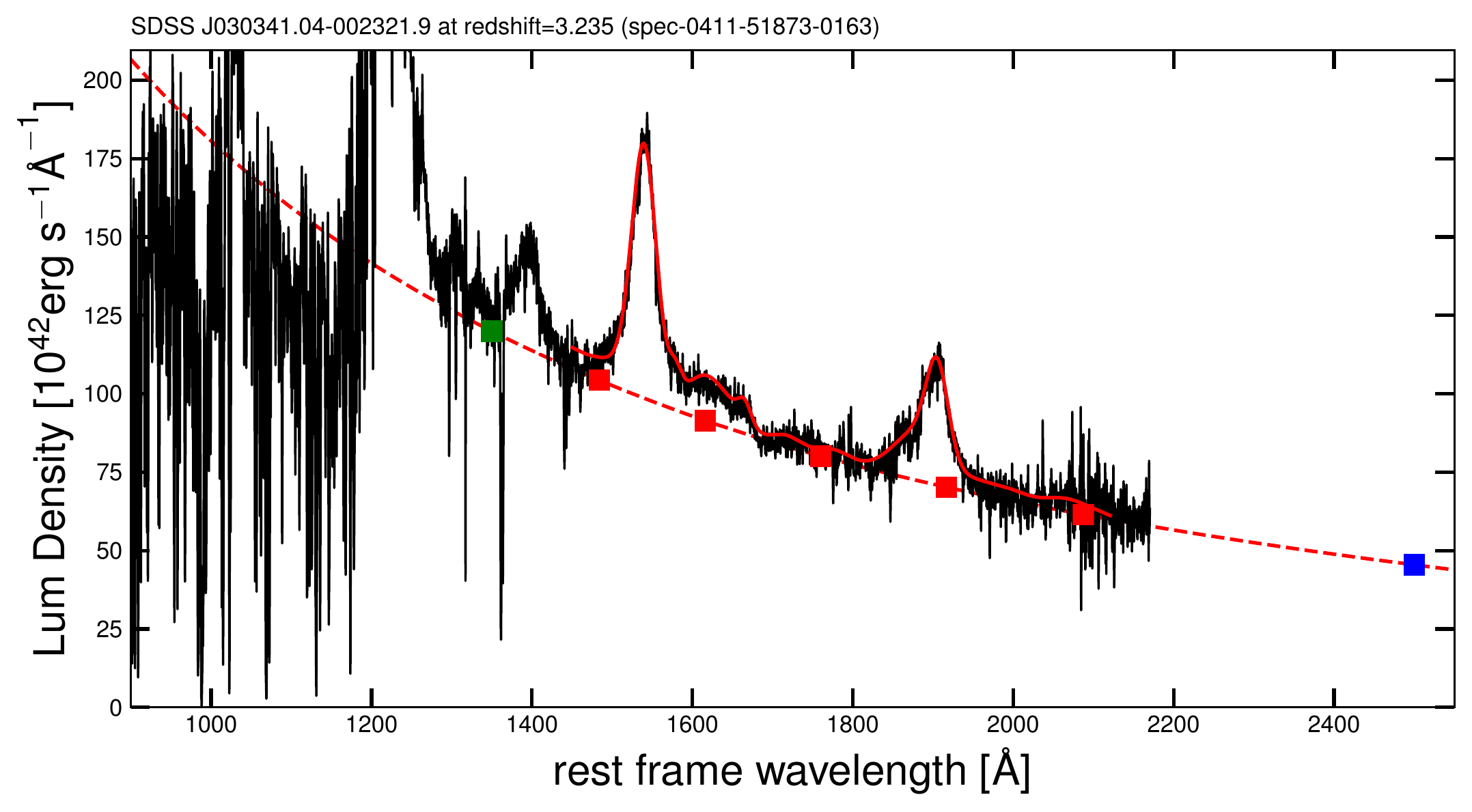}
\includegraphics[width=6cm]{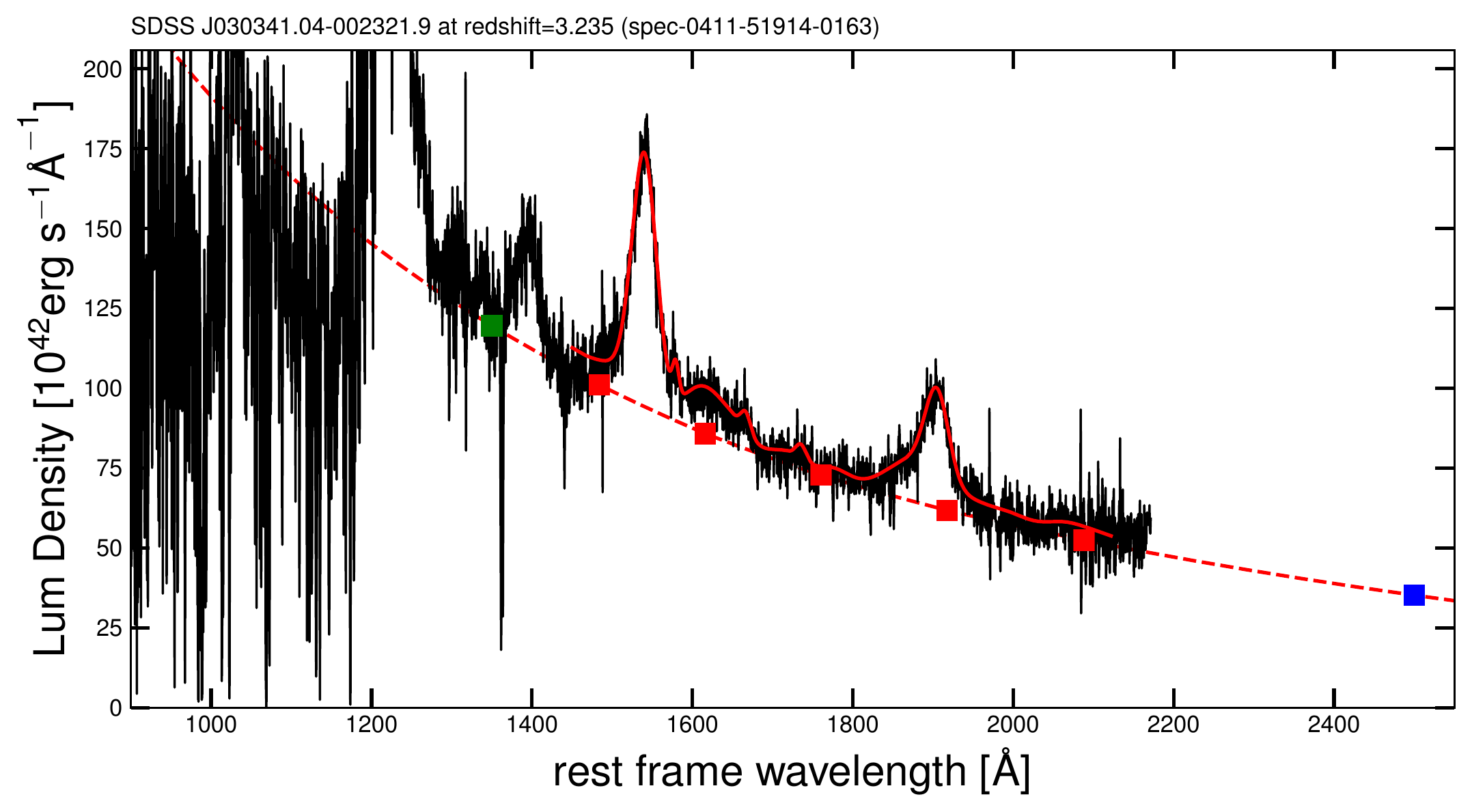}
\includegraphics[width=6cm]{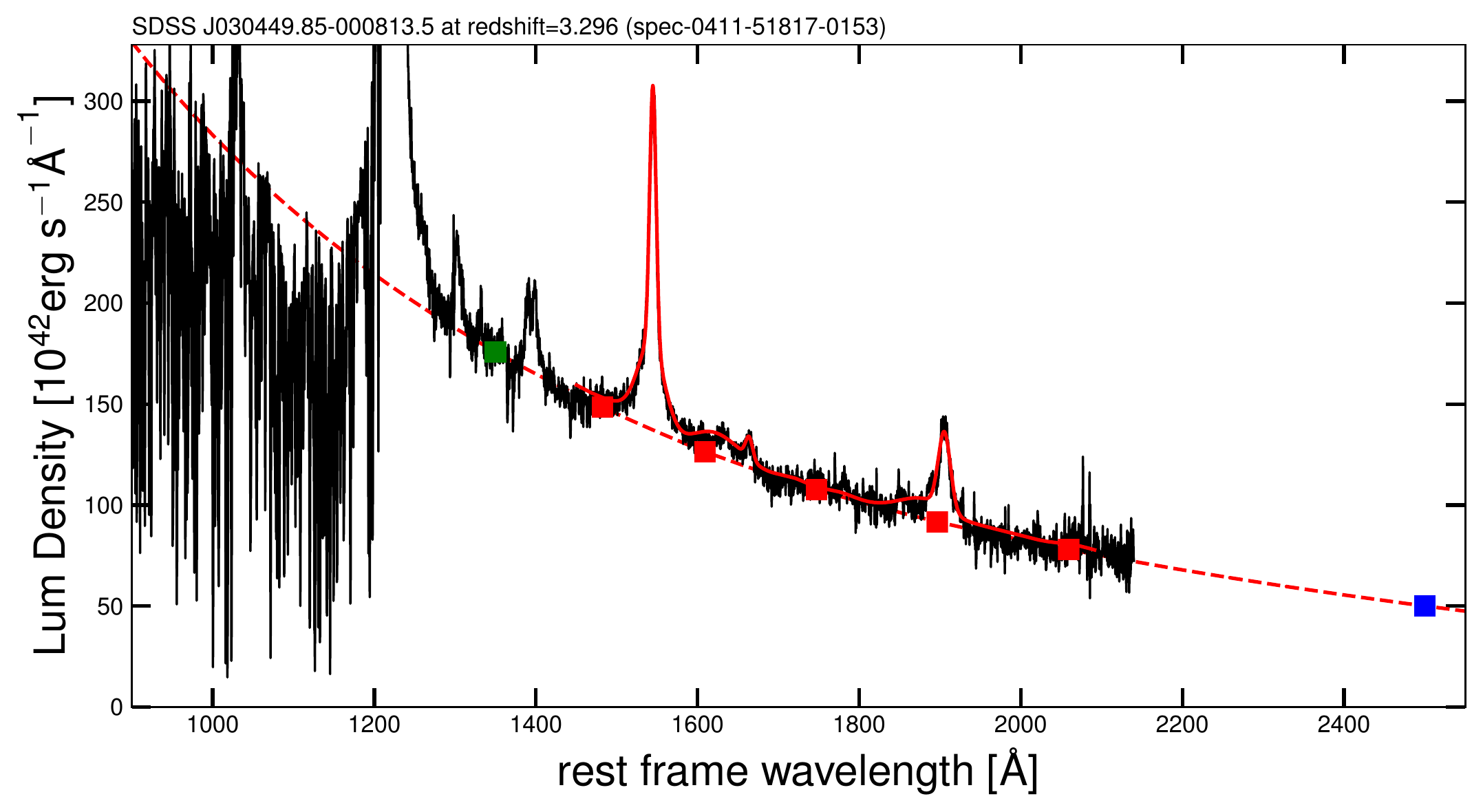}
\includegraphics[width=6cm]{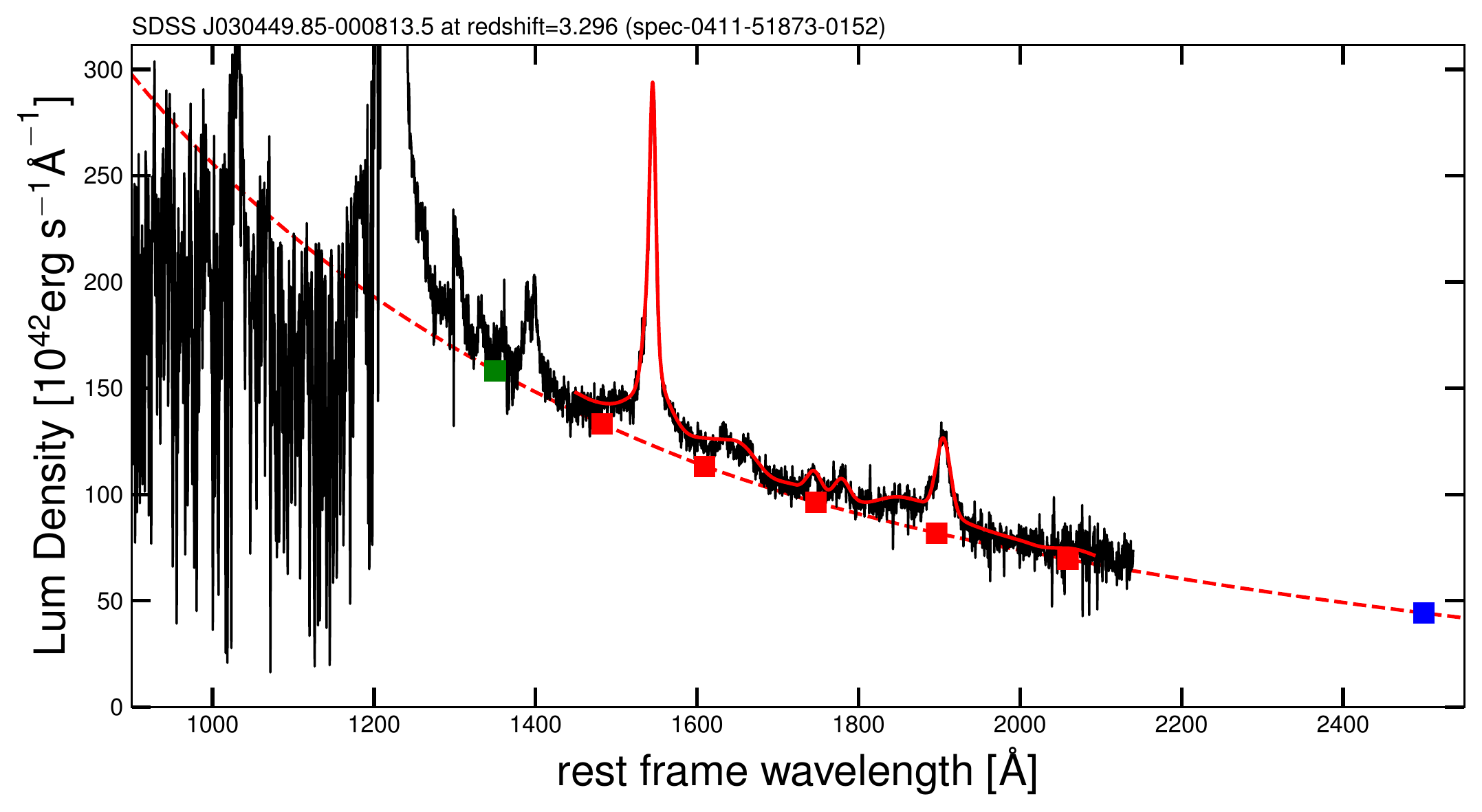}
\includegraphics[width=6cm]{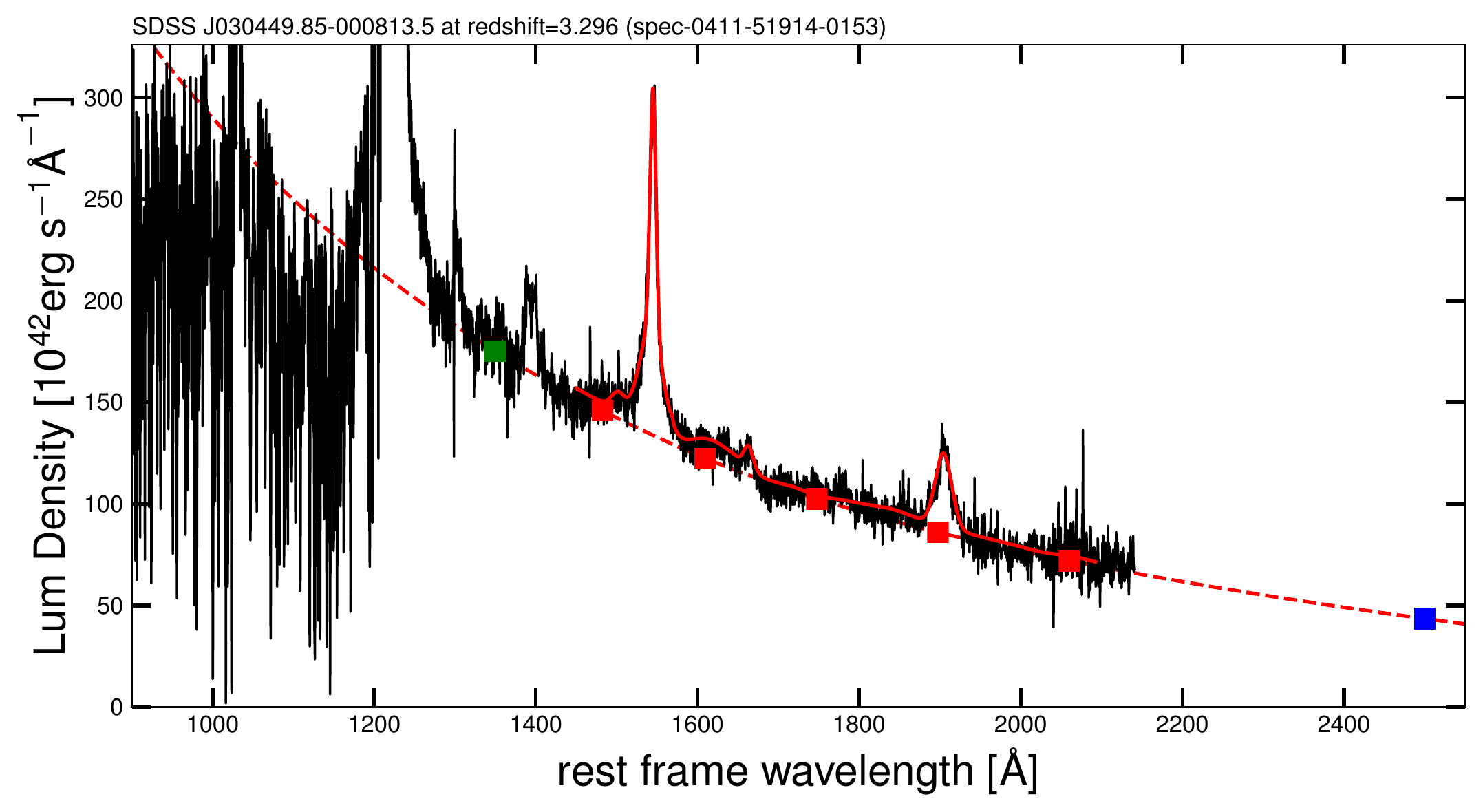}
\includegraphics[width=6cm]{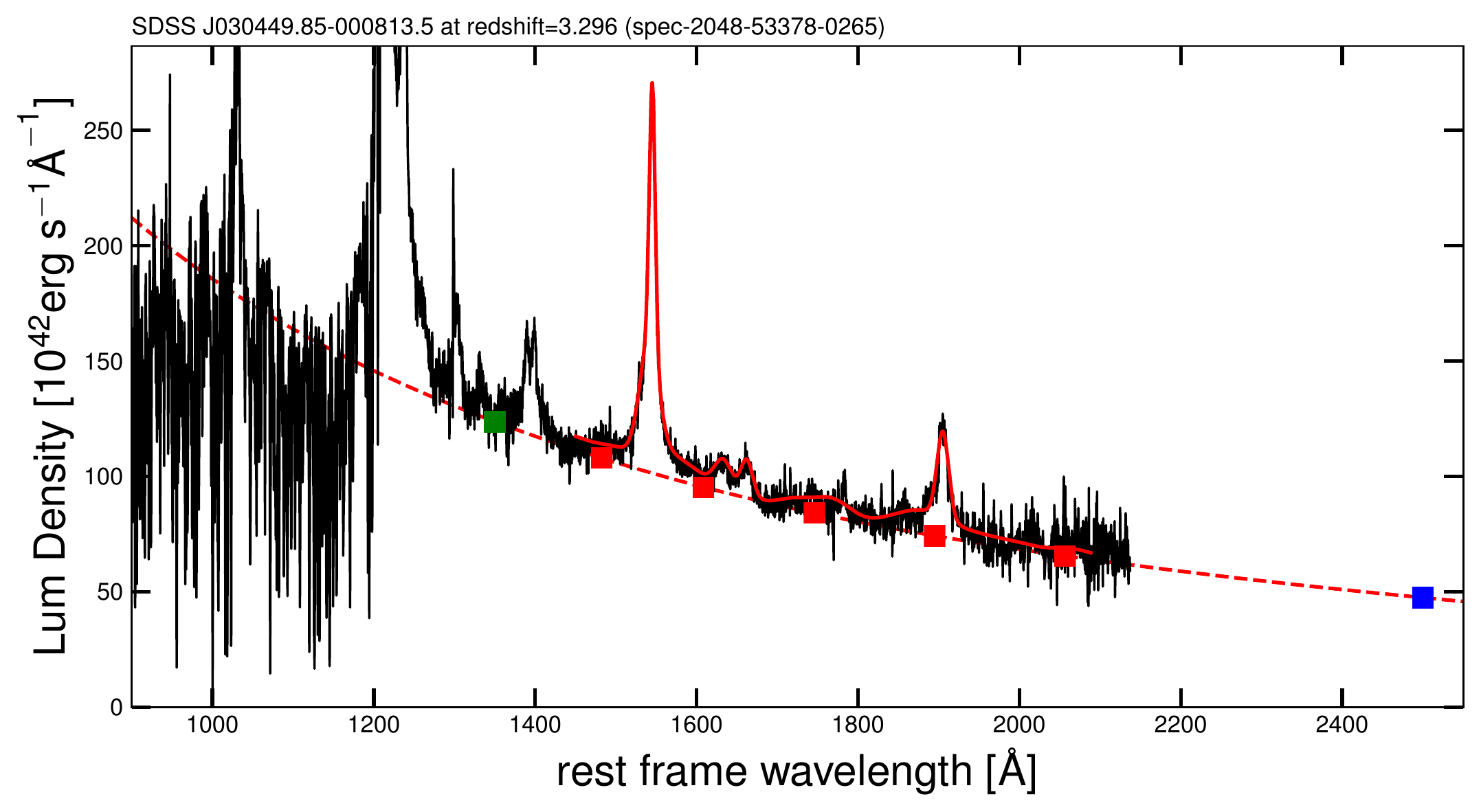}
\includegraphics[width=6cm]{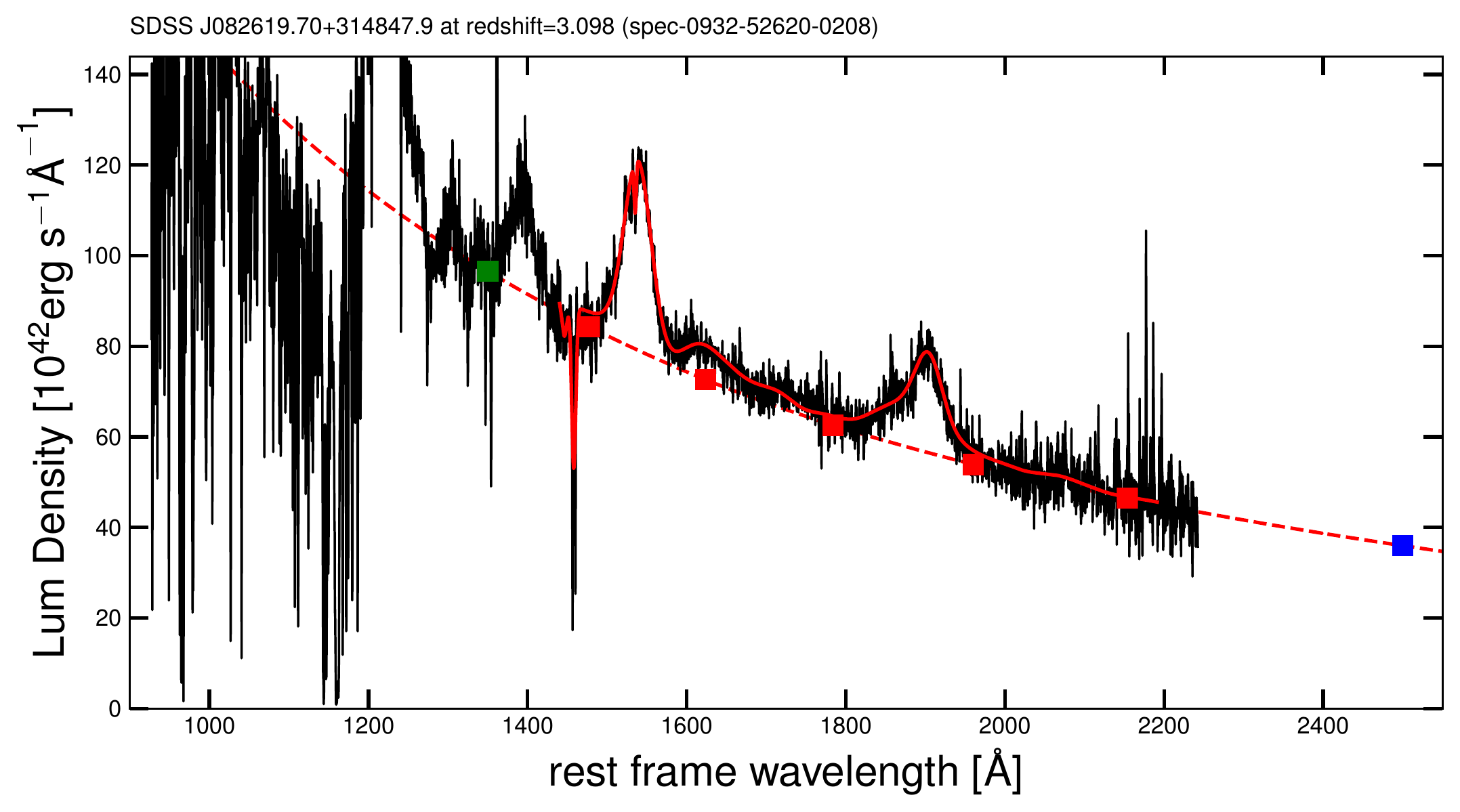}
\includegraphics[width=6cm]{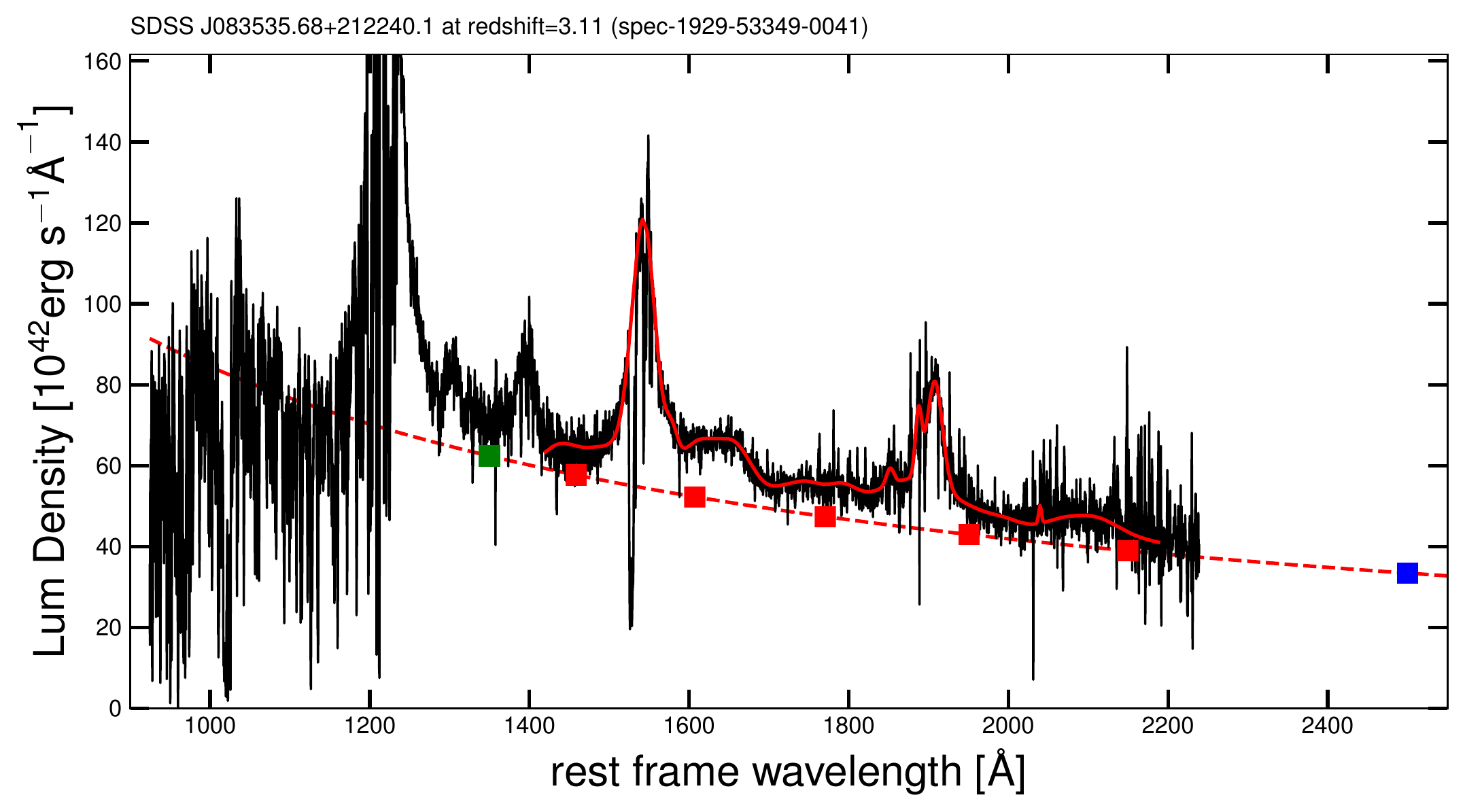}
\includegraphics[width=6cm]{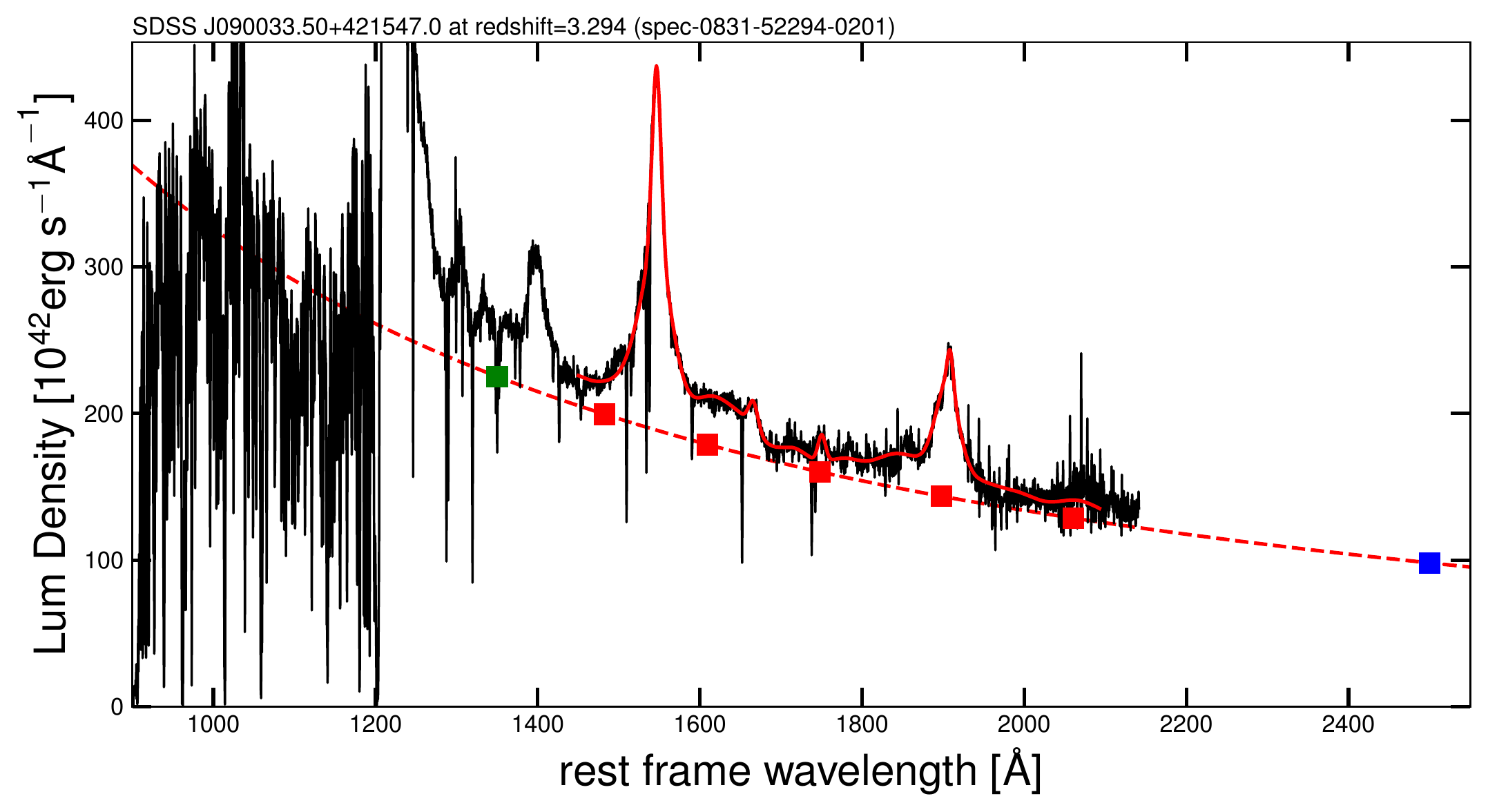}
\includegraphics[width=6cm]{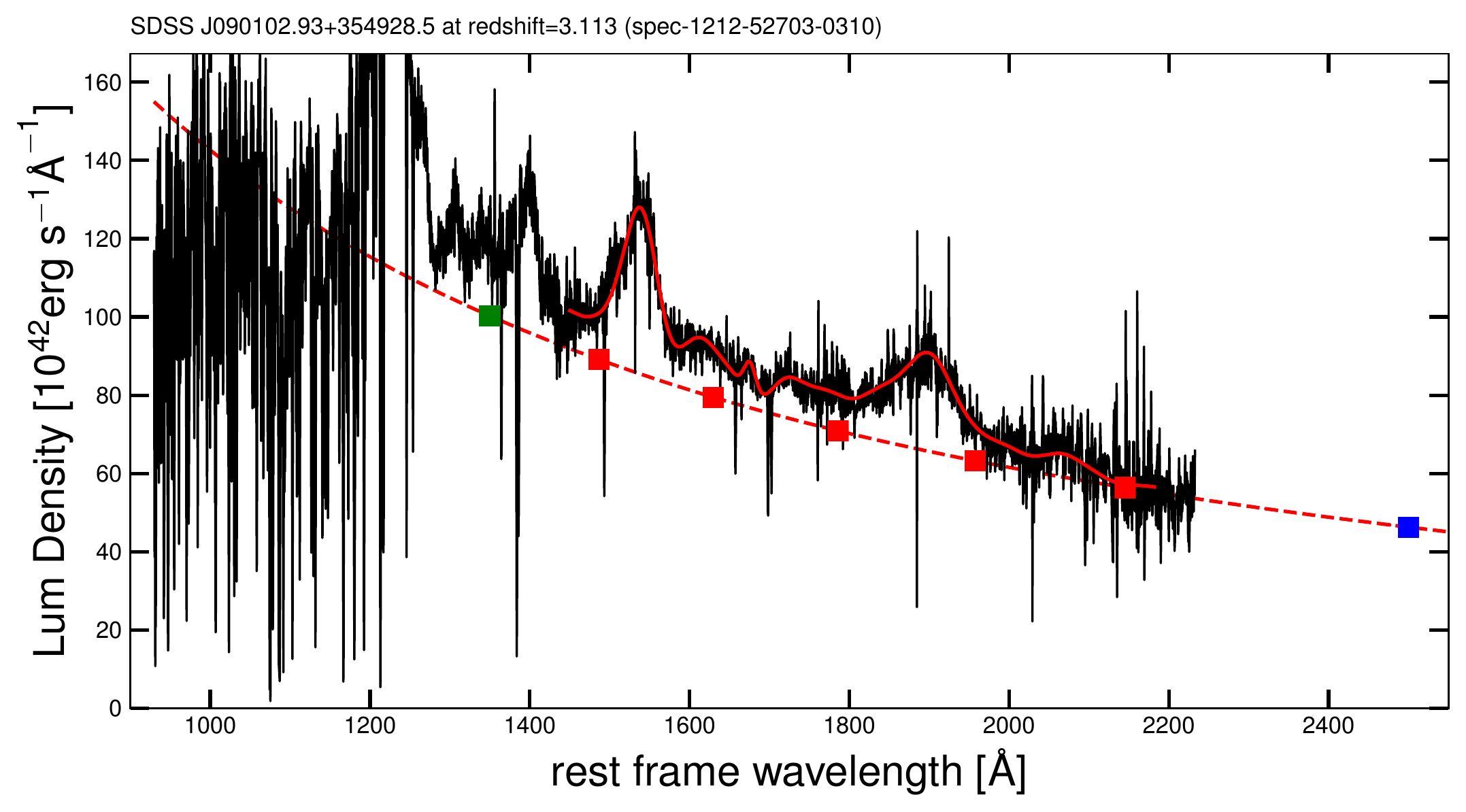}
\includegraphics[width=6cm]{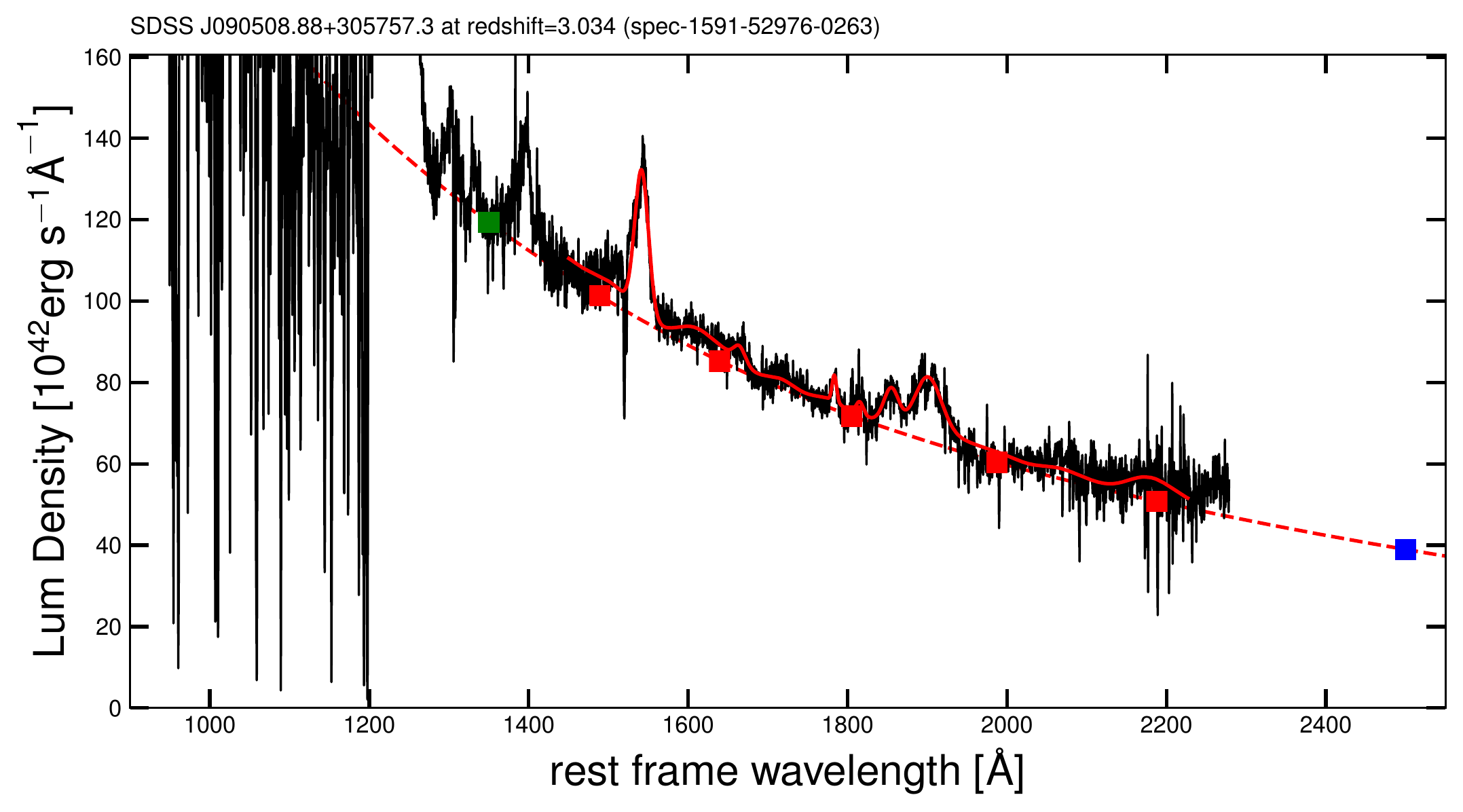}
\includegraphics[width=6cm]{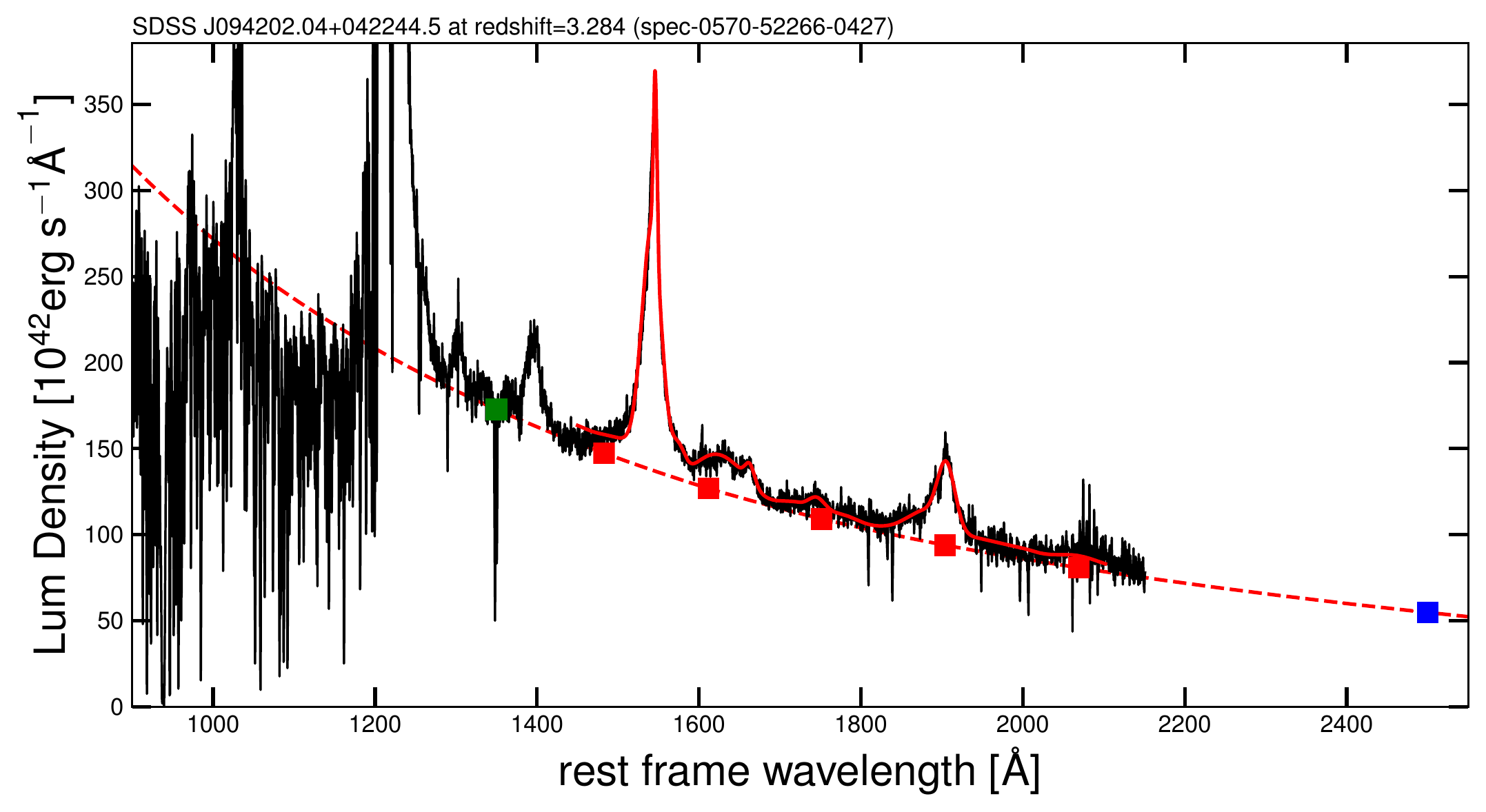}
\includegraphics[width=6cm]{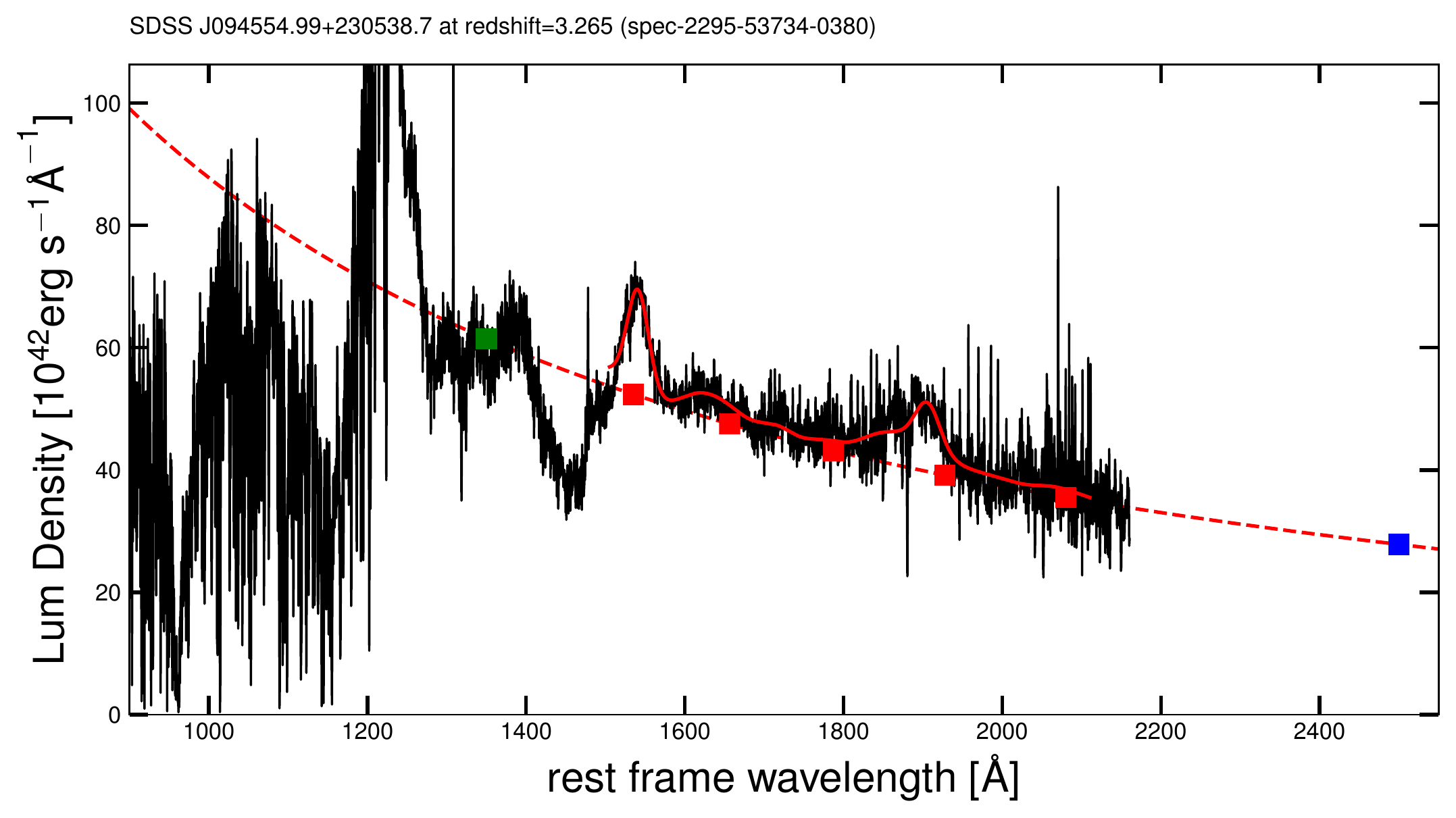}
\includegraphics[width=6cm]{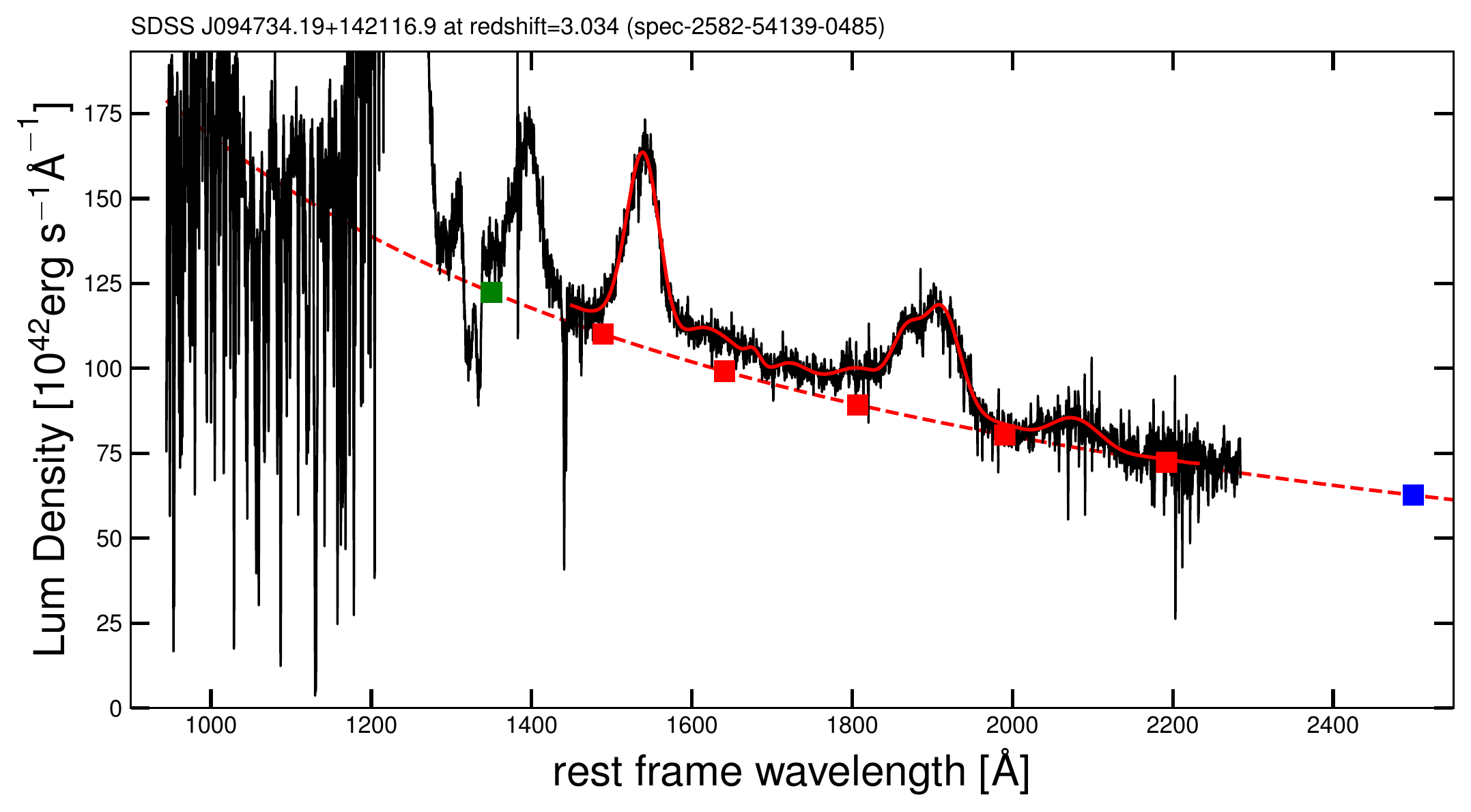}
\includegraphics[width=6cm]{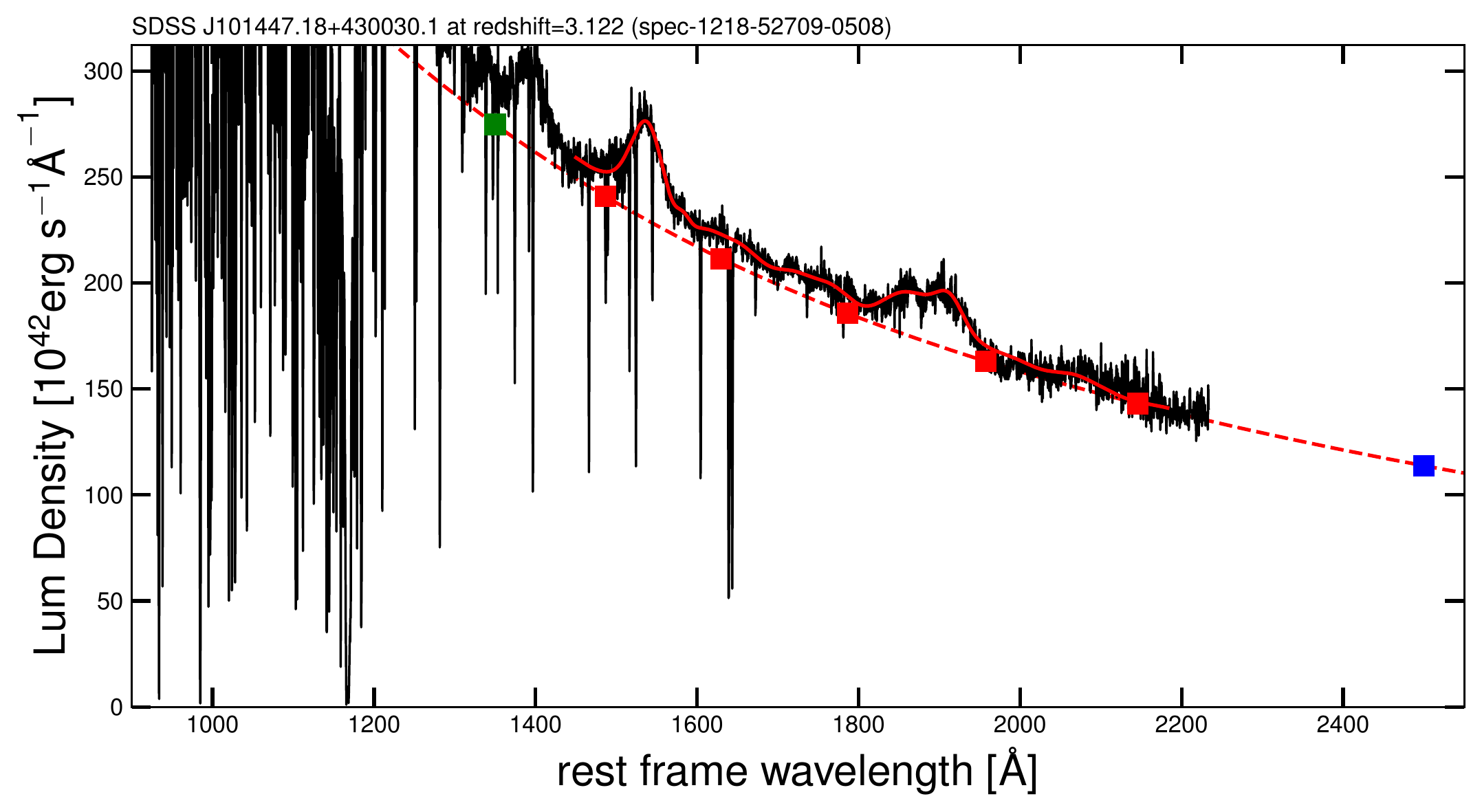}
\includegraphics[width=6cm]{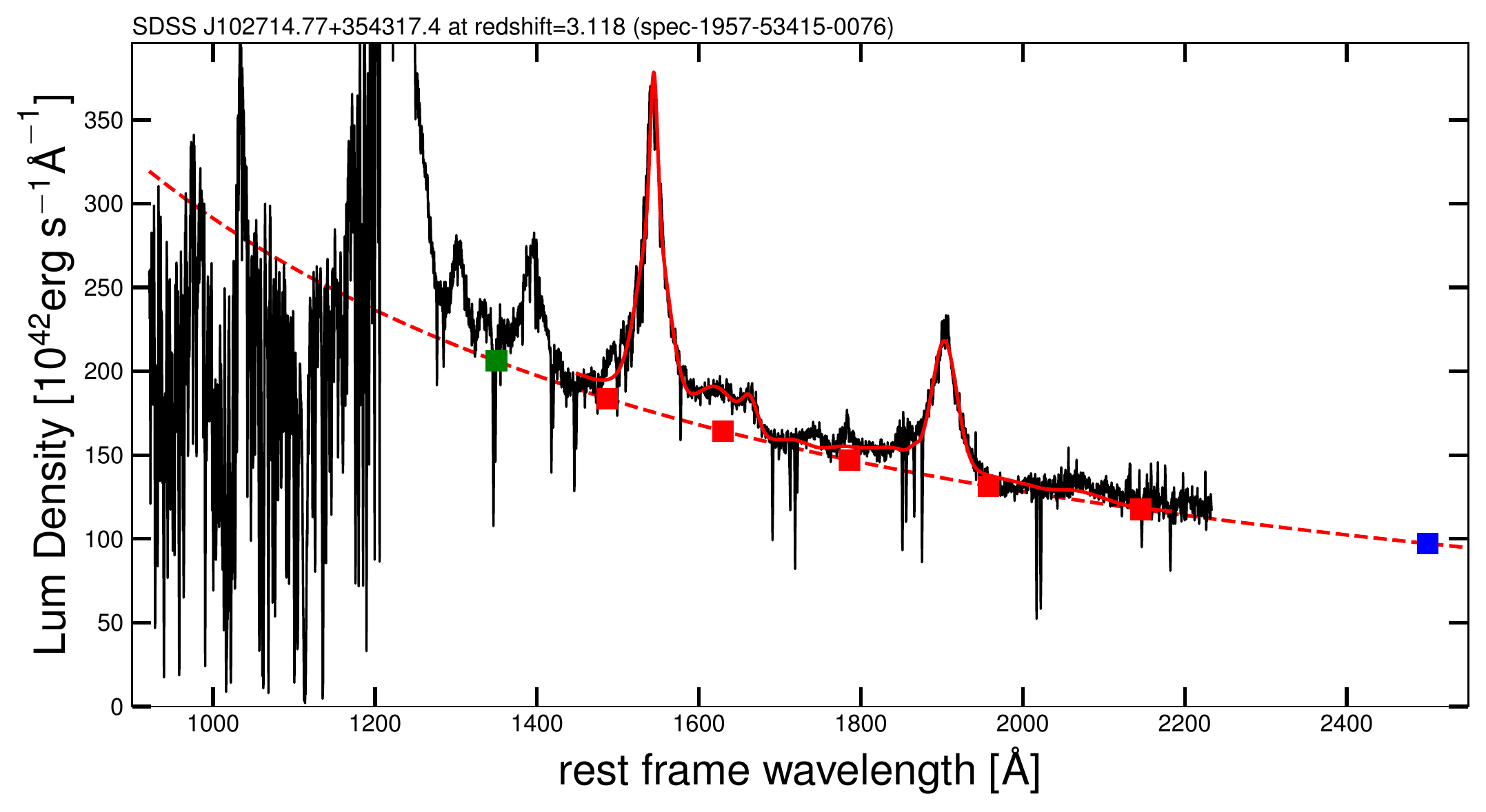}
\includegraphics[width=6cm]{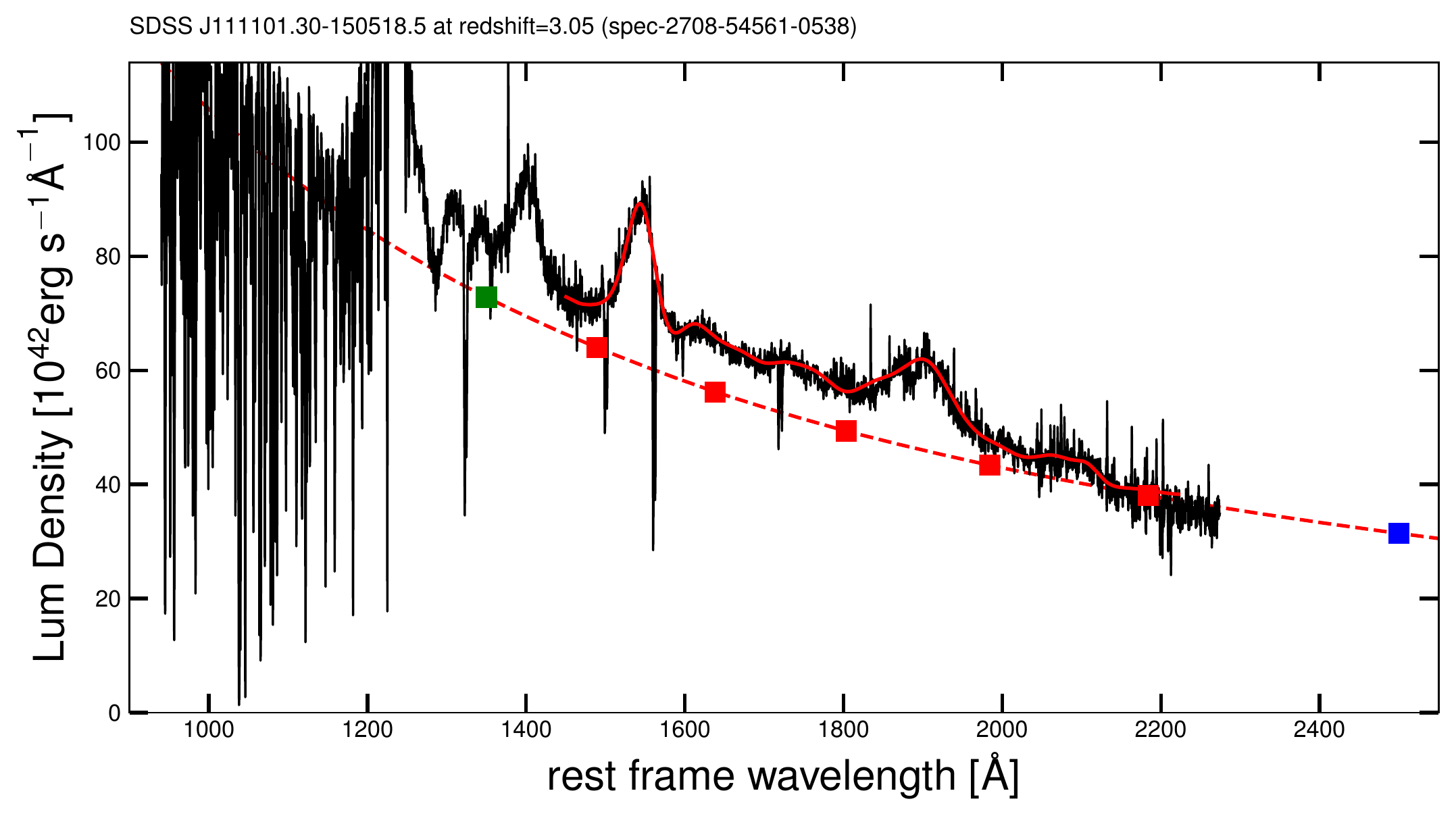}
\includegraphics[width=6cm]{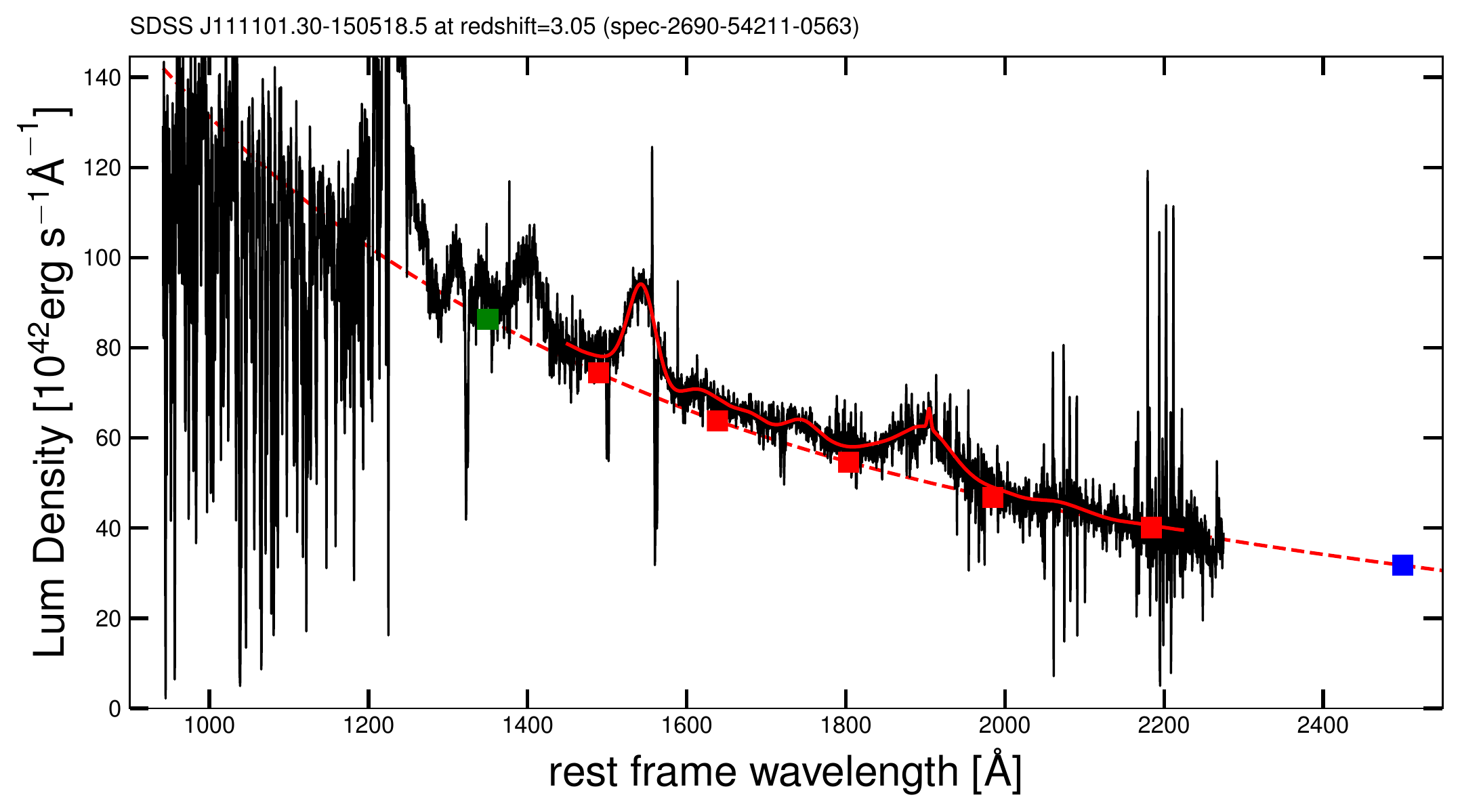}
\caption{Ultraviolet SDSS spectra of the $z$\,$\simeq$\,3 quasar sample (black line). The red line represents the best-fit model of the data using \qsfit. The red squares are the continuum luminosities estimated by \qsfit, whilst the green (1350 \AA) and blue (2500 \AA) squares represent the nuclear continuum luminosities extrapolated from the best fit of the continuum (red dotted line).}
\label{spec}
\end{figure*}

\begin{figure*}
\addtocounter{figure}{-1}
\includegraphics[width=6cm]{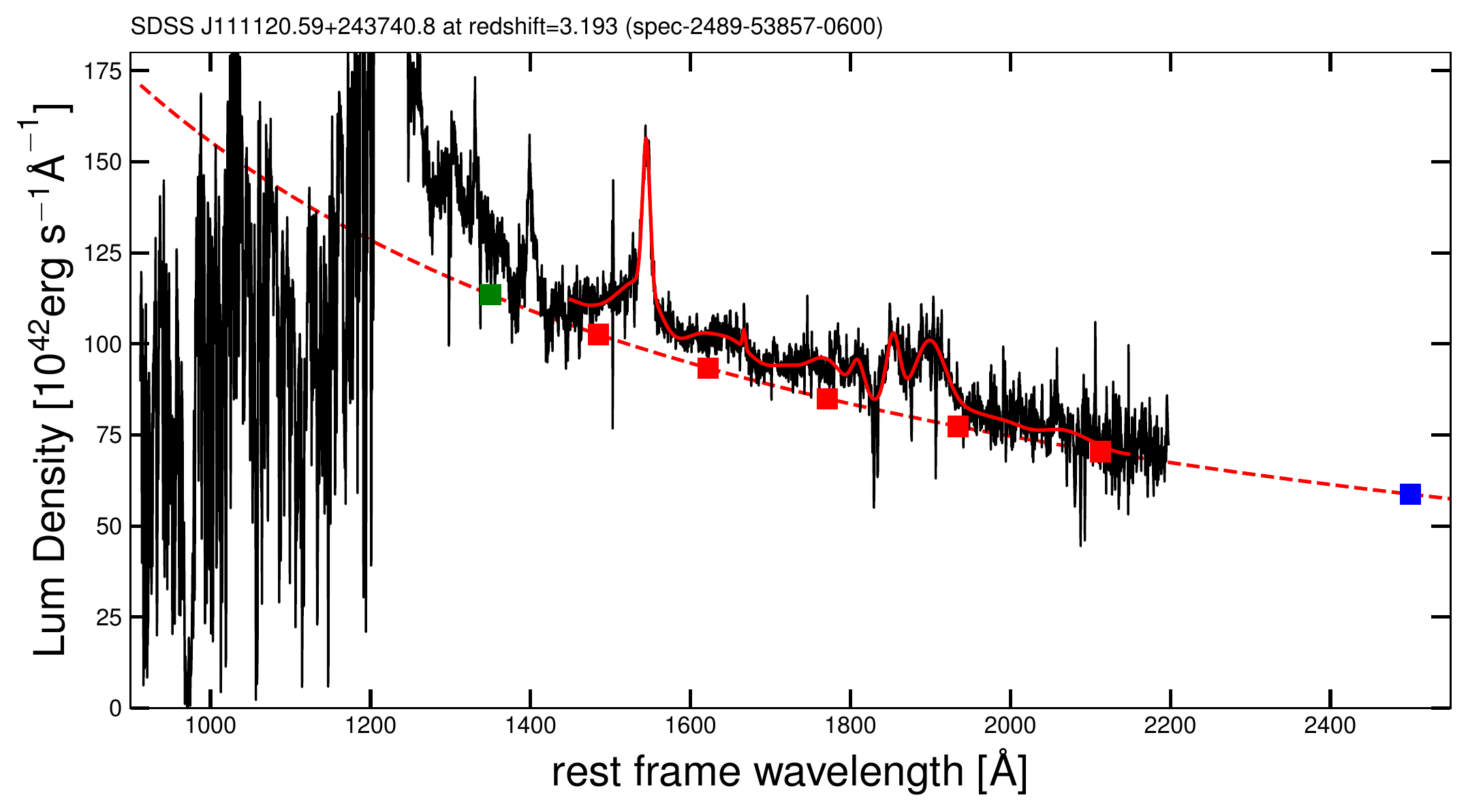}
\includegraphics[width=6cm]{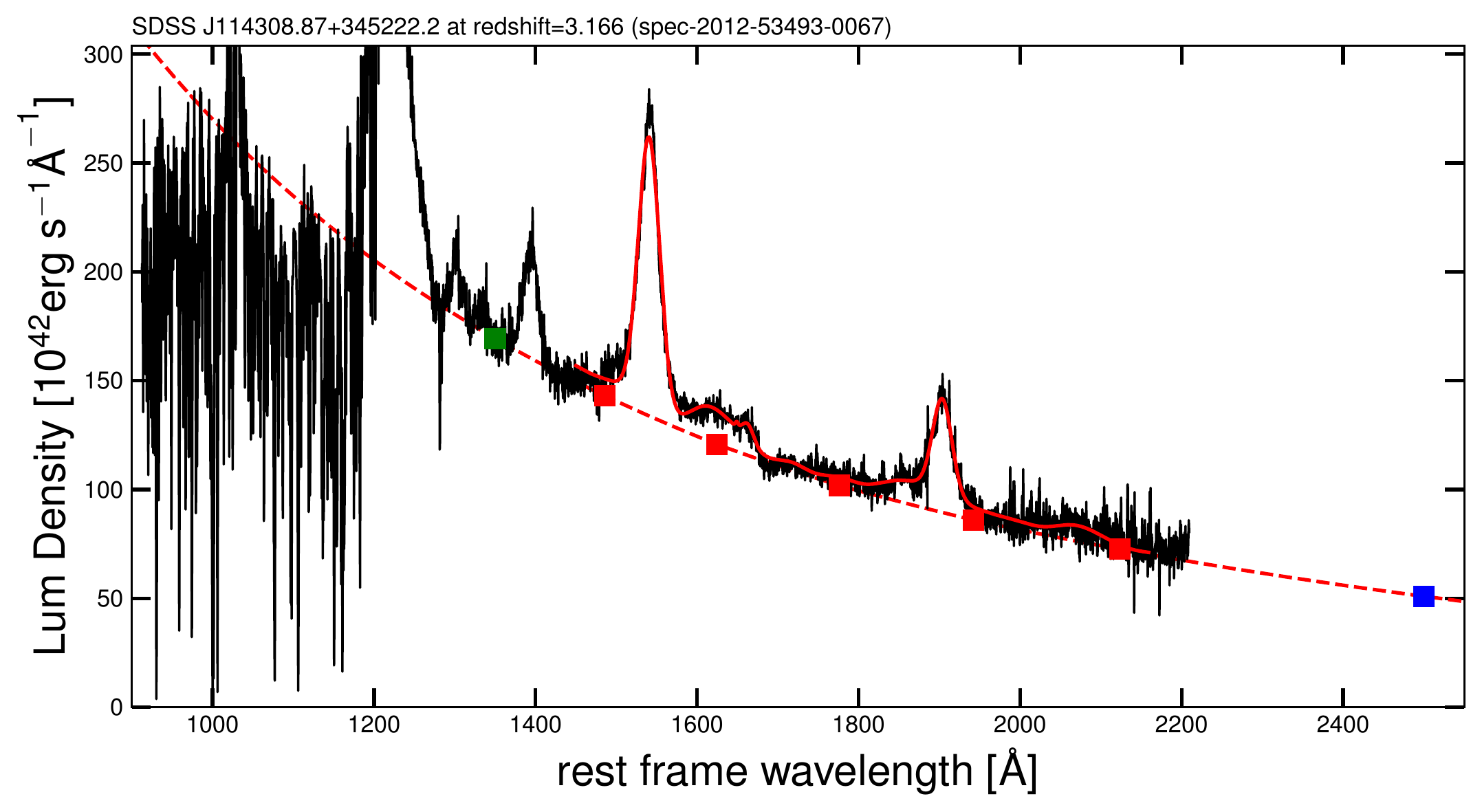}
\includegraphics[width=6cm]{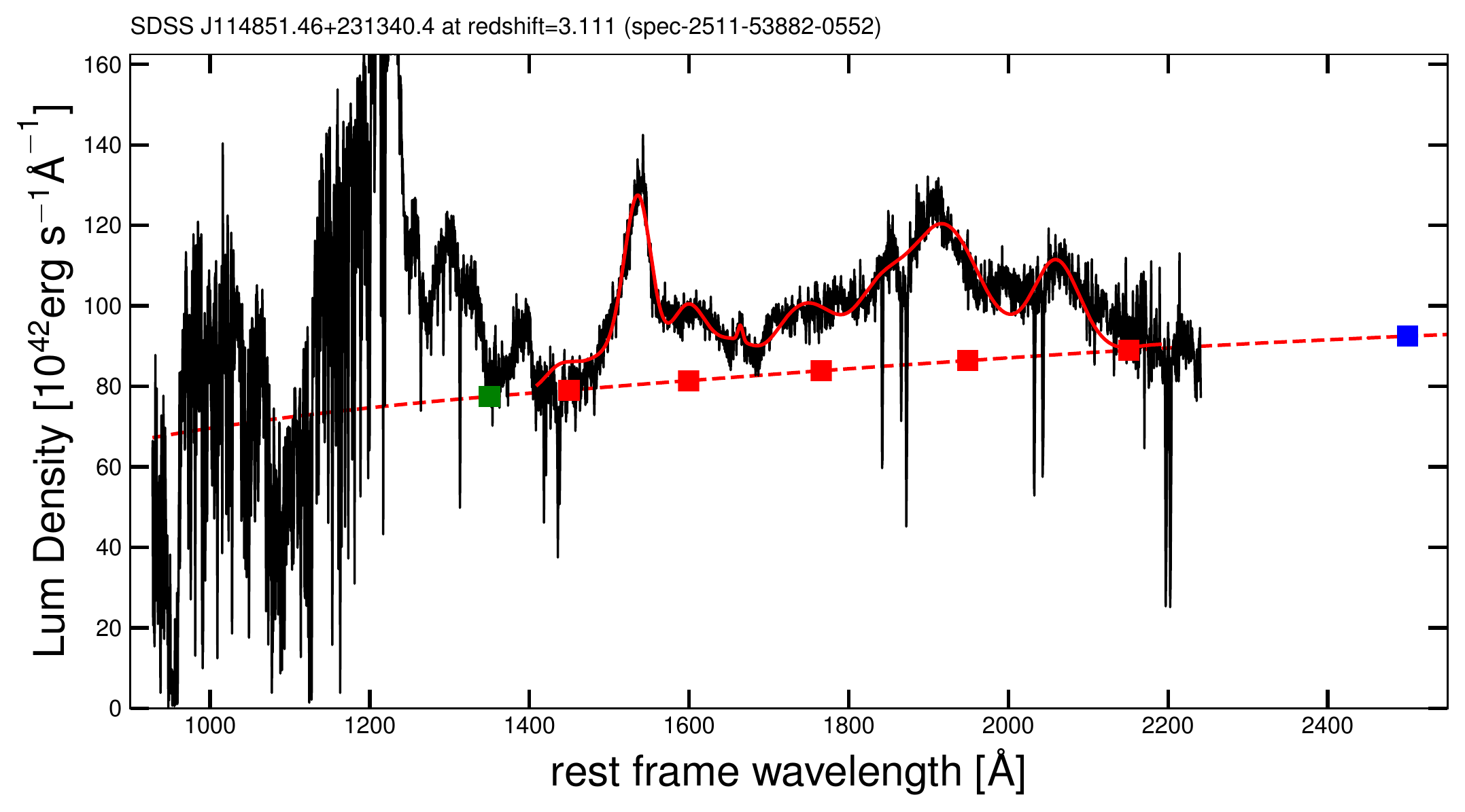}
\includegraphics[width=6cm]{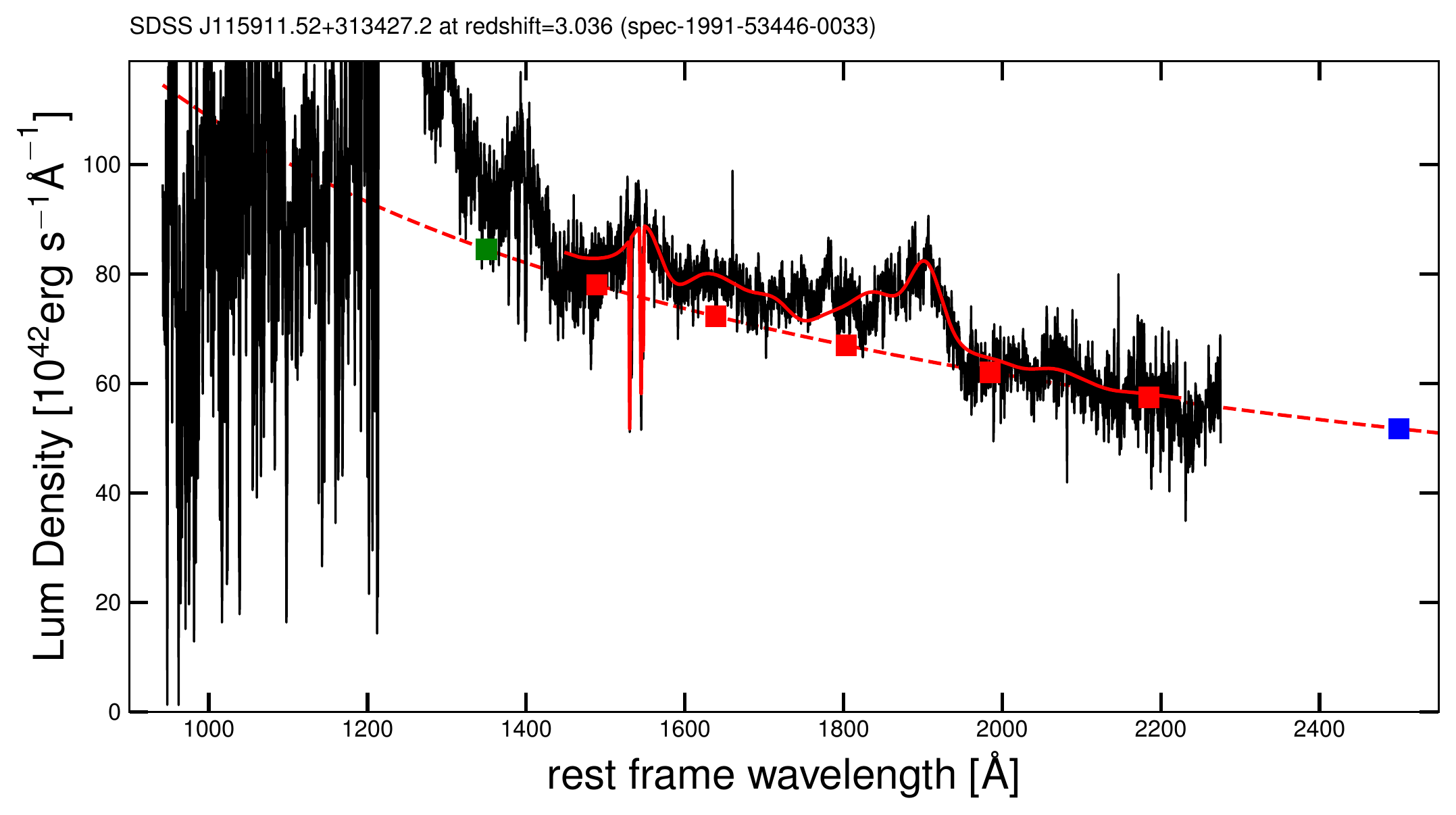}
\includegraphics[width=6cm]{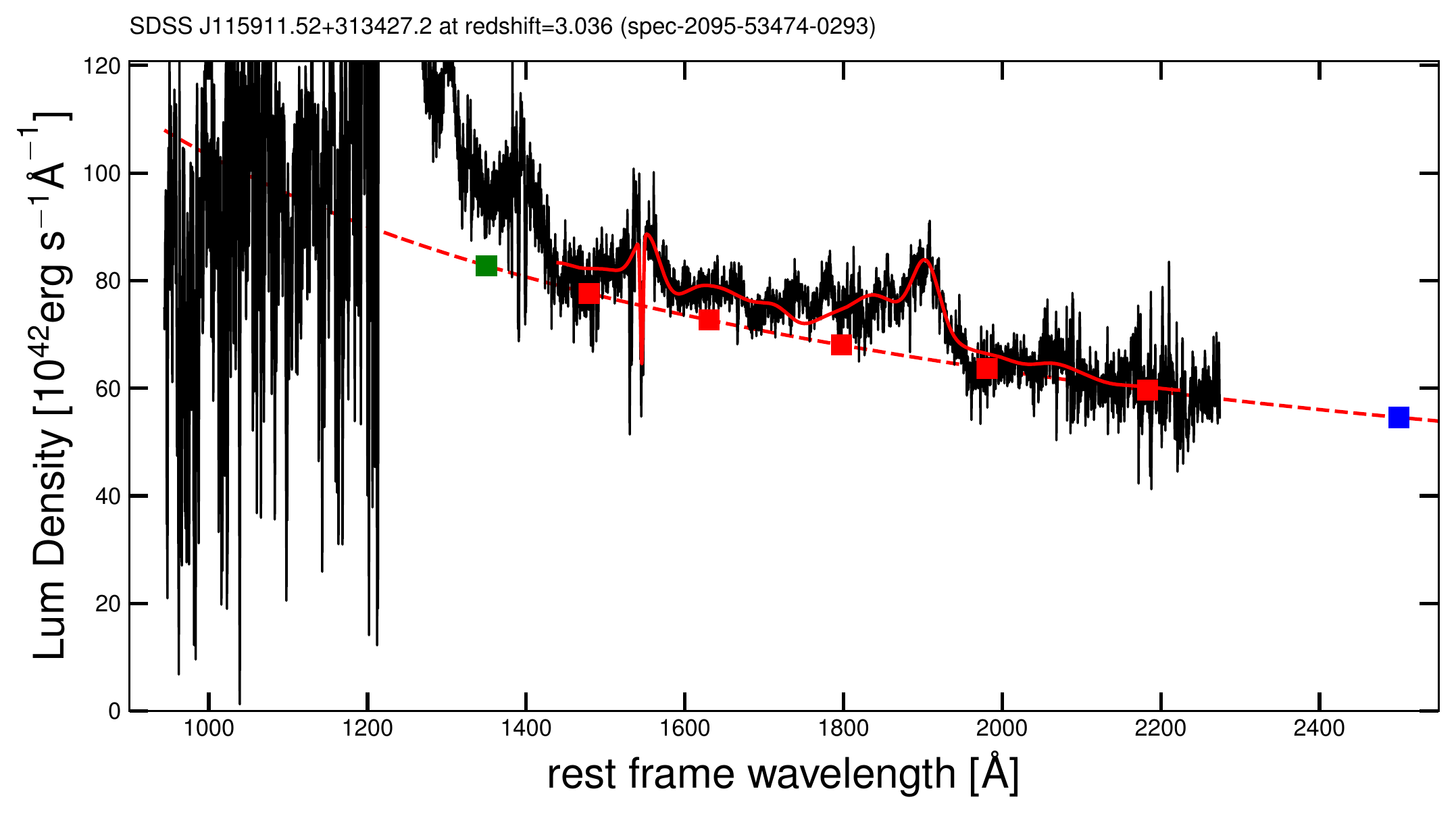}
\includegraphics[width=6cm]{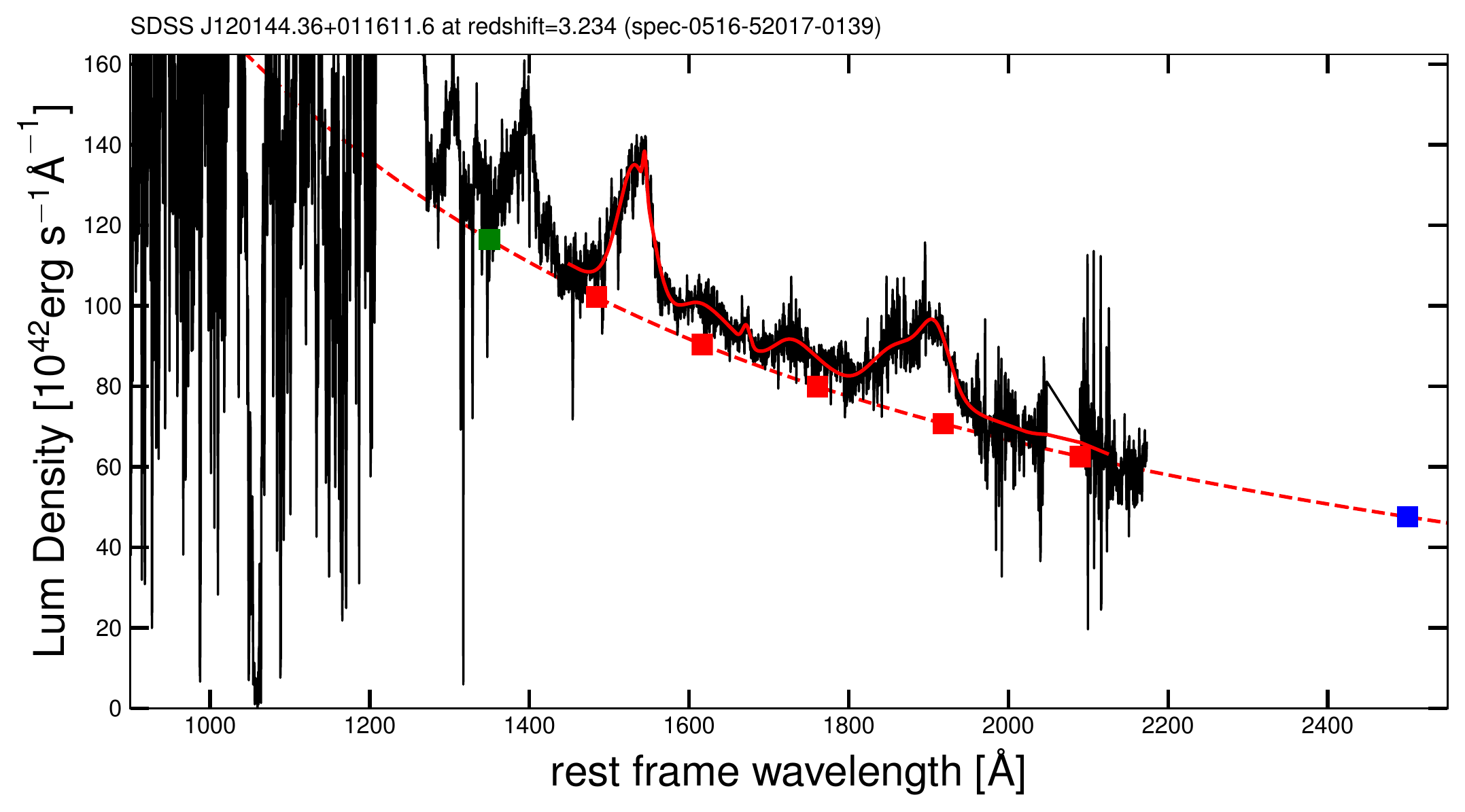}
\includegraphics[width=6cm]{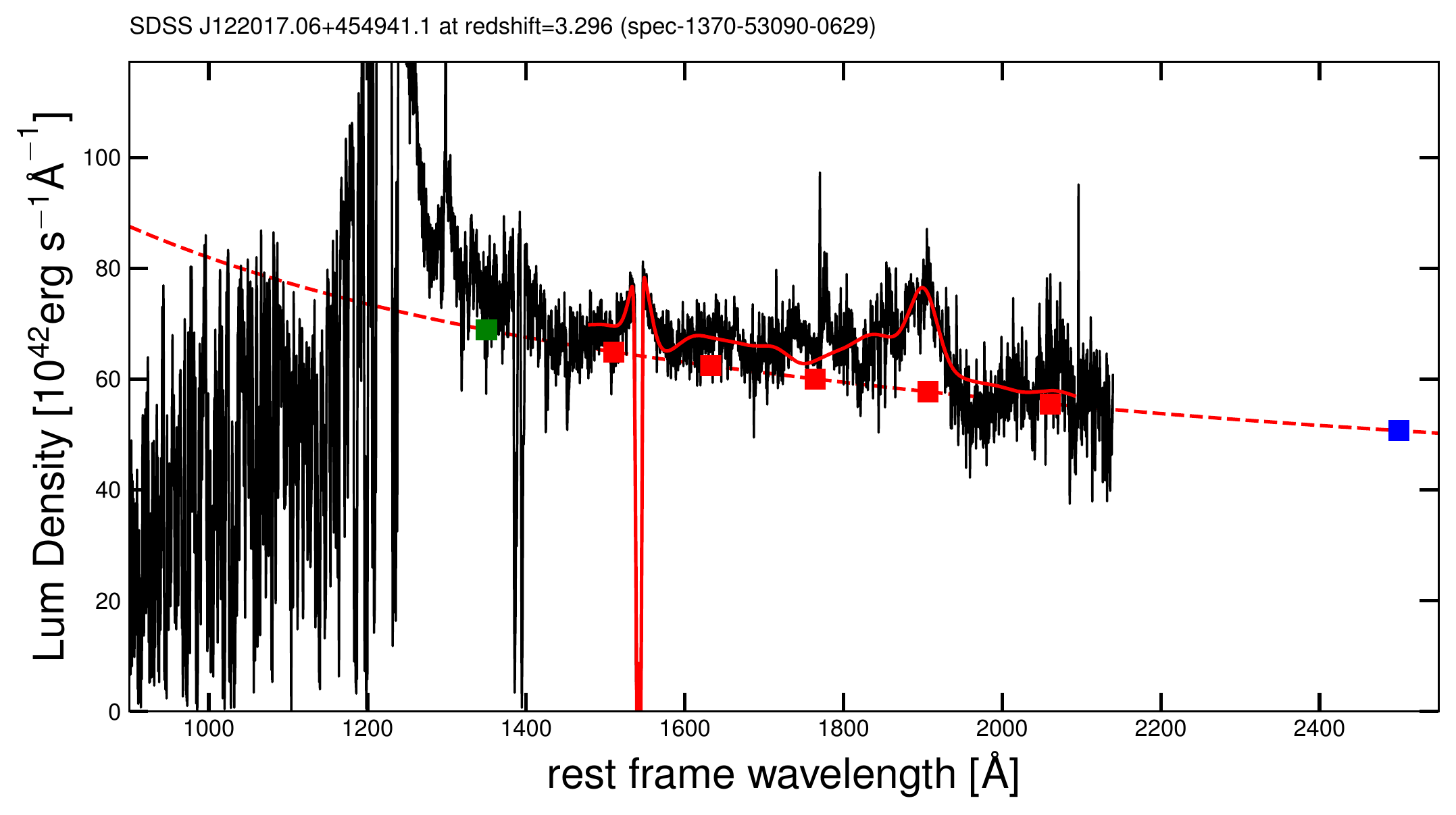}
\includegraphics[width=6cm]{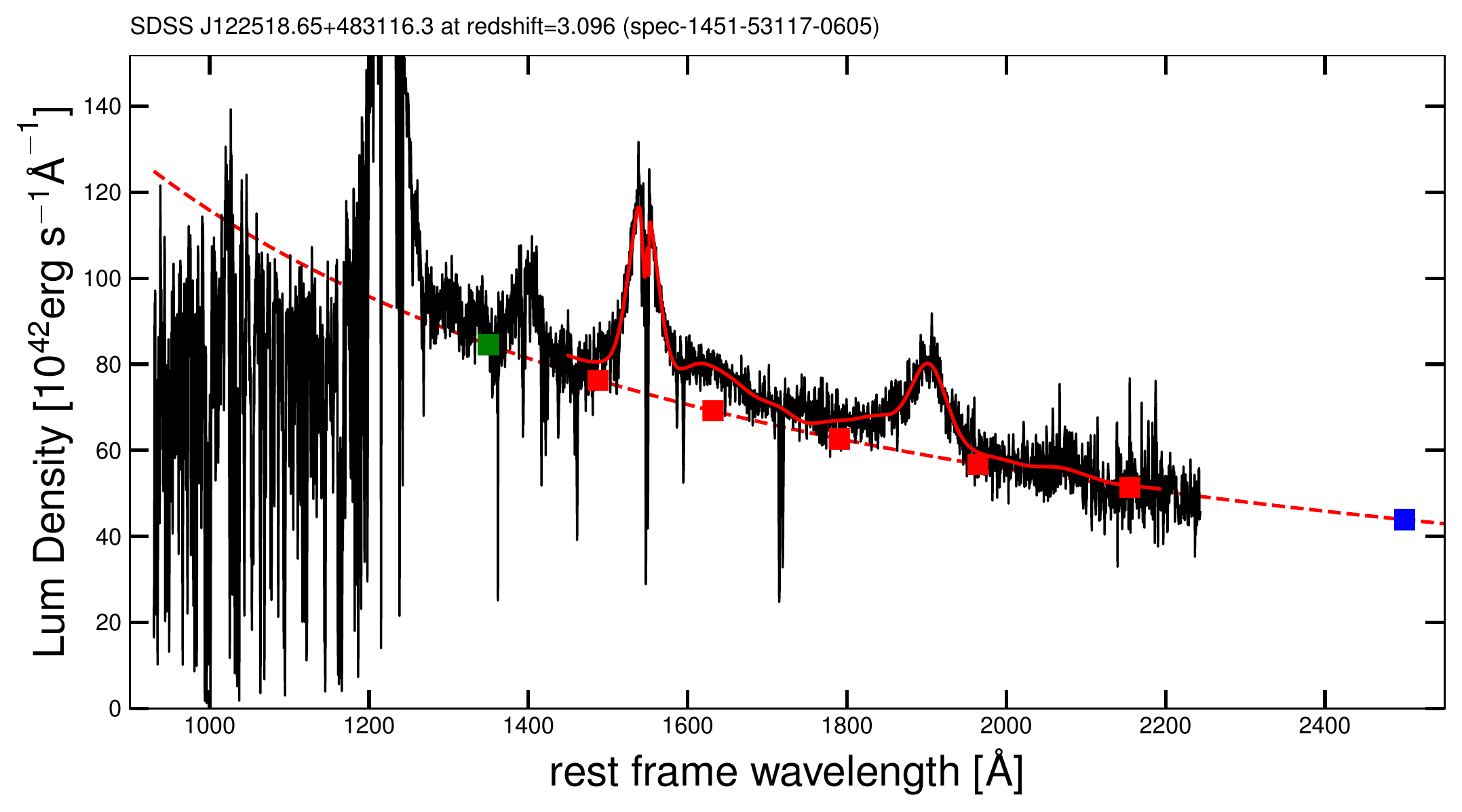}
\includegraphics[width=6cm]{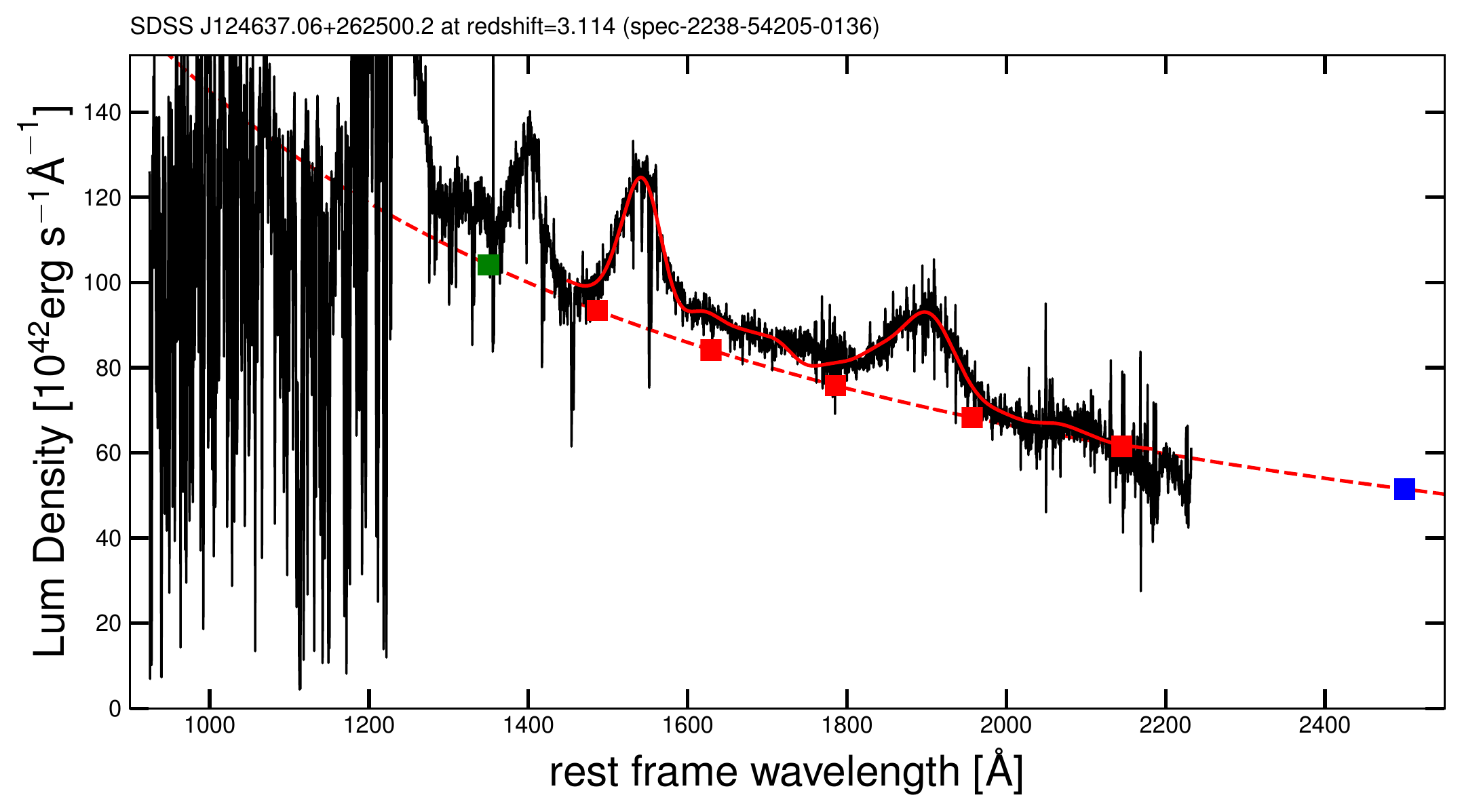}
\includegraphics[width=6cm]{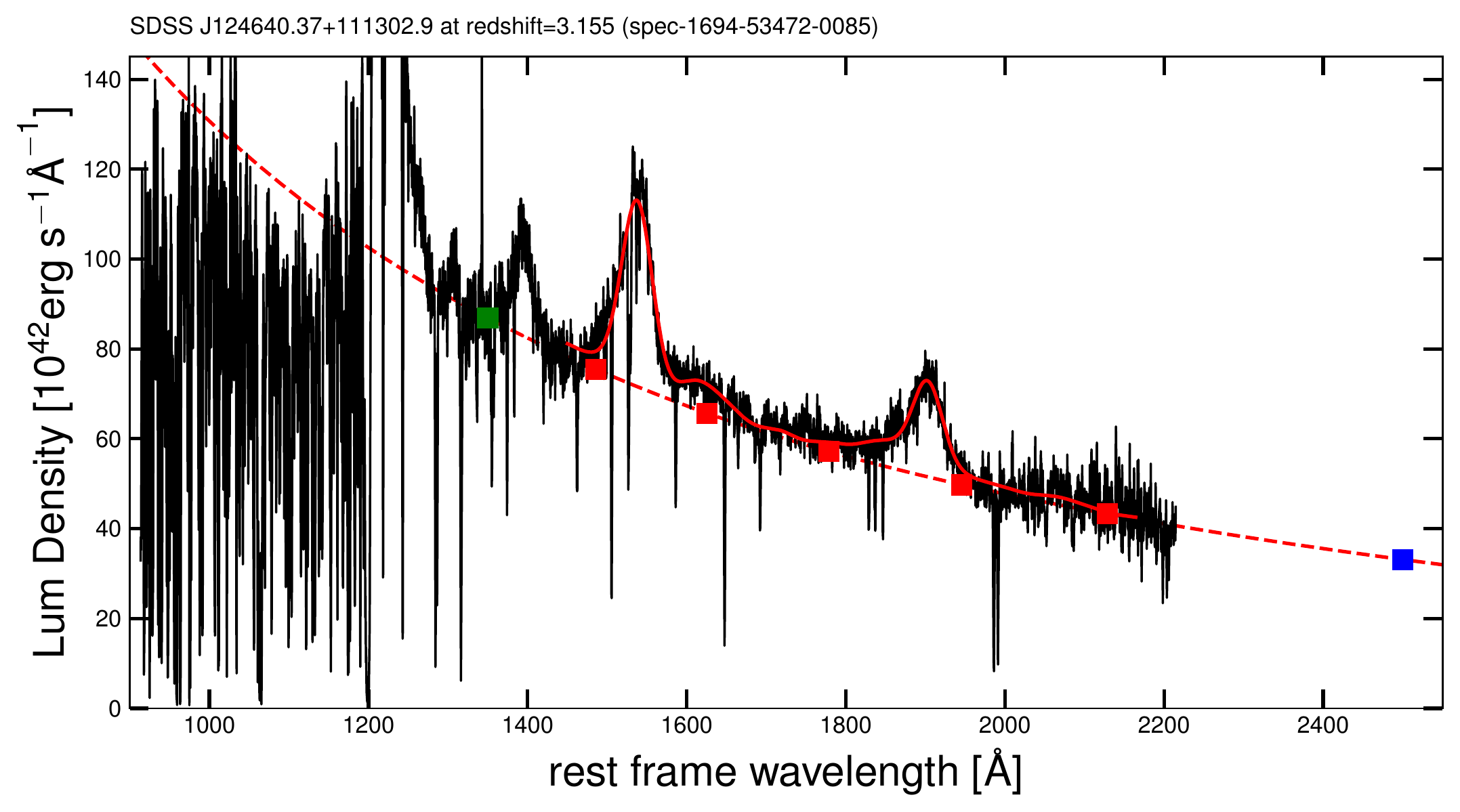}
\includegraphics[width=6cm]{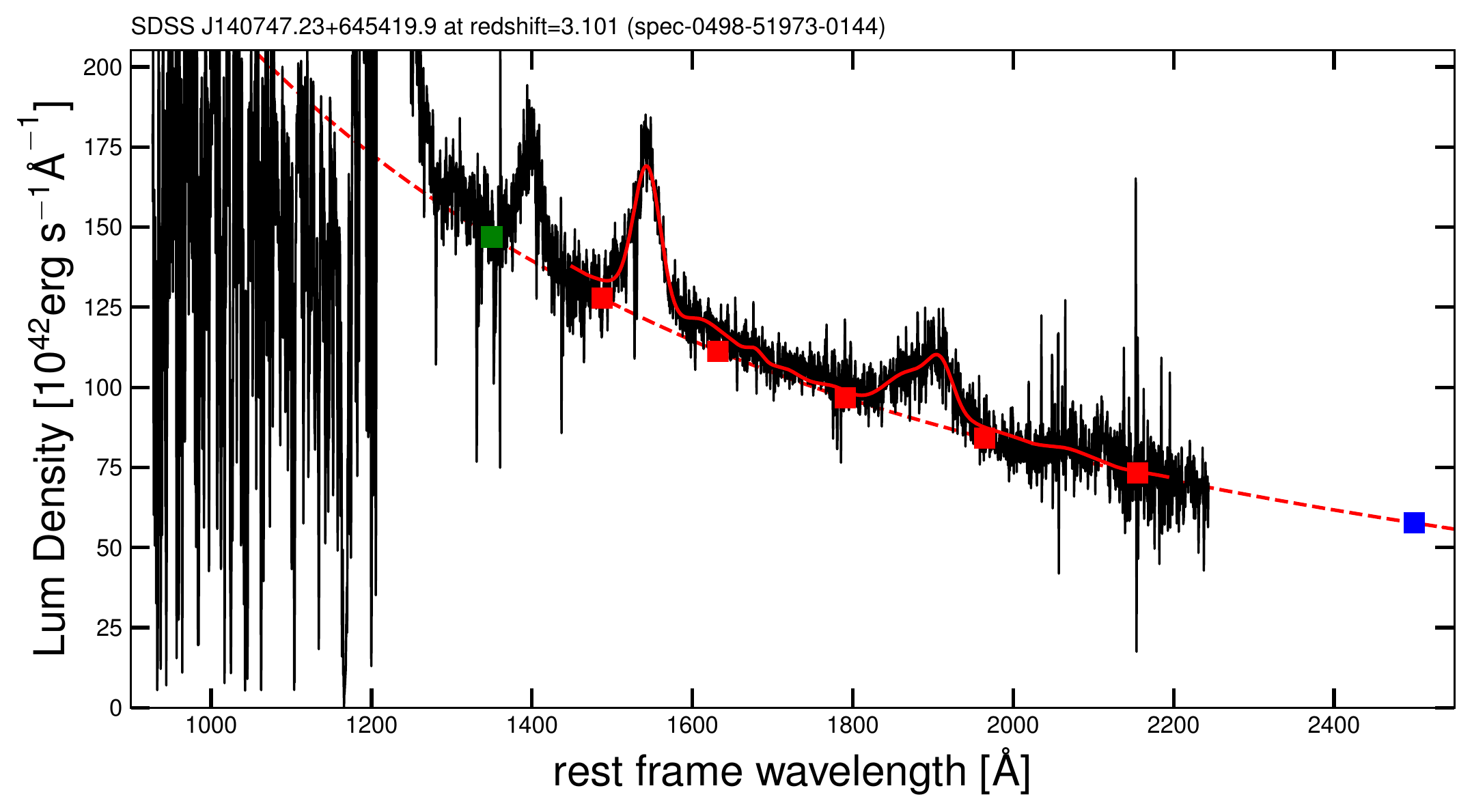}
\includegraphics[width=6cm]{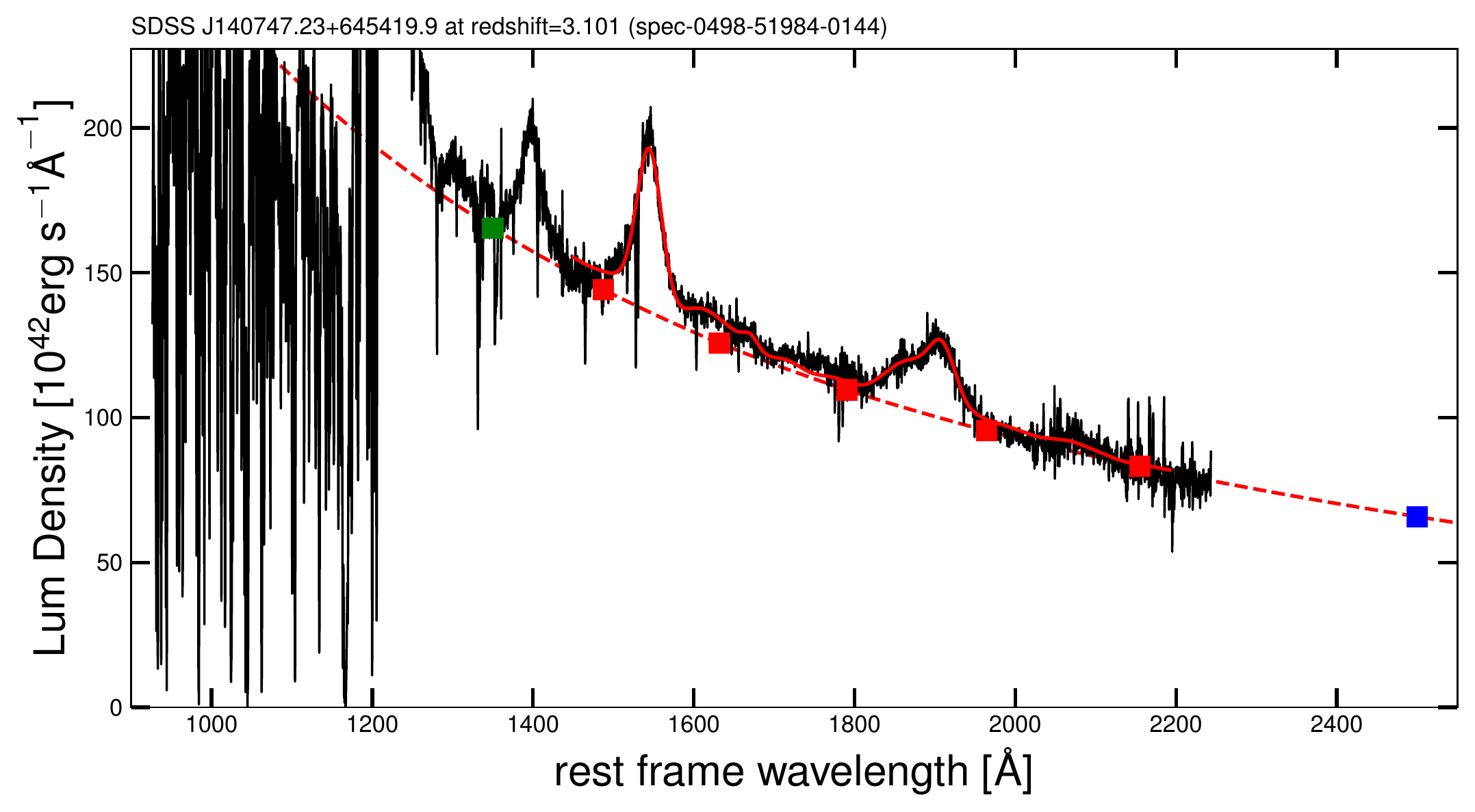}
\includegraphics[width=6cm]{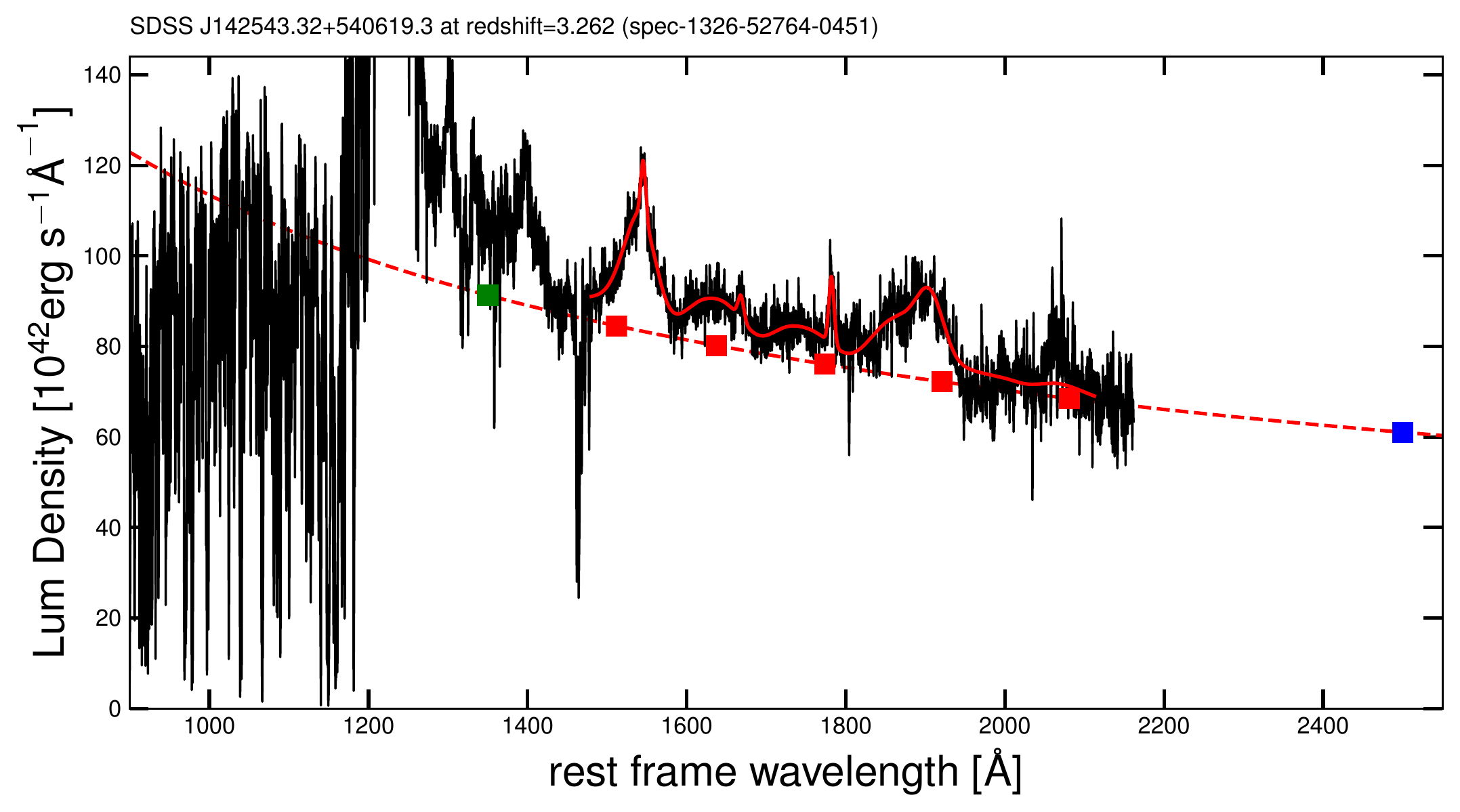}
\includegraphics[width=6cm]{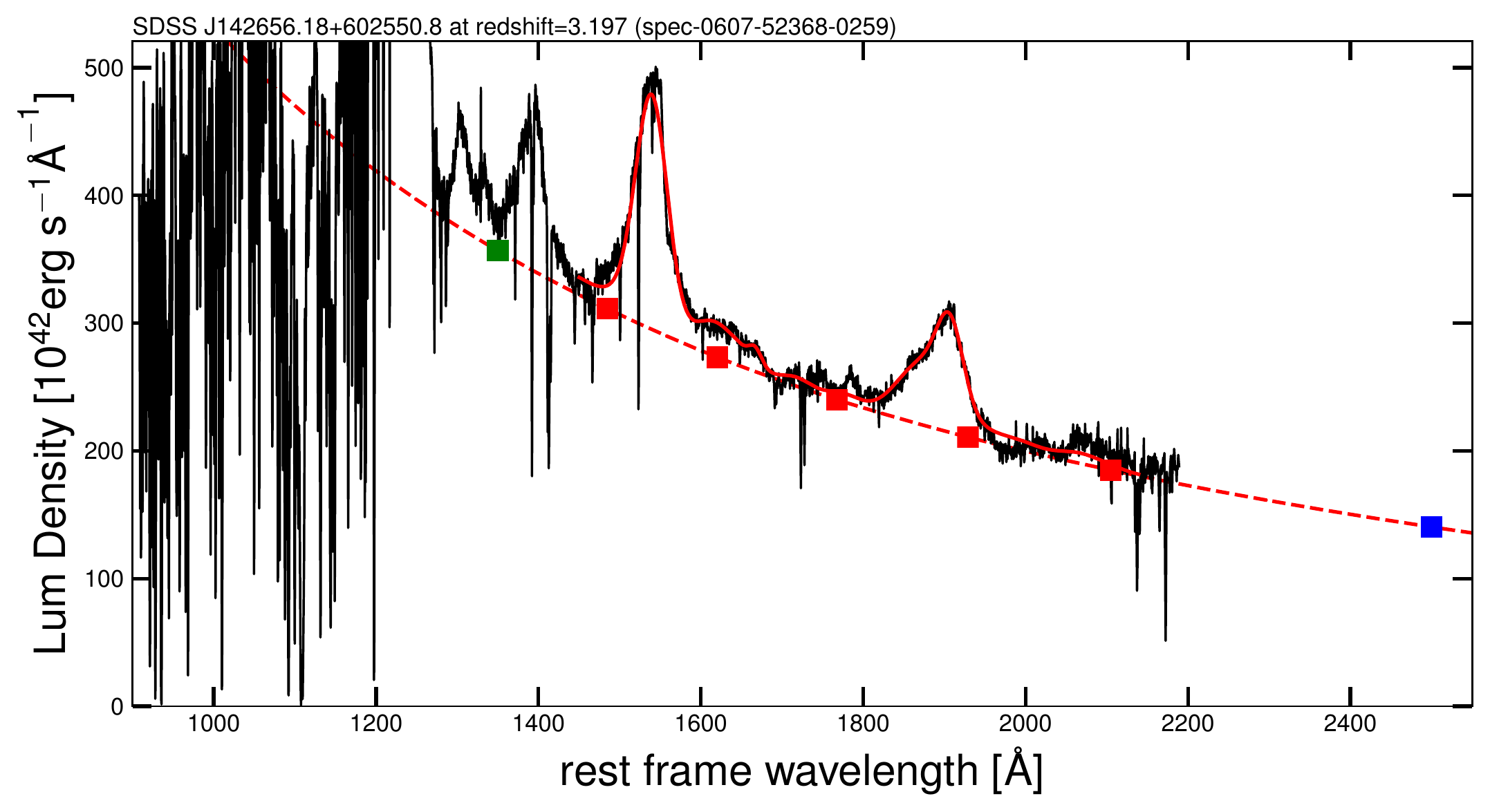}
\includegraphics[width=6cm]{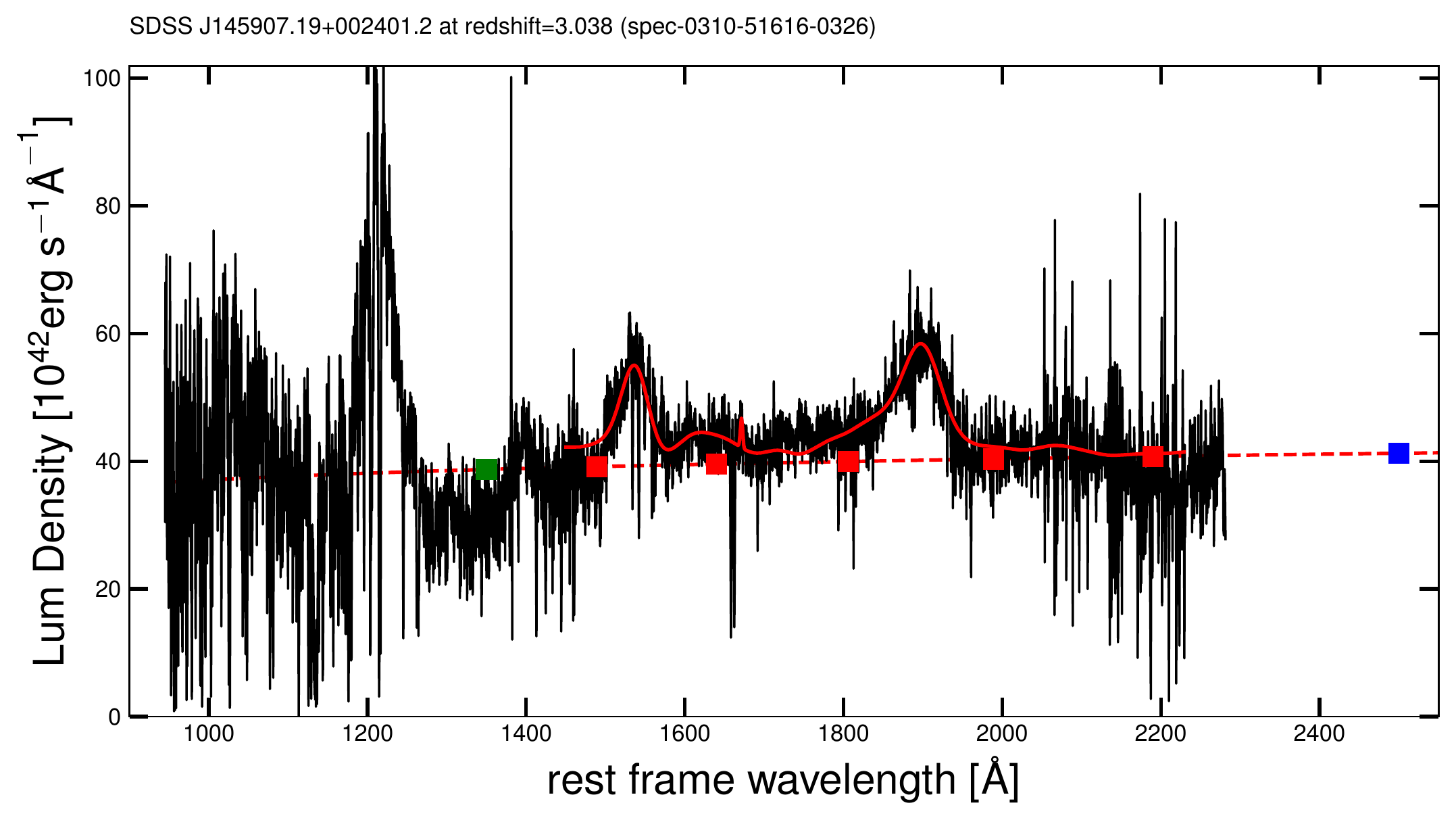}
\includegraphics[width=6cm]{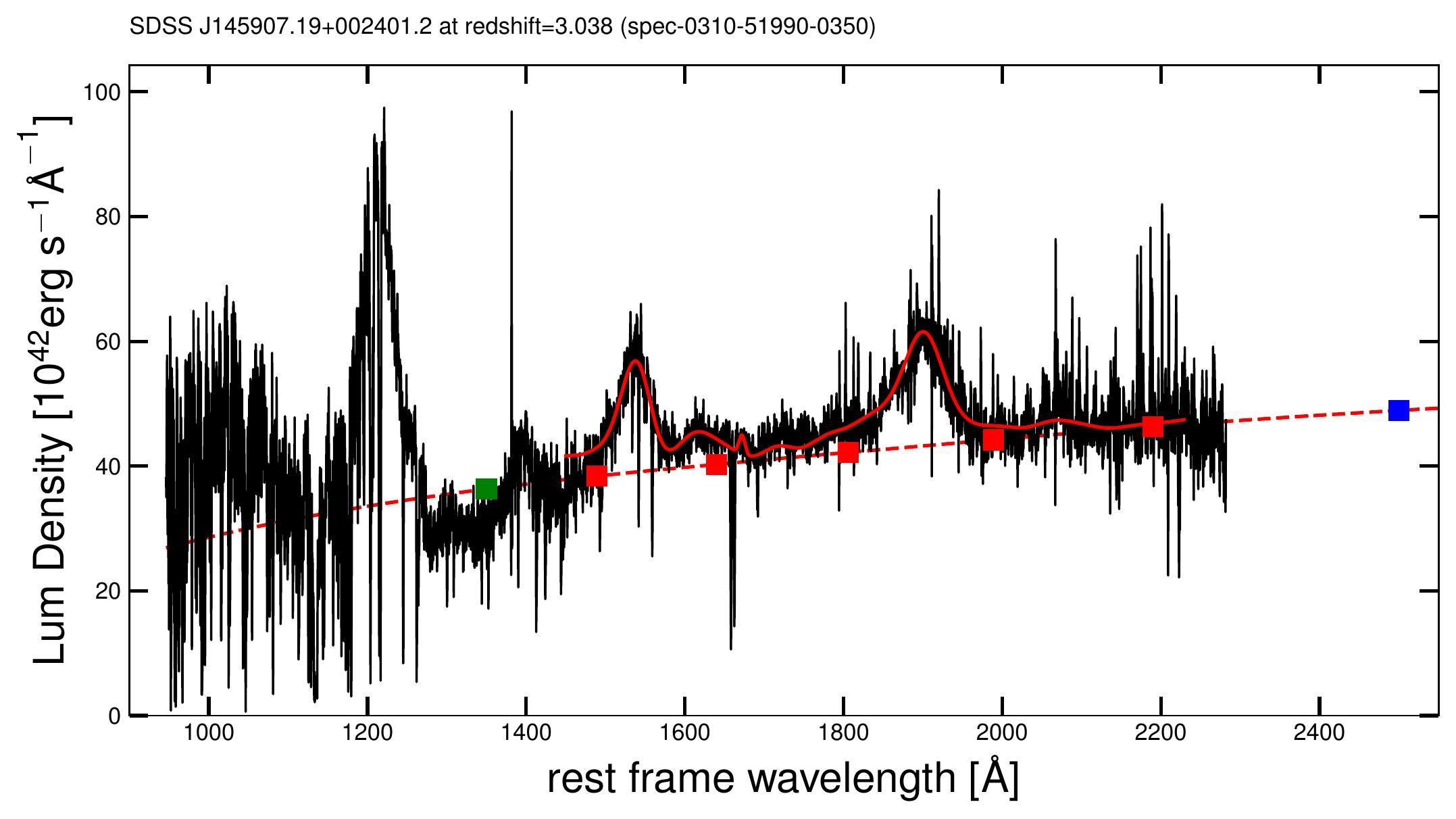}
\includegraphics[width=6cm]{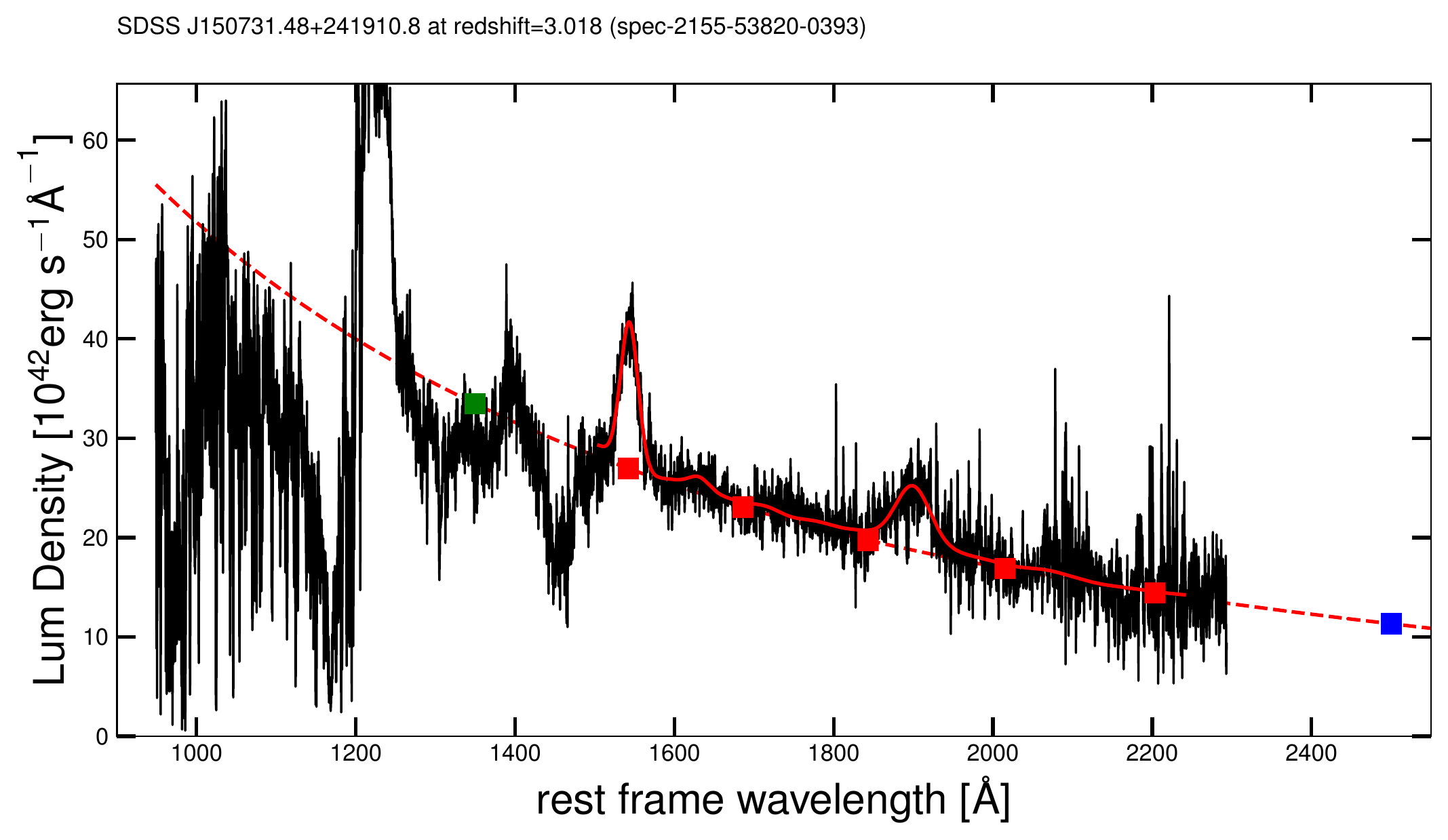}
\includegraphics[width=6cm]{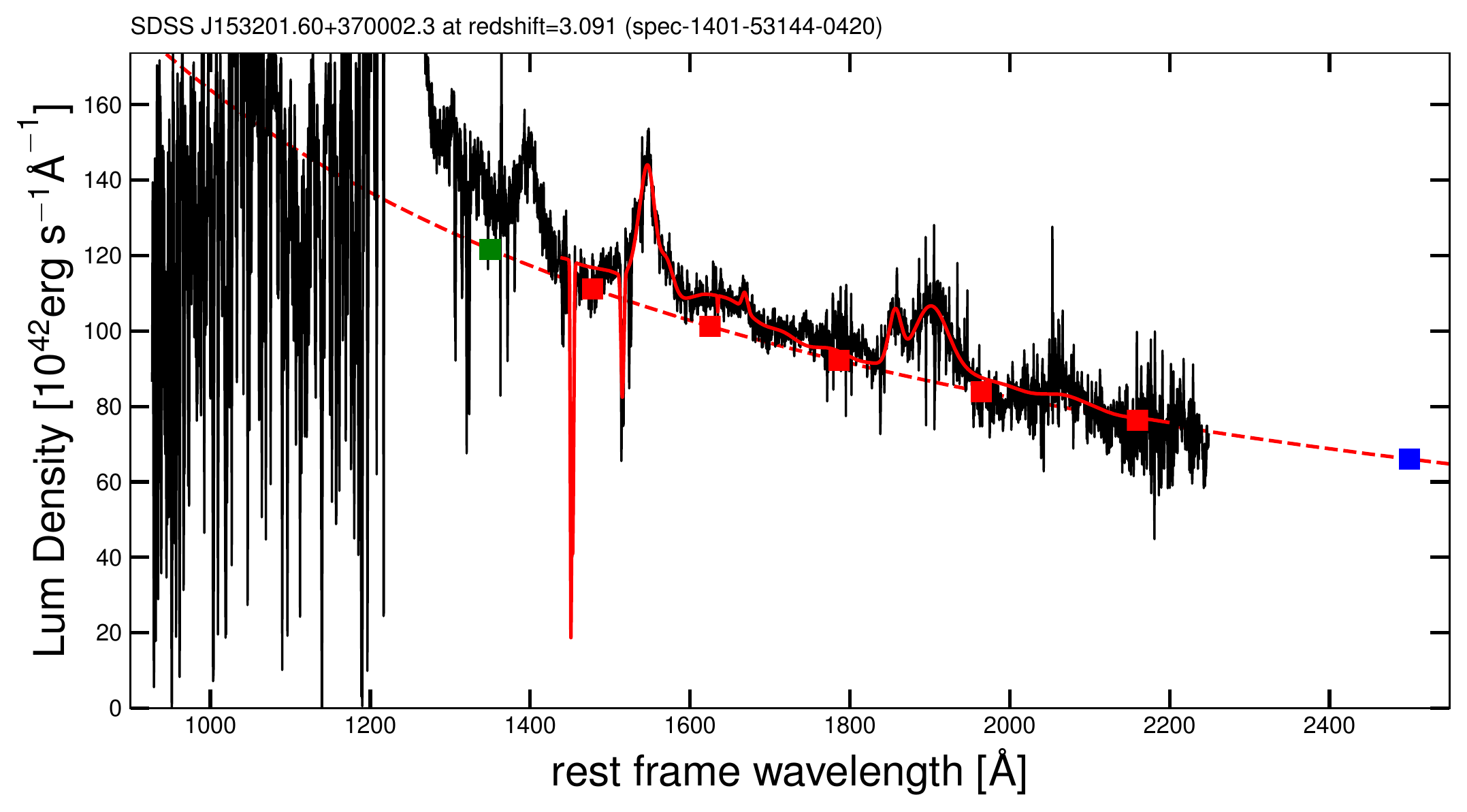}
\includegraphics[width=6cm]{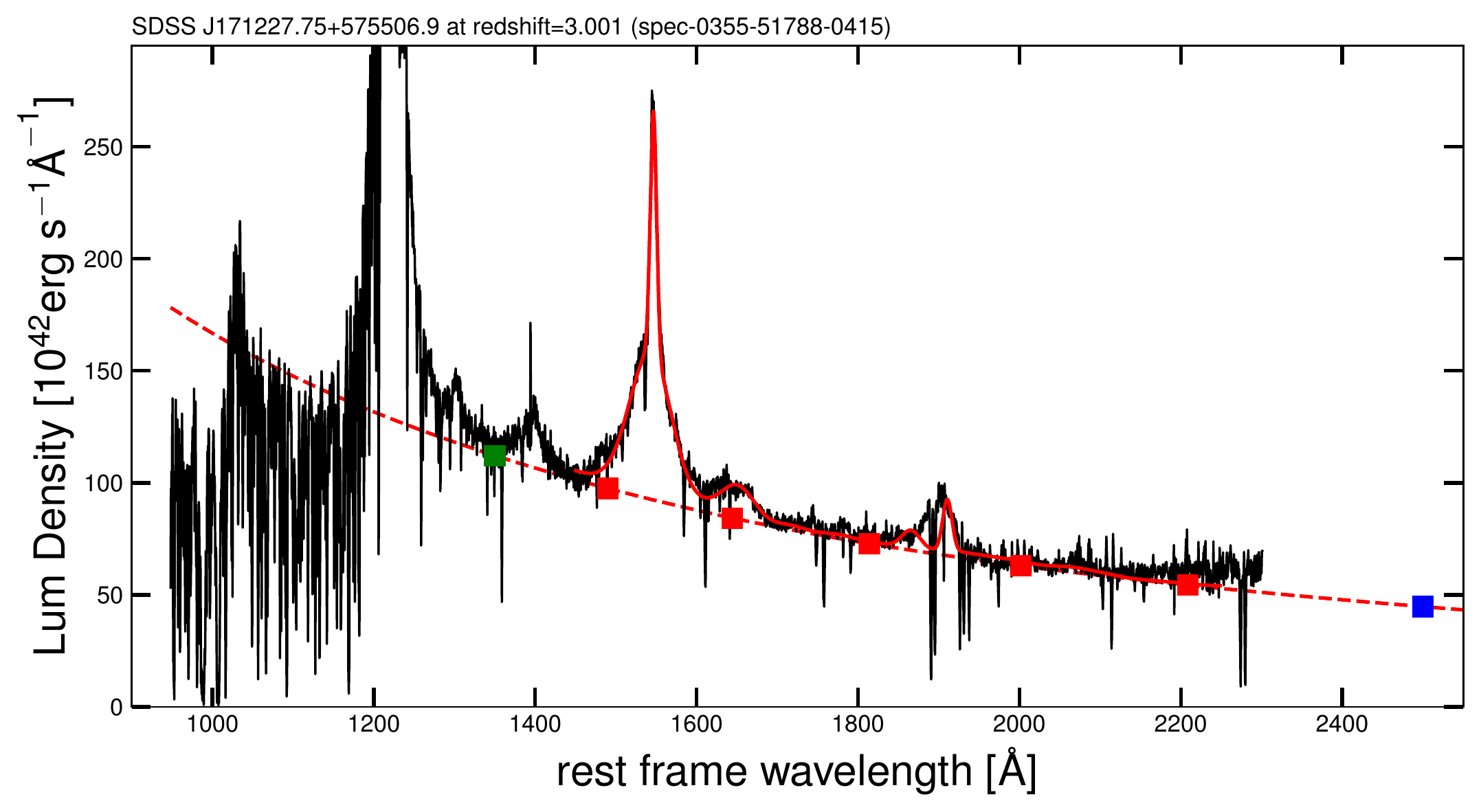}
\includegraphics[width=6cm]{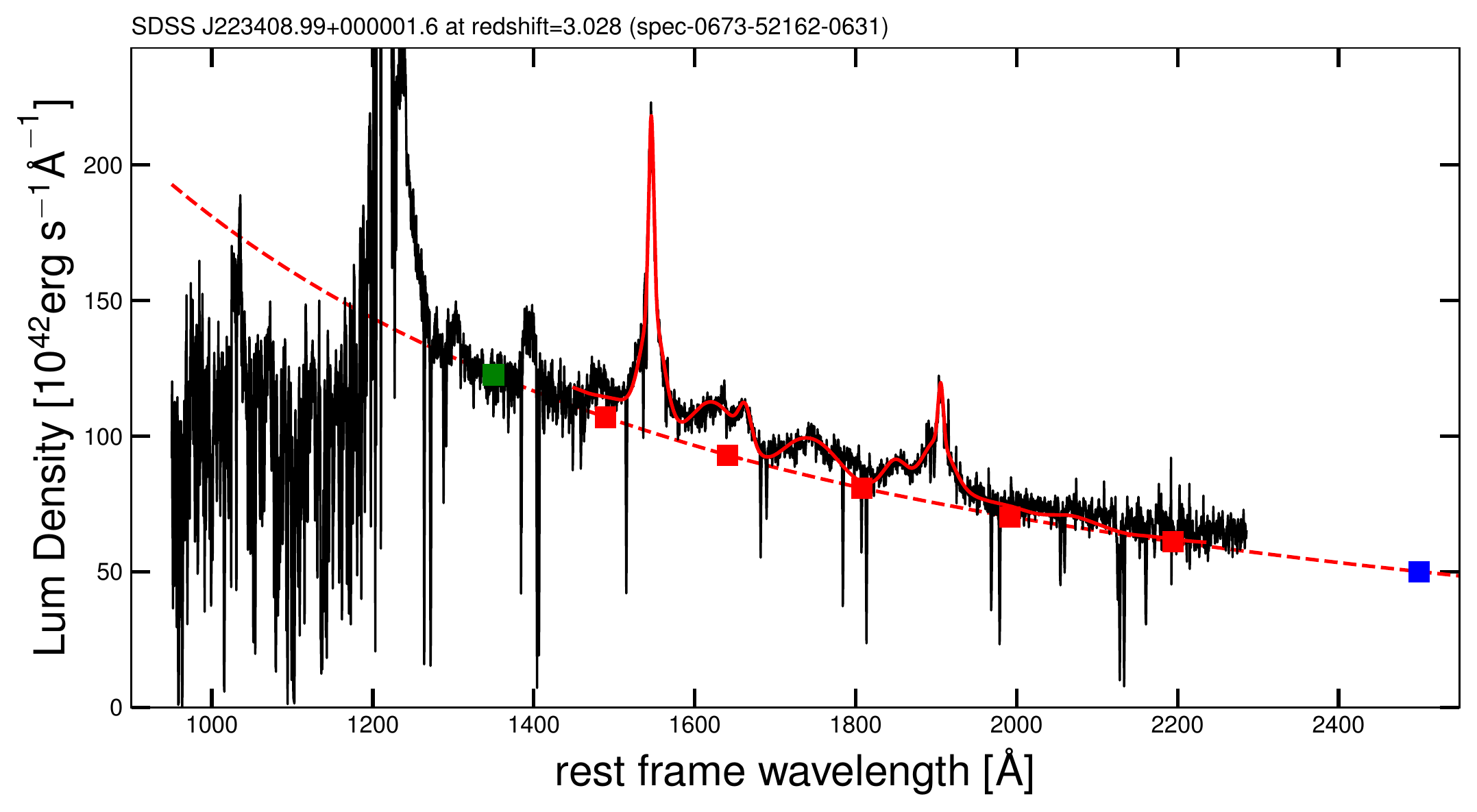}
\caption{{\it Continued}}
\label{spec}
\end{figure*}

\longtab[1]{
\begin{landscape}
\begin{longtable}{lcccccccccc}
\caption{UV spectral properties of the $z$\,$\simeq$\,3 sample.}\\
\label{tbl1}\\
 Name        &        X-ray    &   SDSS spectrum   &      $\alpha_\lambda$  &     FWHM     &     $v_{\rm peak}$    &         $\sigma$ &   centroid shift &        AS   &            $F_{\rm C\,IV}/10^{-17}$   &  EW      \\
 & class & & & km s$^{-1}$ & km s$^{-1}$ & km s$^{-1}$ & km s$^{-1}$ & & erg s$^{-1}$cm$^{-2}$ & \AA\ \\
 \hline
\endhead
\endfoot
\endlastfoot
  030341.04-002321.9 & $N$  &   spec-0411-51873-0163 & $-1.48 \pm 0.03$ & 6358$\pm$  16 &   $1469 \pm 13$  &     $4997\pm 68$ &  $2232\pm   93$  &    $0.76\pm 0.01$  & $ 3970\pm   42$ &     $37.9\pm0.81$ \\
  030341.04-002321.9 & $N$  &   spec-0411-51914-0163 & $-1.75 \pm 0.04$ & 6252$\pm$  23 &   $1266 \pm 21$  &     $5189\pm135$ &  $2556\pm  111$  &    $0.63\pm 0.01$  & $ 3970\pm   78$ &     $39.7\pm1.15$ \\
  030449.85-000813.4 & $N$  &   spec-0411-51817-0153 & $-1.95 \pm 0.03$ & 2138$\pm$  22 &   $ 704 \pm 14$  &     $2920\pm316$ &  $ 848\pm  117$  &    $0.75\pm 0.01$  & $ 3117\pm  122$ &     $22.9\pm0.89$ \\
  030449.85-000813.4 & $N$  &   spec-0411-51873-0152 & $-1.97 \pm 0.04$ & 2274$\pm$  18 &   $ 734 \pm  6$  &     $3232\pm234$ &  $ 700\pm   45$  &    $0.75\pm 0.01$  & $ 3127\pm   81$ &     $24.3\pm0.57$ \\
  030449.85-000813.4 & $N$  &   spec-0411-51914-0153 & $-2.11 \pm 0.04$ & 2167$\pm$   8 &   $ 656 \pm  6$  &     $2644\pm 72$ &  $ 858\pm   46$  &    $0.65\pm 0.01$  & $ 3049\pm   34$ &     $22.6\pm0.37$ \\
  030449.85-000813.4 & $N$  &   spec-2048-53378-0265 & $-1.52 \pm 0.06$ & 2167$\pm$  10 &   $ 666 \pm 12$  &     $2691\pm 49$ &  $ 975\pm   37$  &    $0.69\pm 0.02$  & $ 2988\pm   40$ &     $28.7\pm0.62$ \\
  082619.70+314848.0 & $N$  &   spec-0932-52620-0208 & $-1.50 \pm 0.03$ & 8061$\pm$ 120 &   $ 956 \pm 63$  &     $4890\pm579$ &  $2665\pm  110$  &    $0.45\pm 0.02$  & $ 2562\pm  183$ &     $28.1\pm2.31$ \\
  083535.69+212240.1 & $N$  &   spec-1929-53349-0041 & $-0.93 \pm 0.01$ & 5748$\pm$ 181 &   $1024 \pm 41$  &     $4603\pm188$ &  $ 885\pm   98$  &    $1.00\pm 0.02$  & $ 2930\pm   44$ &     $40.9\pm0.66$ \\
  090033.50+421547.0 & $-$  &   spec-0831-52294-0201 & $-1.37 \pm 0.02$ & 4509$\pm$  52 &   $ 511 \pm 24$  &     $4711\pm292$ &  $ 400\pm  311$  &    $0.80\pm 0.03$  & $ 8242\pm  292$ &     $40.3\pm1.88$ \\
  090102.93+354928.5 & $N$  &   spec-1212-52703-0310 & $-1.48 \pm 0.04$ & 8816$\pm$ 307 &   $2040 \pm134$  &     $5942\pm471$ &  $3372\pm  540$  &    $0.80\pm 0.06$  & $ 2582\pm  141$ &     $24.2\pm1.42$ \\
  090508.88+305757.3 & $N$  &   spec-1591-52976-0263 & $-1.75 \pm 0.05$ & 4442$\pm$  64 &   $ 937 \pm  4$  &     $4114\pm304$ &  $3225\pm  282$  &    $0.40\pm 0.01$  & $ 1483\pm   53$ &     $12.5\pm0.47$ \\
  094202.04+042244.5 & $N$  &   spec-0570-52266-0427 & $-1.74 \pm 0.02$ & 3251$\pm$  13 &   $ 569 \pm 10$  &     $3583\pm 64$ &  $2013\pm   29$  &    $0.54\pm 0.01$  & $ 5591\pm   66$ &     $39.0\pm0.62$ \\
  094554.99+230538.7 & $W$  &   spec-2295-53734-0380 & $-1.03 \pm 0.05$ & 7190$\pm$ 201 &   $1285 \pm 28$   &    $5289\pm130$ &  $ 114\pm  129$  &    $0.63\pm 0.02$  & $ 1039\pm   71$ &     $20.8\pm1.90$ \\
  094734.19+142116.9 & $N$  &   spec-2582-54139-0485 & $-1.15 \pm 0.01$ & 8806$\pm$ 153 &   $1053 \pm292$  &     $4559\pm626$ &  $1813\pm  288$  &    $0.59\pm 0.07$  & $ 3440\pm  236$ &     $24.9\pm2.03$ \\
  101447.18+430030.1 & $N$  &   spec-1218-52709-0508 & $-1.47 \pm 0.01$ & 8409$\pm$ 180 &   $2727 \pm234$  &     $5234\pm499$ &  $2394\pm  655$  &    $1.03\pm 0.25$  & $ 2669\pm  160$ &     $ 9.8\pm0.60$ \\
  102714.77+354317.4 & $N$  &   spec-1957-53415-0076 & $-1.16 \pm 0.07$ & 4964$\pm$  83 &   $1392 \pm109$ &      $4969\pm257$ &  $ 843\pm  358$  &    $1.21\pm 0.17$  & $ 8423\pm  305$ &     $40.5\pm1.61$ \\
  111101.30-150518.5 & $w$  &   spec-2708-54561-0538 & $-1.64 \pm 0.02$ & 8458$\pm$ 431 &   $  75 \pm636$  &     $4600\pm297$ &  $1342\pm  252$  &    $0.56\pm 0.15$  & $ 1467\pm   67$ &     $18.1\pm0.93$ \\
  111101.30-150518.5 & $w$  &   spec-2690-54211-0563 & $-1.68 \pm 0.02$ & 8148$\pm$ 118 &   $ 540 \pm192$  &     $4006\pm159$ &  $1648\pm   95$  &    $0.74\pm 0.06$  & $ 1285\pm   25$ &     $14.5\pm0.33$ \\
  111120.59+243740.8 & $w$  &   spec-2489-53857-0600 & $-1.18 \pm 0.06$ & 2661$\pm$  75 &   $ 792 \pm240$ &      $3224\pm172$ &  $2374\pm  115$  &    $0.75\pm 0.22$  & $ 1186\pm   49$ &     $10.3\pm0.50$ \\
  114308.87+345222.2 & $N$  &   spec-2012-53493-0067 & $-1.74 \pm 0.02$ & 5245$\pm$  11 &   $1285 \pm  8$   &    $3571\pm 75$ &  $1471\pm   45$  &    $0.70\pm 0.01$  & $ 5355\pm   41$ &     $35.6\pm0.49$ \\
  114851.46+231340.4 & $W$  &   spec-2511-53882-0552 & $ 0.43 \pm 0.13$ & 7335$\pm$ 130 &   $2214 \pm 88$   &    $8088\pm549$ &  $ -154\pm  452$  &    $0.81\pm 0.10$  & $ 3529\pm  267$ &     $37.6\pm3.91$ \\
  115911.52+313427.2 & $W$  &   spec-1991-53446-0033 & $-0.85 \pm 0.02$ & 7732$\pm$ 837 &   $1527 \pm324$  &     $5793\pm196$ &  $ -846\pm   90$  &    $1.32\pm 0.19$  & $  997\pm   62$ &     $10.4\pm0.68$ \\
  115911.52+313427.2 & $W$  &   spec-2095-53474-0293 & $-0.79 \pm 0.01$ & 7190$\pm$ 336 &   $1121 \pm243$  &     $5386\pm424$ &  $ -871\pm  325$  &    $1.12\pm 0.30$  & $  866\pm   55$ &     $ 8.9\pm0.59$ \\
  120144.36+011611.6 & $W$  &   spec-0516-52017-0139 & $-1.42 \pm 0.02$ & 8409$\pm$ 241 &   $1140 \pm 58$   &    $6329\pm514$ &  $3021\pm  586$  &    $0.41\pm 0.03$  & $ 2536\pm  178$ &     $24.8\pm2.00$ \\
  122017.06+454941.1 & $W$  &   spec-1370-53090-0629 & $-0.65 \pm 0.04$ & 5709$\pm$ 550 &   $1266 \pm150$  &     $2765\pm407$ &  $1847\pm  528$  &    $0.75\pm 0.17$  & $  491\pm   34$ &     $ 7.5\pm0.60$ \\
  122518.66+483116.3 & $N$  &   spec-1451-53117-0605 & $-1.09 \pm 0.05$ & 6154$\pm$ 404 &   $ 579 \pm218$  &     $4573\pm284$ &  $ -189\pm  344$  &    $0.67\pm 0.06$  & $ 2715\pm  138$ &     $30.5\pm2.21$ \\
  124637.06+262500.2 & $N$  &   spec-2238-54205-0136 & $-1.30 \pm 0.14$ & 1136$\pm$ 143 &   $1585 \pm511$  &     $4967\pm500$ &  $1123\pm  263$  &    $1.04\pm 0.23$  & $ 2424\pm  266$ &     $22.5\pm3.64$ \\
  124640.37+111302.9 & $N$  &   spec-1694-53472-0085 & $-1.65 \pm 0.09$ & 7287$\pm$ 413 &   $1411 \pm311$  &     $4427\pm583$ &  $2520\pm  354$  &    $0.59\pm 0.13$  & $ 2430\pm  242$ &     $30.1\pm4.02$ \\
  140747.23+645419.9 & $N$  &   spec-0498-51973-0144 & $-1.53 \pm 0.01$ & 6474$\pm$  76 &   $1005 \pm323$ &      $4894\pm501$ &  $1348\pm   80$  &    $0.94\pm 0.25$  & $ 2832\pm  212$ &     $19.5\pm1.70$ \\
  140747.23+645419.9 & $N$  &   spec-0498-51984-0144 & $-1.53 \pm 0.01$ & 6483$\pm$ 283 &   $ 850 \pm163$  &     $4857\pm295$ &  $1374\pm  125$  &    $0.85\pm 0.08$  & $ 3281\pm  112$ &     $20.6\pm0.66$ \\
  142543.32+540619.3 & $W$  &   spec-1326-52764-0451 & $-0.83 \pm 0.04$ & 5438$\pm$ 123 &   $ 724 \pm 15$  &     $3574\pm113$ &  $2047\pm  124$  &    $0.38\pm 0.01$  & $ 1242\pm   39$ &     $13.9\pm0.51$ \\
  142656.18+602550.8 & $N$  &   spec-0607-52368-0259 & $-1.59 \pm 0.04$ & 8264$\pm$ 192 &   $1392 \pm433$  &     $5554\pm381$ &  $3457\pm  458$  &    $0.67\pm 0.17$  & $11946\pm  413$ &     $37.5\pm1.69$ \\
  145907.19+002401.2 & $w$  &   spec-0310-51616-0326 & $ 0.28 \pm 0.16$ & 8574$\pm$ 357 &   $2330 \pm483$  &     $7124\pm296$ &  $2063\pm  438$  &    $1.00\pm 0.83$  & $ 1779\pm  159$ &     $42.3\pm5.82$ \\
  145907.19+002401.2 & $w$  &   spec-0310-51990-0350 & $ 0.75 \pm 0.21$ & 8603$\pm$ 186 &   $2185 \pm214$  &     $7090\pm266$ &  $1941\pm  460$  &    $1.00\pm 0.06$  & $ 1747\pm  114$ &     $39.4\pm3.11$ \\
  150731.48+241910.8 & $W$  &   spec-2155-53820-0393 & $-1.46 \pm 0.05$ & 5990$\pm$ 272 &   $ 453 \pm 81$   &    $5196\pm800$ &  $ 463\pm  446$  &    $0.57\pm 0.04$  & $  916\pm  113$ &     $29.9\pm4.73$ \\
  153201.60+370002.3 & $w$  &   spec-1401-53144-0420 & $-1.06 \pm 0.04$ & 5671$\pm$ 503 &   $ 364 \pm 39$   &    $6149\pm719$ &  $ 459\pm  698$  &    $0.93\pm 0.08$  & $ 2385\pm  234$ &     $19.4\pm2.09$ \\
  171227.75+575506.9 & $N$  &   spec-0355-51788-0415 & $-1.44 \pm 0.02$ & 2583$\pm$  30 &   $ 647 \pm  4$   &    $5825\pm 40$ &  $1498\pm   74$  &    $0.98\pm 0.02$  & $ 7180\pm   51$ &     $59.9\pm0.61$ \\
  223408.99+000001.6 & $N$  &   spec-0673-52162-0631 & $-1.27 \pm 0.02$ & 2167$\pm$  26 &   $ 608 \pm 14$  &     $3637\pm129$ &  $1104\pm  144$  &    $0.97\pm 0.04$  & $ 3064\pm  102$ &     $23.6\pm0.92$ \\
\end{longtable}
\end{landscape}
}
\end{appendix}
\end{document}